\documentclass[
    nofootinbib,
    amsmath,
    amssymb,
    aps,
    prb,
    twocolumn,
    floatfix,
    superscriptaddress]{revtex4-2}

\usepackage[hypertexnames=false]{hyperref}
\usepackage{graphicx} 
\usepackage{booktabs}
\usepackage{braket}
\usepackage{bm}
\usepackage[T1]{fontenc}
\usepackage{amsfonts}
\usepackage{mathrsfs}  
\usepackage{dsfont}
\usepackage{enumitem}
\usepackage{algorithm}
\usepackage{algpseudocode}
\usepackage{xcolor}

\newcommand{\ho}[1]{\hat{\mathcal{#1}}}

\newcommand{\vms}{\vec{\mathbf{s}}}

\newcommand{\ren}[1]{\breve{#1}}
\newcommand{\normq}[1]{\mathring{#1}}

\newcommand{\SMnote}{}
\begin{document}
\title{Conditions for Dark-State Coherent Control of Excitons in TMD Monolayers:
A Reduced Intravalley Lindblad Framework for WSe$_2$}

\author{Mauricio Martinez}
\email[Corresponding author: ]{m.martinezguerrero@montana.edu}
\affiliation{Department of Chemistry and Biochemistry, Montana State University, 103 Chemistry and Biochemistry Building, Bozeman, 59717, MT, USA}

\author{Martin A. Mosquera}
\email{martinmosquera@montana.edu}
\affiliation{Department of Chemistry and Biochemistry, Montana State University, 103 Chemistry and Biochemistry Building, Bozeman, 59717, MT, USA}

\begin{abstract}
Dark-state coherence underlies electromagnetically induced transparency (EIT) and stimulated Raman adiabatic passage (STIRAP), and how well it is protected depends on the coupling configuration. We develop a Lindblad framework for driven excitonic three-level systems at fixed center-of-mass momentum, selected by a systematic screening of the exciton spectrum, in which a single exciton couples effectively to the radiation field, which drives it and sets its decay, and to the exchange mean field of a dilute exciton gas, which shifts the energy levels. In the ladder ($\Xi$) system of hBN-encapsulated WSe$_2$, a terahertz pulse preceding a weak optical pulse converts the excitation into the optically dark $2p$ exciton while the lossy bright exciton stays nearly empty; under adiabatic driving, reversing the pulse order suppresses the conversion by more than an order of magnitude. The exciton gas shifts the two-photon resonance; the center of the transparency window measures this shift on the sample itself, and the window is EIT rather than Autler--Townes splitting only below a well-defined control strength. The $V$ system, whose two excited states both decay, offers no such protection. The conversion efficiency and the transparency are upper bounds; sensitivity analyses outside the model show that the conversion keeps half of its value and its dark-state character only if the pure dephasing and the population decay of the $2p$ exciton are both slower than about the inverse pulse width, i.e.\ at picosecond pulses only if its homogeneous width is below about $0.9$~meV or $0.5$~meV, respectively, far below the $7.9$~meV measured so far. The depth of the transparency window and the pulse-order contrast decide whether a sample meets these conditions.
\end{abstract}

\maketitle

\section{Introduction}
\label{sec:intro}

Monolayer transition-metal dichalcogenides (TMDs) such as MoS$_2$, MoSe$_2$, WS$_2$ and WSe$_2$ are direct-gap semiconductors at the monolayer limit~\cite{Mak2010,Splendiani2010} whose optical response is dominated by tightly bound excitons~\cite{Wang2018RMP,Mueller2018npj,Chernikov2014}. Reduced dielectric screening yields binding energies of hundreds of meV and a strongly non-hydrogenic Rydberg series~\cite{Chernikov2014,berkelbach2013,Stier2018}, while broken inversion symmetry and strong spin--orbit coupling lock spin and valley degrees of freedom and make them optically addressable with circularly polarized light~\cite{coupled-spin_Xiao,Xu2014NatPhys}. These properties have made TMDs a versatile platform for optoelectronics, valleytronics and van der Waals heterostructure engineering~\cite{MakShan2016,Manzeli2017}.

Beyond their linear optical properties, TMD monolayers have become a testing ground for excitonic coherence. In hBN-encapsulated samples at cryogenic temperatures the homogeneous linewidth of the bright $1s$ exciton approaches its intrinsic limit~\cite{Moody2015,Jakubczyk2016,Cadiz2017,selig2016excitonic}, which in WSe$_2$ is not purely radiative: phonon emission into lower-lying, momentum-dark K--$\Lambda$ and K--K$'$ excitons adds a width comparable to the radiative one for $1s$ and dominant for $2s$ and higher states~\cite{Brem_intrinsic}. Valley superpositions have been prepared and measured~\cite{Jones2013,Hao2016,Wang2016valley} and rotated on ultrafast timescales by the optical Stark effect~\cite{Sie2015,Kim2014,Ye2017} and by lightwave driving~\cite{Langer2018}; intra-excitonic transitions have been resolved with terahertz and mid-infrared probes~\cite{poellmann2015resonant,Cha2016,Steinleitner2017,Merkl2019} and driven strongly enough to produce Autler--Townes doublets and Floquet dressing~\cite{Yong2019,Kobayashi2023}. EIT-type interference in twisted homobilayers~\cite{Lin2021twist}, control of the nonlinear polarization with shaped pulses~\cite{Meron2025}, tuning of the radiative coupling by coherent feedback~\cite{Rogers2020} and long-lived coherence of single moir\'e excitons~\cite{Wang2024moire} have also been reported. Microscopic theory identifies phonons~\cite{selig2016excitonic,Christiansen2017,Lengers2020} and exciton--exciton scattering~\cite{katsch2018theory,Katsch2020} as the main sources of exciton dephasing.

In atomic, molecular and optical physics, coherence is engineered rather than only observed. Resonant and adiabatic pulse sequences, composite pulses, shortcuts to adiabaticity and optimal control steer quantum systems with high fidelity~\cite{Shore2011,ShapiroBrumer2012,DongPetersen2010,Glaser2015,GueryOdelin2019,Koch2022}, and underlie quantum memories and light--matter interfaces~\cite{Lukin2003,Hammerer2010,Lvovsky2009}, the preparation of ultracold molecules~\cite{ni2008high,Bergmann2019Roadmap} and gate operations in quantum information processing~\cite{Vitanov2017,Koch2022}. Among the most widely used are protocols built on the dark state of a driven three-level system, a coherent superposition that decouples from the lossy intermediate level: used in the stationary regime it gives electromagnetically induced transparency (EIT)~\cite{Boller1991,Harris1997,Fleischhauer2005}, and used dynamically it transfers population adiabatically without populating the intermediate level, in stimulated Raman adiabatic passage (STIRAP)~\cite{Gaubatz1990,Bergmann1998,Vitanov2017}.

Whether a dark state is protected from loss depends on the configuration of the driven couplings (Sect.~\ref{sec:three_level_framework}). In a ladder ($\Xi$) configuration the coherence links the bottom and top states and can bypass a lossy middle state when the top state is long-lived, as in cascade and Rydberg EIT~\cite{GeaBanacloche1995,Mohapatra2007}; in a $V$ configuration it links two excited states that both decay~\cite{Akulshin1998,Fleischhauer2005}. The $\Lambda$ configuration is the canonical geometry of coherent population trapping, EIT and STIRAP~\cite{Arimondo1996,Fleischhauer2005,Vitanov2017}, and the closed-loop $\Delta$ configuration adds a phase-sensitive interference used in cyclic population transfer~\cite{KralShapiro2001,Kral2007}. Dark-state protocols have been carried over to semiconductor nanostructures: EIT through biexciton coherences~\cite{Phillips2003EITGaAsQW} and intersubband coherence~\cite{Serapiglia2000} in quantum wells, coherent population trapping of single spins in quantum dots~\cite{Xu2008CPT,Ramsay2010}, and proposed adiabatic transfer between coupled quantum dots~\cite{Hohenester2000STIRAPQDs}. The internal states of excitons have been driven coherently at terahertz frequencies: intense fields transfer optically dark $1s$ paraexcitons in Cu$_2$O into $2p$~\cite{Leinss2008}, and pulses resonant with $1s\to2p$ in InGaAs quantum wells also populate the dipole-forbidden $2s$ state~\cite{Rice2013}. STIRAP has been suggested for a terahertz-coupled $\Lambda$ system of two exciton polaritons and the dark $2p$ exciton in a microcavity~\cite{Tomaino2012}, and EIT and slow light have been predicted for Rydberg excitons in Cu$_2$O~\cite{ZielinskaRaczynska2016,Ziemkiewicz2020}.

Transferring these protocols to excitons in TMD monolayers faces an obstacle that is absent in dilute atomic gases: the excitonic levels are embedded in a solid and coupled to several dissipative structures at once. The bright exciton decays radiatively on a sub-picosecond timescale~\cite{poellmann2015resonant,Moody2015}; phonons dephase optical coherences and broaden the lines as the temperature rises~\cite{selig2016excitonic,Moody2015,Christiansen2017}; exciton--exciton interactions renormalize the energies and introduce excitation-induced dephasing~\cite{shahnazaryan2017exciton,katsch2018theory,erkensten2021exciton,Katsch2020}; spin- and momentum-dark excitons act as additional reservoirs~\cite{Zhang2015dark,Malic2018dark,Madeo2020}; and Auger processes and exciton--exciton annihilation open further loss channels~\cite{auger_Dezheng,auger_Han}. Each of these can destroy the interference on which a dark state relies, or displace the resonance conditions it requires, so assessing a dark-state protocol calls for an open-system description that treats coherent driving and dissipation on the same footing~\cite{RossiKuhn2002,HaugJauho2008,kira2011semiconductor,BreuerPetruccione}. The excitonic dark-state schemes proposed so far, for a GaAs-based microcavity~\cite{Tomaino2012} and bulk Cu$_2$O~\cite{ZielinskaRaczynska2016,Ziemkiewicz2020}, do not contain the features that shape the problem in a TMD monolayer: a strongly non-hydrogenic Rydberg series in which bright and dark states are interleaved, a sub-picosecond radiative decay of the bright exciton, and the simultaneous coupling to phonon, photon and exciton-gas reservoirs. To our knowledge, a microscopic open-system treatment of EIT and STIRAP protocols for excitons in a TMD monolayer, including the choice of the excitonic three-level system itself, has not been reported, nor have the conditions that a real monolayer must meet for these protocols to work been quantified.

In this work we develop such a treatment and apply it to hBN-encapsulated WSe$_2$. The model couples a single tagged exciton to acoustic phonons, the quantized radiation field and a dilute intravalley exciton gas. A systematic configuration screening first uses the exciton energies, interband brightness, intra-excitonic dipoles and state sizes from a variational Wannier solver to select three-level systems in the $\Xi$ and $V$ configurations; the selected three-level systems are then embedded in a Lindblad framework, derived within the Born--Markov and full-secular approximations, and their driven-dissipative dynamics is solved for EIT-type response and STIRAP-type conversion. The central constraint of the framework is the restriction of the tagged exciton to a fixed center-of-mass momentum $\mathbf{K}_0$, set by the optical geometry: the acoustic deformation-potential coupling then vanishes identically, the exciton gas enters only through a coherent exchange mean field that renormalizes the exciton energies and control detunings, and the dissipator becomes purely radiative. Within this limit, a terahertz pulse preceding a weak optical pulse converts the optical excitation into the dark $2p$ exciton while the bright $1s$ exciton stays nearly empty; the center of the cascade transparency window measures the exchange shift under the conditions of the experiment, and the window is EIT rather than Autler--Townes splitting only below a well-defined control strength; and the $V$ system serves as the radiatively unprotected reference. Because the fixed-$\mathbf{K}_0$ limit generates no pure dephasing and no population decay of $2p$ beyond a negligible radiative decay, these conversion efficiencies and transparencies are upper bounds. Sensitivity analyses outside the model turn them into requirements: the conversion keeps half of its value and its dark-state character only if the pure dephasing and the population decay of $2p$ are both slower than about the inverse pulse width, i.e.\ at picosecond pulses only if the homogeneous width of $2p$ is below about $0.9$~meV or $0.5$~meV, respectively, far below the $7.9$~meV measured so far~\cite{Kusaba2021}. The depth of the cascade window and the pulse-order contrast are the measurements that decide whether a sample meets these conditions (Sect.~\ref{subsec:real-sample}).

The paper is organized as follows. Section~\ref{sec:three_level_framework} reviews the driven three-level framework and the $\Xi$ and $V$ configurations, and Sect.~\ref{sec:multi-env-model} introduces the tagged-exciton and reservoir Hamiltonians and their fixed-$\mathbf{K}_0$ projection. Section~\ref{sec:top_manifold} presents the configuration screening, applies it to WSe$_2$/hBN and constructs the drive Hamiltonians of the selected three-level systems, and Sect.~\ref{sec:lindblad_three_level} derives their rotating-frame Lindblad equations, including the exchange renormalization. Section~\ref{sec:results} presents the results and Sect.~\ref{sec:conclusions} concludes. The intermediate steps of the fixed-$\mathbf{K}_0$ projection, the general derivation of the Lindblad equation, the bath correlators, the variational exciton solver, the numerical integration of the dynamics and the sensitivity to dephasing and population decay of $2p$ are given in Apps.~\ref{app:K0-projection}--\ref{app:dephasing}\SMnote.

\paragraph{Notation:} Boldface symbols (e.g., $\mathbf{v}$) denote two-dimensional vectors, while three-dimensional vectors are represented using both arrow and boldface notation (e.g., $\vec{\mathbf{r}}$). Operators in the Schr\"odinger, interaction, and drive-rotating frames are denoted by a hat, tilde, and bar, respectively: \(\hat{\mathcal A}\), \(\tilde{\mathcal A}\), and \(\bar{\mathcal A}\).
Renormalized scalar quantities and normalized ones are marked by a breve and a ring, respectively: \(\ren y\) and \(\normq y\).
\section{Generic three-level framework for dark-state control}
\label{sec:three_level_framework}

\emph{Dark-state coherence} can be used in two complementary ways. In the
\emph{interference (response) regime}, a strong \emph{control} field makes the
driven pathways interfere destructively, so that a weak \emph{probe} field is
not absorbed: electromagnetically induced transparency
(EIT)~\cite{Fleischhauer2005}. In the \emph{adiabatic (control) regime}, the
same interference is used dynamically: \emph{pump} and \emph{Stokes} pulses
applied in the counterintuitive order, Stokes first, keep the system in a dark
eigenstate that bypasses a lossy intermediate level, as in stimulated Raman
adiabatic passage (STIRAP)~\cite{Bergmann1998,Vitanov2017}. The explicit dark
state of the retained ladder system, and the distinction between coherent
cancellation and dissipative protection, are given in
Sect.~\ref{sec:Xi-component-equations}.

\label{subsec:unified_three_level_ham}
In both regimes the control acts on the slow envelopes and relative timing of
the drives, while the carriers oscillate at optical or terahertz frequencies.
Two standard steps separate these scales~\cite{Steck2012QuantumAtomOptics}. A
selected three-level system $\{\ket{1},\ket{2},\ket{3}\}$ with reference
energies $\epsilon_a$ is driven on a set $\mathcal L$ of directed legs
$(i,j)$, each by a field with signed carrier frequency $\omega_{L,ij}$, real
Rabi envelope $\Omega_{ij}(t)$ and phase $\phi_{ij}$, which includes the phase
of the dipole--field matrix element; the leg label fixes the sign and phase
conventions, not an energy ordering, and the experimentally quoted frequency
is $|\omega_{L,ij}|$. With the signed transition frequencies
$\omega_{ij}=(\epsilon_i-\epsilon_j)/\hbar$ and the one-photon detunings
$\Delta_{ij}=\omega_{L,ij}-\omega_{ij}$, the counter-rotating terms of the
electric-dipole coupling oscillate at $\simeq2|\omega_{ij}|$ and average out
over the envelopes, so they are discarded when
$\Omega_{ij},|\Delta_{ij}|\ll|\omega_{ij}|$: this is the rotating-wave
approximation (RWA). The remaining couplings still carry time-dependent
phases. A diagonal rotation removes them,
\begin{equation}
\ket{\bar\psi(t)}=\hat U^\dagger(t)\ket{\psi(t)},
\qquad
\hat U(t)=\exp\left[
it\sum_a\vartheta_a\ket{a}\bra{a}
\right],
\label{eq:drive-rotation}
\end{equation}
with $\vartheta_1=0$ and $\vartheta_j-\vartheta_i=\omega_{L,ij}$ on each
driven leg, so that in this drive-rotating frame the Hamiltonian depends on
time only through the envelopes. Dark states, adiabatic following and the
two-photon resonance are defined in this frame. After subtracting the
irrelevant scalar $\epsilon_1 I$, the RWA Hamiltonian reads
\begin{equation}
\begin{split}
\bar H^{(\mathrm{RWA})}(t)
&=
\sum_a
(\epsilon_a-\epsilon_1+\hbar\vartheta_a)\ket{a}\bra{a}\\
&+
\frac{\hbar}{2}\sum_{(i,j)\in\mathcal L}
\left[
\Omega_{ij}(t)e^{i\phi_{ij}}\ket{j}\bra{i}
+\mathrm{h.c.}
\right].
\end{split}
\end{equation}
Its diagonal entries, fixed by the configuration-dependent
$\vartheta_a$, give the one- and two-photon detunings; its
off-diagonal entries are the coherent field--matter couplings.

\subsection{Three-level configurations}
\label{subsec:three_level_topologies}
\begin{figure*}[t]
  \centering
  \includegraphics[width=\textwidth]{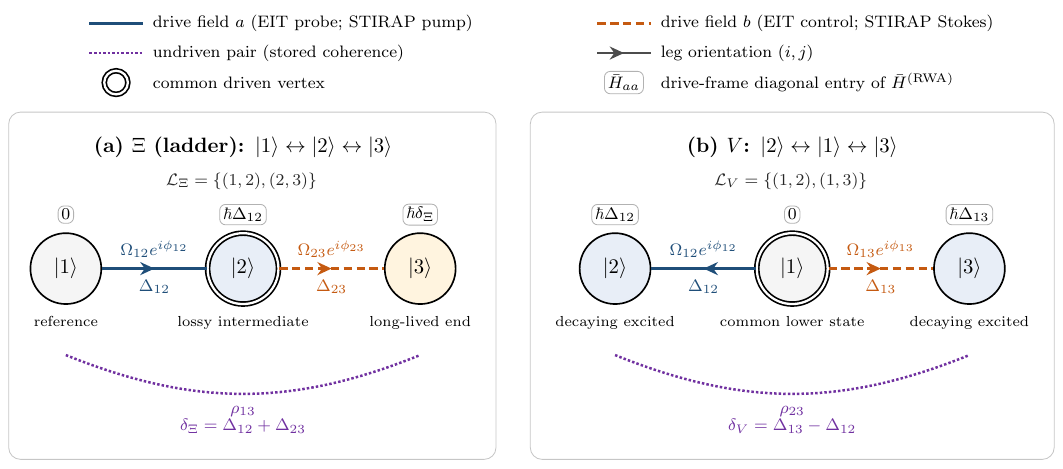}
  \caption{Drive-architecture graphs of the (a) $\Xi$ and (b) $V$
  configurations. Nodes are the retained states $\ket{a}$; edges are the
  driven legs $(i,j)\in\mathcal{L}$, each carrying the Rabi envelope and
  phase $\Omega_{ij}e^{i\phi_{ij}}$ and the one-photon detuning
  $\Delta_{ij}=\omega_{L,ij}-\omega_{ij}$. Solid (dashed) edges are
  addressed by drive field $a$ ($b$), i.e.\ the probe (control) field of
  EIT or the pump (Stokes) field of STIRAP. The arrowhead marks the leg
  orientation $(i,j)$, which fixes the sign convention of
  $\omega_{ij}$, $\Delta_{ij}$ and $\phi_{ij}$; it does not denote an
  energy ordering, and node positions carry no energy information. The
  tag above each node is its diagonal entry in the drive-rotating frame,
  $\bar H_{aa}=\epsilon_a-\epsilon_1+\hbar\vartheta_a$. The dotted arc
  joins the undriven pair that carries the stored coherence,
  $\rho_{13}$ in $\Xi$ and $\rho_{23}$ in $V$; its two-photon detuning is
  the difference of the end-node tags, equivalently the sum of
  $\pm\Delta_{ij}$ along the driven path between the two ends ($+$ when
  a leg is traversed along its orientation), giving
  $\delta_\Xi=\Delta_{12}+\Delta_{23}$ and
  $\delta_V=\Delta_{13}-\Delta_{12}$. The double-ringed node is the
  common driven vertex. Node shading encodes the lifetime hierarchy:
  reference state (gray), radiatively decaying (blue), long-lived
  (cream). The graphs were drawn with TikZ code generated by the Claude Opus 5.5 model to a layout, notation and node
  assignment specified by the authors, who checked every node tag, edge
  orientation and label against the equations of
  Sect.~\ref{sec:three_level_framework}.}
  \label{fig:topology_graphs}
\end{figure*}
Figure~\ref{fig:topology_graphs} represents the $\Xi$ and $V$ configurations~\cite{Fleischhauer2005,Bergmann1998,Vitanov2017} as graphs: the nodes are the retained states, the edges are the driven legs $(i,j)\in\mathcal{L}$, and the stored coherence connects the one pair of states that is not driven directly, so the classification rests on which states are connected, not on how they are ordered in energy. The two driven legs are addressed by two drive fields, $a$ and $b$: the probe and the control in EIT, the pump and the Stokes field in STIRAP. The $\Lambda$ and $\Delta$ configurations are left for future studies.

\subsubsection{\texorpdfstring{$\Xi$}{Xi} (ladder) configuration}
\label{subsub:ladder}
In the ladder or $\Xi$ configuration the three states are connected sequentially, so the driven legs are $|1\rangle\leftrightarrow|2\rangle$ (field $a$) and $|2\rangle\leftrightarrow|3\rangle$ (field $b$) [Fig.~\ref{fig:topology_graphs}(a)], written with the directed legs $(1,2)$ and $(2,3)$ [and $(1,2)$, $(1,3)$ for $V$] so that the same formalism covers both configurations. This configuration is natural in excitonic implementations, where one leg is optical and interband and the other intra-excitonic, in the THz/mid-IR range~\cite{Combescot2015,singh1984dynamics,poellmann2015resonant}. With $\vartheta_2=\omega_{L,12}$ and $\vartheta_3=\omega_{L,12}+\omega_{L,23}$ in Eq.~\eqref{eq:drive-rotation}, the RWA Hamiltonian in the drive-rotating frame is
\begin{equation}
\begin{split}
\bar H_\Xi&^{(\mathrm{RWA})}(t)
=+\hbar\Delta_{12}\ket{2}\bra{2}
+\hbar\delta_\Xi\ket{3}\bra{3}\\
&+\frac{\hbar}{2}\left[
\Omega_{12}(t)e^{i\phi_{12}}\ket{2}\bra{1}
+\Omega_{23}(t)e^{i\phi_{23}}\ket{3}\bra{2}
+\mathrm{h.c.}
\right]\,,
\end{split}
\label{eq:Hdrv_ladder}
\end{equation}
with the two-photon detuning of the outer-state coherence
\begin{equation}
\delta_\Xi
=\Delta_{12}+\Delta_{23}
=\omega_{L,12}+\omega_{L,23}-\omega_{13}.
\end{equation}
Although the ladder is not the canonical dark-state geometry, EIT-like response and adiabatic transfer occur when $\delta_{\Xi}$ is small and the lifetime hierarchy is favorable~\cite{Fleischhauer2005,Bergmann1998,Vitanov2017}. It is the workhorse of cascade and Rydberg EIT, in which a strong control field couples the intermediate state to a long-lived upper state whose lifetime sets the width of the two-photon resonance~\cite{GeaBanacloche1995,Mohapatra2007}. Its essentials, a lower (probe) leg with a large transition dipole, a strong upper (control) leg and a long-lived, weakly dephasing top state, map in the excitonic realization onto the interband brightness of the optically addressed state, the intra-excitonic dipole of the internal leg and the compactness of the dark $p$-like top state that the size penalty of Sect.~\ref{sec:top_manifold} rewards.

\subsubsection{\texorpdfstring{$V$}{V} configuration}
\label{subsub:V}

In a $V$ configuration a common lower state couples to two upper states, so the driven legs are $|1\rangle\leftrightarrow|2\rangle$ (field $a$) and $|1\rangle\leftrightarrow|3\rangle$ (field $b$) [Fig.~\ref{fig:topology_graphs}(b)]. With $\vartheta_2=\omega_{L,12}$ and $\vartheta_3=\omega_{L,13}$ in Eq.~\eqref{eq:drive-rotation}, the RWA Hamiltonian in the drive-rotating frame is
\begin{equation}
\label{eq:Hdrv_V}
\begin{split}
\bar H_V&^{(\mathrm{RWA})}(t)
=+\hbar\Delta_{12}\ket{2}\bra{2}
+\hbar\Delta_{13}\ket{3}\bra{3}\\
&+\frac{\hbar}{2}\left[
\Omega_{12}(t)e^{i\phi_{12}}\ket{2}\bra{1}
+\Omega_{13}(t)e^{i\phi_{13}}\ket{3}\bra{1}
+\mathrm{h.c.}
\right]\,,
\end{split}
\end{equation}
with the two-photon detuning of the coherence between the two excited states
\begin{equation}
\delta_V
=\Delta_{13}-\Delta_{12}
=\omega_{L,13}-\omega_{L,12}-\omega_{23}.
\end{equation}
The coherence of interest is carried by the two \emph{excited} states, so its lifetime is bounded by their combined radiative width. When the two states are quasidegenerate, the driven pathways overlap within a linewidth, and such systems can display electromagnetically induced \emph{absorption} rather than transparency, as observed for Zeeman sublevels~\cite{Akulshin1998}; when their splitting greatly exceeds both radiative widths, as in the excitonic three-level system retained here (Sect.~\ref{sec:V-component-equations}), no transparency window exists between the two arms, and the accessible signature is the excited-state quantum beat at the splitting frequency, damped by the combined radiative width. Because both driven states must be optically bright, no radiatively protected combination exists, and the screening of Sect.~\ref{sec:top_manifold} retains $V$ only as a comparison case, quantifying how coherent control degrades when the stored coherence cannot be shielded.

\subsection{Connection to the excitonic implementation}
\label{subsec:connection_excitonic_implementation}

On projection onto a selected excitonic three-level system, the retained states are Wannier excitons at fixed center-of-mass momentum $\bm K_0$, together with the crystal ground state when an interband leg is present; the Rabi envelopes follow from the projected exciton--photon coupling, and the reference energies and detunings are renormalized in the open-system treatment of Sect.~\ref{sec:lindblad_three_level}. Interband optical legs connect the crystal ground state to bright excitons, whereas intra-excitonic THz/mid-IR legs connect bound states~\cite{Combescot2015,Cha2016,Steinleitner2017,Kobayashi2023}. In the retained three-level systems field $a$ is the interband optical field in both configurations, while field $b$ is a THz/mid-IR field in $\Xi$ and a second optical field in $V$; the common driven vertex, the double-ringed node of Fig.~\ref{fig:topology_graphs}, is the intermediate exciton in $\Xi$ and the crystal ground state in $V$. Because $s$-like envelopes are optically bright and $p$-like ones interband-dark, we classify three-level systems by their \emph{drive architecture}, i.e.\ by which legs are interband-optical or intra-excitonic and which state is the common driven vertex, rather than by energy ordering: a state is \emph{bright/lossy} when it is $s$-like and radiatively coupled to the crystal ground state, and \emph{dark/long-lived} when it is $p$-like and dipole-forbidden from it. Whether a configuration is viable then depends on interband brightness, intra-excitonic dipole strength, spectral separation and reservoir-induced decoherence, quantified in Sects.~\ref{sec:multi-env-model} and~\ref{sec:top_manifold}.

\section{Open quantum system model for a 2D exciton}
\label{sec:multi-env-model}
This section specifies the Hamiltonian ingredients and the subsystem–reservoir partition defining the microscopic open-system model used in the subsequent analysis of dark-coherence control. The free and interaction Hamiltonians, standard in the theory of semiconductor excitons, are written out in App.~\ref{app:model-hamiltonians}\SMnote; Sect.~\ref{subsec:K0-projection} contains the model-specific fixed-$\bm K_0$ reduction used throughout the work. 

We consider a tagged 2D Wannier exciton as the subsystem of interest, which is later restricted to an excitonic three-level system adapted to a chosen control configuration. This subsystem is coupled to three distinct environments---a phonon reservoir, a quantized external electromagnetic reservoir, and a dilute intravalley exciton gas---through the total Hamiltonian
\begin{equation}
\hat{\mathcal H}=\hat{\mathcal H}_X+\hat{\mathcal H}_{\rm ph}+\hat{\mathcal H}_{\gamma}+\hat{\mathcal H}_{X_g}
+\hat{\mathcal H}_{X\text{-}ph}+\hat{\mathcal H}_{X\text{-}\gamma}+\hat{\mathcal H}_{X\text{-}X_g},
\label{eq:H-total-schematic}
\end{equation}
where the 2D free exciton Hamiltonian $\hat{\mathcal H}_X$ acts on the tagged-exciton subsystem, $\hat{\mathcal H}_{\rm ph}$, $\hat{\mathcal H}_{\gamma}$, and $\hat{\mathcal H}_{X_g}$ describe the three free Hamiltonian reservoirs, and the interaction terms $\hat{\mathcal H}_{X\text{-}\alpha}$ couple the tagged exciton to each reservoir $\alpha\in\{{\rm ph},{\gamma},X_g\}$. We restrict the electronic dynamics to a single valley and neglect spin degrees of freedom. Moreover, we assume that the reservoirs remain approximately stationary on the subsystem timescales and neglect explicit bath--bath couplings in the reduced dynamics. In the open-system treatment below, the tagged exciton is restricted to a fixed center-of-mass momentum sector $\bm K_0$. The Hamiltonians, written in App.~\ref{app:model-hamiltonians}\SMnote{} without this kinematic restriction, are the following. The tagged exciton is a Wannier exciton with internal state $n$ and center-of-mass wavevector $\mathbf K$, of energy $\epsilon_n(\mathbf K)=E_g-E_{b,n}+\hbar^2\mathbf K^2/2M$, where $E_g$ is the quasiparticle gap, $E_{b,n}$ are the binding energies of the Wannier equation (Table~\ref{tab:appD_states}) and $M$ is the total electron--hole mass. It couples to acoustic phonons through the intravalley deformation potential; to the quantized radiation field, written in the multipolar gauge at electric-dipole order, through an intra-excitonic term that conserves the number of tagged excitons and an interband term that creates or annihilates one; and to a dilute intravalley exciton gas, with mean inter-exciton spacing $d_X\gg a_B$, through the Coulomb exciton--exciton interaction, separated into direct and exchange parts. The tagged exciton and the gas share the same excitonic modes and dispersion but occupy different Fock-space sectors, the optically prepared one-exciton sector and the complementary many-exciton sector. Accordingly, the populations $\bar\rho_{ii}(t)$ of Sect.~\ref{sec:lindblad_three_level} describe the optically prepared sector of a dilute photoexcited population, not an isolated exciton in an otherwise empty crystal. The kinematic restriction is imposed on the interaction Hamiltonians in Sect.~\ref{subsec:K0-projection}.

\subsection{Fixed-\texorpdfstring{$\bm K_0$}{K0} projection of the interaction Hamiltonians}
\label{subsec:K0-projection}
The Hamiltonians of App.~\ref{app:model-hamiltonians}\SMnote{} are written without any kinematic restriction on the tagged exciton. We now restrict the tagged exciton to a fixed center-of-mass momentum sector $\bm K_0$ by projecting the subsystem operators entering the interaction Hamiltonians onto that sector. Two cases must be distinguished. The exciton--phonon, intra-excitonic exciton--photon, and exciton--exciton couplings conserve the number of tagged excitons, whereas the interband exciton--photon coupling creates or annihilates one. The corresponding projectors are, respectively,
\begin{subequations}
\label{eq:fixed-K0-projector}
    \begin{align}
    \hat{\mathcal P}_{\bm K_0}^{(1X)}
    &=
    \sum_n
    |n;\bm K_0\rangle
    \langle n;\bm K_0|\,,
    \label{eq:fixed-K0-one-exciton-projector}
    \\
    \hat{\mathcal P}_{\bm K_0}^{(0\oplus 1X)}
    &=
    |0\rangle\langle0|
    +
    \sum_n
    |n;\bm K_0\rangle
    \langle n;\bm K_0|\,,
    \label{eq:fixed-K0-zero-one-exciton-projector}
    \end{align}
\end{subequations}
where $\ket{0}$ is the crystal ground state and $\ket{n;\bm K_0}=\hat X^\dagger_{n,\bm K_0}\ket{0}$. In the following, $\hat{\mathcal P}_{\bm K_0}$ denotes whichever of Eqs.~\eqref{eq:fixed-K0-projector} is appropriate for the interaction under consideration. It acts exclusively on the subsystem Hilbert space, with the identity operator on the reservoir Hilbert spaces understood implicitly. The fixed-$\bm K_0$ interaction Hamiltonians are defined by
\begin{equation}
    \ho{H}_{X-\alpha}^{(\bm K_0)}
    \equiv
    \hat{\mathcal P}_{\bm K_0}\,
    \ho{H}_{X-\alpha}\,
    \hat{\mathcal P}_{\bm K_0}\,,
    \qquad
    \alpha\in\{{\rm ph},\gamma,X_g\}.
    \label{eq:generic-fixed-K0-definition}
\end{equation}
The free Hamiltonians require no separate treatment. The reservoir Hamiltonians $\ho{H}_{\rm ph}$, $\ho{H}_{\gamma}$, and $\ho{H}_{X_g}$ act on the complementary tensor factors, while $\ho{H}_X$, Eq.~\eqref{eq:Exc-ham}\SMnote, is diagonal in the center-of-mass momentum, so that $[\ho{H}_X,\hat{\mathcal P}_{\bm K_0}]=0$ and its restriction is simply
\begin{equation}
    \ho{H}_X^{(\bm K_0)}
    \equiv
    \hat{\mathcal P}_{\bm K_0}\,
    \ho{H}_X\,
    \hat{\mathcal P}_{\bm K_0}
    =
    \sum_n
    \epsilon_n(\bm K_0)\,
    |n;\bm K_0\rangle
    \langle n;\bm K_0|\,,
    \label{eq:fixed-K0-free-exciton}
\end{equation}
with the crystal ground state taken as the zero of energy. Because $\hat{\mathcal P}_{\bm K_0}$ commutes with the free evolution generated by $\ho{H}_X$ and by the reservoir Hamiltonians, performing the projection in the Schr\"odinger picture and subsequently transforming to the interaction and drive-rotating frames of App.~\ref{app:general_rotating_lindblad} is equivalent to projecting in those frames. For each interaction, the projection converts the momentum sums of the interaction Hamiltonians into Kronecker constraints on the momentum transferred at the vertex; the intermediate steps are given in App.~\ref{app:K0-projection}.
 
\paragraph{Exciton--phonon coupling:} Projecting the exciton--phonon Hamiltonian with $\hat{\mathcal P}^{(1X)}_{\bm K_0}$ enforces $\bm K=\bm K_0$ and a vanishing transferred phonon momentum, $\bm q=\bm 0$. At $\bm q=\bm 0$ the orthonormality of the internal wavefunctions gives $F_{n'n}(\bm 0)=\delta_{n'n}$, so that no internal transition can be induced, while the remaining diagonal acoustic deformation-potential coupling vanishes because the deformation-potential matrix element scales as $g^{n'n}_{\bm q,\nu}\propto\sqrt{|\bm q|}$. Hence
\begin{equation}
    \ho{H}_{X-{\rm ph}}^{(\bm K_0)}
    =
    \hat{\mathcal P}_{\bm K_0}^{(1X)}
    \ho{H}_{X-\mathrm{ph}}
    \hat{\mathcal P}_{\bm K_0}^{(1X)}
    =0\,.
    \label{eq:fixed-K0-phonon-Hamiltonian-zero}
\end{equation}
This result holds at the Hamiltonian level. The vanishing follows from the combined assumptions of a sharply defined center-of-mass momentum, strict intravalley momentum conservation, and a deformation-potential coupling that vanishes at zero momentum transfer. It is not a statement that phonons are irrelevant in monolayer semiconductors, where phonon-induced scattering contributes significantly to excitonic dephasing~\cite{selig2016excitonic,himami2018}. Finite-momentum scattering, intervalley processes, phonon sidebands, disorder-assisted transitions, and the finite momentum width of an excitonic wave packet lie outside the present reduction; their consequences for the scope of the results are discussed in Sects.~\ref{sec:lindblad_three_level} and~\ref{subsec:scope}.
 
\paragraph{Exciton--photon coupling:} The intra-excitonic coupling conserves the tagged-exciton number and is projected with $\hat{\mathcal P}^{(1X)}_{\bm K_0}$, which enforces $\bm s_\parallel=\bm 0$. The interband coupling creates or annihilates the tagged exciton and is projected with $\hat{\mathcal P}^{(0\oplus1X)}_{\bm K_0}$, which requires the photon to carry the in-plane momentum of the created exciton, $\bm s_\parallel=\bm K_0$. The projected couplings read
\begin{widetext}
\begin{subequations}
\label{eq:fixed-K0-X-photon}
\begin{align}
\ho{H}_{X-\gamma}^{{\rm intra},(\bm K_0)}
&=
-\sum_{n',n}
|n';\bm K_0\rangle
\langle n;\bm K_0|\,
\bm{\mathcal D}_{n'n}^{\rm intra}
\cdot
\left[
\hat{\bm E}_{\bm 0}^{(+)}
+
\hat{\bm E}_{\bm 0}^{(-)}
\right],
\label{eq:projected-intraband-length-gauge}
\\
\ho{H}_{X-\gamma}^{{\rm inter},(\bm K_0)}
&=
-\sum_n
\bigg\{
|n;\bm K_0\rangle\langle0|\,
\bm{\mathcal D}_{n0}^{\rm inter}(\bm K_0)
\cdot
\left[
\hat{\bm E}_{\bm K_0}^{(+)}
+
\hat{\bm E}_{-\bm K_0}^{(-)}
\right]
\nonumber\\
&\qquad\qquad
+
{\rm H.c.}
\bigg\},
\label{eq:projected-interband-length-gauge}
\end{align}
\end{subequations}
\end{widetext}
with the dipole form factors $\bm{\mathcal D}^{\rm intra}_{n'n}$ and $\bm{\mathcal D}^{\rm inter}_{n0}$ of App.~\ref{app:model-hamiltonians}\SMnote, the interband one evaluated at $\bm s_\parallel=\bm K_0$. The fixed-$\bm K_0$ exciton--photon Hamiltonian is $\ho{H}_{X-\gamma}^{(\bm K_0)}=\ho{H}_{X-\gamma}^{{\rm intra},(\bm K_0)}+\ho{H}_{X-\gamma}^{{\rm inter},(\bm K_0)}$. It supplies both the coherent drive, obtained in Sect.~\ref{subsec:explicit_drives} by replacing the field operators with prescribed coherent amplitudes, and the coupling to the quantized radiation fluctuations treated in App.~\ref{app:bath-corr}. The projection fixes only the in-plane photon momentum. The out-of-plane component $s_\perp$, and with it the photon frequency, remains unrestricted, so that the photon reservoir is not reduced to a single mode. The additional restriction of the intra-excitonic THz/mid-IR sector to a single linear polarization, introduced in Sect.~\ref{subsec:explicit_drives}, leaves these momentum constraints unchanged.
 
\paragraph{Exciton--exciton coupling:} Projecting the tagged-exciton pair operator $\hat X^\dagger_{n',\bm K+\bm q}\hat X_{n,\bm K}$ of the exciton--exciton interaction with $\hat{\mathcal P}^{(1X)}_{\bm K_0}$ enforces $\bm K=\bm K_0$ and $\bm q=\bm 0$, while the gas momentum $\bm Q$ remains unrestricted. At zero transferred momentum the gas exciton stays in its center-of-mass sector, although its internal state may change, $(n,\bm K_0)+(l,\bm Q)\longrightarrow(n',\bm K_0)+(l',\bm Q)$. The two equivalent assignments of the identical exciton lines to the tagged subsystem and to the gas contribute equally and cancel its prefactor $1/2$. Of its two parts, the direct intralayer term $G^{n'l';ln}_{\rm dir}(\bm q)$ vanishes as $\bm q\to\bm 0$ because of the neutral-exciton form factors~\cite{erkensten2021exciton}, whereas the exchange term cross-couples the internal wavefunctions of the two excitons through the Coulomb kernel and remains finite, $G^{n'l';ln}_{\rm exch}(\bm K_0,\bm Q,\bm 0)\neq0$~\cite{ciuti1998,shahnazaryan2017exciton,erkensten2021exciton}. The fixed-$\bm K_0$ exciton--gas Hamiltonian therefore reads
\begin{equation}
\begin{split}
    \ho{H}_{X-X_g}^{(\bm K_0)}
    =&
    \sum_{n',n}
    |n';\bm K_0\rangle
    \langle n;\bm K_0|\\
    &\otimes \frac{1}{A}
    \sum_{l',l,\bm Q}
    \mathcal G_{\mathrm{exch}}^{n'l';ln}
        (\bm K_0,\bm Q,\bm 0)\,
    \hat X^\dagger_{l',\bm Q}
    \hat X_{l,\bm Q}\,,
\end{split}
    \label{eq:final-projected-Xg-Hamiltonian}
\end{equation}
where $\hat X^\dagger_{n',\bm K_0}\hat X_{n,\bm K_0}$ has been written as $|n';\bm K_0\rangle\langle n;\bm K_0|$ on the one-exciton sector, and $\mathcal G_{\rm exch}\equiv A\,G_{\rm exch}$ is the area-rescaled exchange vertex. Its carrier--carrier and electron--hole contributions are derived, with independent internal-state labels on all four exciton lines, in App.~\ref{app:xx-internal-state-exchange}. The Hamiltonian~\eqref{eq:final-projected-Xg-Hamiltonian} describes number-conserving Coulomb scattering that preserves the center-of-mass momenta of both excitons. Its diagonal internal-state block is elastic, whereas its off-diagonal exchange channels transfer internal excitation between the tagged exciton and the gas while conserving the total exciton number and, for resonant channels, the total energy. Its first-order treatment in the stationary gas, which yields the exchange mean field renormalizing the control detunings, is given in Sect.~\ref{sec:VE-renormalization} and App.~\ref{app:exciton-gas-mean-field}.
 
Equations~\eqref{eq:fixed-K0-free-exciton}--\eqref{eq:final-projected-Xg-Hamiltonian}, together with the unprojected reservoir Hamiltonians and the separation of the photon field into coherent and fluctuating parts, Eq.~\eqref{eq:pho_drive-fluct}, constitute the kinematically restricted Hamiltonian, 
\begin{equation}
\begin{split}   
\ho{H}^{(\bm K_0)}(t)=&\ho{H}_X^{(\bm K_0)}+\sum_{\alpha}\ho{H}_{\mathcal B_\alpha}+\ho{H}_{\rm drv}^{(\bm K_0)}(t)+\ho{H}^{(\bm K_0)}_{X-ph}\\&+\ho{H}^{(\bm K_0)}_{X-X_g}\,.
\end{split}
\label{eq:H_total_proj}
\end{equation}
The three-level reduction and the Lindblad equation of Sect.~\ref{sec:lindblad_three_level} and App.~\ref{app:general_rotating_lindblad} are constructed from this reduced Hamiltonian. Within it, the projection removes the phonon coupling, reduces the exciton gas to an exchange mean field that shifts the energy levels, and leaves the radiation field as the only source of dissipation, with in-plane photon momentum fixed to $\bm s_\parallel=\bm K_0$ on interband legs and $\bm s_\parallel=\bm 0$ on intra-excitonic ones. We emphasize that some features of this limit are not necessarily representative of the physics of a real monolayer, notably the absence of pure dephasing and the $A^{-1}$ suppression of the intra-excitonic rates. The limitations of the model and their implications for the results are discussed in Sect.~\ref{subsec:scope}.

\section{Selection of excitonic three-level systems}
\label{sec:top_manifold}

This section introduces a systematic procedure for selecting the three-level systems used in the remainder of this work. A viable configuration must combine adequate optical accessibility, sufficiently strong intra-excitonic connectivity, and energy separations compatible with the interference and adiabatic-passage mechanisms of its configuration. The screening takes as its only inputs the exciton energies and matrix elements of the variational solver (App.~\ref{app:rayleigh-ritz}) and returns a short list of candidate exciton triplets ranked by configuration-dependent scores. The choice of states therefore follows from the exciton spectrum alone and is reproducible, and the same procedure applies to other materials and dielectric environments, where the ordering of bright and dark states, and hence the viable three-level systems, may differ~\footnote{Analogous pathway-screening strategies based on transition strengths and wave-function overlaps have been used in ultracold-molecule STIRAP~\cite{ni2008high,warner2023efficient,Borsalino2014KRb,Bergmann2019Roadmap}}.
 
Following the drive-architecture classification of Sect.~\ref{subsec:connection_excitonic_implementation}, each configuration imposes its own set of physical requirements, and the score encodes them quantitatively so that the three-level systems best suited to a given protocol rise to the top.
Throughout, we retain the three-level labeling of
Sect.~~\ref{subsec:three_level_topologies}, with $\lvert1\rangle=\lvert0\rangle$ denoting
the crystal ground state in both configurations.
The common driven vertex is $\lvert2\rangle$ for $\Xi$
and $\lvert1\rangle$ for $V$.

\subsection{Scoring policies for selecting three-level systems}
 \label{subsec:policies}
\paragraph{Screening primitives:}

The ranking is built from three quantities. Interband
brightness measures optical accessibility from the crystal
ground state and enters the scores of both retained
configurations. Intraband dipole matrix elements determine
the coupling strength of the internal leg in the $\Xi$
configuration. Exciton size provides a soft robustness
prior that mildly disfavors overly extended Rydberg-like
states.
 
For every exciton eigenstate $\ket{n}$ we extract the following quantities from the solver output:
\begin{enumerate}
\item the binding energy $E_{b,n}$, which fixes the leg splittings
      $\Delta E_{ab}\equiv|\epsilon_a-\epsilon_b|$ used throughout this section;
\item the interband brightness proxy
      $B_n := |\varphi_n(\bm r{=}0)|^2$,
      proportional to the interband oscillator strength and nonzero only for
      $s$-like ($m{=}0$) states; and
\item the intraband dipole magnitudes
      $D_{nm}:=|e\langle n|\bm r|m\rangle|$,
      which govern THz/mid-IR transitions between exciton internal states and
      are nonzero only for $\Delta m=\pm 1$.
\end{enumerate}
In addition, we compute the mean-square radius $\langle r^2\rangle_n$ and define the size-penalty factor
\begin{equation}
  \Pi_n
  =
  \frac{1}{1+\left(\langle r^2\rangle_n/R_{\mathrm{ref}}^2\right)^{\varsigma}},
  \label{eq:size_penalty}
\end{equation}
with $R_{\mathrm{ref}}^2=\langle r^2\rangle_{1s}$ and $\varsigma=1$. The factor $\Pi_n$ is used here as a mild geometric robustness prior against overly extended, Rydberg-like states. Accordingly, $\Pi_n$ only regularizes the configuration score by mildly favoring compact states.

\paragraph{Normalized figures of merit:}
Because the raw brightness and dipole matrix elements span widely different numerical scales, we define the normalized quantities
\begin{equation}
  \normq{B}_n \;=\; \frac{B_n}{\max_k B_k}\,,
  \qquad
  \normq{D}_{nm} \;=\; \frac{D_{nm}}{\max_{i<j}D_{ij}}\,.
  \label{eq:normalized_primitives}
\end{equation}
These are combined into configuration-dependent scores below. Here the subscripts $n,m$ run over the exciton eigenstates that fill the three-level slots $\ket{1},\ket{2},\ket{3}$ of a given candidate three-level system.
 
\paragraph{Scoring policy:} Each configuration is screened by a product of normalized
factors representing its required optical couplings.
This product suppresses candidates with weak optical
accessibility or, where required, a weak intraband
transition. The exponents in Eqs.~\eqref{eq:score_xi} and \eqref{eq:score_v} determine
the relative weights: $w_B=w_D=1$ for the brightness and dipole factors, while $w_S=0.3$ makes the size penalty a weaker regularizer. Since the factors are normalized, the scores determine a relative ranking rather than an absolute measure of protocol performance. The $\Xi$ score combines interband brightness with intraband dipole strength, whereas the $V$ score requires both excitonic states to be optically bright.
 
\paragraph{Ladder $\Xi$:}
The crystal ground $\ket{1}$ is coupled optically into a bright exciton $\ket{2}$, which is then driven internally to $\ket{3}$ by a THz/mid-IR field (legs $\ket{1}\!\leftrightarrow\!\ket{2}$ and $\ket{2}\!\leftrightarrow\!\ket{3}$, as in Sect.~\ref{subsub:ladder}). Two requirements must hold together: $\ket{2}$ must be optically accessible from the ground (interband brightness $\normq B_2$), and the internal $\ket{2}\!\leftrightarrow\!\ket{3}$ leg must be strongly dipole-coupled (intraband dipole $\normq D_{23}$). The score enforces both as near-hard criteria, with the size penalty a mild regularizer:
\begin{equation}
  \mathcal S_{\Xi}
  \;=\;
  \normq{B}_2^{\,w_B}\;
  \normq{D}_{23}^{\,w_D}\;
  \bigl(\Pi_2\,\Pi_3\bigr)^{w_S}.
  \label{eq:score_xi}
\end{equation}
No hard energy-window factor is included in the ladder
score. The two driven legs occupy the interband optical
and intraband THz/mid-IR sectors. Their experimental
addressability must nevertheless be checked against the
chosen pulse bandwidths, Rabi frequencies, and nearby
omitted transitions.
 
\paragraph{$V$ configuration:}
Both arms are interband-optical transitions from the crystal ground $\ket{1}$ into two bright excitons $\ket{2},\ket{3}$ (legs $\ket{1}\!\leftrightarrow\!\ket{2}$ and $\ket{1}\!\leftrightarrow\!\ket{3}$), so the single dominant requirement is that \emph{both} target excitons be optically bright; a three-level system with one dark arm cannot be driven as a $V$ and is disqualified. The score therefore rewards the product of the two brightnesses,
\begin{equation}
  \mathcal S_{V}
  \;=\;
  \bigl(\normq{B}_2\,\normq{B}_3\bigr)^{w_B}\;
  \bigl(\Pi_2\,\Pi_3\bigr)^{w_S}.
  \label{eq:score_v}
\end{equation}

\paragraph{Architecture-driven classification:}
The retained $\Xi$ system contains one interband leg
and one intraband leg, sharing the excitonic vertex
$\lvert2\rangle$. The $V$ system contains two interband
legs, sharing the crystal ground state
$\lvert1\rangle=\lvert0\rangle$.
 
\subsection{Application to WSe$_2$/hBN}
\label{subsub:selected-top}
We apply the screening procedure to hBN-encapsulated monolayer WSe$_2$~\footnote{The choice of substrate is motivated by the fact that Auger processes~\cite{auger_Dezheng,auger_binder,auger_Han} are
suppressed in hBN-encapsulated monolayers~\cite{wse2_hBN_zipfel}.
In general, our current model includes neither Auger nor
$X-X$ annihilation processes; their neglect is justified
here by the dilute exciton-gas regime considered.}, using the eleven lowest exciton eigenstates obtained from the variational solver with the material parameters listed in App.~\ref{app:rayleigh-ritz}. A complete list of the output data needed for the scoring classification is detailed in App.~\ref{app:eigenstate_inputs}. Tables~\ref{tab:rank_ladder_wse2}--\ref{tab:rank_v_wse2} collect the top-three candidates for each configuration together with their score components; in each row the exciton fillers are listed in slot order $(\ket{1},\ket{2},\ket{3})$, with $g\equiv\ket{0}$ denoting the crystal ground state.
 
The ladder $\Xi$ baseline is $(\ket{2},\ket{3})=(1s,2p)$, combining maximum interband brightness ($\normq{B}_{1s}=1$) with a strong $s\to p$ intraband dipole at $\Delta E=113$~meV. The $V$ baseline $(1s,2s)$ selects the brightest pair of optically accessible states.
 
\begin{widetext}
 
\begin{table}[h]
\centering
\caption{Top-3 ladder~$\Xi$ candidates for WSe$_2$/hBN. The optical leg addresses the bright state $\ket{2}$ from the crystal ground $\ket{1}=g$, while the THz/mid-IR leg couples $\ket{2}\leftrightarrow\ket{3}$. Under the drive-architecture classification the internal leg is not constrained to be ascending, so $\ket{3}$ may lie below $\ket{2}$ in energy (e.g.\ rank~3). Weights: $w_B=w_D=1$, $w_S=0.3$.}
\label{tab:rank_ladder_wse2}
\begin{tabular}{c l c c c c c}
\hline\hline
Rank & $(\ket{1},\ket{2},\ket{3})$ & $\mathcal S_{\Xi}$ & $\normq{B}_2$ &
  $\normq{D}_{23}$ & $(\Pi_2,\,\Pi_3)$ & $\Delta E_{23}$\,[meV] \\
\hline
1 & $(g,\,1s,\,2p)$ & $0.01101$ & $1.000$ & $0.026$ &
  $(0.500,\,0.108)$ & $113.2$ \\
2 & $(g,\,1s,\,3p)$ & $0.00196$ & $1.000$ & $0.008$ &
  $(0.500,\,0.017)$ & $142.5$ \\
3 & $(g,\,2s,\,2p)$ & $0.00191$ & $0.105$ & $0.085$ &
  $(0.055,\,0.108)$ & $11.0$ \\
\hline\hline
\end{tabular}
\end{table}
 
\begin{table}[h]
\centering
\caption{Top-3 $V$ candidates for WSe$_2$/hBN. Both legs are interband optical transitions from the crystal ground $\ket{1}=g$ to bright excitons $\ket{2}$ and $\ket{3}$; the common lower vertex is the crystal ground state. Weights: $w_B=1$, $w_S=0.3$.}
\label{tab:rank_v_wse2}
\begin{tabular}{c l c c c c c}
\hline\hline
Rank & $(\ket{1},\ket{2},\ket{3})$ & $\mathcal S_V$ & $\normq{B}_2$ &
  $\normq{B}_3$ & $(\Pi_2,\,\Pi_3)$ &
  $(E_b(\ket{2}),\,E_b(\ket{3}))$\,[meV] \\
\hline
1 & $(g,\,1s,\,2s)$ & $0.03572$ & $1.000$ & $0.105$ &
  $(0.500,\,0.055)$ & $(161.6,\,37.4)$ \\
2 & $(g,\,1s,\,3s)$ & $0.00624$ & $1.000$ & $0.030$ &
  $(0.500,\,0.011)$ & $(161.6,\,16.1)$ \\
3 & $(g,\,1s,\,4s)$ & $0.00181$ & $1.000$ & $0.012$ &
  $(0.500,\,0.003)$ & $(161.6,\,8.9)$ \\
\hline\hline
\end{tabular}
\end{table}
\end{widetext}
 
\subsection{Photon source model and drive Hamiltonians for the selected configurations}
\label{subsec:explicit_drives}
Once a candidate three-level system has been selected, the microscopic exciton--photon coupling in Eq.~\eqref{eq:X-gam_ham}\SMnote{} is projected first onto the fixed-$\bm K_0$ Hilbert space and afterwards onto the chosen three-level subspace, yielding an effective low-dimensional control Hamiltonian with source-model-dependent Rabi couplings and detunings. 

To this end, we introduce a two-color semiclassical drive (TCSD) together with a quantized radiation reservoir, valid for any three-level configuration $\mathcal T$. The transverse electric field is decomposed as
\begin{equation}
\begin{split}
\hat{\bm E}_{\rm T}(t)
&=
\bm E_{\rm drv}(t)
+
\delta\hat{\bm E}_{\rm T}(t),
\\
\bm E_{\rm drv}(t)
&=
\bm E_{\rm opt}(t)
+
\bm E_{\rm THz}(t).
\label{eq:pho_drive-fluct}
\end{split}
\end{equation}
Here $\bm E_{\rm opt}$ and $\bm E_{\rm THz}$ are prescribed coherent
fields driving the interband and intra-excitonic transitions,
respectively, while $\delta\hat{\bm E}_T$ denotes the corresponding
quantized radiation fluctuations responsible for radiative emission and
absorption and for the photonic Lamb-shift corrections; see App.~\ref{app:bath-corr}. After the fixed-$\bm K_0$ projection, the optical interband sector has
$\bm s_\parallel=\bm K_0$, whereas the intra-excitonic THz/mid-IR
sector has $\bm s_\parallel=\bm0$; in both cases $s_\perp$ remains
unrestricted, see Sect.~\ref{subsec:K0-projection}.

The two frequency sectors are assigned different polarization structures
in the reduced photonic model.  The optical interband reservoir retains
both transverse photon polarizations.  By contrast, the intra-excitonic
THz/mid-IR sector is restricted to a single in-plane linear polarization
$\bm e_d$, common to its coherent drive and quantum fluctuations.  Thus,
all intra-excitonic tones belonging to a given control protocol are taken
to have their electric field parallel to $\bm e_d$.  This restriction
selects one real linear combination of each $m=\pm1$ $p$ doublet and
suppresses its orthogonal partner in the $\Xi$ and $V$ configurations. 

The coherent-drive Hamiltonian is obtained by replacing the quantized electric field by its prescribed coherent component in the fixed-$\bm K_0$ $X-\gamma$ Hamiltonian,
\begin{equation}
\label{eq:Hdrv_def}
\begin{split}
\ho{H}_{\rm drv}(t)
&:=
\ho{H}_{X-\gamma}^{(\bm K_0)}
\!\left[
\hat{\bm E}_{\rm T}\rightarrow\bm E_{\rm drv}(t)
\right]
\\
&=
\ho{H}_{\rm drv}^{\rm inter}(t)
+
\ho{H}_{\rm drv}^{\rm intra}(t).
\end{split}
\end{equation}
The two contributions inherit the projected intra-excitonic and interband structures derived in Sect.~\ref{subsec:K0-projection}, see Eqs.~\eqref{eq:fixed-K0-X-photon}. The explicit parameterization of the coherent drive is introduced below. 

\subsubsection{Leg-resolved drive form}
\label{subsubsec:leg_resolved_drive}
Within the TCSD model, every dipole-allowed driven leg of a given configuration
$\mathcal{T}\in\{\Xi,V\}$ is assigned to either the optical
or the THz/mid-IR component of the field according to whether it is
an interband ground-to-exciton transition or an intra-excitonic
bound-to-bound transition. Denoting by $L^{\rm opt}_\mathcal{T}$ and
$L^{\rm THz}_\mathcal{T}$ the corresponding sets of driven legs, the
total coherent drive takes the leg-resolved form 
\begin{widetext}
\begin{equation}
\bm E_{\rm drv}^{(\mathcal T)}(t)
=
\frac{1}{2}
\sum_{(0,n)\in L_{\mathcal T}^{\rm opt}}
\left[
\bm{E}_{0n}^{\rm opt}\,
f_{0n}(t)\,
e^{-i\omega_{0n}^{\rm opt}t+i\varphi_{0n}^{\rm opt}}
+
{\rm c.c.}
\right]\,+\,
\frac{1}{2}
\sum_{(n',n)\in L_{\mathcal T}^{\rm THz}}
\left[
\bm{E}_{n'n}^{\rm THz}\,
f_{n'n}(t)\,
e^{-i\omega_{n'n}^{\rm THz}t+i\varphi_{n'n}^{\rm THz}}
+
{\rm c.c.}
\right]\,,
\label{eq:modelA_legresolved}
\end{equation}
\end{widetext}
with the leg-assignment table
\begin{equation}
\begin{aligned}
L^{\rm opt}_\Xi   &=\{(1,2)\},      & \qquad L^{\rm THz}_\Xi   &=\{(2,3)\},\\
L^{\rm opt}_V     &=\{(1,2),(1,3)\},& \qquad L^{\rm THz}_V     &=\varnothing,
\end{aligned}
\label{eq:legassignment}
\end{equation}
where the leg labels follow the configuration-specific identifications of
Sect.~\ref{subsubsec:topology_spec}. The general rule is that
interband ground-to-exciton legs are driven by $\bm E^{\rm opt}$ and
bound-exciton-to-bound-exciton legs are driven by $\bm E^{\rm THz}$,
which yields the two experimental architectures of one-color optical pump combined with a THz/mid-IR
control field ($\Xi$) and two-color optical control ($V$). 

All interband tones are taken co-circularly polarized, with the helicity that
couples to $\mathbf d^{K}_{cv}$ of Eq.~\eqref{eq:bloch-interband-dipole} and not to its time-reversed partner
$\mathbf d^{K'}_{cv}$, so that $|\mathbf e\cdot\mathbf d^{K}_{cv}|=|\mathbf d^{K}_{cv}|$
and both retained three-level systems are addressed within the single valley $\xi=K$
assumed in Sect.~\ref{sec:multi-env-model}; the intra-excitonic tone of $\Xi$ is linearly polarized
along $\mathbf e_d$.

Equation~\eqref{eq:modelA_legresolved} is written for a normally
incident optical beam, i.e.\ $\bm s_\parallel\simeq\bm 0$ for the
interband tones. Because the fixed-$\bm K_0$ projection of the interband
coupling requires the photon to carry the in-plane center-of-mass
momentum of the created exciton,
$\bm s_\parallel=\bm K_0$ [Eq.~\eqref{eq:projected-interband-exciton-operator}],
normal incidence forces $\bm K_0\simeq\bm 0$ in every three-level system
containing an interband leg, that is, in the $\Xi$ and $V$ configurations.
This is consistent with the radiative light-cone condition
${\omega^{\rm B}_{\mu}}>c|\bm K_0|$ derived in
App.~\ref{app:photon-reservoir}, which is then satisfied with a large
margin, and with the evaluation of the exciton-gas exchange vertices at
$\bm K_0=\bm 0$ in App.~\ref{app:offdiag}. The intraband leg of the retained $\Xi$ system requires
$s_\parallel=0$, whereas its interband leg satisfies
$\bm s_\parallel=\bm K_0$. Thus, the optical geometry
determines the retained exciton momentum.

The leg-resolved form~\eqref{eq:modelA_legresolved} accommodates two
distinct physical regimes through the choice of envelope $f_\ell(t)$,
which are operationally not interchangeable. The
\emph{interference/response regime}, in which the system reaches a
stationary or quasi-stationary response under sustained illumination
and the relevant observables are spectral quantities such as the
linear susceptibility $\chi(\omega_{\rm probe})$, transparency
contrast, and group delay, is naturally addressed by a finite-duration
plateau envelope
\begin{equation}
\begin{split}
f^{\rm plateau}_\ell(t)
&=
\tfrac{1}{2}\!\left[1+\tanh\!\left(\frac{t-t_{\rm on}}{\tau_{\rm on}}\right)\right]\\
&\times\,\tfrac{1}{2}\!\left[1-\tanh\!\left(\frac{t-t_{\rm off}}{\tau_{\rm off}}\right)\right].
\end{split}
\label{eq:envelope_plateau}
\end{equation}
The ramp times $\tau_{\rm on},\tau_{\rm off}$ suppress the broadband transient of an abrupt switch-on while remaining short compared with the relaxation timescales of the three-level system, so that $f^{\rm plateau}_\ell(t)\simeq1$ throughout the observation window; a plateau $T_{\rm pulse}=t_{\rm off}-t_{\rm on}$ longer than all relaxation timescales gives the true steady-state response, a shorter one the quasi-stationary response of short-window pump--probe configurations.

In the \emph{adiabatic control regime}, the
relevant observables are time-domain populations and coherences at the
end of a finite pulse sequence, and the dynamics relies
explicitly on the time-ordered overlap of the applied pulses; this
regime is naturally addressed by Gaussian envelopes
\begin{equation}
f^{\rm G}_\ell(t)
=
\exp\!\left[-\frac{(t-t_\ell)^2}{2\sigma_\ell^2}\right],
\label{eq:envelope_G}
\end{equation}
parameterized by a center time $t_\ell$ and a width $\sigma_\ell$. For the retained $\Xi$ system, the counterintuitive ordering $t_{23}<t_{12}$ is the candidate sequence for converting the excitation created by a weak optical leg into the $2p$ exciton by dark-state following, while suppressing occupation of the intermediate $1s$ state; the opposite ordering, $t_{12}<t_{23}$, describes sequential excitation. For $\Xi$, plateau envelopes address the cascade response and Gaussian envelopes the finite-pulse conversion; for $V$, a plateau addresses the optical response and a synchronous Gaussian pair, $t_{12}=t_{13}$, the free excited-state quantum beat of Sect.~\ref{subsub:V}.

The Gaussian widths are bounded from below by spectral selectivity, since each tone must address only its intended leg, and, for the $V$ beat, from above by the radiative decay of the coherence it prepares (App.~\ref{app:drive-windows}\SMnote): $\sigma_\ell\gtrsim0.067$~ps for the terahertz leg of $\Xi$, set by the $1s\to3p$ transition; $\sigma_\ell\gtrsim0.093$~ps for the upper tone of $V$, set by the $0\to3s$ transition; and $0.093\lesssim\sigma_\ell\lesssim0.18$~ps for the $V$ beat, for which we use $\sigma_\ell=0.12$~ps with $\hbar\Omega_{12},\hbar\Omega_{13}\lesssim1$~meV. These bounds settle only the bandwidth; the requirement that the Rabi energies remain small compared with the same detunings depends on the field amplitude and is assessed separately.

The two envelope families also call for different control strengths. The cascade dip is an interference feature only below a control strength of order the half-width of the vertex coherence, $\hbar\Omega_{23}\sim1.6$~meV, and an Autler--Townes doublet above it~\cite{AbiSalloum2010,Anisimov2011}, whereas the conversion benefits from the largest $\Omega_{23}$ that selectivity permits; the two analyses are therefore quoted at different control strengths, and a transparency claim is made only where the interference description applies (Sect.~\ref{subsec:Xi-response}).

\subsubsection{From the microscopic exciton--photon coupling to configuration-specific driven Hamiltonians}
\label{subsubsec:RWA_form}

After projection onto the fixed-$\bm K_0$ sector and subsequently onto a selected three-level system, the Rabi frequencies entering the effective Hamiltonians follow directly from the electric-dipole matrix elements discussed in App.~\ref{app:K0-projection-photon}.

\paragraph{Interband and intra-excitonic Rabi frequencies:}
For an interband optical leg between the crystal ground state $\ket{0}$ and a bright exciton $\ket{n}$, the projected dipole matrix element is
\begin{equation}
\bm{\mathcal D}^{\rm inter}_{n0}
=
\sqrt{A}\,
\varphi_n^{*}(\bm r=\bm0)\,
\bm d_{cv}^{\xi},
\label{eq:interband-dipole-drive}
\end{equation}
where the momentum-independent band-edge interband dipole $\bm d_{cv}^{\xi}$ is defined in App.~\ref{app:K0-projection-photon}. For a phasor $\bm{\mathcal E}^{\rm opt}_{0n}$, the corresponding real Rabi envelope is~\footnote{The area $A$ entering $\bm{\mathcal D}^{\rm inter}_{n0}$ through
Eq.~\eqref{eq:interband-dipole-drive} is not a free or ambiguous scale:
it is the same physical monolayer normalization area fixed by the
Wannier-envelope normalization of footnote~\ref{fn:normalization}\SMnote,
$\int_A d^2\bm r\,|\varphi_{n\xi}(\bm r)|^2=1$, and is set once a
sample is chosen---an experimentally accessible quantity (e.g.\ the
illuminated flake area), not a fitted or tunable parameter.
Equation~\eqref{eq:Omega_inter_def} is accordingly read with $A$ fixed
at this physical value; the Rabi frequency $\Omega_{n0}^{\rm amp}$ is
instead controlled through the external field amplitude
$\bm E^{\rm opt}_{0n}$, the physically meaningful and experimentally
adjustable design parameter (laser intensity/power) for a normally
incident drive of the form~\eqref{eq:modelA_legresolved}.}
\begin{equation}
\Omega_{n0}(t)
=
\Omega_{n0}^{\rm amp}f_{0n}(t),
\qquad
\Omega_{n0}^{\rm amp}
=
\frac{1}{\hbar}
\left|
\bm{E}^{\rm opt}_{0n}
\cdot
\bm{\mathcal D}^{\rm inter}_{n0}
\right|.
\label{eq:Omega_inter_def}
\end{equation}
These amplitudes must satisfy three conditions together: (i) $|\Omega|\ll|\Delta|$
against the nearest omitted transition ($\hbar\Omega_{13}\ll21.3$~meV in $V$,
29.3~meV in $\Xi$) and the RWA; (ii) the drives must leave the
tagged sector almost empty, $\max_t(1-\bar\rho_{11})\ll1$, because the
one-exciton Fock space of Sect.~\ref{sec:multi-env-model} holds no second
tagged exciton; this requires small interband pulse areas and, in $\Xi$, an
optical leg that is weak against the 3.21~meV interband linewidth of
Eq.~\eqref{eq:1s-lifetime-check}, $\Omega^{\rm amp}_{n0}\ll\gamma^{(\gamma,{\rm inter})}_{1s\to0}$; and (iii) $n_X$ inside the window where the
mean-field exchange treatment of App.~\ref{app:offdiag} is validated. Their
verification for every run is reported in Sect.~\ref{sec:results}.

For an intra-excitonic THz/mid-IR leg $\ket{n}\leftrightarrow\ket{n'}$, the relevant dipole matrix element is
\begin{equation}
\bm{\mathcal D}^{\rm intra}_{n'n}
=
-e\langle n'|\bm r|n\rangle
=
-e\int d^2\bm r\,
\varphi_{n'}^{*}(\bm r)\,
\bm r\,
\varphi_n(\bm r),
\label{eq:intra-dipole-drive}
\end{equation}
and the Rabi envelope is
\begin{equation}
\Omega_{n'n}(t)
=
\Omega_{n'n}^{\rm amp}f_{n'n}(t),
\qquad
\Omega_{n'n}^{\rm amp}
=
\frac{1}{\hbar}
\left|
\bm{E}^{\rm THz}_{n'n}
\cdot
\bm{\mathcal D}^{\rm intra}_{n'n}
\right|.
\label{eq:Omega_intra_def}
\end{equation}
For each driven leg $(i,j)$, we write the resonant
coupling as
\begin{equation}
\bra{j}\hat H_{\mathrm{drv}}^{(\mathrm{RWA})}(t)\ket{i}
=
\frac{\hbar}{2}\Omega_{ij}(t)
e^{+i\omega_{L,ij}t+i\phi_{ij}},
\end{equation}
following the convention of Sect.~\ref{subsec:unified_three_level_ham}.
Here $\Omega_{ij}(t)\geq0$ is the Rabi envelope, and
$\phi_{ij}$ contains the phase of the dipole--field
coupling, including the minus sign of the
electric-dipole interaction~\footnote{The real field in Eq.~\eqref{eq:modelA_legresolved} contains two complex-conjugate
components. The RWA retains the component resonant
with the chosen transition, using
$\omega_{ij}=(\epsilon_i-\epsilon_j)/\hbar$.
If this is the complex-conjugate component, its field
phase changes sign. Thus, $\phi_{ij}$ must be obtained
from the retained coupling rather than copied directly
from the field expression.}.

Interband accessibility is governed by $\varphi_n(\bm r=0)$ and the
valley-dependent polarization projection of $\bm d_{cv}^{\xi}$, whereas
intra-excitonic dipole transitions obey the envelope selection rule
$\Delta m=\pm1$.  For the latter, the linearly polarized THz/mid-IR
field selects the real $p$-state combination whose dipole is parallel to
$\bm e_d$. 

\subsubsection{Specialization to the top-ranked WSe$_2$/hBN three-level systems}
\label{subsubsec:topology_spec}

As in Sect.~\ref{sec:three_level_framework}, $\omega_{ab}\equiv(\epsilon_a-\epsilon_b)/\hbar$ is the signed transition frequency of a driven leg and $\omega_{L,ij}$ the signed drive frequency, with $|\omega_{L,ij}|$ the experimentally quoted carrier frequency. Positive Bohr frequencies entering spectral densities and dissipative rates are denoted by
$\omega^{\rm B}_{ab}\equiv|\epsilon_a-\epsilon_b|/\hbar$. The baseline three-level systems identified above are driven as follows.

\paragraph{Ladder $\Xi$:}
We choose $\ket{1}=\ket{0}$ (crystal ground state), $\ket{2}=\ket{1s}$, and $\ket{3}=\ket{2p}$, with an optical tone driving $\ket{0}\leftrightarrow\ket{1s}$ and a THz/mid-IR tone driving $\ket{1s}\leftrightarrow\ket{2p}$. The drive Hamiltonian is Eq.~\eqref{eq:Hdrv_ladder} of Sect.~\ref{subsub:ladder}, with $\Omega_{12}\equiv\Omega_{1s,0}$ an interband coupling~\eqref{eq:Omega_inter_def} and $\Omega_{23}\equiv\Omega_{2p,1s}$ an intra-excitonic coupling~\eqref{eq:Omega_intra_def}; the one-photon detunings are $\Delta_{12}=\omega_{L,12}-\omega_{12}$ and $\Delta_{23}=\omega_{L,23}-\omega_{23}$, and the two-photon detuning is $\delta_\Xi$ of the same subsection.

\paragraph{$V$ system:}
We choose $\ket{1}=\ket{0}$, $\ket{2}=\ket{1s}$, and $\ket{3}=\ket{2s}$, with two optical tones driving $\ket{0}\leftrightarrow\ket{1s}$ and $\ket{0}\leftrightarrow\ket{2s}$. The drive Hamiltonian is Eq.~\eqref{eq:Hdrv_V} of Sect.~\ref{subsub:V}, with $\Omega_{12}\equiv\Omega_{1s,0}$ and $\Omega_{13}\equiv\Omega_{2s,0}$ interband couplings~\eqref{eq:Omega_inter_def}; the one-photon detunings are $\Delta_{12}=\omega_{L,12}-\omega_{12}$ and $\Delta_{13}=\omega_{L,13}-\omega_{13}$, and the two-photon detuning is $\delta_V$ of the same subsection.

\section{Reduced Lindblad dynamics of the three-level systems}
\label{sec:lindblad_three_level}

After the configuration selection and source-model construction of Sect.~\ref{sec:top_manifold}, the excitonic problem reduces to
a driven three-level system of the tagged exciton, with $\bar\rho_{ii}(t)$ below understood
throughout as populations of the tagged-exciton sector defined in
Sect.~\ref{sec:multi-env-model}, distinct from the dilute exciton-gas
reservoir. Of the three reservoirs of the model, the fixed-$\bm K_0$ projection of
Sect.~\ref{subsec:K0-projection} leaves two active: the photon field, which is the only
source of dissipation, and the exciton gas, which enters as an exchange mean field that
shifts the energy levels; the phonon coupling drops out (items~\ref{item3} and~\ref{item4}
below). The open-system structure is common to the $\Xi$ and $V$
configurations; their driven legs, allowed transitions, and
dissipative channels are specified separately below.

For any of the selected configurations
$\mathcal T\in\{\Xi,V\}$, the subsystem is restricted to the
corresponding three-level Hilbert space
$\mathcal H_{\mathcal T}=\operatorname{span}
\{\ket{1},\ket{2},\ket{3}\}$ at fixed exciton center-of-mass
momentum $\bm K_0$.  After the projection onto this three-level system, the
Born--Markov approximation and the full-secular approximation derived
in App.~\ref{app:general_rotating_lindblad} yield the generic rotating-frame
master equation
\begin{equation}
\frac{d\bar{\rho}_{\mathcal T}}{dt}
=
-\frac{i}{\hbar}
\left[
\bar{\mathcal H}^{(\mathrm{RWA})}_{\mathcal T,\mathrm{eff}}(t) +\bar{\mathcal H}^{(\gamma,\mathrm{th})}_{LS,\mathcal T}
,
\bar{\rho}_{\mathcal T}
\right]
+
\bar{\mathcal D}^{(\gamma)}_{\mathcal T}
\!\left[\bar{\rho}_{\mathcal T}\right].
\label{eq:generic-topology-lindblad}
\end{equation}
Here $\bar{\mathcal H}^{(\mathrm{RWA})}_{\mathcal T,\mathrm{eff}}(t)$
is the configuration-dependent coherent Hamiltonian of
Eqs.~\eqref{eq:Hdrv_ladder}--\eqref{eq:Hdrv_V}, with the exciton-gas mean-field renormalizations incorporated into its level energies and detunings. In addition, $\bar{\mathcal H}^{(\gamma,\mathrm{th})}_{LS,\mathcal T}$ is the Lamb shift Hamiltonian due to the fluctuations of the photon reservoir; its explicit calculation is provided in App.~\ref{app:photon-reservoir}. 

The only contributing reservoir to the dissipator in Eq.~\eqref{eq:generic-topology-lindblad} is generated by the fluctuating photon field. To write this contribution in a form common to both
retained configurations, let $\mu=(h_\mu,\ell_\mu)$ denote a dipole-allowed
photon-coupled leg whose higher- and lower-energy states both
belong to $\mathcal H_{\mathcal T}$, and define
\begin{equation}
L_\mu
=
\ket{\ell_\mu}\bra{h_\mu}.
\end{equation}
Within the full-secular approximation, distinct nondegenerate
transition legs contribute independently, and the photon
dissipator takes the generic form
\begin{equation}
\bar{\mathcal D}^{(\gamma)}_{\mathcal T}
[\bar{\rho}_{\mathcal T}]
=
\sum_{\mu\in\mathcal L_{\mathcal T}^{(\gamma)}}
\left[
\gamma_{h_\mu\to\ell_\mu}^{(\gamma)}
\mathcal L_{L_\mu}[\bar{\rho}_{\mathcal T}]
+
\gamma_{\ell_\mu\to h_\mu}^{(\gamma)}
\mathcal L_{L_\mu^\dagger}[\bar{\rho}_{\mathcal T}]
\right],
\label{eq:generic-photon-dissipator}
\end{equation}
where the Lindblad superoperators are 
\begin{equation}
\mathcal L_L[\rho]
=
L\rho L^\dagger
-\frac{1}{2}
\left\{
L^\dagger L,\rho
\right\},
\label{eq:generic-lindblad-superoperator}
\end{equation}
and $\mathcal L_{\mathcal T}^{(\gamma)}$ denotes the set of
photon-coupled legs retained by the configuration.  The downward and
upward rates are evaluated from the momentum-constrained photon
spectral densities as described in App.~\ref{app:bath-corr}~\footnote{The photon occupation entering
Eqs.~\eqref{eq:photon-up-down}\SMnote{} is the Bose--Einstein distribution
$n_\gamma(\omega)=\left[e^{\hbar\omega/k_{\rm B}T_\gamma}-1\right]^{-1}$ at the
Bohr frequency of the leg. For the interband legs,
$\hbar{\omega^{\rm B}_{\mu}}\simeq1.7$~eV, it is $\approx10^{-29}$ even at
$T_\gamma=300$~K, so these legs are purely emissive; for the intraband leg of
$\Xi$, $\hbar\omega^{\rm B}_\mu=113.19$~meV, it is $2.4\times10^{-143}$,
$3.9\times10^{-8}$ and $1.27\times10^{-2}$ at $4$, $77$ and $300$~K. Both upward
rates are nevertheless retained formally, so that the same generator applies at
any $T_\gamma$ (item~\ref{item6}).}. Consistency checks of the Lindblad equations are discussed in App.~\ref{app:consistency}.

Before specializing to a particular three-level configuration, we list the assumptions and structural properties of the Lindblad equation~\eqref{eq:generic-topology-lindblad}:
\begin{enumerate}
    \item \label{item1} The finite-temperature photon Lamb shift is set to zero,
$\bar H_{\mathrm{LS},\mathcal T}^{(\gamma,\mathrm{th})}\simeq0$, in
Eq.~\eqref{eq:generic-topology-lindblad} and in the numerical analysis of
Sect.~\ref{sec:results}; its zero-temperature free-space part is already
absorbed into the reference exciton energies (App.~\ref{app:photon-reservoir}).
The interband contribution is algebraically suppressed as
$(k_BT_\gamma/\hbar\omega^{\rm B}_\mu)^2$, and the intra-excitonic contribution
carries the $A^{-1}$ prefactor of the corresponding fixed-$\bm K_0$ radiative
rate, Eq.~\eqref{eq:photon-intra-lamb}\SMnote{}, and vanishes in the thermodynamic
plane-wave limit. The photon-induced emission and absorption rates are kept.
    \item \label{item2} The projection of Sect.~\ref{subsec:K0-projection}
constrains the center-of-mass momentum of the tagged exciton but does not
collapse the reservoirs to a single mode. Momentum conservation fixes the
in-plane photon momentum to $\bm s_\parallel=\bm 0$ for an intra-excitonic
transition and to $\bm s_\parallel=\bm K_0$ for an interband one, leaving
$s_\perp$ unrestricted; this encodes the normally incident drive of
Sect.~\ref{subsubsec:leg_resolved_drive}. Setting the transferred in-plane
momentum to zero is a kinematic consequence of the fixed-$\bm K_0$ restriction,
whereas setting the reservoir momentum itself to zero is a separate modeling
choice tied to the spatial mode (App.~\ref{app:K0-projection}). Every rate
carrying an explicit $A^{-1}$ below inherits its magnitude from that choice.
\item \label{item3} For the acoustic deformation-potential interaction considered
here, the fixed-$\bm K_0$ projection of the exciton--phonon matrix
element~\eqref{eq:exciton-phonon-matrix-element-forward}\SMnote{} vanishes
identically, because conservation of the tagged-exciton center-of-mass momentum
enforces zero transferred phonon momentum (Sect.~\ref{subsec:K0-projection} and
App.~\ref{app:K0-projection-phonons}). Phonons therefore generate neither a
dissipator nor a dynamical Lamb shift within the retained three-level system
(Sect.~\ref{subsec:K0-projection}); the phonon
reservoir drawn in Fig.~\ref{fig:cartoon} is part of the Hamiltonian but not of
the reduced generator.
\item \label{item4} The exciton--exciton interaction is retained to first order
in the stationary one-body density matrix of the dilute, internally incoherent
exciton gas. It therefore enters as a coherent exchange mean field and generates
neither a dissipator nor a dynamical Lamb shift. The composition of the gas is a further,
material-specific limitation. In monolayer WSe$_2$, spin-forbidden dark excitons
lie about $40$~meV below the bright intravalley $1s$
exciton~\cite{Zhang2015dark,Wang2017dark}, and momentum-forbidden intervalley
excitons are also energetically favorable~\cite{Malic2018dark}; both are
populated on sub-picosecond to picosecond timescales after optical
injection~\cite{Selig2018,Madeo2020}. A gas that has relaxed at the $4$~K
baseline therefore resides largely in these dark species, whose exchange
coupling to the tagged exciton requires spin- and valley-resolved vertices that
the construction of App.~\ref{app:K0-projection-exciton-gas} does not contain.
Throughout this work $n_X$ accordingly denotes the areal density of the bright
intravalley $1s$ population sharing the spin and valley of the tagged exciton,
treated as a stationary input scanned over $n_X\lesssim10^{11}$~cm$^{-2}$. The
dark reservoir is a named omission of the present model.
\item\label{item5} The $p$-like states of the selected $\Xi$ system are
twofold degenerate, $m=\pm1$ (Sect.~\ref{subsub:selected-top}). We take the
coherent THz/mid-IR drive and the fluctuating photon field in the same
linear-polarization channel, which selects the real combination whose
intra-excitonic dipole is aligned with that polarization; the orthogonal member
of the doublet is neither driven nor populated by any retained process and is
excluded. Together with the nondegeneracy of the retained photon-coupled Bohr
frequencies, this ensures that the full-secular approximation acts on
nondegenerate transition blocks.
\item\label{item6} The three-level truncation omits the $2s$ state, which lies
$10.97$~meV above the target $2p$ and is dipole-connected to it with
$|d_{2s,2p}|=2.740\,a_B$, i.e.\ $3.21$ times the retained $|d_{1s,2p}|$
(Table~\ref{tab:appD_dipoles}). This channel and the retained $2p\to1s$ channel
are both intra-excitonic, so they share the $A^{-1}$ scaling and their ratio is
$A$-independent; in the fixed-$\bm K_0$ reservoir the larger dipole of the
omitted channel almost exactly compensates its smaller frequency, and the two
spectral densities agree to $0.2\%$ (App.~\ref{app:omitted-channels}\SMnote{}).
This near equality is an artifact of the fixed-$\bm K_0$ limit, whose
one-dimensional photon continuum makes both spectral densities linear in
frequency; with in-plane recoil allowed the ratio would be $9.4\times10^{-3}$.
Within the present reduction the two channels therefore differ only in thermal
occupation: the omitted upward channel is $1.5\times10^{-14}$ of the retained
downward one at $T_\gamma=4$~K, $0.24$ at $77$~K and $1.86$ at $300$~K. We adopt
$T_\gamma=4$~K as the baseline, where the reduction is exact for every practical
purpose, provided the far-infrared radiation reaching the sample is thermalized
with the cold stage; $77$~K remains usable at the level of a $24\%$ correction,
and room-temperature results are not reported. Within the present generator the
photon occupations are the only temperature-dependent input, and they change the
observables of Sect.~\ref{sec:results} by less than $10^{-7}$ even at $300$~K.
The choice of temperature rests instead on three conditions that hold
independently of the fixed-$\bm K_0$ limit. The projection removes
exciton--phonon scattering identically, so the dissipator is a radiative-only
loss budget, which is closest to complete at low temperature, where the
phonon-absorption channels that broaden the measured lines from a few meV below
$10$~K to several tens of meV at $300$~K freeze out~\cite{selig2016excitonic};
even there the budget is incomplete in WSe$_2$, because phonon emission into
lower-lying dark excitons does not freeze out (Sect.~\ref{subsec:scope}). The
$\Delta\to0$ reduction of the exchange vertices is cleanest at the same
temperature (App.~\ref{app:offdiag}), and the magnetoexciton Rydberg data that
fix the screening parameters of App.~\ref{app:rayleigh-ritz} were acquired
there~\cite{Stier2018}. Intra-excitonic $1s$--$2p$ transitions were resolved
at room temperature~\cite{Merkl2019}, so the restriction is one of the present
loss model rather than of the experimental platform; extending the predictions
to $T_\gamma\gtrsim77$~K requires a phonon-assisted dissipator and a
finite-$\mathbf{K}$ treatment, and restoring the $2s$ state would promote $\Xi$
to a four-level problem.
\item \label{item7} A second class of omitted exit is opened by the drive. The
THz/mid-IR tone, resonant with the $1s\to2p$ transition at
$\hbar\omega_{L,23}=113.19$~meV, lies
$E_b(1s)-2E_b(2p)=64.80$~meV above the ionization edge of the target state and
can carry the $2p$ exciton into the electron--hole continuum at electric-dipole
order, whereas the $1s$ state is safe from it by exactly $E_b(2p)$. This channel
is neither thermally nor $A^{-1}$ suppressed, but it is a bound-to-continuum
process outside the bound-state basis of App.~\ref{app:rayleigh-ritz}. A
golden-rule estimate within the same Rytova--Keldysh model
(App.~\ref{app:ionization}\SMnote{}) gives $\hbar\Gamma_{\rm ion}=2.9~\mu$eV at
the EIT/Autler--Townes crossover $\hbar\Omega_{23}\simeq1.6$~meV and
$28~\mu$eV, i.e.\ $0.9\%$ of $\hbar\gamma^{(\gamma,{\rm inter})}_{1s\to0}$, at
$\hbar\Omega_{23}=5$~meV. Evaluated to first order on the solution of the reduced
dynamics, without modifying the generator, the loss per converted $2p$ exciton
in the conversion of Sect.~\ref{subsec:Xi-transfer} is $0.5$--$1.6\%$ at
$\hbar\Omega_{23}=5$~meV and $0.2$--$0.7\%$ at $3$~meV, more than thirty times
smaller than the emission from the $1s$ vertex. It bears on the achievable
conversion efficiency, not on the structure of the generator.
\end{enumerate}

\subsection{On the renormalization of exciton energies and control detunings}
\label{sec:VE-renormalization}

The driven three-level dynamics depends on the energy differences of the
retained states. In the rotating frame, the reservoir-induced energy shifts
appear as corrections to the one- and two-photon detunings that control EIT-
and STIRAP-like interference, and within the present reduction the only such
correction is the coherent exchange mean field of the dilute exciton gas
(items~\ref{item1}--\ref{item4}).

The strict fixed-$\bm K_0$, zero-transfer projection of the
exciton--exciton interaction~\eqref{eq:gen-XX-Hamiltonian}\SMnote{} gives, to first order in the stationary
one-body density matrix of the gas, the Hermitian self-energy
$\Sigma_{ab}^{(X_g,0)}(\bm K_0)$ (App.~\ref{app:exciton-gas-mean-field}). Its
diagonal elements shift the retained levels, whereas its off-diagonal elements
describe coherent mixing, not dephasing. In the ladder $\Xi$ the only excitonic
off-diagonal element, $\Sigma^{(X_g,0)}_{1s,2p}$, connects $m=0$ to $|m|=1$ and
vanishes identically by the angular selection rule~\eqref{eq:selrule}, so the
exchange mean field is \emph{exactly} diagonal in the retained $\Xi$ block. In
$V$ the element $\Sigma^{(X_g,0)}_{1s,2s}$ is symmetry-allowed and nonzero, but
it connects states separated by $124.16$~meV and, under the drive
rotation~\eqref{eq:drive-rotation}, acquires the phase
$e^{i(\omega_{L,13}-\omega_{L,12})t}$, so it is non-secular on every timescale
of the reduced dynamics. We neglect it: its size is controlled by
$\chi_{1s,2s}=0.012$ at $n_X=10^{11}$~cm$^{-2}$, and the $V$ results remain
accurate to better than $10\%$ for $n_X\lesssim1.4\times10^{11}$~cm$^{-2}$
(App.~\ref{app:offdiag}). We therefore write the renormalized energy level of an exciton as
\begin{equation}
\ren\epsilon_a
=\epsilon_a+\Sigma_a^{(X_g,0)},
\qquad
\Sigma_a^{(X_g,0)}
\equiv\Sigma_{aa}^{(X_g,0)}(\bm K_0).
\label{eq:VE-renormalized-energies}
\end{equation}
Here, $\epsilon_a$ denotes the bare exciton energy obtained from the solution of the two-dimensional Wannier equation, see App.~\ref{app:rayleigh-ritz}. If the crystal ground state belongs to the three-level system, its self-energy vanishes because it contains no tagged exciton. 

The renormalized frequencies,
\begin{equation}
\ren\omega_{ab}
=\omega_{ab}
+\frac{\Sigma_a^{(X_g,0)}-\Sigma_b^{(X_g,0)}}{\hbar},
\label{eq:VE-renormalized-frequencies}
\end{equation}
with $\omega_{ab} = (\epsilon_a-\epsilon_b)/\hbar$. Thus, the one-photon detuning becomes
\begin{equation}
\begin{split}
\ren\Delta_{ab}
&=\omega_{L,ab}-\ren\omega_{ab}=\Delta_{ab}
-\frac{\Sigma_a^{(X_g,0)}-\Sigma_b^{(X_g,0)}}{\hbar},
\end{split}
\label{eq:VE-renormalized-detunings}
\end{equation}
where $\Delta_{ab}=\omega_{L,ab}-\omega_{ab}$. Only differences of level shifts affect the detunings.
The same principle applies to the two-photon condition.
For each configuration $\mathcal T\in\{\Xi,V\}$, let
$(a_{\mathcal T},b_{\mathcal T})$ denote the ordered pair
of outer states whose coherence defines this condition,
with $(a_{\mathcal T},b_{\mathcal T})=(1,3)$ for
$\mathcal T=\Xi$ and $(2,3)$ for $\mathcal T=V$.
The renormalized two-photon detuning then takes the
common form
\begin{equation}
\ren\delta_{\mathcal T}
=
\delta_{\mathcal T}
+
\frac{
\Sigma_{b_{\mathcal T}}^{(X_g,0)}
-
\Sigma_{a_{\mathcal T}}^{(X_g,0)}
}{\hbar}.
\label{eq:VE-two-photon-detunings}
\end{equation}
Here $\delta_{\mathcal T}$ is the corresponding signed
combination of the unrenormalized one-photon detunings
defined in Sec.~\ref{sec:three_level_framework}. The self-energy of the common or
intermediate state cancels, so the two-photon resonance
is displaced only by the differential mean-field shift
of the two outer states. In the $\Xi$ and $V$ systems,
$\Sigma_1^{(X_g,0)}=0$ because $\ket{1}$ is the crystal
ground state.

The magnitude of these corrections follows from the exchange vertices of
App.~\ref{app:offdiag}. Writing
$\Sigma^{(X_g,0)}_a=n_X\,\zeta_{aa}$ with $\zeta$ in units of $E_0a_B^2$, the
$1s$ level is displaced by
\begin{equation}
\frac{\Sigma^{(X_g,0)}_{1s}}{n_X}
= \zeta_{1s,1s}\,E_0a_B^2 = +2.12\;\mu\mathrm{eV}\,\mu\mathrm{m}^2 ,
\label{eq:gXX}
\end{equation}
i.e. $\Sigma^{(X_g,0)}_{1s}=+2.12$~meV at $n_X=10^{11}$~cm$^{-2}$. Within the
spinless, single-valley model of Sect.~\ref{sec:multi-env-model} the retained
exchange mean field therefore produces a \emph{blue} shift. It is the small
difference of two larger contributions of the carrier exchange, $-12.3$~meV
from the lines between like carriers and $+14.3$~meV from the electron--hole
lines, with $+0.08$~meV from the interband exchange (Table~\ref{tab:offdiag});
in the two-dimensional hydrogenic limit the same calculation reproduces the
exchange constant of Ciuti \emph{et al.}~\cite{ciuti1998} and the exchange
integrals of Glazov \emph{et al.}~\cite{Glazov2009} (check item~(v) of
App.~\ref{app:offdiag}). Because only differences of level shifts enter the
dynamics, the relevant quantities are smaller still. At the same density, in the
$\Xi$ system $\ren\Delta_{12}$ is displaced by $+2.12$~meV, $\ren\Delta_{23}$
by $-0.28$~meV and the two-photon detuning $\ren\delta_\Xi$ by $+1.84$~meV; in
the $V$ system $\ren\delta_V$ is displaced by only $+0.42$~meV, the two
interband shifts largely cancelling. All of these scale linearly with $n_X$ and
stay below the $3.21$~meV linewidth of the interband leg throughout the
validated window, but $\ren\delta_\Xi$ exceeds the detuning that the conversion
of Sect.~\ref{sec:results} tolerates, about $0.25$~meV or half the bandwidth
$\hbar/\sigma=0.44$~meV of its pulses (Sect.~\ref{subsec:Xi-robustness}), for
$n_X\gtrsim1.3\times10^{10}$~cm$^{-2}$. The two-photon resonance required by
Eq.~\eqref{eq:xi-dark-state} must therefore be set at the density of the exciton
gas present during the control sequence, not at the bare level spacing, and the
exchange mean field must be carried explicitly in the detunings rather than
absorbed into the bare energies.

That density is not supplied by the control drive. The weak Gaussian optical
pulse of the $\Xi$ conversion, $\hbar\Omega_{12}=0.080$~meV with $\sigma=1.5$~ps,
creates at most $\max_t(1-\bar\rho_{11})=2.3\times10^{-3}$ excitons per
normalization area, i.e.\ $2.4\times10^{4}$ excitons per cm$^2$ at
$A=9.62~\mu$m$^2$ (Sect.~\ref{subsec:operating-point}), six orders of magnitude
below the densities considered here. The gas is therefore a separate population,
created by an excitation that precedes or accompanies the control sequence and
treated as a stationary input (item~\ref{item4}). The field amplitudes and the
carrier frequencies thus remain independent settings, linked only by the
requirement that the latter be calibrated at the actual $n_X$. The numerical
determination of the operating point, and a way to perform this calibration on
the sample itself, are given in Sects.~\ref{subsec:operating-point},
\ref{subsec:Xi-robustness} and~\ref{subsec:Xi-response}.

\begin{figure*}[t]
  \centering
  \includegraphics[width=\textwidth]{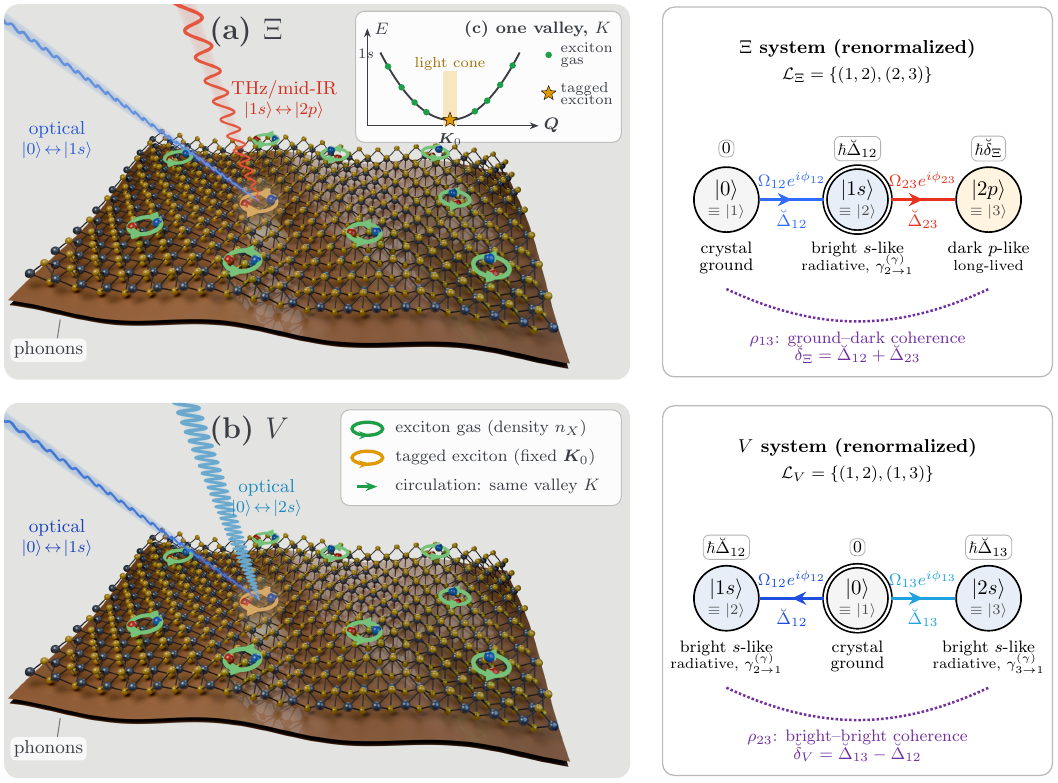}
  \caption{Reduced model of Sects.~\ref{sec:multi-env-model}--\ref{sec:VE-renormalization}.
  Left: cartoon of monolayer WSe$_2$ with a tagged exciton (gold ring) embedded
  in a dilute exciton gas (green rings) of density $n_X$ and mean spacing
  $d_X\gg a_B$. Every exciton carries the same circulation arrow: all of them
  belong to one valley, $K$ (item~\ref{item4} of
  Sect.~\ref{sec:lindblad_three_level}). The sheet ripple represents the phonon
  reservoir, which decouples at fixed $\bm K_0$ (item~\ref{item3}). (a) $\Xi$
  configuration: an interband optical tone drives $\ket{0}\leftrightarrow\ket{1s}$
  and a THz/mid-IR tone drives $\ket{1s}\leftrightarrow\ket{2p}$. (b) $V$
  configuration: two optical tones of different frequency drive
  $\ket{0}\leftrightarrow\ket{1s}$ and $\ket{0}\leftrightarrow\ket{2s}$. The model
  assumes normally incident plane-wave drives
  (Sect.~\ref{subsubsec:leg_resolved_drive}); the inclined, converging beams and
  the localized tagged exciton are drawn this way for visual clarity only and do
  not represent the drive geometry of the model. (c) Exciton
  band of the single valley, common to both configurations: the tagged exciton
  occupies the mode $\bm Q=\bm K_0$ inside the light cone, and the gas occupies
  other modes of the same band. Right: connectivity graphs of the selected
  three-level systems, drawn as in Fig.~\ref{fig:topology_graphs} but with the
  renormalized detunings $\ren\Delta_{ij}$ and $\ren\delta_{\mathcal T}$;
  the exchange renormalization of the level energies is detailed in
  Sect.~\ref{sec:VE-renormalization}. Each edge has the color of the laser that
  drives it. Bright $s$-like states decay radiatively at rates
  $\gamma^{(\gamma)}_{2\to1}$ and $\gamma^{(\gamma)}_{3\to1}$, while the dark
  $p$-like state is long-lived. The dotted arc marks the stored coherence and
  its renormalized two-photon detuning. The original figure was created using Blender software~\cite{Blender} and subsequently modified with Claude Opus 5.5 model, which was used to generate and refine selected graphical elements. The scientific content, geometry, labels, and physical interpretation of the AI-assisted elements were specified and verified by the authors.}
  \label{fig:cartoon}
\end{figure*}

Figure~\ref{fig:cartoon} summarizes the reduced model. The tagged exciton, at
fixed $\bm K_0\simeq0$ and addressed by the two tones of its configuration, shares
spin, valley and band with the dilute bright $1s$ gas, which acts through the
diagonal exchange self-energy alone. The connectivity of each three-level system is
therefore unchanged, and the bare detunings are replaced by
Eqs.~\eqref{eq:VE-renormalized-detunings} and~\eqref{eq:VE-two-photon-detunings},
the only difference from the graphs of Fig.~\ref{fig:topology_graphs}. The
dissipation is radiative: the bright $s$-like states decay to the crystal ground
state, whereas the dark $2p$ end of the ladder has no interband decay channel.

\subsection{Rotating-frame Lindblad equation for the ladder $\Xi$ system}
\label{sec:Xi-component-equations}

For the ladder system
$\{\ket{0},\ket{1s,\bm K_0},\ket{2p,\bm K_0}\}$, the sequential legs
$\ket{1}\leftrightarrow\ket{2}$ and $\ket{2}\leftrightarrow\ket{3}$ are driven by
an interband optical field and an intra-excitonic THz/mid-IR field,
respectively, and the direct leg $\ket{1}\leftrightarrow\ket{3}$ is absent
because $\ket{3}\equiv\ket{2p,\bm K_0}$ is interband-dark. The coherent
Hamiltonian is Eq.~\eqref{eq:Hdrv_ladder} with the detunings replaced by their
renormalized counterparts $\ren\Delta_{12}$ and $\ren\delta_\Xi$, and the
dissipator~\eqref{eq:generic-photon-dissipator} contains the interband leg
$2\to1$ and the intra-excitonic leg $3\to2$. The resulting equations for the
three populations and three coherences,
Eqs.~\eqref{eq:Xi-component-system}\SMnote{}, have three features that organize
the results. The vertex population and the two coherences that involve the
vertex, $\bar\rho_{12}$ and $\bar\rho_{23}$, are damped at rates set by its
radiative width $\gamma^{(\gamma)}_{2\to1}$. The two-photon coherence
$\bar\rho_{13}$, which carries the protocol, is damped only by the
$A^{-1}$-suppressed $3\to2$ rate and the negligible upward rate, and is
therefore long-lived on every timescale of the protocol. The exchange mean field
enters only through the phases of the coherences, via the renormalized
detunings.

\paragraph{The coherent dark state of the ladder.} The retained $\Xi$ target is
reached through an instantaneous eigenstate of the coherent Hamiltonian that
carries no amplitude on the lossy vertex. At renormalized two-photon resonance,
$\ren\delta_\Xi=0$, and for any nonvanishing total drive strength
$\Omega_{12}^2+\Omega_{23}^2>0$, the normalized state
\begin{equation}
|D_\Xi(t)\rangle=
\frac{\Omega_{23}(t)\,|1\rangle-\Omega_{12}(t)\,
e^{i(\phi_{12}+\phi_{23})}\,|3\rangle}
{\sqrt{\Omega_{12}^2(t)+\Omega_{23}^2(t)}}
\label{eq:xi-dark-state}
\end{equation}
satisfies $\bar H^{\rm (RWA)}_{\Xi,\rm eff}(t)\,|D_\Xi(t)\rangle=0$. The two
couplings of Eq.~\eqref{eq:Hdrv_ladder} generate $|2\rangle$ amplitudes
$\tfrac{\hbar}{2}\Omega_{12}e^{i\phi_{12}}$ from the $|1\rangle$ component and
$\tfrac{\hbar}{2}\Omega_{23}e^{-i\phi_{23}}$ from the $|3\rangle$ component,
and Eq.~\eqref{eq:xi-dark-state} is precisely the combination for which these
cancel. The cancellation is exact and, crucially, holds for
\emph{arbitrary} one-photon detuning $\ren\Delta_{12}$: only the two-photon
condition $\ren\delta_\Xi=0$ is required, because $\ren\Delta_{12}$
multiplies $\sigma_{22}$, on which $|D_\Xi\rangle$ has no support. With the
counterintuitive ordering $t_{23}<t_{12}$ of Sect.~\ref{subsec:explicit_drives},
$|D_\Xi\rangle$ interpolates continuously from $|1\rangle=|0\rangle$ to
$|3\rangle=|2p\rangle$ up to a phase, which is the candidate route from the
crystal ground state to the dipole-dark $2p$ exciton. With the weak optical leg
used in Sect.~\ref{sec:results} the mixing angle stays small, and the route
converts only the excitation that the optical pulse creates while the
terahertz dressing lasts (Sect.~\ref{subsec:Xi-transfer}).

Three qualifications must be attached to Eq.~\eqref{eq:xi-dark-state}, and they
are what separates this statement from a claim of protection.
First, the cancellation is a property of the coherent Hamiltonian alone. The
dark state is not a stationary state of the full generator: the photon
dissipator does not annihilate it, since its $|3\rangle$ component decays into
the vertex through $\gamma^{(\gamma)}_{3\to2}$, so
$\bar{\mathcal D}^{(\gamma)}_\Xi[\,|D_\Xi\rangle\langle D_\Xi|\,]\neq0$. A
coherent dark state and a dissipatively protected state are distinct objects,
and only the former is established here.
Second, exact adiabatic following would keep $\bar\rho_{22}(t)\equiv0$, but a
finite pulse sequence follows $|D_\Xi(t)\rangle$ only approximately, so the
vertex occupation is suppressed rather than eliminated; the residual
$\bar\rho_{22}(t)$ and the integrated vertex emission
$\int\!dt\,\gamma^{(\gamma)}_{2\to1}\bar\rho_{22}(t)$ are the quantities that
decide whether the mechanism is useful, and they are reported in
Sect.~\ref{sec:results}.
Third, $\ren\delta_\Xi$ is the \emph{renormalized} two-photon detuning of
Eq.~\eqref{eq:VE-two-photon-detunings}, which carries the density-dependent exchange mean
field. The resonance condition that Eq.~\eqref{eq:xi-dark-state} requires is therefore
set at the density of the exciton gas present during the sequence, not at
the bare level spacing; see the closing remark of
Sect.~\ref{sec:VE-renormalization}.

\subsection{Rotating-frame Lindblad equation for the $V$ system}
\label{sec:V-component-equations}

For the $V$ system $\{\ket{0},\ket{1s,\bm K_0},\ket{2s,\bm K_0}\}$, two
interband optical fields drive the legs $\ket{1}\leftrightarrow\ket{2}$ and
$\ket{1}\leftrightarrow\ket{3}$, and there is no direct electric-dipole coupling
between the two excited states. The coherent Hamiltonian is
Eq.~\eqref{eq:Hdrv_V} with the renormalized detunings, the dissipator contains
the two interband legs, and the component equations are
Eqs.~\eqref{eq:V-component-system}\SMnote{}. Both retained excited states are
interband-bright, and this is the structural difference between $V$ and $\Xi$.
From Table~\ref{tab:appD_states},
$|\varphi_{2s}(\bm 0)|^2=0.02985\,a_B^{-2}$, i.e.\ $10.5\%$ of the $1s$
brightness, so the upper state of the $V$ system carries its own vertex decay,
$\gamma^{(\gamma,\mathrm{inter})}_{2s\to0}=5.48\times10^{11}\,\mathrm{s^{-1}}$,
a radiative lifetime of $1.82$~ps, whereas the $\Xi$ target $2p$ has no
interband channel at all. The stored coherence of $V$ therefore decays on a
timescale comparable to any plausible pulse duration, whereas that of $\Xi$
does not. The two $V$ arms are moreover separated by $124.16$~meV against
radiative widths $\hbar\gamma^{(\gamma)}_{2\to1}=3.21$~meV and
$\hbar\gamma^{(\gamma)}_{3\to1}=0.361$~meV, so they are spectrally isolated and
no transparency window exists between them. The accessible signature of the
bright--bright coherence $\bar\rho_{23}$ is therefore a quantum beat at
$\omega_{23}=188.6$~rad\,ps$^{-1}$, a period of $33.3$~fs, damped at
$\tfrac12(\gamma^{(\gamma)}_{2\to1}+\gamma^{(\gamma)}_{3\to1})$ and hence
surviving about eleven beat periods. It is prepared by the synchronous Gaussian
pair of Sect.~\ref{subsubsec:leg_resolved_drive}, whose width must fall inside
the window given there. This is the quantitative content of the statement in
Sect.~\ref{subsub:V} that no radiatively protected combination exists in a $V$
configuration, and it places the $V$ comparison in a different experimental
regime from the picosecond pulses the $\Xi$ conversion requires. The one channel
the $V$ truncation omits, the intra-excitonic $2s\to2p$ transition, carries
$A^{-1}$ while the retained interband exit of the same state does not, so the
$V$ truncation is better controlled than the $\Xi$ one at every temperature
(App.~\ref{app:omitted-channels}\SMnote{}).

\section{Results}
\label{sec:results}

We now solve the reduced dynamics of Eqs.~\eqref{eq:Xi-component-system}
and~\eqref{eq:V-component-system}\SMnote{} for the retained $\Xi$ and $V$ systems of
hBN-encapsulated WSe$_2$. The numerical method, its validation against
independent analytic and numerical results, and the use of AI tools in
writing the simulation and analysis code are described in
App.~\ref{app:num_Lindblad}. Section~\ref{subsec:operating-point} fixes the
operating point and verifies, for every run, the conditions under which the
reduced model applies. Sections~\ref{subsec:Xi-transfer}
and~\ref{subsec:Xi-robustness} address the conversion of a weak optical
excitation into the dipole-dark $2p$ exciton by dark-state following, and its
robustness, including the density calibration imposed by the exchange mean field.
Section~\ref{subsec:Xi-response} addresses the stationary cascade response and
the distinction between electromagnetically induced transparency and
Autler--Townes splitting. Section~\ref{subsec:V-comparison} presents the $V$
system as the radiatively unprotected comparison,
Sect.~\ref{subsec:scope} states the scope of the results, and
Sect.~\ref{subsec:real-sample} the conditions that a real monolayer must meet.

\subsection{Operating point and admissibility}
\label{subsec:operating-point}

All calculations are performed at the baseline $T_\gamma=4$~K adopted in
Sect.~\ref{sec:lindblad_three_level}, item~\ref{item6}. They start from the crystal ground state,
$\bar\rho(0)=|1\rangle\langle1|$, with all drive phases $\phi_{ij}=0$; in the
open configurations considered here constant phases are a gauge choice. The
control inputs are the renormalized detunings $\hbar\ren\Delta$ and the
Rabi energies $\hbar\Omega$. Unless stated otherwise, every renormalized
detuning vanishes, i.e.\ each tone is tuned to its mean-field-renormalized
resonance at the operating density. The exciton density then enters only
through the conversion of the renormalized detunings into carrier frequencies,
Eqs.~\eqref{eq:VE-renormalized-detunings}--\eqref{eq:VE-two-photon-detunings}, and its physical effect is
examined in Sect.~\ref{subsec:Xi-robustness} by deliberately miscalibrating
that conversion. Field amplitudes follow from the Rabi energies through
Eqs.~\eqref{eq:Omega_inter_def} and~\eqref{eq:Omega_intra_def}. For the optical legs
they are quoted at $A=9.62~\mu$m$^2$, the core area of the single-mode fiber
over which the sample of Ref.~\cite{Stier2018} was mounted. The
largest Rabi energy used, $\hbar\Omega_{23}=5$~meV, is on the intra-excitonic
leg, where it corresponds to $E^{\rm THz}=49$~kV/cm, independent of $A$. The
interband legs are driven weakly. In the $\Xi$ conversion the optical leg has
$\hbar\Omega_{12}\le0.080$~meV, i.e.\ $E^{\rm opt}\le1.8$~V/cm ($4.4$~mW/cm$^2$,
vacuum-equivalent), and the largest interband Rabi energy, $1$~meV in the $V$
beat, corresponds to $E^{\rm opt}=23$~V/cm ($0.68$~W/cm$^2$) on the
$0\to1s$ leg. These amplitudes and the interband rate
$\hbar\gamma^{(\gamma,{\rm inter})}_{1s\to0}=3.21$~meV are those of the
monolayer in vacuum (App.~\ref{app:photon-reservoir}). In a dielectric stack
the field at the sheet and the radiative rate change by sample-dependent
factors~\cite{Fang2019}; the vacuum rate lies within about $30\%$ of the
microscopic radiative width computed for hBN-encapsulated
WSe$_2$~\cite{Brem_intrinsic}, and the results below carry over to other
environments by rescaling at fixed $\Omega/\gamma^{(\gamma,{\rm inter})}_{1s\to0}$,
$\sigma\gamma^{(\gamma,{\rm inter})}_{1s\to0}$ and $\mathcal A$ (App.~\ref{app:photon-reservoir}). The protocol families and their run parameters are collected
in Table~\ref{tab:protocols}.

\begin{table*}
\caption{Protocol families of Sect.~\ref{sec:results} and their run
parameters. Detunings are the renormalized detunings of
Eqs.~\eqref{eq:VE-renormalized-detunings}--\eqref{eq:VE-two-photon-detunings}; the miscalibration runs
of Sect.~\ref{subsec:Xi-robustness} are labeled separately. All runs at
$T_\gamma=4$~K, starting from $\bar\rho(0)=|1\rangle\langle1|$.}
\label{tab:protocols}
\begin{ruledtabular}
\begin{tabular}{llll}
Protocol & Envelope & $\hbar\Omega$ (meV) & Scanned detunings (meV) \\
\hline
$\Xi$ conversion & Gaussian, $\sigma=1.1$--$1.9$~ps,
 & $\Omega_{23}=1$--$5$, $\Omega_{12}\le0.08^{\,b}$ & $\ren\delta_\Xi\in[-1,1]$; \\
 & $\tau_d=(1.0$--$1.5)\,\sigma$ & & miscalibration $|\ren\delta_\Xi|\le1.84$ \\
$\Xi$ response & plateau, $\tau_{\rm on}=0.5$~ps, $T_{\rm pulse}=50$~ps$^{\,a}$
 & $\Omega_{23}=0.25$--$5$, $\Omega_{12}\le0.08^{\,b}$
 & $\ren\Delta_{12}\in[-15,15]$ \\
$V$ response & plateau, $\tau_{\rm on}=0.2$~ps, $T_{\rm pulse}\ge20$~ps$^{\,a}$
 & $\Omega_{12}=0.1$, $\Omega_{13}=0.01$ & $\ren\delta_V\in[-5,5]$ \\
$V$ beat & synchronous Gaussians, $\sigma=0.093$--$0.18$~ps
 & $\Omega_{12}=\Omega_{13}=1$ & -- \\
\end{tabular}
\end{ruledtabular}
\begin{flushleft}
\footnotesize
$^{a}$ Stationary responses are taken from the null space of the generator;
finite plateaus are used to quantify the approach to it (App.~\ref{app:num_Lindblad}).\\
$^{b}$ Weak optical leg $\hbar\Omega_{12}=0.05\min(\hbar\Omega_{23},
\hbar\gamma^{(\gamma,{\rm inter})}_{1s\to0}/2)$, scaled by $0.25$--$2$ in the
linearity control of Sect.~\ref{subsec:Xi-transfer}; the coherent-population-trapping
run uses $\hbar\Omega_{12}=0.05$~meV and $\hbar\Omega_{23}=1$~meV.
\end{flushleft}
\end{table*}

\begin{table}
\caption{Worst case, over all runs of each protocol family of
Table~\ref{tab:protocols}, of the admissibility ratios of
Sect.~\ref{subsec:explicit_drives}. RWA: $\hbar\Omega/\hbar\omega_L$ on the
interband (inter) and intra-excitonic (intra) legs. Selectivity:
$\hbar\Omega/|\Delta|$ against the nearest omitted transition ($0\to2s$ or
$0\to1s$ for the optical tones, $0\to3s$ for the $0\to2s$ tone, $1s\to3p$ for
the THz tone), with $|\Delta|$ reduced by the largest detuning scanned on that
leg and corrected for the mean-field shift of the competing state. Bandwidth:
$\sigma/\sigma_{\min}$ or $\tau_{\rm on}/\tau_{\min}$ for the bounds of
Sect.~\ref{subsec:explicit_drives} ($n=3$). A dash marks a leg that the
protocol does not use.}
\label{tab:admissibility}
\footnotesize
\begin{ruledtabular}
\begin{tabular}{lccccc}
 & \multicolumn{2}{c}{RWA} & \multicolumn{2}{c}{Selectivity} & Bandwidth \\
\cline{2-3}\cline{4-5}
Protocol & inter & intra & inter & intra & \\
\hline
$\Xi$ conversion$^{\,a}$ & $4.6\times10^{-5}$ & $0.045$ & $6.5\times10^{-4}$ & $0.18$ & $\ge16$ \\
$\Xi$ response & $4.7\times10^{-5}$ & $0.044$ & $7.3\times10^{-4}$ & $0.17$ & $4.9$ \\
$V$ response & $5.8\times10^{-5}$ & -- & $8.1\times10^{-4}$ & -- & $1.4$ \\
$V$ beat & $5.8\times10^{-4}$ & -- & $0.047$ & -- & $1.00$ \\
\end{tabular}
\end{ruledtabular}
\begin{flushleft}
\footnotesize
$^{a}$ Weak optical leg, $\Omega_{12}/\gamma^{(\gamma,{\rm inter})}_{1s\to0}\le0.025$;
$\max_t(1-\bar\rho_{11})\le2.3\times10^{-3}$ with the photon dissipator and
$1.2\times10^{-2}$ without it (Sect.~\ref{subsec:operating-point}).
\end{flushleft}
\end{table}

Table~\ref{tab:admissibility} gives, for each protocol family, the worst
case over all its runs of the admissibility ratios of
Sect.~\ref{subsec:explicit_drives}. These ratios test the validity of the
reduced model itself, not the performance of a protocol. The
rotating-wave ratio stays below $0.045$ and the selectivity ratio below
$0.18$. Both maxima occur at the $5$~meV ceiling of the terahertz leg of the conversion protocol,
where the $1s\to3p$ transition at $29.3$~meV sets the limit already identified
in Sect.~\ref{subsec:explicit_drives}. All pulse widths and switching times lie
at or above the bandwidth bounds of the same section; the $V$ beat at
$\sigma=0.093$~ps sits at its bound by construction. The exciton density
stays below $1.1\times10^{11}$~cm$^{-2}$ in every run except the bare-laser
scan of Fig.~\ref{fig:R2}(c), which continues to $2\times10^{11}$~cm$^{-2}$ to
show how the uncalibrated efficiency decays. Otherwise, values above
$10^{11}$~cm$^{-2}$ occur only in the $+5\%$ and $+10\%$ miscalibration runs
at $n_{X}=10^{11}$~cm$^{-2}$ of Sect.~\ref{subsec:Xi-robustness}, which are
designed to probe the edge of the window. Over the whole range
$n_X\lambda\le0.038$, so the composite-boson terms neglected in
App.~\ref{app:offdiag} remain at the per-cent level. The reduced model also requires the drives to leave the tagged sector
almost empty, $\max_t(1-\bar\rho_{11})\ll1$ (Sect.~\ref{subsec:explicit_drives}),
because the one-exciton Fock space of Sect.~\ref{sec:multi-env-model} holds no
second tagged exciton. This condition is met with a wide margin in $\Xi$, where
$\max_t(1-\bar\rho_{11})\le2.3\times10^{-3}$ over all conversion runs with the
photon dissipator ($5.6\times10^{-3}$ in the linearity control of
Sect.~\ref{subsec:Xi-transfer}, at twice the optical Rabi energy),
$1.2\times10^{-2}$ in the coherent-only controls, and
$2.5\times10^{-3}$ on the coherent-population-trapping plateau, and it is
tightest in the $V$ beat, $0.04$--$0.11$ at $1$~meV
(Sect.~\ref{subsec:V-comparison}). Per normalization area this is at most
$1.3\times10^{5}$~cm$^{-2}$ in $\Xi$ and $1.1\times10^{6}$~cm$^{-2}$ in $V$ at
$A=9.62~\mu$m$^2$, at least four orders of magnitude below the gas density;
the drive therefore does not perturb the stationary reservoir assumed in
Sect.~\ref{sec:lindblad_three_level}. Finally, the probability that a $2p$
exciton is ionized during a pulse, $\int dt\,\Gamma_{\rm ion}(t)$, stays below
$0.15$, and the ionization loss of Sect.~\ref{sec:lindblad_three_level},
item~\ref{item7}, below $1.6\%$ of the converted $2p$ population, so its
first-order treatment in Eq.~\eqref{eq:2p-ionization-num}\SMnote{} remains adequate.

Because the optical leg of every $\Xi$ run is weak against the interband
linewidth, $\Omega_{12}/\gamma^{(\gamma,{\rm inter})}_{1s\to0}\le0.025$, the
$1s$ vertex never undergoes resolved Rabi oscillations on that leg. It acts
instead as a damped intermediate state dressed by the terahertz leg, and the
parameter that controls the conversion is built from $\Omega_{23}$, the pulse
width and the interband rate. It is introduced in Sect.~\ref{subsec:Xi-transfer}.

\subsection{Weak-excitation conversion to the dark $2p$ exciton}
\label{subsec:Xi-transfer}

The counterintuitive pulse sequence converts the excitation created by a weak
optical pulse into the dipole-dark $2p$ exciton by dark-state following, and
the lossy $1s$ vertex is populated only transiently. Because the one-exciton
description holds only while the drives leave the tagged sector almost empty
(Sect.~\ref{subsec:operating-point}), the optical leg is kept weak,
$\hbar\Omega_{12}=0.05\min(\hbar\Omega_{23},\hbar\gamma^{(\gamma,{\rm inter})}_{1s\to0}/2)$,
and the sequence is characterized by what becomes of the small excitation it
creates. Part of it ends in $2p$, $\bar\rho_{33}(t_f)$; the rest is emitted by
the vertex, $N_{\rm emit}=\int dt\,\gamma^{(\gamma)}_{2\to1}\bar\rho_{22}$,
which returns the exciton to $|1\rangle$. We use three figures of merit. The
conversion efficiency
\begin{equation}
\eta=\frac{\bar\rho_{33}(t_f)}{\bar\rho_{33}(t_f)+N_{\rm emit}}
\label{eq:conversion-efficiency}
\end{equation}
is the fraction of the excitation that reaches $2p$ rather than being
re-emitted. The vertex suppression $\max_t\bar\rho_{22}/\bar\rho_{33}(t_f)$ is
the peak occupation of the lossy state per converted exciton, and
$P_{\rm ion}/\bar\rho_{33}(t_f)$ is the ionization loss per converted exciton
(Sect.~\ref{sec:lindblad_three_level}, item~\ref{item7}). All three are ratios
of quantities that are quadratic in $\Omega_{12}$, and they do not depend on it:
changing $\Omega_{12}$ by a factor of eight, from $0.25$ to $2$ times the value
above, changes $\eta$ by less than $10^{-3}$ of itself. The results therefore
apply to any optical leg that is weak in this sense.

\begin{figure*}[!t]
\includegraphics[width=\textwidth]{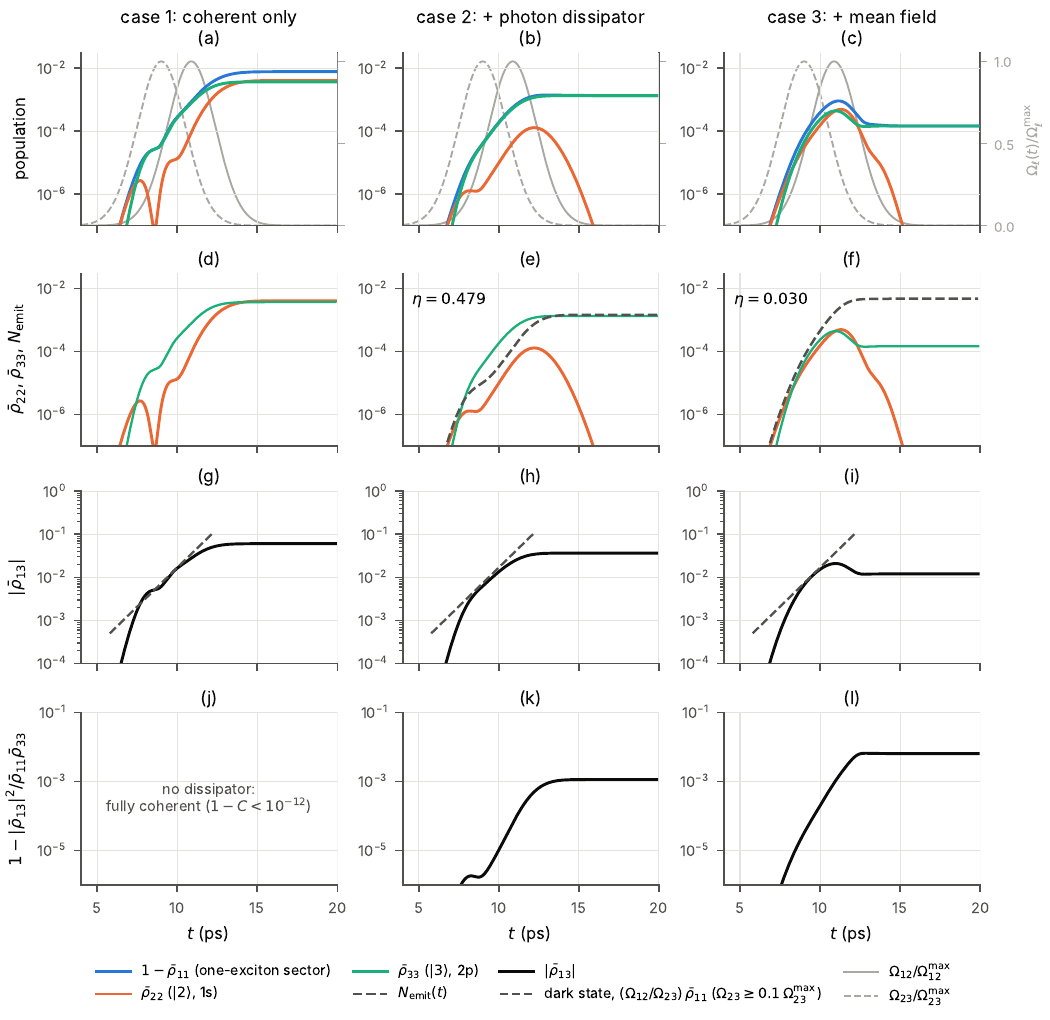}
\caption{Time evolution of the $\Xi$ conversion at the baseline
(counterintuitive Gaussian pulses, $\sigma=1.5$~ps, $\tau_d=1.25\,\sigma$,
$\hbar\Omega_{23}=5$~meV, weak optical leg $\hbar\Omega_{12}=0.080$~meV,
$T_\gamma=4$~K). Columns: case~1, coherent dynamics only; case~2, with the
photon dissipator; case~3, with the photon dissipator and the exchange mean
field at $n_X=10^{11}$~cm$^{-2}$, lasers at the bare resonances
($\hbar\ren\delta_\Xi=1.84$~meV). Colors: blue, excitation $1-\bar\rho_{11}$;
orange, $|2\rangle$ ($1s$ vertex); green, $|3\rangle$ ($2p$). All panels on a
logarithmic scale.
(a)--(c)~Excitation and populations; thin gray lines show the envelopes
$\Omega_{12}/\Omega_{12}^{\max}$ (solid) and $\Omega_{23}/\Omega_{23}^{\max}$
(dashed), right axis.
(d)--(f)~Vertex and $2p$ populations and accumulated number of emitted vertex
photons $N_{\rm emit}$ (dashed; zero in case~1), with the conversion
efficiency $\eta$ of Eq.~\eqref{eq:conversion-efficiency}.
(g)--(i)~Outer coherence $|\bar\rho_{13}|$ (black) and its dark-state value
$(\Omega_{12}/\Omega_{23})\bar\rho_{11}$ (dashed), shown while
$\Omega_{23}\ge0.1\,\Omega_{23}^{\max}$.
(j)--(l)~Incoherent fraction $1-C=1-|\bar\rho_{13}|^2/\bar\rho_{11}\bar\rho_{33}$ of
the converted population, where $\bar\rho_{33}>10^{-8}$; in case~1 it stays
below $10^{-12}$.
The sequential-order control is shown in Fig.~\ref{fig:R1b}.
The data in this figure and in Figs.~\ref{fig:R1b}--\ref{fig:R5} were computed
with the code described in App.~\ref{app:num_Lindblad} and plotted with Python
scripts written with the assistance of Claude Opus 5.5 model to the authors' specification; the authors checked
the plotted quantities against the run outputs.}
\label{fig:R1}
\end{figure*}

Figure~\ref{fig:R1} shows the time evolution at the baseline of
Table~\ref{tab:protocols}: Gaussian pulses with $\sigma=1.5$~ps,
$\tau_d=1.25\,\sigma$, $\hbar\Omega_{23}=5$~meV and $\hbar\Omega_{12}=0.080$~meV
(pulse area $0.46$~rad) at $T_\gamma=4$~K. To separate the mechanisms, the same
sequence is run in three cases. Case~1 retains the coherent dynamics only.
Case~2 adds the photon dissipator of Eq.~\eqref{eq:generic-photon-dissipator}.
Case~3 adds, in addition, the diagonal exchange mean field at
$n_X=10^{11}$~cm$^{-2}$ with the lasers tuned to the bare resonances, so that
the renormalized detunings of
Eqs.~\eqref{eq:VE-renormalized-detunings} and~\eqref{eq:VE-two-photon-detunings} are
$\hbar\ren\Delta_{12}=2.12$~meV, $\hbar\ren\Delta_{23}=-0.28$~meV and
$\hbar\ren\delta_\Xi=1.84$~meV. With the lasers calibrated at the same density,
case~3 reproduces case~2 exactly, because the mean field only shifts the
levels. In all three cases the outer coherence follows the dark state of
Eq.~\eqref{eq:xi-dark-state}, $|\bar\rho_{13}|\simeq(\Omega_{12}/\Omega_{23})\bar\rho_{11}$,
while the terahertz pulse is strong, and freezes as that pulse decays
[Fig.~\ref{fig:R1}(g)--(i)]. In case~1 the sequence leaves an excitation
$1-\bar\rho_{11}=7.8\times10^{-3}$, shared between $2p$
($\bar\rho_{33}=3.7\times10^{-3}$) and the vertex
($\bar\rho_{22}=4.1\times10^{-3}$) [Fig.~\ref{fig:R1}(a), (d)]: the part of the
optical pulse that arrives after the terahertz dressing has decayed excites
the vertex directly. In case~2 that vertex population is emitted. The vertex
occupation peaks at $1.3\times10^{-4}$ and ends at $5\times10^{-10}$, while
$\bar\rho_{33}(t_f)=1.34\times10^{-3}$ and $N_{\rm emit}=1.46\times10^{-3}$,
so that $\eta=0.479$ and the vertex suppression is $0.096$
[Fig.~\ref{fig:R1}(b), (e)]. Per incident photon the same run converts a
fraction $Y=\bar\rho_{33}(t_f)/N_{\rm in}=0.33$ into $2p$. Here
$N_{\rm in}=\int dt\,\Omega_{12}(t)^2/2\gamma^{(\gamma,{\rm inter})}_{1s\to0}=4.05\times10^{-3}$
is the number of photons incident from one side on the normalization area,
calibrated by the perfect reflection of a radiatively limited monolayer at
resonance. A mirror-symmetric sheet driven from one side couples only to the
symmetric half of the incident field, so $Y\le1/2$. More than $90\%$ of the emission occurs after the
peak of the optical pulse, $80\%$ of it between $10.9$ and $13.4$~ps, while
the terahertz pulse, centered at $9.0$~ps, decays. The converted population is
coherent with the crystal ground state. We characterize this by the coherence
fraction $C=|\bar\rho_{13}|^2/(\bar\rho_{11}\bar\rho_{33})$, which equals one
for any pure superposition of $|1\rangle$ and $|3\rangle$. Emission returns
population to $|1\rangle$ without a phase relation to $|3\rangle$, but the
population it returns is itself of order $\bar\rho_{33}$, so it lowers the
coherence fraction only to $1-C=1.1\times10^{-3}$, of the order of $N_{\rm emit}$
[Fig.~\ref{fig:R1}(k)]. This value of $C$ is an upper bound for a real sample
(Sect.~\ref{subsec:scope}). In case~3 the
two-photon detuning, $|\ren\delta_\Xi|\sigma=4.2$, breaks the dark state:
$\bar\rho_{33}(t_f)=1.5\times10^{-4}$, $N_{\rm emit}=4.7\times10^{-3}$,
$\eta=0.030$, and the vertex suppression rises to $3.4$, with
$1-C=6.4\times10^{-3}$ [Fig.~\ref{fig:R1}(c), (f), (l)].

\begin{figure*}
\includegraphics[width=\textwidth]{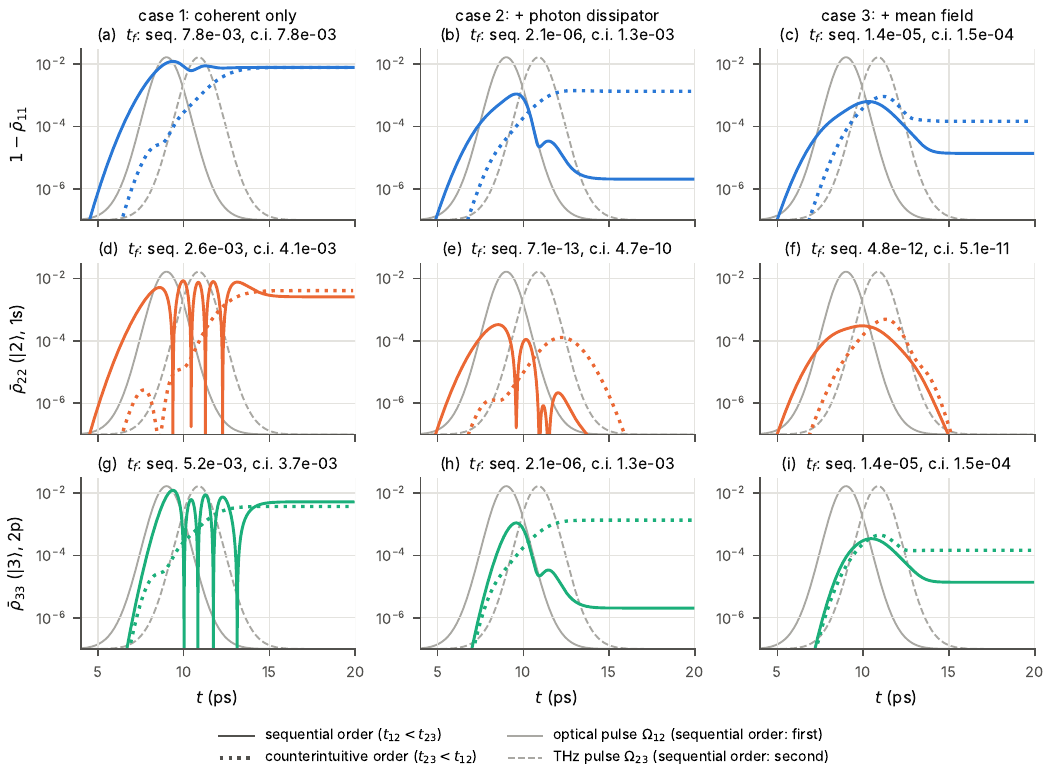}
\caption{Pulse-order control of the $\Xi$ conversion, with the parameters and
the three cases of Fig.~\ref{fig:R1}. Rows: excitation $1-\bar\rho_{11}$
[(a)--(c)] and populations $\bar\rho_{22}$ [(d)--(f)] and $\bar\rho_{33}$
[(g)--(i)], with the colors of Fig.~\ref{fig:R1}, on a logarithmic scale.
Solid lines: sequential order, $t_{12}<t_{23}$; dotted lines:
counterintuitive order, $t_{23}<t_{12}$ (the curves of Fig.~\ref{fig:R1}).
Both orders use the same two Gaussian envelopes at the same times, with the
legs exchanged; thin gray lines show the early (solid) and late (dashed) pulse
of the sequential order, each normalized to its peak. The panel titles give
the final values in the sequential (seq.) and counterintuitive (c.i.) orders.
AI-assisted plotting code, as stated in the caption of Fig.~\ref{fig:R1}.}
\label{fig:R1b}
\end{figure*}

Three controls identify the mechanism. First, the three cases isolate its
ingredients. Without dissipation the vertex keeps the population that the
optical pulse excites outside the dressing; the photon dissipator removes it
by emission, which is the loss channel of the conversion; an uncompensated
mean-field shift lowers $\eta$ sixteen-fold. Second, reversing the pulse order
suppresses the conversion (Fig.~\ref{fig:R1b}). In the sequential order,
$t_{12}<t_{23}$, the optical pulse comes first and excites the vertex before
the terahertz pulse acts. Without dissipation the terahertz pulse then drives
Rabi-type oscillations between $1s$ and $2p$ [Fig.~\ref{fig:R1b}(d), (g)], and
the final populations, $\bar\rho_{33}=5.2\times10^{-3}$ and
$\bar\rho_{22}=2.6\times10^{-3}$, are set by the pulse area rather than by
dark-state following. With the dissipator most of the vertex population is
emitted before the terahertz pulse can convert it: the sequential order gives
$\bar\rho_{33}(t_f)=2.1\times10^{-6}$, $\eta=7.3\times10^{-4}$ and a vertex
suppression of $162$, against $0.479$ and $0.096$ in the counterintuitive
order [Fig.~\ref{fig:R1b}(e), (h)]; in case~3 it gives $\eta=2.8\times10^{-3}$.
Over the operating window ($\hbar\Omega_{23}=3$--$5$~meV, $\sigma=1.1$--$1.9$~ps,
$\tau_d=(1.0$--$1.5)\,\sigma$) the sequential order gives $\eta\le0.027$. Third,
case~3 shows that the two-photon condition must be met at the renormalized
resonance. The density calibration this requires is quantified in
Sect.~\ref{subsec:Xi-robustness}.

\begin{figure*}
\includegraphics[width=\textwidth]{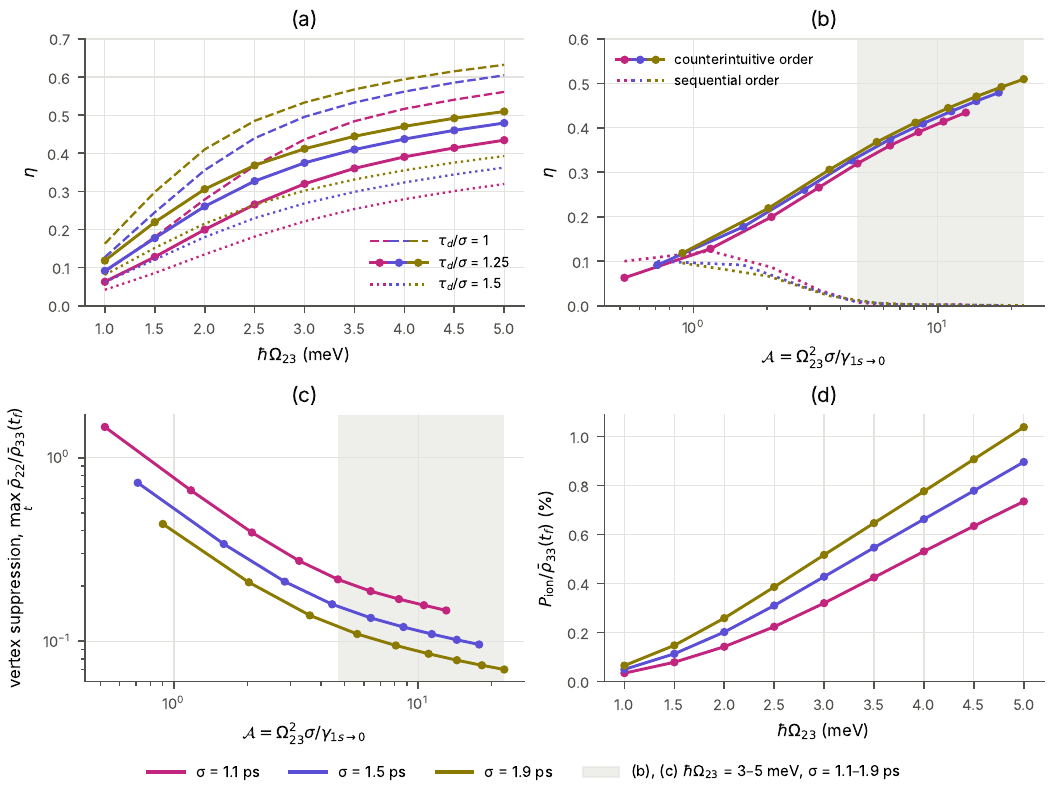}
\caption{Adiabaticity and loss channels of the $\Xi$ conversion
(counterintuitive Gaussian pulses, weak optical leg
$\hbar\Omega_{12}=0.05\min(\hbar\Omega_{23},\hbar\gamma^{(\gamma,{\rm inter})}_{1s\to0}/2)$,
$\ren\delta_\Xi=0$, $T_\gamma=4$~K). Colors give the pulse width $\sigma$
(legend below the figure); line styles and markers give the quantity (legend
in each panel, whose keys show the style in every color present in the panel).
(a)~Conversion efficiency $\eta$ [Eq.~\eqref{eq:conversion-efficiency}] for
$\hbar\Omega_{23}=1$--$5$~meV and delays $\tau_d/\sigma=1$ (dashed), $1.25$
(solid with dots) and $1.5$ (dotted).
(b)~$\eta$ at $\tau_d=1.25\,\sigma$ against $\mathcal A$
[Eq.~\eqref{eq:adiabaticity-A}], in the counterintuitive (solid) and
sequential (dotted) order.
(c)~Vertex suppression $\max_t\bar\rho_{22}/\bar\rho_{33}(t_f)$ against
$\mathcal A$, at $\tau_d=1.25\,\sigma$.
(d)~Ionization loss per converted exciton, $P_{\rm ion}/\bar\rho_{33}(t_f)$, at
$\tau_d=1.25\,\sigma$.
Shading in (b) and (c) marks the operating window $\hbar\Omega_{23}=3$--$5$~meV,
$\sigma=1.1$--$1.9$~ps, i.e.\ $\mathcal A=4.7$--$22.5$. AI-assisted plotting code, as stated in the caption of Fig.~\ref{fig:R1}.}
\label{fig:R3}
\end{figure*}

The conversion is controlled by the terahertz leg. To first order in
$\Omega_{12}$ the crystal ground state keeps unit amplitude, and the
amplitudes of $1s$ and $2p$ obey a two-level equation in which the vertex
carries the complex energy $-i\hbar\gamma^{(\gamma,{\rm inter})}_{1s\to0}/2$,
the terahertz leg couples it to $2p$, and the optical leg acts as a source.
This equation reproduces $\eta$ of the Lindblad runs to $2\times10^{-3}$ of
its value (App.~\ref{app:num_Lindblad}). In the weak-excitation regime used
throughout, the conversion is thus linear response: what the framework supplies
is not the form of this response but its inputs, namely the fixed-$\bm K_0$
radiative rates, the exchange shift of the two-photon resonance and the
conditions under which the reduction holds. The controlling parameter of the
amplitude equation is
\begin{equation}
\mathcal A=\frac{\Omega_{23}^2\,\sigma}{\gamma^{(\gamma,{\rm inter})}_{1s\to0}} ,
\label{eq:adiabaticity-A}
\end{equation}
the pulse width measured in units of the time
$\gamma^{(\gamma,{\rm inter})}_{1s\to0}/\Omega_{23}^2$ in which the dressed
vertex pumps population into the dark state, i.e.\ of the inverse width of the
cascade transparency window of Sect.~\ref{subsec:Xi-response}. On this
variable the efficiencies for the three pulse widths collapse onto a single
curve to within $0.03$ [Fig.~\ref{fig:R3}(b)], whereas at equal $\Omega_{23}$
they differ by up to $0.11$. $\mathcal A$ is the terahertz-leg analogue of the
parameter $a^2/g$ that controls stimulated Raman adiabatic passage through a
decaying intermediate state~\cite{Vitanov1997}. The efficiency rises
monotonically with $\mathcal A$, from $0.09$ at $\mathcal A=0.7$ to $0.48$ at
the baseline, $\mathcal A=17.8$ ($\tau_d=1.25\,\sigma$). Unlike STIRAP with two
strong pulses, it does not approach unity. With a weak optical leg the mixing
angle $\tan\theta=\Omega_{12}/\Omega_{23}$ stays small while the terahertz
pulse is strong, so the dark state carries the $2p$ amplitude
$\Omega_{12}/\Omega_{23}$ only until the terahertz dressing becomes too weak to
keep the vertex adiabatically eliminated. The $2p$ amplitude then freezes
[Fig.~\ref{fig:R1}(h)], and the rest of the optical pulse is absorbed by the
undressed vertex and re-emitted. This freeze-out is the writing step of EIT
light storage~\cite{Fleischhauer2000,Fleischhauer2005}, here into the $2p$
state of a single monolayer rather than into a collective excitation of an
optically dense medium. In such media the optimal storage efficiency depends
only on the optical depth and is reached by shaping the pulses~\cite{Gorshkov2007};
in a monolayer, whose coupling to light is fixed by the radiative width, the
corresponding limit is $Y\le1/2$. A larger $\mathcal A$ delays this freeze-out,
and a shorter delay increases the overlap of the optical pulse with the
dressing: $\tau_d=\sigma$ gives the highest efficiency at every point of the
grid, up to $\eta=0.63$ at $\sigma=1.9$~ps and $\hbar\Omega_{23}=5$~meV, and
$\tau_d=1.5\,\sigma$ the lowest [Fig.~\ref{fig:R3}(a)]. The pulse order matters
only once the dressing is effective: the counterintuitive order overtakes the
sequential one near $\mathcal A\simeq1$ and leads by more than an order of
magnitude for $\mathcal A\gtrsim4$ [Fig.~\ref{fig:R3}(b)].

The excitation that is not converted is lost to vertex emission, and the other
two figures of merit follow the same variable. The vertex suppression falls
with $\mathcal A$, to $0.04$--$0.37$ over the operating window
($\mathcal A=4.7$--$22.5$) [Fig.~\ref{fig:R3}(c)]: the vertex never holds more
than about a third of the population that is finally converted, and at the
baseline a tenth. Drive-induced ionization
(Sect.~\ref{sec:lindblad_three_level}, item~\ref{item7}) grows with
$\Omega_{23}$ and $\sigma$, as the rate of
Eq.~\eqref{eq:2p-ionization-num}\SMnote{} implies, and reaches
$P_{\rm ion}/\bar\rho_{33}(t_f)=0.9\%$ at the baseline and at most $1.6\%$ over
the grid [Fig.~\ref{fig:R3}(d)]. It is evaluated to first order and is not
contained in $\bar\rho_{33}(t_f)$. Per converted exciton the baseline thus
emits $1.09$ vertex photons and loses $0.009$ excitons to ionization, and
vertex emission dominates the loss everywhere in the window, by a factor of
$37$ or more. The longest delay minimizes ionization, because the $2p$
population then builds up while the terahertz pulse, which comes first, is
already decaying ($0.7\%$ against $1.6\%$ at $\sigma=1.9$~ps and $5$~meV), but
it lowers the efficiency (from $0.63$ to $0.39$), so the choice of delay is
governed by the emission.

\subsection{Robustness and density calibration}
\label{subsec:Xi-robustness}

\begin{figure*}
\includegraphics[width=\textwidth]{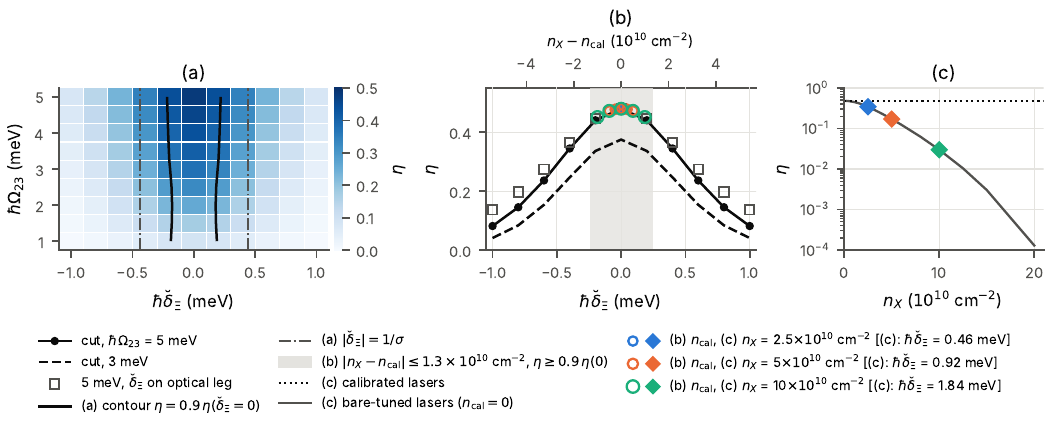}
\caption{Two-photon detuning and density calibration of the $\Xi$ conversion
(counterintuitive Gaussian pulses, $\sigma=1.5$~ps, $\tau_d=1.25\,\sigma$,
weak optical leg, $T_\gamma=4$~K). (a)~Conversion efficiency $\eta$ against
the renormalized two-photon detuning, carried by the terahertz leg
($\ren\Delta_{12}=0$), and the terahertz Rabi energy. Solid line: contour
$\eta=0.9\,\eta(\ren\delta_\Xi=0)$; dash-dotted lines: $|\ren\delta_\Xi|=1/\sigma$.
(b)~Cuts of (a) at $\hbar\Omega_{23}=5$~meV (solid line with dots) and 3~meV
(dashed); open squares: 5~meV with the detuning carried by the optical leg
($\ren\Delta_{23}=0$). Open circles: lasers calibrated at
$n_{\rm cal}=2.5\times10^{10}$, $5\times10^{10}$ and $10^{11}$~cm$^{-2}$
(color and size), with the gas at $n_X=n_{\rm cal}(1\pm0.05)$ and
$n_{\rm cal}(1\pm0.10)$; runs with the same density error $n_X-n_{\rm cal}$
coincide. Top axis: the density error that produces $\ren\delta_\Xi$,
Eqs.~\eqref{eq:gXX}, \eqref{eq:VE-renormalized-detunings} and~\eqref{eq:VE-two-photon-detunings}. Shading:
density errors $|n_X-n_{\rm cal}|\le1.3\times10^{10}$~cm$^{-2}$, within which
$\eta\ge0.9\,\eta(0)$. (c)~Lasers at the bare resonances ($n_{\rm cal}=0$):
$\eta$ against the gas density $n_X$ (solid line), logarithmic scale;
diamonds: $n_X=2.5\times10^{10}$, $5\times10^{10}$ and $10^{11}$~cm$^{-2}$
(colors as in (b)), at which the mean field produces
$\hbar\ren\delta_\Xi=0.46$, $0.92$ and $1.84$~meV. Dotted line: calibrated
lasers, $\eta=0.479$ at every $n_X$. AI-assisted plotting code, as stated in the caption of Fig.~\ref{fig:R1}.}
\label{fig:R2}
\end{figure*}

The conversion tolerates a two-photon detuning of about half the pulse
bandwidth, nearly independently of the drive [Fig.~\ref{fig:R2}(a)].
Figure~\ref{fig:R2} scans the renormalized two-photon detuning over
$|\hbar\ren\delta_\Xi|\le1$~meV at $\sigma=1.5$~ps and
$\tau_d=1.25\,\sigma$, with the detuning carried by the terahertz leg
($\ren\Delta_{12}=0$). The map is symmetric in $\ren\delta_\Xi$.
At $\hbar\Omega_{23}=5$~meV the efficiency falls from $0.479$ at resonance to
$0.32$ at $|\ren\delta_\Xi|=1/\sigma$ ($\hbar/\sigma=0.44$~meV), a relative loss
of $32\%$, and to $0.082$ at $1$~meV [Fig.~\ref{fig:R2}(b)]. The additional
loss is again vertex emission: $N_{\rm emit}/\bar\rho_{33}(t_f)$ grows from
$1.09$ to $11.2$ over the same range. The half-width at which $\eta$ has
dropped by $10\%$ of its resonant value is $0.18$--$0.22$~meV for
$\hbar\Omega_{23}=1$--$5$~meV. In STIRAP with two strong pulses a finite
detuning couples the dark state to bright states that lie at
$\pm\hbar\Omega_0/2$, and a stronger drive widens the tolerance. With a weak
optical leg the $2p$ amplitude is instead accumulated from the optical source
during the conversion, over a time of order $\sigma$, and a two-photon
detuning dephases it against that source, so the tolerance is set mainly by
the pulse duration. The efficiency depends on how the two-photon detuning is
split between the legs, not on $\ren\delta_\Xi$ alone: with the detuning
carried by the optical leg instead, it is up to $0.055$ higher
[Fig.~\ref{fig:R2}(b), open squares].

The detuning tolerance sets a requirement on the exciton density, and the
requirement concerns the absolute density error, not the relative one. If the
carrier frequencies are calibrated at a density $n_{\rm cal}$ while the gas
has density $n_X$, Eqs.~\eqref{eq:gXX}, \eqref{eq:VE-renormalized-detunings} and~\eqref{eq:VE-two-photon-detunings} give
renormalized detunings proportional to $n_X-n_{\rm cal}$, in particular
$\hbar\ren\delta_\Xi=\Sigma^{(Xg,0)}_{2p}(n_X)-\Sigma^{(Xg,0)}_{2p}(n_{\rm cal})
=+1.84~{\rm meV}\times(n_X-n_{\rm cal})/10^{11}~{\rm cm}^{-2}$, of which a
fraction $1.15$ sits on the optical leg and $-0.15$ on the terahertz leg.
Calibration errors with the same $n_X-n_{\rm cal}$ therefore give the same
conversion: a $10\%$ error at $n_{\rm cal}=5\times10^{10}$~cm$^{-2}$ and a $5\%$
error at $10^{11}$~cm$^{-2}$ both give $\eta=0.472$ [Fig.~\ref{fig:R2}(b)].
All miscalibrated runs agree with a scan of the density error at the same
split of the detuning to $5\times10^{-4}$, and lie up to $0.011$ above the cut
of the map because most of the detuning sits on the optical leg. The
efficiency stays within $10\%$ of its resonant value for
$|n_X-n_{\rm cal}|\lesssim1.3\times10^{10}$~cm$^{-2}$, i.e.\
$|\hbar\ren\delta_\Xi|\lesssim0.25$~meV$\,\simeq0.56\,\hbar/\sigma$, and it is
halved at $|n_X-n_{\rm cal}|\simeq3.9\times10^{10}$~cm$^{-2}$. The tolerance is
$13\%$ of the density at $n_{\rm cal}=10^{11}$~cm$^{-2}$, $27\%$ at
$5\times10^{10}$~cm$^{-2}$ and $53\%$ at $2.5\times10^{10}$~cm$^{-2}$, so a
$10\%$ calibration error, which lowers $\eta$ by at most $0.028$ (to $0.451$ at
$10^{11}$~cm$^{-2}$), stays inside it at every density considered. Without
calibration, with the lasers at the bare resonances, the efficiency falls to
$0.342$, $0.170$ and $0.030$ at $n_X=2.5\times10^{10}$, $5\times10^{10}$ and
$10^{11}$~cm$^{-2}$, i.e.\ to $0.71$, $0.35$ and $0.062$ of the calibrated
value [Fig.~\ref{fig:R2}(c)], since $|\hbar\ren\delta_\Xi|=0.46$, $0.92$ and
$1.84$~meV is then $1.0$, $2.1$ and $4.2$ times $\hbar/\sigma$. The two-photon
resonance must therefore be set at the density of the exciton gas present
during the sequence, as anticipated in Sect.~\ref{sec:VE-renormalization}, once
$n_X\gtrsim1.3\times10^{10}$~cm$^{-2}$, where bare-tuned lasers fall below
$0.9$ of the calibrated efficiency; the calibration then needs an accuracy of
about $1.3\times10^{10}$~cm$^{-2}$. Section~\ref{subsec:Xi-response}
shows that the center of the cascade transparency window measures
$\ren\delta_\Xi$ directly and can serve as this calibration.

\subsection{Cascade response: EIT versus Autler--Townes splitting}
\label{subsec:Xi-response}

\begin{figure*}
\includegraphics[width=\textwidth]{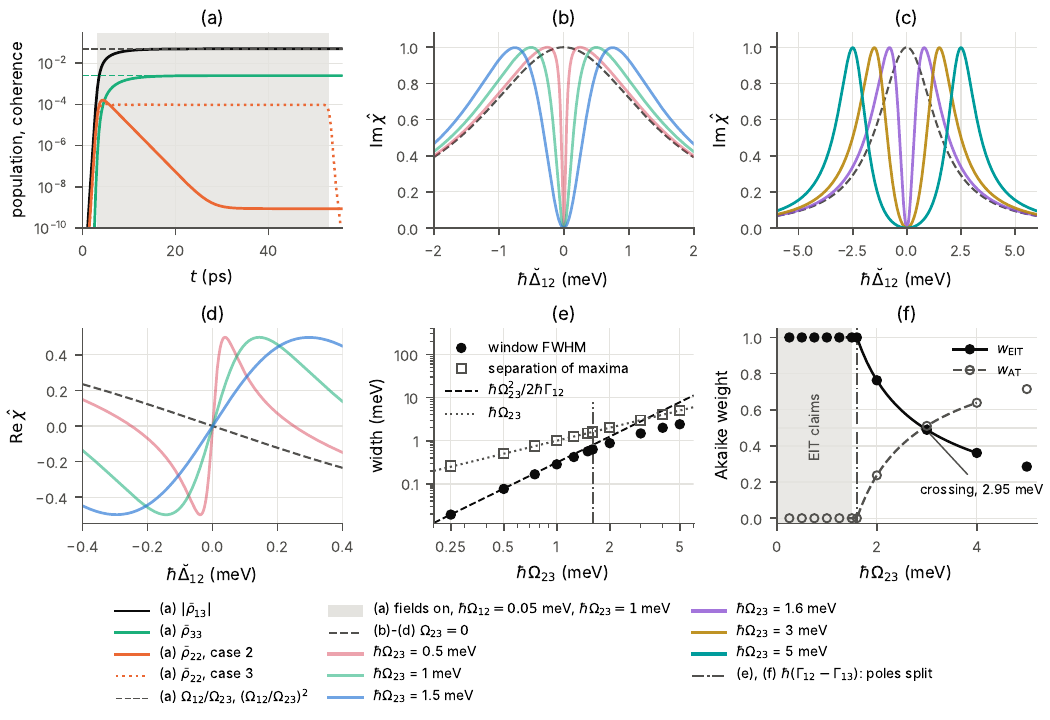}
\caption{Cascade response of the $\Xi$ system ($T_\gamma=4$~K, photon
dissipator, renormalized detunings $\ren\Delta_{23}=0$).
(a)~Coherent population trapping on a $50$~ps plateau,
$\hbar\Omega_{12}=0.05$~meV and $\hbar\Omega_{23}=1$~meV, both legs resonant,
logarithmic scale: outer coherence $|\bar\rho_{13}|$ (black), $2p$ population
$\bar\rho_{33}$ (green) and vertex population $\bar\rho_{22}$ (orange), with
$\bar\rho_{22}$ of case~3 (orange, dotted); dashed lines:
$\Omega_{12}/\Omega_{23}$ and $(\Omega_{12}/\Omega_{23})^2$; shading: fields on. (b)~Absorption ${\rm Im}\,\hat\chi$ against the probe
detuning for $\hbar\Omega_{23}=0.5$, $1$ and $1.5$~meV (interference regime)
and (c)~for $1.6$, $3$ and $5$~meV (crossover and Autler--Townes regime);
dashed: $\Omega_{23}=0$. (d)~Dispersion ${\rm Re}\,\hat\chi$ near the window
center for the spectra of (b). Weak probe,
$\hbar\Omega_{12}=0.05\min(\hbar\Omega_{23},\hbar\Gamma_{12})$; color encodes
$\hbar\Omega_{23}$ in (b)--(d). (e)~Window width (FWHM of the absorption
hole) and separation of the absorption maxima against $\hbar\Omega_{23}$,
with $\Omega_{23}^2/2\Gamma_{12}$ (dashed) and $\Omega_{23}$ (dotted).
(f)~Per-point Akaike weights of the interference (EIT) and doublet (AT) line
shapes of Ref.~\cite{Anisimov2011}; dots: control strengths of (b)--(e);
lines: $0.1$~meV steps. Shading: control strengths at which transparency is
claimed. Dash-dotted line in (e) and (f): $\hbar(\Gamma_{12}-\Gamma_{13})=1.603$~meV,
above which the poles of the response split. AI-assisted plotting code, as stated in the caption of Fig.~\ref{fig:R1}.}
\label{fig:R4}
\end{figure*}

Under a sustained drive the dark state of Eq.~\eqref{eq:xi-dark-state} is reached
by dissipation rather than by adiabatic following. Figure~\ref{fig:R4}(a)
shows coherent population trapping on the plateau of Table~\ref{tab:protocols}:
both legs resonant, $\hbar\Omega_{12}=0.05$~meV and $\hbar\Omega_{23}=1$~meV,
switched simultaneously by the envelope of Eq.~\eqref{eq:envelope_plateau} for
$50$~ps. The small bright component of the ground state is driven into the
$1s$ vertex and returned by emission until the population has been pumped into
$|D\rangle$: $\bar\rho_{22}$ falls to $1\%$ of its peak $10.5$~ps after
switch-on, and at the end of the plateau the population outside $|D\rangle$ is
$1.8\times10^{-9}$, so the photon dissipator prepares a pure state. The
simultaneous switch-off keeps the mixing angle, $\tan\theta=\Omega_{12}/\Omega_{23}$,
and leaves the coherent superposition $\cos\theta\,|1\rangle-\sin\theta\,|3\rangle$
(run details in App.~\ref{app:num_Lindblad}\SMnote). Without the
dissipator (case~1) the bright component, $\sin^2\theta=2.5\times10^{-3}$, is
not removed and stays outside $|D\rangle$. With the mean field and
bare-tuned lasers (case~3) the two-photon detuning prevents the pumping into
the dark state: the vertex stays populated, emission continues
($N_{\rm emit}=0.023$), and only $\bar\rho_{33}=4.8\times10^{-6}$ reaches $2p$.

The weak-probe spectrum shows the same dark state as a transparency window
[Fig.~\ref{fig:R4}(b)--(d)]. We scan the probe detuning
$\ren\Delta_{12}$ at $\ren\Delta_{23}=0$ with
$\hbar\Omega_{12}=0.05\min(\hbar\Omega_{23},\hbar\Gamma_{12})$ and take the
stationary state from the null space of the generator. The response is
expressed by the normalized susceptibility
$\hat\chi=2\Gamma_{12}\bar\rho_{12}/\Omega_{12}$, whose imaginary part is
the absorption with the two-level peak normalized to one; here
$\Gamma_{12}=1.60$~meV$/\hbar$ and
$\Gamma_{13}=5.6\times10^{-7}$~meV$/\hbar$ are the damping rates of
$\bar\rho_{12}$ and $\bar\rho_{13}$ in Eqs.~\eqref{eq:Xi-component-system}\SMnote{}.
The spectra have the standard form of the driven
ladder~\cite{GeaBanacloche1995,Fleischhauer2005}. Because the outer coherence
is almost undamped, the transparency is complete, with an absorption of at
most $6\times10^{-5}$ at the window center over the scan; the window has the
width $\Omega_{23}^2/2\Gamma_{12}$ for weak control and narrows below it as the
control grows, and the dispersion at the center is steep and of opposite sign
to the two-level one (App.~\ref{app:num_Lindblad}\SMnote).

A transparency window alone does not identify EIT: for every control strength
the absorption maxima lie at $\pm\hbar\Omega_{23}/2$
[Fig.~\ref{fig:R4}(e)]. The regime is fixed instead by the poles of the
response~\cite{AbiSalloum2010}. For $\Omega_{23}<\Gamma_{12}-\Gamma_{13}=1.603$~meV$/\hbar$ they lie
at the same detuning and the spectrum is exactly a broad Lorentzian minus a
narrow one, the interference form; above, they split and the spectrum
becomes an Autler--Townes doublet. The
model-selection test of Anisimov \emph{et al.}~\cite{Anisimov2011} agrees
below the threshold, where the interference model is selected with Akaike
weight $1-10^{-12}$, but places the change of preference higher: the weight
of the doublet model passes $0.5$ only at $\hbar\Omega_{23}=2.95$~meV and is
$0.71$ at $5$~meV [Fig.~\ref{fig:R4}(f)]. Between $1.6$ and $2.95$~meV the
spectrum is a crossover with split poles but an interference-like line
shape~\cite{AbiSalloum2010}. We therefore call the dip electromagnetically
induced transparency only for $\hbar\Omega_{23}\le1.5$~meV, as anticipated
in Sect.~\ref{subsec:explicit_drives}, which is below the control strength the
conversion of Sect.~\ref{subsec:Xi-transfer} uses.

The window center measures the renormalized two-photon detuning and can
therefore serve as the density calibration required in
Sect.~\ref{subsec:Xi-robustness}. With the lasers calibrated at $n_{\rm cal}$
and the gas at $n_X$, the window appears displaced from the calibrated probe
resonance by $-\ren\delta_\Xi$, i.e. at the probe detuning that restores
two-photon resonance. For all eighteen calibration and miscalibration settings
of Sect.~\ref{subsec:Xi-robustness}, including bare-tuned lasers with
$|\hbar\ren\delta_\Xi|$ up to $1.84$~meV, the computed center agrees
with $-\ren\delta_\Xi$ to $2\times10^{-4}$~meV at
$\hbar\Omega_{23}=0.5$ and $1$~meV. One window width corresponds to a density
error of $4\times10^{9}$~cm$^{-2}$ at $0.5$~meV, a third of the tolerance of
$1.3\times10^{10}$~cm$^{-2}$ that the conversion requires, and of
$1.5\times10^{10}$~cm$^{-2}$ at $1$~meV, comparable to it; since the center of
a complete window can be located to a small fraction of its width, both
control strengths can provide the calibration. The narrow window forms slowly, however: at $0.5$~meV the population is
pumped into the dark state in about $17$~ps. After a $50$~ps plateau the absorption at the center
is still $0.07$ instead of $1.4\times10^{-5}$, whereas at $1.5$ and $5$~meV
the plateau and the null space agree to $10^{-9}$. A calibration at
$\hbar\Omega_{23}\simeq1$~meV, where pumping takes $4$~ps, avoids this
limitation at the cost of a window four times wider.

\subsection{The $V$ system as the unprotected comparison}
\label{subsec:V-comparison}

\begin{figure*}
\includegraphics[width=\textwidth]{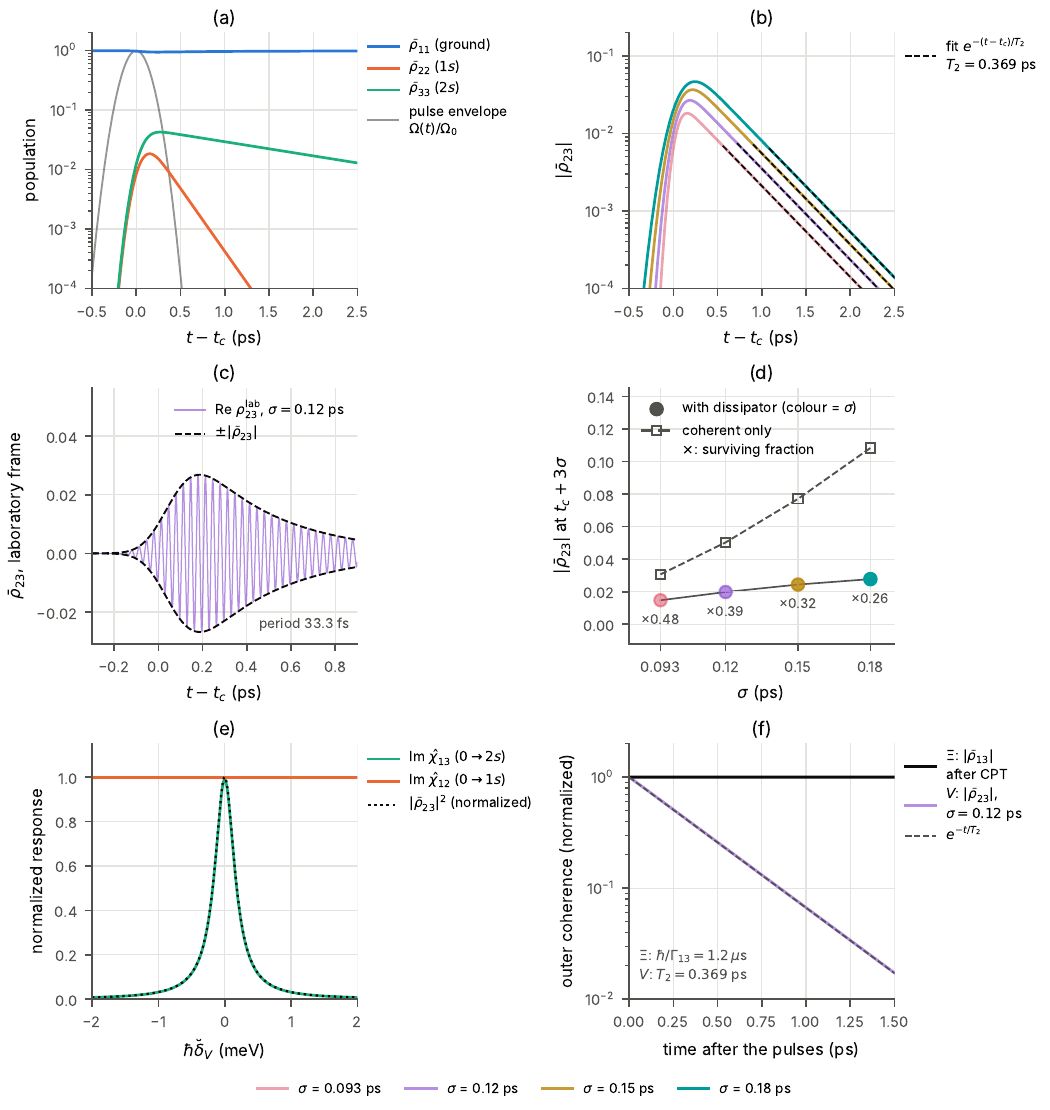}
\caption{The $V$ system ($|0\rangle$, $1s$, $2s$; $T_\gamma=4$~K, photon
dissipator, both legs resonant). (a)~Populations after a synchronous Gaussian
pair, $\sigma=0.12$~ps, $\hbar\Omega_{12}=\hbar\Omega_{13}=1$~meV, logarithmic
scale: ground state (blue), $1s$ (orange), $2s$ (green); gray: pulse envelope.
(b)~Excited-state coherence $|\bar\rho_{23}|$ for $\sigma=0.093$, $0.12$,
$0.15$ and $0.18$~ps (red, purple, ochre, teal; opacity increasing with $\sigma$); dashed:
exponential fits with $T_2=0.369$~ps. (c)~Laboratory-frame coherence
${\rm Re}\,\rho^{\rm lab}_{23}$ at $\sigma=0.12$~ps and its envelope
$\pm|\bar\rho_{23}|$ (dashed); period $33.3$~fs. (d)~$|\bar\rho_{23}|$ at the
end of the pulses, $t_c+3\sigma$, with (dots, colors of (b)) and without (open squares) the
dissipator; numbers: surviving fraction. (e)~Stationary response under weak
plateau drive ($\hbar\Omega_{12}=0.1$~meV, $\hbar\Omega_{13}=0.01$~meV)
against the two-photon detuning: absorption of the $0\to2s$ arm,
${\rm Im}\,\hat\chi_{13}$ (green), and of the $0\to1s$ arm,
${\rm Im}\,\hat\chi_{12}$ (orange), each normalized to its two-level peak, and
$|\bar\rho_{23}|^2$ normalized to its maximum (dotted). (f)~Outer coherence
after the pulses, normalized to its value at the end of the pulses: $\Xi$,
$|\bar\rho_{13}|$ after the coherent-population-trapping plateau of
Fig.~\ref{fig:R4}(a) (black); $V$, $|\bar\rho_{23}|$ at $\sigma=0.12$~ps
(purple); dashed: $e^{-t/T_2}$. AI-assisted plotting code, as stated in the caption of Fig.~\ref{fig:R1}.}
\label{fig:R5}
\end{figure*}

In the $V$ system both excited states radiate, so the coherence between them
is lost on the radiative time scale, and the observable is a quantum beat
rather than a transparency window. Figure~\ref{fig:R5}(a)--(c) shows the
dynamics prepared by the synchronous Gaussian pair of Table~\ref{tab:protocols}
($\sigma=0.12$~ps, $\hbar\Omega_{12}=\hbar\Omega_{13}=1$~meV, both legs
resonant). The pulses transfer $4\%$ of the population to $2s$ and $2\%$ to
$1s$, which then decay at their own radiative rates [Fig.~\ref{fig:R5}(a)]. The
coherence $|\bar\rho_{23}|$ decays exponentially with $T_2=0.369$~ps for every
pulse width [Fig.~\ref{fig:R5}(b)], the inverse of
$\Gamma^{V}_{23}=\tfrac12(\gamma^{(\gamma)}_{2\to1}+\gamma^{(\gamma)}_{3\to1})=1.783$~meV$/\hbar$
of Eq.~\eqref{eq:V-component-system}\SMnote{}, and in the laboratory frame it
oscillates at the $1s$--$2s$ splitting with a period of $33.3$~fs
[Fig.~\ref{fig:R5}(c)]. Without the dissipator (case~1) the pulses leave a pure
state that does not decay. The exchange mean field at $n_X=10^{11}$~cm$^{-2}$
with bare-tuned lasers (case~3) lowers $|\bar\rho_{23}|$ at the end of the
pulses by $16\%$, but leaves $T_2$ unchanged, since the mean field shifts the
levels and does not enter the damping.

The damping limits how much coherence a pulse pair can prepare, which is the
origin of the narrow pulse-width window of Sect.~\ref{subsec:explicit_drives}.
At the end of the pulses, $t_c+3\sigma$, the coherence grows with $\sigma$
from $0.015$ to $0.028$, but the fraction that survives the excitation,
relative to the coherent-only pulses, falls from $0.48$ at $\sigma=0.093$~ps
to $0.26$ at $0.18$~ps [Fig.~\ref{fig:R5}(d)]. The window
$0.093\lesssim\sigma\lesssim0.18$~ps is thus not a technical restriction but
the time-domain expression of the missing radiative protection: a longer pulse
spends a larger part of the excitation inside $T_2$. These pulses create the
largest excitation of all runs, $\max_t(1-\bar\rho_{11})\le0.11$, still within
the one-exciton description (Sect.~\ref{subsec:operating-point}) to about
$10\%$ at $1$~meV and $2\%$ at $0.5$~meV (App.~\ref{app:num_Lindblad}\SMnote).

The stationary response shows no interference. Under a weak plateau drive
($\hbar\Omega_{12}=0.1$~meV, $\hbar\Omega_{13}=0.01$~meV) the absorption of the
$0\to2s$ arm, ${\rm Im}\,\hat\chi_{13}$, is a single Lorentzian of half-width
$\Gamma^{V}_{13}=0.18$~meV$/\hbar$, and that of the resonant $0\to1s$ arm is
flat [Fig.~\ref{fig:R5}(e)]. The two arms are
$124$~meV apart, so no transparency window can form between them, as
anticipated in Sect.~\ref{subsub:V}; the information on the
$V$ coherence is carried by the beat. The $V$ results also carry the
density-dependent uncertainty of the neglected off-diagonal exchange element
(Sect.~\ref{sec:VE-renormalization} and App.~\ref{app:offdiag}): at
$10^{11}$~cm$^{-2}$ it moves the response width $\Gamma^{V}_{13}$ by up to $7\%$,
the beat amplitude by about $4\%$ and $T_2$ by at most $1.4\%$, linearly in
$n_X$.

The comparison with the ladder is direct [Fig.~\ref{fig:R5}(f)]. After the
coherent-population-trapping plateau of Sect.~\ref{subsec:Xi-response}, the
$\Xi$ outer coherence $|\bar\rho_{13}|$ changes by $1.3\times10^{-6}$ in
$1.5$~ps, the decay at
$\Gamma^{\Xi}_{13}=5.6\times10^{-7}$~meV$/\hbar$ of
Eq.~\eqref{eq:Xi-component-system}\SMnote{}, i.e. a lifetime of $1.2~\mu$s. In the same time
the $V$ coherence falls to $1.7\%$ of its initial value. The ratio of the two
outer-coherence damping rates is $3.2\times10^{6}$. It measures the radiative
protection of the dipole-dark $2p$ state, whose only retained exit is the
intra-excitonic $2p\to1s$ channel suppressed as $A^{-1}$; what this lifetime
means for a real monolayer is discussed in Sect.~\ref{subsec:scope}.

\subsection{Scope of the results}
\label{subsec:scope}

The results above rest on a dissipator that is radiative only. The reduced
model contains no channel beyond those generated by the fixed-$\mathbf K_0$
projection of its reservoirs, which leaves the photon field as the only
dissipative one, and in particular no pure dephasing of any coherence. The contributions it leaves out, each of which can only add
damping, are:
\begin{enumerate}[label=(\roman*)]
\item pure dephasing of the coherences by exciton--phonon scattering, which
requires a finite momentum transfer and vanishes identically at fixed
$\mathbf K_0$ (Sect.~\ref{sec:lindblad_three_level}, item~\ref{item3};
App.~\ref{app:K0-projection})~\cite{selig2016excitonic};
\item phonon emission from the bright states into the momentum- and
spin-dark K--$\Lambda$ and K--K$'$ excitons, which in WSe$_2$ lie below the
bright $1s$ exciton~\cite{Berghauser2018,Brem_intrinsic} and give it a
non-radiative width even at $T\to0$, computed to be $3.9$~meV~\cite{Brem_intrinsic};
\item the relaxation cascade of the excited states into lower-lying
ones~\cite{brem2018exciton}, which adds exits to the $A^{-1}$-suppressed
$2p\to1s$ radiative channel retained here;
\item excitation-induced dephasing and collision broadening by the exciton
gas, which enters here only as a stationary mean field of the bright
intravalley $1s$ population
(Sect.~\ref{sec:lindblad_three_level}, item~\ref{item4})~\cite{katsch2018theory,Katsch2020};
\item scattering by residual free carriers, which the model does not contain
and which Ref.~\cite{Kusaba2021} proposes as the origin of the measured $2p$ width;
\item dielectric and strain disorder, i.e.\ inhomogeneous broadening of the
two-photon resonance: across an hBN-encapsulated WSe$_2$ flake the $1s$
energy varies by about $25$~meV, while on areas of a few micrometers it
fluctuates by an order of magnitude less than the homogeneous
width~\cite{Boule2020}, about $0.3$~meV. This is already the two-photon
tolerance $|\hbar\ren\delta_\Xi|\lesssim0.25$~meV of
Sect.~\ref{subsec:Xi-robustness}, and it is a lower bound for the relevant
scale: the two-photon resonance is set by the $2p$ energy, and under
dielectric disorder the shifts of the $n=2$ excitons are correlated with
those of $n=1$ and larger, by a factor of $2.2$ for monolayers on
hBN~\cite{Raja2019};
\item a finite in-plane photon phase space, a finite beam waist and a
distribution of center-of-mass momenta, which restore recoil and the phonon
channels and modify, though not remove, the $A^{-1}$-suppressed rates.
\end{enumerate}
The measured widths of hBN-encapsulated WSe$_2$ give the scale of these
contributions. The homogeneous width of $1s$ extrapolates to $2.96$~meV (full
width at half maximum) at $T\to0$~\cite{Boule2020}, comparable to the radiative
width of $3.21$~meV used here; the $2s$ line is narrower, $1.7$~meV in
reflectance at $5$~K~\cite{Boule2020} and $2.45$~meV extrapolated to
$T\to0$~\cite{Kusaba2021}; and the dark $2p$ state, observed by sum-frequency
generation, has a width of $7.9$~meV at $T\to0$~\cite{Kusaba2021}, against
$\hbar\gamma^{(\gamma)}_{3\to2}=1.1$~neV in the model. The conversion
efficiencies, coherence fractions and coherence times reported here are
therefore upper bounds; the conditions that a real sample must meet are
given in Sect.~\ref{subsec:real-sample}. Lifting the fixed-$\mathbf K_0$ restriction,
which restores contributions (i)--(iv) and (vii), is the subject of separate
work.
The lifetime $\hbar/\Gamma^{\Xi}_{13}=1.2~\mu$s of the $\Xi$ outer coherence
is the $A^{-1}$-suppressed $2p\to1s$ radiative rate at $A=9.62~\mu$m$^2$, not a
prediction of the $2p$ coherence time of a real monolayer; with in-plane
recoil allowed, the free-space rate of the same dipole would give
$3.5~\mu$s (App.~\ref{app:photon-reservoir}), an artifact of
the fixed-$\mathbf K_0$ limit. The protection
established for $\Xi$ is protection against the lossy $1s$ vertex, not against
every channel. The $V$ results are much less exposed: their coherence is
already limited by radiation, at $\hbar\Gamma^{V}_{23}=1.78$~meV, and a pure
dephasing small compared with this rate changes them little.

The remaining limitations are those of the reduced model: the $4$~K baseline,
the $2p\leftrightarrow2s$ and $2s\to2p$ channels omitted by the three-level
truncation, and the first-order treatment of the ionization of $2p$
(Sect.~\ref{sec:lindblad_three_level}, items~\ref{item6} and~\ref{item7}).

\subsection{Requirements for a real monolayer}
\label{subsec:real-sample}

The protocols of Sects.~\ref{subsec:Xi-transfer}--\ref{subsec:Xi-response}
rely on a $2p$ state whose only exit is the $A^{-1}$-suppressed radiative
channel. The measured width of $2p$, $7.9$~meV at
$T\to0$~\cite{Kusaba2021}, implies pure dephasing of the $0$--$2p$ coherence
at a rate $\gamma_\phi$, population decay of $2p$ at a rate
$\gamma_{\rm rel}$, or both, with
$\hbar\gamma_{\rm rel}/2+\hbar\gamma_\phi=3.96$~meV if the width is
homogeneous. Sensitivity analyses outside the fixed-$\mathbf K_0$ model, in
which these terms are added to the $\Xi$ generator by hand
(App.~\ref{app:dephasing}\SMnote), give the conditions that the transparency
window and the conversion impose.

Within the fixed-$\mathbf K_0$ limit, the model predicts how deep the cascade
window is at a given terahertz Rabi energy, through its own $0$--$2p$ decay
rate $\Gamma^{\Xi}_{13}$. From the weak-probe susceptibility of
Sect.~\ref{subsec:Xi-response}, the residual absorption at the window center
is $\Gamma_{12}\Gamma^{\Xi}_{13}/(\Gamma_{12}\Gamma^{\Xi}_{13}+\Omega_{23}^2/4)$.
Because $\Gamma^{\Xi}_{13}$ is radiative only, this is $3.6\times10^{-6}$ at
$\hbar\Omega_{23}=1$~meV: the predicted window is essentially complete. A pure
dephasing replaces $\Gamma^{\Xi}_{13}$ by $\Gamma^{\Xi}_{13}+\gamma_\phi$, so
that the window depth is set by $\gamma_\phi$: for
$\hbar\gamma_\phi=3.96$~meV the residual absorption would be $0.96$ at
$\hbar\Omega_{23}=1$~meV and $0.50$ at $5$~meV. For terahertz Rabi energies of
$1$--$5$~meV and pulse widths of $1.1$--$1.9$~ps, the conversion efficiency
falls to half its value when the dephasing time $1/\gamma_\phi$ is about the
pulse width $\sigma$, while the $0$--$2p$ coherence of the converted
population is lost about eight times earlier [Fig.~\ref{fig:dephasing}(a)
and (b)]. Conversely, since the model predicts a complete window, a shallower
measured window at a known $\Omega_{23}$ quantifies the $0$--$2p$ dephasing
that the model does not contain, a quantity not measured so far.

Population decay of $2p$ is more damaging than pure dephasing. When a quarter
or more of the measured width is population decay, no $2p$ population
survives the pulses ($\eta<10^{-12}$). If $2p$ relaxes into non-radiating
states, the dark-sector fraction, i.e.\ the share of the excitation that ends
in $2p$ or in those states rather than being re-emitted
(App.~\ref{app:dephasing}\SMnote), is at most $0.14$ at
$\hbar\Omega_{23}=5$~meV ($0.05$ at $2.5$~meV and $0.009$ at $1$~meV), against
$\eta=0.48$ without the width. The optical-first order gives more ($0.21$),
so this is incoherent relaxation, not dark-state conversion; if $2p$ relaxes
back to $1s$, the excitation is re-emitted. Below the measured width, population decay acts twice. After the transfer,
the converted population decays freely: measured a time $k\sigma$ after the
optical pulse, it halves at $\gamma_{\rm rel}\sigma\simeq\ln2/k$
[Fig.~\ref{fig:dephasing}(c) for $k=6$], so it must outlive the time it is
used. During the transfer, at $\hbar\Omega_{23}=5$~meV and $\sigma=1.5$~ps,
the pulse-order contrast of the dark-sector fraction falls from $655$ without
the width to $2$ at $\gamma_{\rm rel}\sigma\simeq1$, where the dark-sector
fraction still keeps $83\%$ of its value, but less and less of it by
dark-state transfer.

For the conversion to keep half of its value and its dark-state character at
$\sigma=1.5$~ps, the conditions are therefore $\hbar\gamma_\phi\lesssim0.44$~meV
and $\hbar\gamma_{\rm rel}\lesssim0.47$~meV, both close to $\hbar/\sigma$,
i.e.\ a homogeneous $2p$ width (full width at half maximum) of at most about
$0.9$~meV for pure dephasing or $0.5$~meV for population decay. The measured
width bounds the homogeneous width from above, and meeting these conditions
would require $89$--$94\%$ of it to be inhomogeneous. In comparable samples,
however, the $1s$ line is close to its homogeneous limit ($3.12$~meV
extrapolated to $T\to0$~\cite{Kusaba2021}, against a homogeneous width of
$2.96$~meV~\cite{Boule2020}), and the $2s$ population decays in $0.75$~ps,
dominated by non-radiative processes~\cite{Boule2020}. Present samples
therefore miss the conditions. Reference~\cite{Kusaba2021} attributes the
excess of the $2p$ width over the $1s$ and $2s$ widths to residual carriers
[item~(v) of Sect.~\ref{subsec:scope}] and infers a similar phonon-emission
rate; in gated, charge-neutral samples the $2p$ width would then approach the
$1s$ and $2s$ values, about $2.5$--$3$~meV, still well above the conditions.

Two measurements decide whether a sample meets them: the depth of the cascade
window at a known $\Omega_{23}$, which measures the $0$--$2p$ dephasing, and
the pulse-order contrast of the dark-sector fraction, which at
$\hbar\Omega_{23}=5$~meV is $14$ under pure dephasing with
$\hbar\gamma_\phi=3.96$~meV but $0.6$--$0.7$, below one, when a quarter or
more of the width is population decay.

\section{Conclusions}
\label{sec:conclusions}

We have developed a configuration-resolved framework for dark-state coherent control of excitons in two-dimensional semiconductors and applied it to hBN-encapsulated WSe$_2$. Its first ingredient is a systematic screening of the exciton spectrum that ranks candidate three-level systems by interband brightness, intra-excitonic dipole strength and state compactness, and so makes the choice of three-level systems explicit and reproducible. The dielectric screening of a TMD monolayer produces a strongly non-hydrogenic Rydberg series, in which optically bright $s$-like and interband-dark $p$-like states are interleaved and intra-excitonic splittings fall in the terahertz and mid-infrared range. The screening therefore classifies three-level systems by their drive architecture, that is, by which legs are interband optical and which are intra-excitonic, rather than by the energy ordering familiar from atomic quantum optics. For WSe$_2$ it selects a $\Xi$ system, in which an optical leg addresses the brightest exciton and a terahertz leg connects it to a compact, optically dark state, together with a bright--bright $V$ system as the natural comparison.

The selected three-level systems are then described by a Lindblad equation, derived within the Born--Markov and full-secular approximations from a model in which a single exciton couples to acoustic phonons, the quantized radiation field and a dilute exciton gas. The central assumption is the restriction to a fixed exciton center-of-mass momentum $\mathbf{K}_0$. Its consequences are structural rather than numerical. Momentum conservation forbids any momentum transfer to the lattice, so the deformation-potential coupling to acoustic phonons vanishes identically and produces neither dephasing nor a Lamb shift. The exciton gas enters only through a Hermitian exchange mean field, which renormalizes the exciton energies and the control detunings linearly in the density without introducing dissipation. The photon reservoir reduces to a one-dimensional continuum; this yields an interband radiative rate fixed by the Wannier envelope and the band-edge dipole, which for the monolayer in vacuum lies within about $30\%$ of the microscopic radiative width computed for hBN-encapsulated WSe$_2$~\cite{Brem_intrinsic}, while it suppresses intra-excitonic emission and renders the thermal photon Lamb shifts negligible. The resulting dissipator is purely radiative, and it is most complete at the low temperature adopted here.

Within this framework the two three-level systems behave in opposite ways. In the $\Xi$ system, a weak optical pulse preceded by an overlapping terahertz pulse, the counterintuitive order of STIRAP, converts the excitation it creates into the optically dark exciton by dark-state following, while the lossy bright intermediate exciton is only weakly populated: about half of the excitation is converted at the baseline ($\eta=0.48$, up to $0.63$ in the operating window), a fraction $0.33$ of the incident photons against the bound $1/2$ of a single monolayer, and reversing the pulse order suppresses the conversion by more than an order of magnitude over the operating window. The excitation that is not converted is emitted by that intermediate state rather than lost to non-adiabatic coupling, and the conversion efficiency is controlled by a single dissipative adiabaticity parameter of the terahertz leg, the analogue of the one that governs STIRAP through a decaying intermediate state~\cite{Vitanov1997}. The exchange mean field, a small blue shift of the exciton levels, displaces the two-photon resonance by more than the conversion tolerates already at moderate densities, so the drives must be calibrated at the actual exciton density. The cascade transparency window supplies this calibration on the sample itself: in the model it is complete, its center measures the renormalized two-photon detuning directly, and it qualifies as EIT rather than Autler--Townes splitting only below a control strength of about $1.5$~meV~\cite{AbiSalloum2010,Anisimov2011}. The $V$ system, whose two excited states both radiate, offers no comparable protection. Its coherence decays on the radiative timescale and is accessible only as a damped quantum beat, which establishes it as the radiatively unprotected reference.

These results also delimit the scope of the present reduction, and its natural extensions follow from relaxing its assumptions. Within the intravalley regime, a finite momentum transfer would restore exciton--phonon scattering, collisional broadening by the exciton gas and a finite in-plane photon phase space. All three reservoirs would then generate pure dephasing and additional Lamb-shift renormalizations of the detunings~\cite{selig2016excitonic,himami2018}, so the conversion efficiencies and $\Xi$ coherence times reported here are upper bounds with respect to those channels. Conversely, the sensitivity analyses outside the model (Sect.~\ref{subsec:real-sample} and App.~\ref{app:dephasing}\SMnote{}) turn these bounds into requirements that a real sample must meet. At picosecond pulses the conversion keeps half of its value only for a pure dephasing of $2p$ with $\hbar\gamma_\phi\lesssim0.44$~meV, nine times below the half-width of the $2p$ line in hBN-encapsulated WSe$_2$, $3.96$~meV~\cite{Kusaba2021}, and keeps its dark-state character only for a population decay of $2p$ with $\hbar\gamma_{\rm rel}\lesssim0.47$~meV; the $0$--$2p$ coherence of the converted exciton and a complete transparency window require dephasing rates two to three orders of magnitude below that half-width. Present samples miss these conditions, and gated, charge-neutral samples would still be expected to miss them. The depth of the cascade transparency window at a known terahertz Rabi energy and the pulse-order contrast are the measurements that decide whether a sample meets them. Spin- and momentum-dark excitons, which dominate a relaxed exciton gas in tungsten-based monolayers~\cite{Zhang2015dark,Wang2017dark,Malic2018dark,Selig2018,Madeo2020}, require spin- and valley-resolved exchange vertices. The screening methodology will likewise need refinement for other materials and dielectric environments, where Auger processes and exciton--exciton annihilation~\cite{auger_Dezheng,auger_binder,auger_Han}, trions in doped samples, or a different ordering of bright and dark states change which three-level systems are viable; such effects enter naturally as additional factors in the configuration scores, and so does the lifetime of the final state, which the requirements found here make the decisive factor. Beyond the intravalley regime, intervalley scattering and optical-phonon branches~\cite{Jin2014} add strongly coupled reservoirs with memory, already visible in the phonon sidebands and non-Lorentzian line shapes of TMD monolayers~\cite{groll2026phonon,Christiansen2017,Lengers2020}, for which a Lindblad generator is no longer sufficient and non-Markovian methods are required~\cite{deVega2017,BreuerPetruccione,RivasHuelga2012,NazirMcCutcheon2016,Tanimura2020,Weiss2012,Strathearn2018}, from projection-operator and reaction-coordinate master equations to hierarchical equations of motion and tensor-network techniques.

In summary, this work provides a microscopically derived, material-specific framework in which EIT and STIRAP protocols for excitons in two-dimensional semiconductors can be designed and quantified, and it states in numbers the conditions that a real transition-metal dichalcogenide monolayer must meet for them to work, together with the measurements that decide whether it does.

\section*{Acknowledgments}
We thank N. Borys for useful discussions. We thank the MonArk collaboration for their constructive feedback during a preliminary presentation of this work, which helped to improve the development of this research project. We thank J.~Guerrero for his assistance with the Blender software~\cite{Blender} used to prepare the original version of Fig.~\ref{fig:cartoon}. This work was supported by the National Science Foundation (NSF) Q-AMASE-i program within the framework of the MonArk NSF Quantum Foundry, under the NSF award DMR-1906383. During the preparation of this work the authors used the large language models ChatGPT (OpenAI; GPT-4.1, 5.1, 5.2, 5.5 and 5.6 Sol) and Claude (Anthropic; Opus 4.8, 5 and 5.5) to support and review the derivations, identify errors and typographical errors, and improve the clarity of explanations throughout the manuscript. The use of AI tools in writing the numerical code, in the dephasing analysis and in preparing the figures is described in Apps.~\ref{app:num_Lindblad} and~\ref{app:dephasing}\SMnote{} and in the captions of Figs.~\ref{fig:topology_graphs}, \ref{fig:cartoon}, \ref{fig:R1} and~\ref{fig:dephasing}. The authors directed this work, verified all AI-assisted contributions, and take full responsibility for the content of the paper.

\section*{Data availability}
The codes that generate the data supporting the findings of this article are
openly available in a public repository~\cite{code_repo_note}. Data underlying
the results presented in this paper are not publicly available at this time;
they may be obtained from the authors upon reasonable request, and their release
is at the discretion of the authors.

\onecolumngrid
\renewcommand{\thefootnote}{\arabic{footnote}}

\appendix 

\section{Hamiltonians of the model}
\label{app:model-hamiltonians}
This appendix collects the free and interaction Hamiltonians summarized in Sect.~\ref{sec:multi-env-model}, before the fixed-$\bm K_0$ restriction; their projections are derived in App.~\ref{app:K0-projection}.

\subsection{Free Hamiltonians}
\label{subsec:free}
\paragraph{Tagged exciton:} The full 2D exciton state carrying center-of-mass wavevector $\mathbf{K}$ is
$\Psi_{n}(\mathbf{r},\mathbf{K})=e^{i\mathbf{K}\cdot\mathbf{R}}\,\varphi_{n}(\mathbf{r})$,
where $\mathbf{R}=\alpha_e\mathbf{r}_e+\alpha_h\mathbf{r}_h$ is the center-of-mass coordinate and $\alpha_{e(h)}=m_{e(h)}/M$ are the electron and hole mass fractions.
In second quantization the exciton creation operator is related to the underlying electron and hole operators through~\footnote{A similar definition holds for the annihilation operator.}
\begin{equation}
    \label{eq:X-carriertrans}
    \hat{X}^\dagger_{n,\mathbf{K}}=\sum_{\mathbf{k}}\varphi_n(\mathbf{k})\,
\hat{c}^\dagger_{e,\alpha_e\mathbf{K}+\mathbf{k}}\,\hat{c}^\dagger_{h,\alpha_h\mathbf{K}-\mathbf{k}}\,,
\end{equation}
where $\varphi_n(\mathbf{k})$ is the Fourier transform of $\varphi_n(\mathbf{r})$ and $\hat{c}^\dagger_{e(h)}$ creates an electron (hole)~\footnote{
The wave functions $\varphi_n(\bm k)$ and $\varphi_n(\bm r)$ satisfy the following normalization conditions, respectively
\begin{equation*}
\begin{split}
\sum_{\bm k}
\left|
\varphi_{n}(\bm k)
\right|^2
&=
1\,,\\
\varphi_{n}(\bm r)
&=
\frac{1}{\sqrt{A}}
\sum_{\bm k}
e^{i\bm k\cdot\bm r}
\varphi_{n}(\bm k),
\quad
\int_{A}d^2\bm r\,
|\varphi_{n}(\bm r)|^2=1\,,
\end{split}
\end{equation*}
where $A$ is the monolayer normalization area.
\label{fn:normalization}}. By neglecting spin and valley dynamics, the free exciton Hamiltonian reads~\cite{singh1984dynamics}
\begin{equation}
    \label{eq:Exc-ham}
    \ho{H}_X=\sum_{\mathbf{K},n}\epsilon_{n}(\mathbf{K})\,\hat{X}_{n,\mathbf{K}}^\dagger\,
    \hat{X}_{n,\mathbf{K}}\,,
\end{equation}
where $\hat{X}_{n,\mathbf{K}}^\dagger$ ($\hat{X}_{n,\mathbf{K}}$) is the exciton creation (annihilation) operator and $\epsilon_n(\mathbf{K})$ is the exciton energy. The energy $\epsilon_\nu$ in the previous expression is with dispersion
\begin{equation}
    \label{eq:energy-X}
    \epsilon_{n}(\mathbf{K})=E_g-E_{b,n}+\frac{\hbar^2 \bm K^2}{2M}\,,
    \qquad
    E_{b,n}\equiv-E_n>0\,,
\end{equation}
where $M=m_e+m_h$ is the total electron-hole pair mass, $E_g$ is the quasiparticle gap, $E_n<0$ are the eigenenergies of the Wannier equation
and $E_{b,n}\equiv-E_n>0$ the corresponding binding energies tabulated in
Table~\ref{tab:appD_states}~\footnote{We solve the Wannier equation using a variational Rayleigh–Ritz algorithm, whose implementation and validation are detailed in App.~\ref{app:rayleigh-ritz}.} and $\bm K$ is the center-of-mass momentum. 
\paragraph{Phonon Hamiltonian:} The free phonon Hamiltonian is
\begin{equation}
    \label{eq:phonon_ham}
    \ho{H}_{ph}=\sum_{\mathbf{q},\nu}\hbar\,\omega_{\mathbf{q},\nu}\,\hat{b}_{\mathbf{q},\nu}^\dagger\hat{b}_{\mathbf{q},\nu}\,,
\end{equation}
where $\hat{b}_{\mathbf{q},\nu}^\dagger$ and $\hat{b}_{\mathbf{q},\nu}$ create and annihilate a phonon of wavevector $\mathbf{q}$ in branch $\nu$, and $\omega_{\mathbf{q},\nu}$ is the dispersion relation. We restrict to acoustic phonons for the deformation-potential coupling adopted below.

\paragraph{Photon Hamiltonian:} We describe the photon field in the
multipolar gauge representation
\cite{PowerZienau1959,Woolley1971,BabikerLoudon1983,
TodorovSirtori2012,TaghizadehPedersen2018}. This representation is
particularly convenient for the few-level reduction performed below. The quantized
electric field reads as~\footnote{Here the subindex $\bm T$ denotes the transversality with respect to the photon wavevector, i.e. $\mathbf{e}_{\vms, \lambda}\cdot \vms=0$.}
\begin{equation}
\hat{\bm E}_{\bm T}(\vec{\mathbf{r}})
=
i\sum_{\vec s,\lambda}
\mathscr{A}_{\vms}\left(
    \mathbf{e}_{\vms, \lambda}\,\hat{a}_{\vms,\lambda}\,e^{i\,\vms\cdot\vec{\mathbf{r}}} -h.c.\right)\,,
\label{eq:quantized-electric-field}
\end{equation}
where $\hat{a}_{\vms,\lambda}$ and $\hat{a}^\dagger_{\vms,\lambda}$ are photon annihilation and creation operators, $\omega_{\vms}$ is the photon frequency, $\mathbf{e}_{\vms,\lambda}$ is a polarization vector transverse to the propagation direction, and $\mathscr{A}_{\vms}=\left(
\left[\hbar\omega_{\vec s}\right]/\left[2\epsilon_0\,V\right]\right)^{1/2}$~\footnote{The quantized volume is parametrized as $V=A L_{\gamma}$, where $A$ is the monolayer normalization area and $L_{\gamma}$ is
the quantization length perpendicular to the layer.}. The free photon Hamiltonian reads as
\begin{equation}
    \label{eq:photon_ham}
    \ho{H}_\gamma = \sum_{\vms,\lambda}\hbar\,\omega_{\vms}
    \left(\hat{a}_{\vms,\lambda}^\dagger\hat{a}_{\vms,\lambda}+\frac{1}{2}\right)\,.
\end{equation}
The photon wavevector is decomposed into its component parallel to the
monolayer, $\bm s_\parallel$, and its out-of-plane component,
$s_\perp$, i.e. $\vec s=(\bm s_\parallel,s_\perp)$. Thus, the photon
dispersion relation reads
$\omega_{\vec s}
=
c\sqrt{|\bm s_\parallel|^2+s_\perp^2}$.
Equations~\eqref{eq:quantized-electric-field} and~\eqref{eq:photon_ham} describe the full microscopic electromagnetic
field and, therefore, retain both transverse polarizations~\footnote{The
polarization-selective reduction introduced in Sect.~\ref{subsec:explicit_drives} is applied only to the
intra-excitonic THz/mid-IR sector of the reduced fixed-$\bm K_0$
open-system model; the interband optical photon sector retains the full
transverse-polarization reservoir.}. 

\paragraph{Dilute exciton gas:} We assume a dilute intravalley exciton gas with mean inter-exciton spacing $d_X\gg a_B$, where $a_B$ is the exciton Bohr radius, such that wave-function-overlap effects, composite-boson corrections, and inter-exciton correlations remain perturbative. Although the tagged exciton and the excitons forming the gas are physically indistinguishable, the subsystem--reservoir separation is performed at the level of excitonic mode sectors rather than particle labels: the subsystem corresponds to the optically prepared and monitored one-exciton sector, whereas the gas occupies the complementary many-exciton reservoir sector. 
The free exciton-gas Hamiltonian is
\begin{equation}
\label{eq:Xg-Ham}
\hat{\mathcal H}_{X_g}=\sum_{n,\bm{Q}} \epsilon_{n}(\bm{Q})\, \hat X_{n,\bm{Q}}^\dagger \hat X_{n,\bm{Q}},
\end{equation}
where $\epsilon_{n}(\bm Q)$ is the exciton dispersion given by
Eq.~\eqref{eq:energy-X}, evaluated at the gas center-of-mass momentum
$\bm Q$. We stress that Eq.~\eqref{eq:energy-X} defines
$\epsilon_n(\bm K)$; the symbols $E_n$ and $E_{b,n}$ appearing there are
the Wannier eigenvalue and the corresponding binding energy, and no
separate dispersion $E_n(\bm Q)$ is introduced anywhere in this work. The tagged exciton and the exciton gas share the same underlying excitonic modes and dispersion, while differing in the Fock-space sectors assigned to the system and reservoir~\footnote{In the dilute regime ($n_Xa_B^2\ll1$), deviations of the composite-exciton commutation relations from canonical bosonic behavior and initial exchange correlations between the subsystem and reservoir sectors are neglected. This mode-space partition then permits the two sectors to be represented on separate tensor factors in the reduced open-system description, while carrier-exchange effects associated with the fermionic substructure of excitons remain explicitly contained in the microscopic exciton--exciton interaction vertex. This partition is what the populations $\bar\rho_{ii}(t)$ of the
reduced density matrix in Sect.~\ref{sec:lindblad_three_level} refer
to: under the normally incident, macroscopically illuminating drives
of Sect.~\ref{subsec:explicit_drives}, the whole sample is excited
coherently and uniformly, so $\bar\rho_{ii}(t)$ describes the state of
the optically prepared tagged-exciton sector of a dilute photoexcited
population, not an isolated single exciton in an otherwise empty
crystal; the exciton gas, created by a separate excitation that precedes or
accompanies the control sequence (Sect.~\ref{sec:VE-renormalization}),
populates the complementary many-exciton reservoir sector already
introduced above and enters the dynamics only through the mean-field
exchange self-energy $\Sigma^{(Xg,0)}_{ab}$ of
Sect.~\ref{sec:VE-renormalization}. }.

\subsection{Interaction Hamiltonians}
\label{subsec:int_ham}
Here, we introduce the general effective interaction Hamiltonians coupling the tagged exciton to the reservoirs. Their projections onto the fixed-$\bm K_0$ sector adopted throughout this work are given in Sect.~\ref{subsec:K0-projection}.
\paragraph{Exciton--phonon Hamiltonian:} The intravalley exciton--phonon Hamiltonian reads as~\cite{singh1984dynamics}
\begin{equation}
    \ho{H}_{X-\mathrm{ph}}
    =
    \sum_{n',n}
    \sum_{\bm K,\bm q,\nu}
    g_{\bm q,\nu}^{n'n}\,
    \hat X_{n',\bm K+\bm q}^{\dagger}
    \hat X_{n,\bm K}
    \left(
        \hat b_{\bm q,\nu}
        +
        \hat b_{-\bm q,\nu}^{\dagger}
    \right),
    \label{eq:X-ph_Ham}
\end{equation}
where $\nu$ labels the phonon branch. For an acoustic deformation-potential
coupling, the internal-state-resolved matrix element is~\cite{singh1984dynamics}
\begin{equation}
    g_{\bm q,\nu}^{n'n}
    =
    \sqrt{
        \frac{\hbar |\bm q|}
             {2\rho_m v_\nu A}
    }
    \left[
        D_\nu^{e}
        F_{n'n}(\alpha_h\bm q)
        -
        D_\nu^{h}
        F_{n'n}(-\alpha_e\bm q)
    \right],
    \label{eq:exciton-phonon-matrix-element-forward}
\end{equation}
where $\alpha_{e(h)}=m_{e(h)}/M$ is the electron (hole) mass fraction, $D_\nu^i$ ($i$=$\{$electron (e), hole (h)$\}$) is the deformation-potential constant, $v_\nu$ is the sound velocity in branch $\nu$, $\rho_m$ is the 2D mass density, and $A$ is the total sample area. The excitonic form factor is $F_{n'n}(\bm q) = \langle n'|
    e^{i\bm q\cdot\bm r}
    |n\rangle$.

\paragraph{Exciton--photon Hamiltonian:} 
In the multipolar gauge, the linear exciton--photon ($X-\gamma$) interaction follows from
$\ho{H}_{X-\gamma}=-\int_V d^3\vec r\,
\hat{\bm P}^{\rm pol}(\vec{\bm r})\cdot
\hat{\bm E}_{\bm T}(\vec{\bm r})$~\cite{Loudon}~\footnote{The complete multipolar Hamiltonian also contains the photon-independent polarization self-energy $\sim\left[\hat{\mathbf P}^{\rm pol}\right]^{2}$. In the present effective excitonic description, $\ho{H}_X$ is understood as the renormalized matter Hamiltonian, so that photon-independent self-energy contributions are absorbed into the chosen reference exciton energies. We therefore retain explicitly only the linear photon-dependent interaction. This should be understood as a renormalization convention and does not imply that the Wannier equation itself contains the electromagnetic polarization self-energy.}. The total $X-\gamma$ Hamiltonian can be decomposed into its intra-excitonic and interband contributions~\cite{Combescot2015}, i.e. $\ho{H}_{X-\gamma}=\ho{H}^{\rm intra}_{X-\gamma}+\ho{H}^{\rm inter}_{X-\gamma}$ with
\begin{subequations}
\label{eq:X-gam_ham}
\begin{align}
\ho{H}_{X-\gamma}^{\rm intra}
&=
-\sum_{\bm K,\bm s_\parallel}
\sum_{n',n}
\hat X_{n',\bm K+\bm s_\parallel}^{\dagger}
\hat X_{n,\bm K}\,
\bm{\mathcal D}_{n'n}^{\rm intra}
\cdot
\left[
\hat{\bm E}_{\bm s_\parallel}^{(+)}
+
\hat{\bm E}_{-\bm s_\parallel}^{(-)}
\right],
\label{eq:intra-X-gamma_ham}
\\
\ho{H}_{X-\gamma}^{\rm inter}
&=
-\sum_{\bm s_\parallel,n}
\bigg\{
\hat X_{n,\bm s_\parallel}^{\dagger}\,
\bm{\mathcal D}_{n0}^{\rm inter}(\bm s_\parallel)
\cdot
\left[
\hat{\bm E}_{\bm s_\parallel}^{(+)}
+
\hat{\bm E}_{-\bm s_\parallel}^{(-)}
\right]
+
{\rm H.c.}
\bigg\}.
\label{eq:inter-X-gamma_ham}
\end{align}
\end{subequations}
Throughout the present work, the light–matter interaction is treated at electric-dipole order with respect to the internal electron–hole coordinate~\footnote{Accordingly, the spatial variation of the electromagnetic field across the exciton is neglected, while its temporal dependence and its spatial dependence on the exciton center-of-mass coordinate are retained. This long-wavelength approximation does not constrain the photon frequency and therefore allows resonant THz/mid-IR transitions between distinct internal exciton states.}. In Eqs.~\eqref{eq:X-gam_ham}, the electric field~\eqref{eq:quantized-electric-field} is decomposed into positive- and negative-frequency components carrying in-plane momentum $\bm s_\parallel$,
\begin{subequations}
\begin{align}
\hat{\bm E}_{\bm s_\parallel}^{(+)}
&=
i\sum_{s_{\perp},\lambda}
\mathscr{A}_{(\bm s_\parallel,s_{\perp})}
\bm e_{(\bm s_\parallel,s_{\perp}),\lambda}
\hat a_{(\bm s_\parallel,s_{\perp}),\lambda},
\label{eq:E-positive-q}
\\
\hat{\bm E}_{-\bm s_\parallel}^{(-)}
&=
\left[
\hat{\bm E}_{\bm s_\parallel}^{(+)}
\right]^{\dagger}.
\label{eq:E-negative-q}
\end{align}
\end{subequations}
The corresponding intra-excitonic and interband electric-dipole form factors are
\begin{subequations}
\label{eq:exciton-polarization-general}
\begin{align}
\bm{\mathcal D}_{n'n}^{\rm intra}
&=
-e\langle n'|\bm r|n\rangle
=
-e\int d^2\bm r\,
\varphi_{n'}^{*}(\bm r)\,
\bm r\,
\varphi_n(\bm r),
\label{eq:intra-exciton-polarization-general}
\\
\bm{\mathcal D}_{n0}^{\rm inter}(\bm s_\parallel)
&=
\sum_{\bm k}
\varphi_n^{*}(\bm k)\,
\bm d_{cv}^{\xi}
(\bm k-\alpha_h\bm s_\parallel),
\label{eq:interband-exciton-polarization-general}
\end{align}
\end{subequations}
where $\bm r=\bm r_e-\bm r_h$ is the electron--hole relative coordinate and
$\bm d_{cv}^{\xi}(\bm k)$ is the interband Bloch electric-dipole matrix element for states in the valley $\xi$, with $\bm k$ measured relative to the corresponding valley center. The first expression~\eqref{eq:intra-exciton-polarization-general} governs bound-to-bound intra-excitonic transitions, whereas the second~\eqref{eq:interband-exciton-polarization-general} describes optical creation or annihilation of a Wannier exciton from the crystal ground state.

\paragraph{Exciton--exciton Hamiltonian:} Exciton - exciton ($X$-$X$) interactions have been extensively investigated~\cite{ivanov1993self,tassone1999exciton,shahnazaryan2017exciton,erkensten2021exciton,katsch2018theory,erkensten2023electrically}. To retain processes that may change the internal states of either exciton, we use the internal-state-resolved extension of the standard effective two-exciton Hamiltonian~\cite{erkensten2023electrically}
\begin{equation}
    \ho{H}_{X-X}
    =
    \frac{1}{2}
    \sum_{\substack{n',n,l',l\\
                    \bm K,\bm Q,\bm q}}
    G_{\bm K,\bm Q,\bm q}^{n'l';ln}\,
    \hat X^\dagger_{n',\bm K+\bm q}
    \hat X^\dagger_{l',\bm Q-\bm q}
    \hat X_{l,\bm Q}
    \hat X_{n,\bm K}.
    \label{eq:gen-XX-Hamiltonian}
\end{equation}
Here $\mathbf K$ and $\mathbf Q$ denote the incoming center-of-mass momenta of the two interacting excitons, while $\mathbf q$ is the momentum transferred between them. Accordingly, Eq.~\eqref{eq:gen-XX-Hamiltonian} describes the scattering process $
(n,\mathbf K)+(l,\mathbf Q)
\longrightarrow
(n',\mathbf K+\mathbf q)+(l',\mathbf Q-\mathbf q)$. The matrix element $G_{\bm K,\bm Q,\bm q}^{n'l';ln}$ in Eq.~\eqref{eq:gen-XX-Hamiltonian} is decomposed into its direct and exchange contributions as follows
\begin{equation}
    G_{\bm K,\bm Q,\bm q}^{n'l';ln}
    =
    G_{\mathrm{dir}}^{n'l';ln}(\bm q)
    +
    G_{\mathrm{exch}}^{n'l';ln}
        (\bm K,\bm Q,\bm q),
    \label{eq:direct-exchange-decomposition-general}
\end{equation}
where the definition of the exchange contribution includes the conventional numerical prefactors associated with the chosen effective bosonic Hamiltonian. The decomposition into direct and exchange contributions follows the microscopic exciton--exciton framework of Ref.~\cite{erkensten2023electrically}. Here the indices $(n,n',l,l')$ are retained as independent internal-state labels. The fixed-$\bm K_0$ reduction of Eq.~\eqref{eq:gen-XX-Hamiltonian} is given in Sect.~\ref{subsec:K0-projection}, and the explicit internal-state-resolved exchange vertex in App.~\ref{app:xx-internal-state-exchange}.

\section{Fixed \texorpdfstring{$\bm{K}_0$}{K0} projection of interacting Hamiltonians}
\label{app:K0-projection}
This appendix collects the intermediate steps of the fixed-$\bm K_0$ projection of the interaction Hamiltonians whose results are summarized in Sect.~\ref{subsec:K0-projection}. Mathematically, the fixed-$\bm K_0$ restriction is implemented by explicitly projecting the tagged-exciton operators appearing in the interaction Hamiltonians onto the fixed-$\bm K_0$ sector, as described below.

A sufficiently general subsystem--environment interaction for the Hamiltonians considered here can be written as
\begin{equation}
    \ho{H}_{X-\alpha}
    =
    \sum_{\mu,\bm K,\bm Q, \bm q}
    \mathcal W_{\mu}^{(\alpha)}
        (\bm K,\bm Q,\bm q)\,
    \hat{\mathcal A}_{\mu}^{(\alpha)}
        (\bm K,\bm q)
    \otimes
    \hat{\mathcal B}_{\mu}^{(\alpha)}
        (\bm Q,\bm q).
    \label{eq:generic-momentum-resolved-interaction}
\end{equation}
Here $\alpha$ labels the environment, $\mu$ is a compound index containing
all remaining discrete labels, $\bm K$ is the center-of-mass momentum
associated with the tagged exciton, $\bm Q$ denotes an independent
environmental momentum, and $\bm q$ parameterizes the momentum exchanged
at the interaction vertex. The quantity
$\mathcal W_{\mu}^{(\alpha)}(\bm K,\bm Q,\bm q)$ is the microscopic
coupling matrix element, while
$\hat{\mathcal A}_{\mu}^{(\alpha)}$ and
$\hat{\mathcal B}_{\mu}^{(\alpha)}$ act on the subsystem and environmental
Hilbert spaces, respectively. Hermitian-conjugate contributions are
understood to be included in Eq.~\eqref{eq:generic-momentum-resolved-interaction}.

The projectors onto the fixed-$\bm K_0$ sector, Eqs.~\eqref{eq:fixed-K0-projector}, and the projected interactions, Eq.~\eqref{eq:generic-fixed-K0-definition}, are defined in Sect.~\ref{subsec:K0-projection}; $\hat{\mathcal P}_{\bm K_0}$ denotes whichever projector is appropriate for the interaction under consideration.

Applying the projector to the subsystem operator generates the reservoir-specific Kronecker constraints required by momentum conservation. Since these constraints depend on whether the interaction conserves exciton number or creates/annihilates an exciton, we derive them explicitly for each microscopic interaction below.

\subsection{Projected exciton--phonon Hamiltonian}
\label{app:K0-projection-phonons}
The exciton-phonon Hamiltonian~\eqref{eq:X-ph_Ham} preserves the exciton number and therefore requires the fixed-$\bm K_0$ projector~\eqref{eq:fixed-K0-one-exciton-projector}. Applying this projector to Eq.~\eqref{eq:X-ph_Ham} gives
\begin{equation}
    \hat{\mathcal P}_{\bm K_0}^{(1X)}
    \hat H_{X-\mathrm{ph}}
    \hat{\mathcal P}_{\bm K_0}^{(1X)}
    =
    \sum_{n',n,\nu}
    g_{\bm 0,\nu}^{n'n}
    |n';\bm K_0\rangle
    \langle n;\bm K_0|
    \left(
        \hat b_{\bm 0,\nu}
        +
        \hat b_{\bm 0,\nu}^{\dagger}
    \right).
    \label{eq:projected-exciton-phonon-Hamiltonian}
\end{equation}
In the fixed-$\bm K_0$ limit, orthonormality of the internal exciton
wavefunctions implies $F_{n'n}(\bm q)\bigl|_{\bm q = \bm 0}=\delta_{n'n}$ and thus, zero-momentum transfer cannot change the internal exciton
state. Moreover, in this regime, the diagonal acoustic deformation-potential matrix element vanishes since $g_{\bm q,\nu}^{n'n}\propto\sqrt{|\bm q|}$ as
$\bm q\rightarrow\bm 0$, as follows directly from Eq.~\eqref{eq:exciton-phonon-matrix-element-forward}. Consequently, the projected exciton--phonon Hamiltonian vanishes identically, Eq.~\eqref{eq:fixed-K0-phonon-Hamiltonian-zero}. The physical content and the limitations of this result are discussed in Sect.~\ref{subsec:K0-projection}.

\subsection{Projected exciton--photon Hamiltonian}
\label{app:K0-projection-photon}

We now project the exciton--photon Hamiltonians~\eqref{eq:X-gam_ham} onto the fixed-$\bm K_0$ sector. Since intra-excitonic processes conserve exciton number whereas interband processes create or annihilate an exciton, we must use the projectors~\eqref{eq:fixed-K0-one-exciton-projector} and~\eqref{eq:fixed-K0-zero-one-exciton-projector}, respectively.

\paragraph{Fixed-$\bm K_0$ limit of the intra-excitonic Hamiltonian:}
For the number-conserving exciton operator entering Eq.~\eqref{eq:intra-X-gamma_ham},
\begin{equation}
\hat{\mathcal P}_{\bm K_0}^{(1X)}
\hat X_{n',\bm K+\bm s_\parallel}^{\dagger}
\hat X_{n,\bm K}
\hat{\mathcal P}_{\bm K_0}^{(1X)}
=
\delta_{\bm K,\bm K_0}
\delta_{\bm s_\parallel,\bm 0}
|n';\bm K_0\rangle
\langle n;\bm K_0|.
\label{eq:projected-intraband-exciton-operator}
\end{equation}
Hence the fixed-$\bm K_0$ condition enforces vanishing in-plane photon momentum,
$\bm s_\parallel=\bm 0$, while leaving $s_\perp$ unrestricted. The projected intra-excitonic interaction is therefore Eq.~\eqref{eq:projected-intraband-length-gauge} of Sect.~\ref{subsec:K0-projection}, with the dipole form factor~\eqref{eq:intra-exciton-polarization-general}.

This matrix element obeys the usual envelope electric-dipole selection
rule $\Delta m=\pm1$.  For a degenerate $p$ state with $m=\pm1$,
a fixed in-plane linear polarization $\bm e_d$ selects the real linear
combination $\ket{p_d}$ whose transition dipole is parallel to
$\bm e_d$, while the orthogonal combination $\ket{p_\perp}$ has a
dipole perpendicular to it.  In the reduced open-system model we retain
this single polarization branch only for the intra-excitonic THz/mid-IR
photon sector.  This additional polarization-sector restriction does not
modify the fixed-$\bm K_0$ momentum constraint
$\bm s_\parallel=\bm0$ derived above; its implementation in the photon
correlators is given in App.~\ref{app:bath-corr}.

\paragraph{Fixed-$\bm K_0$ limit of the interband Hamiltonian:}
For the exciton creation operator entering Eq.~\eqref{eq:inter-X-gamma_ham},
\begin{equation}
\hat{\mathcal P}_{\bm K_0}^{(0\oplus1X)}
\hat X_{n,\bm s_\parallel}^{\dagger}
\hat{\mathcal P}_{\bm K_0}^{(0\oplus1X)}
=
\delta_{\bm s_\parallel,\bm K_0}
|n;\bm K_0\rangle
\langle0|,
\label{eq:projected-interband-exciton-operator}
\end{equation}
and therefore the photon must carry the in-plane center-of-mass momentum of the retained exciton sector,
$\bm s_\parallel=\bm K_0$. The projected interband interaction is therefore Eq.~\eqref{eq:projected-interband-length-gauge} of Sect.~\ref{subsec:K0-projection}, where the projected interband dipole form factor is
\begin{equation}
\bm{\mathcal D}_{n0}^{\rm inter}
=
\bm d_{cv}^{\xi}
\sum_{\bm k}\varphi_n^{*}(\bm k)\,
\bm d_{cv}^{\xi}
(\bm k-\alpha_h\bm K_0).
\label{eq:projected-inter-dipole}
\end{equation}
For band-edge transitions, we approximate the Bloch interband
electric-dipole matrix element by a momentum-independent complex
vector within the selected valley~\cite{wang_interlayer,Brem_intrinsic,plexitons}. Using the
$k\cdot p$ parametrization adopted in Refs.~\cite{coupled-spin_Xiao,tight-banding_yan},
the corresponding dipole at the $K$ valley is
\begin{equation}
\bm d_{cv}^{K}
=
i\frac{e\hbar}{2\sqrt{\mu E_g}}\,(1,i),
\label{eq:bloch-interband-dipole}
\end{equation}
where $E_g$ is the direct band gap and
$\mu=m_e^\ast m_h^\ast/(m_e^\ast+m_h^\ast)$ is the electron--hole
reduced mass. The opposite valley carries the time-reversed
complex polarization. Consequently, the projected interband dipole form factor reduces to
\begin{equation}
\bm{\mathcal D}_{n0}^{\rm inter}
=
\bm d_{cv}^{K}
\sum_{\bm k}\varphi_n^{*}(\bm k)
=
\sqrt{A}\,
\varphi_n^{*}(\bm r=\bm0)\,
\bm d_{cv}^{\xi},
\label{eq:projected-inter-dipole-local}
\end{equation}
where
$\varphi_n(\bm r)=A^{-1/2}\sum_{\bm k}e^{i\bm k\cdot\bm r}\varphi_n(\bm k)$. For the opposite valley, $\bm d_{cv}^{K'}$ is the time-reversed counterpart of $\bm d_{cv}^{K}$~\eqref{eq:bloch-interband-dipole}.

\subsection{Projected exciton--exciton Hamiltonian}
\label{app:K0-projection-exciton-gas}
We now apply the fixed-center-of-mass-momentum reduction to the
exciton--exciton interaction. We restrict the discussion to intravalley
excitons in a single monolayer, suppress the fixed valley label, and neglect
spin. The internal states of the tagged and gas excitons are denoted by
$n,n'$ and $l,l'$, respectively. In addition,  their center-of-mass dispersions are taken
to be given by Eq.~\eqref{eq:energy-X}, with the same total effective mass $M$ for all internal states. By applying the fixed-$\bm K_0$ projector~\eqref{eq:fixed-K0-one-exciton-projector} to the effective Hamiltonian~\eqref{eq:gen-XX-Hamiltonian}, one obtains
\begin{equation}
    \hat{\mathcal P}_{\bm K_0}^{(1X)}
    \hat X^\dagger_{n',\bm K+\bm q}
    \hat X_{n,\bm K}
    \hat{\mathcal P}_{\bm K_0}^{(1X)}
    =
    \delta_{\bm K,\bm K_0}
    \delta_{\bm q,\bm 0}
    |n';\bm K_0\rangle
    \langle n;\bm K_0|.
    \label{eq:fixed-K0-exciton-exciton-projection}
\end{equation}
Thus, the fixed-$\bm K_0$ restriction enforces $\bm q=\bm 0$ but leaves the exciton-gas momentum $\bm Q$ unrestricted. Together with the cancellation of the prefactor $1/2$ of Eq.~\eqref{eq:gen-XX-Hamiltonian} and the vanishing of the direct term at zero momentum transfer, both discussed in Sect.~\ref{subsec:K0-projection}, this yields the projected exciton--gas Hamiltonian, Eq.~\eqref{eq:final-projected-Xg-Hamiltonian}.  

\subsubsection{Projected internal-state-resolved exchange vertex}
\label{app:xx-internal-state-exchange}

The projected Hamiltonian~\eqref{eq:final-projected-Xg-Hamiltonian}
requires the exchange matrix element
$G_{\rm exch}^{\,n'l';ln}(\bm K_0,\bm Q,\bm 0)$ with independent
internal states on the incoming and outgoing tagged- and exciton gas lines. This quantity is decomposed into two contributions, the carrier-carrier and the electron-hole exchanges~\cite{erkensten2023thesis}. In the following, we provide explicit expressions for both of these terms in the fixed-$\bm K_0$ projection. The carrier--carrier vertex is derived below rather than taken over, because the cited form is specialized to a single internal state. 

\paragraph{Carrier--carrier exchange.} We derive the labeled vertex directly from the monolayer carrier Hamiltonian by carrying out explicitly the independent internal state labels and thus, extending previous calculations~\cite{erkensten2023electrically}.

Writing the exciton operator~\eqref{eq:X-carriertrans} in terms of the
valence electron rather than the hole,
\begin{equation}
\hat X^{\dagger}_{n,\bm K}
=\sum_{\bm p}\varphi_n(\bm p+\alpha_h\bm K)\,
 \hat c^{\dagger}_{\bm p+\bm K}\,\hat v_{\bm p},
\label{eq:exciton-valence-form}
\end{equation}
and evaluating the two-exciton Coulomb matrix element between the incoming
pair and the outgoing pair with one carrier (electron or hole) exchanged,
the exchange contribution is, with $\bm\Delta\equiv\bm K_0-\bm Q$,
\begin{equation}
\mathcal E_{\rm cc}^{\,n'l';ln}(\bm K_0,\bm Q)
=\mathcal E_{\rm cc,r}^{\,n'l';ln}(\bm K_0,\bm Q)
+\mathcal E_{\rm cc,a}^{\,n'l';ln}(\bm K_0,\bm Q),
\label{eq:appA-xx-cc-oriented}
\end{equation}
with
\begin{equation}
\begin{split}
\mathcal E_{\rm cc,r}^{\,n'l';ln}(\bm K_0,\bm Q)
= -\sum_{\bm k,\bm k'}\Big\{
&\big[V_{\bm k-\bm k'+\alpha_e\bm\Delta}
     +V_{\bm k-\bm k'-\alpha_h\bm\Delta}\big]\,
  \varphi^*_{n'}(\bm k)\varphi_n(\bm k)\,
  \varphi^*_{l'}(\bm k')\varphi_l(\bm k')
\\[1mm]
+\;&V_{\bm k-\bm k'-\alpha_h\bm\Delta}\,
  \varphi^*_{n'}(\bm k'+\alpha_h\bm\Delta)\varphi_l(\bm k')\,
  \varphi^*_{l'}(\bm k-\alpha_h\bm\Delta)\varphi_n(\bm k)
\\[1mm]
+\;&V_{\bm k-\bm k'+\alpha_e\bm\Delta}\,
  \varphi^*_{n'}(\bm k'-\alpha_e\bm\Delta)\varphi_l(\bm k')\,
  \varphi^*_{l'}(\bm k+\alpha_e\bm\Delta)\varphi_n(\bm k)\Big\}\,,
\label{eq:appA-xx-cc-rep}
\end{split}
\end{equation}
\begin{equation}
\begin{split}
\mathcal E_{\rm cc,a}^{\,n'l';ln}(\bm K_0,\bm Q)
= +\sum_{\bm k,\bm k'}V_{\bm k-\bm k'}\Big\{
&\big[\varphi^*_{n'}(\bm k-\alpha_e\bm\Delta)\varphi_n(\bm k-\alpha_e\bm\Delta)
 +\varphi^*_{n'}(\bm k+\alpha_h\bm\Delta)\varphi_n(\bm k+\alpha_h\bm\Delta)\big]
 \varphi_l(\bm k)\,\varphi^*_{l'}(\bm k')
\\[1mm]
+\;&\varphi^*_{n'}(\bm k')\varphi_n(\bm k)
 \big[\varphi^*_{l'}(\bm k+\alpha_e\bm\Delta)\varphi_l(\bm k+\alpha_e\bm\Delta)
 +\varphi^*_{l'}(\bm k-\alpha_h\bm\Delta)\varphi_l(\bm k-\alpha_h\bm\Delta)\big]\Big\}\,.
\label{eq:appA-xx-cc-att}
\end{split}
\end{equation}
In every exchange process the Coulomb line joins carriers that belong to
different incoming excitons. In $\mathcal E_{\rm cc,r}$ it joins two like
carriers (conduction--conduction or valence--valence) through the repulsive
kernel~\eqref{eq:RK-momentum-space}, and the fermionic exchange sign makes
the contribution negative. The first line of Eq.~\eqref{eq:appA-xx-cc-rep}
is the contraction in which the envelopes of each exciton pair with
themselves, the second and third lines are those in which the envelopes
cross-pair between the tagged exciton and the gas. In
$\mathcal E_{\rm cc,a}$ the line joins the electron of one exciton to the
hole of the other through the attractive conduction--valence interaction,
and the same exchange sign makes the contribution positive; in each of its
terms three envelopes share a momentum and the fourth, an outgoing envelope,
is summed against the kernel. Both parts are of the same order in $A^{-1}$.
Here the interaction is evaluated in the grouping of carriers of the
incoming pair; the exchange overlap weighted by the incoming energies
cancels identically against the intra-exciton kinetic and Coulomb energies,
because the incoming product state is an eigenstate of the two isolated
excitons, so that no separate non-orthogonality correction arises at this
order. As a check on the bookkeeping, the direct (Hartree) pieces of the
three Coulomb channels enter with weights $+1$, $+1$ and $-2$ and cancel
identically, reproducing the absence of a direct interaction between two
neutral excitons at zero momentum transfer.

At equal center-of-mass momenta, $\bm\Delta=\bm 0$ --- which covers the case
$\bm K_0=\bm 0$ with the gas treated in the low-center-of-mass limit used in
App.~\ref{app:offdiag} --- Eqs.~\eqref{eq:appA-xx-cc-rep}
and~\eqref{eq:appA-xx-cc-att} collapse to
\begin{equation}
\begin{split}
\mathcal E_{\rm cc}^{\,n'l';ln}\Big|_{\bm\Delta=\bm 0}
= -2\sum_{\bm k,\bm k'} V_{\bm k-\bm k'}
\Big[&\varphi^*_{n'}(\bm k)\varphi_n(\bm k)\,
      \varphi^*_{l'}(\bm k')\varphi_l(\bm k')
+\varphi^*_{n'}(\bm k')\varphi_l(\bm k')\,
      \varphi^*_{l'}(\bm k)\varphi_n(\bm k)\Big]\\
+2\sum_{\bm k,\bm k'} V_{\bm k-\bm k'}
\Big[&\varphi^*_{n'}(\bm k)\varphi_n(\bm k)\,\varphi_l(\bm k)\,
      \varphi^*_{l'}(\bm k')
+\varphi^*_{n'}(\bm k')\,\varphi^*_{l'}(\bm k)\varphi_n(\bm k)\varphi_l(\bm k)\Big]\,,
\label{eq:appA-xx-cc-zero-transfer}
\end{split}
\end{equation}
which is the form used in App.~\ref{app:offdiag}. The two like-carrier
contractions in the first line carry equal weight, and both are symmetric
under $(n'\leftrightarrow n,\;l'\leftrightarrow l)$. In the second line the
kernel acts on a single outgoing envelope, and the Wannier equation obeyed
by every envelope, $\varepsilon_{\bm k}\varphi_a(\bm k)-\sum_{\bm k'}
V_{\bm k-\bm k'}\varphi_a(\bm k')=E_a\varphi_a(\bm k)$ with
$\varepsilon_{\bm k}=\hbar^2k^2/2\mu$ and $E_a=-E_b(a)$
[Eq.~\eqref{eq:radial} in momentum space], reduces it to a single momentum
sum,
\begin{equation}
\mathcal E_{\rm cc,a}^{\,n'l';ln}\Big|_{\bm\Delta=\bm 0}
=2\sum_{\bm k}\varphi^*_{n'}(\bm k)\varphi^*_{l'}(\bm k)
\varphi_n(\bm k)\varphi_l(\bm k)\,
\big[2\varepsilon_{\bm k}-E_{n'}-E_{l'}\big].
\label{eq:appA-xx-cc-att-wannier}
\end{equation}
This term carries no Coulomb kernel and therefore no additional factor
$A^{-1}$: it is of the same order as the first line of
Eq.~\eqref{eq:appA-xx-cc-zero-transfer} and of opposite sign. The energies
in the bracket are those of the outgoing states, because the inter-exciton
interaction was evaluated in the incoming grouping of carriers; evaluating
it in the outgoing grouping replaces $E_{n'}+E_{l'}$ by $E_n+E_l$. The two
forms coincide on the energy shell, and hence for every diagonal element,
but differ off the diagonal, the familiar in/out ambiguity of the
composite-boson reduction~\cite{Combescot2015}. We adopt the Hermitian
average, $E_{n'}+E_{l'}\to\tfrac12(E_n+E_l+E_{n'}+E_{l'})$, which makes the
exciton-gas self-energy built from Eq.~\eqref{eq:appA-xx-cc-zero-transfer}
Hermitian.

In Eqs.~\eqref{eq:appA-xx-cc-oriented} and~\eqref{eq:appA-xx-cc-zero-transfer},
$\alpha_{e\,(h)}=m_{e\,(h)}/M,$ is the electron (hole) mass fraction and we use $\varphi_n(\bm k)$ for the normalized relative-motion exciton envelope with
$\sum_{\bm k}|\varphi_n(\bm k)|^2=1$. Moreover, $V_{\bm p}^{\lambda\lambda'}$ denotes the screened Coulomb
interaction between carriers in bands
$\lambda,\lambda'\in\{c,v\}$.  For the long-range single-monolayer interaction considered here, the band dependence of the ordinary
carrier--carrier Coulomb kernel is neglected, and thus, $V_{\bm p}^{cc}= V_{\bm p}^{vv}=V_{\bm p}^{cv}
\equiv V_{\bm p}$.
For consistency with the exciton states used throughout this work, the
screened interaction $V_{\bm p}$ is evaluated with the same effective
Rytova--Keldysh model employed in
Appendix~\ref{app:rayleigh-ritz}.  The two-dimensional Fourier transform
of the real-space Rytova--Keldysh potential
in Eq.~\eqref{eq:keldysh} gives~\footnote{Notice that the momentum $\bm p$ in
Eq.~\eqref{eq:RK-momentum-space} is the momentum carried by the microscopic
carrier--carrier Coulomb interaction; in
Eq.~\eqref{eq:appA-xx-cc-oriented} it is $\bm p=\bm k-\bm k'$.
It must therefore not be identified with the center-of-mass momentum
transferred between the two excitons, which has already been set to zero by
the fixed-$\bm K_0$ projection.}
\begin{equation}
    V_{\bm p}
    =
    \frac{e^2}
    {2\varepsilon_0 A |\bm p|
    \left(\kappa_{\rm env}+r_0|\bm p|\right)}\,,
    \label{eq:RK-momentum-space}
\end{equation}
where $A$ is the monolayer normalization area.  Within this effective screening parametrization, the dielectric function reads from the previous expression, namely $ \varepsilon_{\bm p}^{\rm RK}=\kappa_{\rm env}+r_0|\bm p|$. For the $\mathrm{WSe_2}/hBN$ material considered here we use
$\kappa_{\rm env}=4.5$ and $r_0=4.5\,{\rm nm}$, consistent with the
exciton-spectrum calculation of Appendix~\ref{app:rayleigh-ritz}
and Ref.~\cite{Stier2018}.  

The quantity
$\mathcal E_{\rm cc}^{\,n'l';ln}(\bm K_0,\bm Q)$ in
Eq.~\eqref{eq:appA-xx-cc-oriented} corresponds to one of the two possible
assignments of the excitonic lines in the parent two-exciton Hamiltonian:
the first line is identified with the tagged exciton and the second with the
exciton gas.  Since the sums in the original $X-X$ Hamiltonian~\eqref{eq:gen-XX-Hamiltonian} run over both excitonic lines, the opposite assignment is also present and is obtained by
the simultaneous interchange $(n',n,\bm K_0)
    \longleftrightarrow
    (l',l,\bm Q)$. The tagged exciton - exciton gas cross interaction therefore receives the two contributions: $\mathcal E_{\rm cc}^{\,n'l';ln}(\bm K_0,\bm Q)$ and $\mathcal E_{\rm cc}^{\,l'n';nl}(\bm Q,\bm K_0)$. Because the excitonic operators in the effective Hamiltonian~\eqref{eq:gen-XX-Hamiltonian} are treated as
bosonic, these two terms multiply the same tagged--exciton gas operator product. Combining them with the overall factor $1/2$ of the generic
exciton--exciton Hamiltonian~\eqref{eq:gen-XX-Hamiltonian} gives the carrier--carrier exchange coefficient
entering Eq.~\eqref{eq:final-projected-Xg-Hamiltonian},
\begin{equation}
G_{\rm cc}^{\,n'l';ln}(\bm K_0,\bm Q,\bm 0)
=
\frac{1}{2}
\Big\{
\mathcal E_{\rm cc}^{\,n'l';ln}(\bm K_0,\bm Q)
+
\mathcal E_{\rm cc}^{\,l'n';nl}(\bm Q,\bm K_0)
\Big\}.
\label{eq:appA-xx-cc-symmetrized}
\end{equation}
Since the Coulomb matrix element in Eq.~\eqref{eq:RK-momentum-space} scales as $V_{\bm p}\propto A^{-1}$,
the exchange matrix elements inherit the same normalization-area
dependence. It is therefore convenient to introduce the area-independent interaction
vertices
\begin{equation}
\mathcal G_{\rm cc}^{\,n'l';ln}(\bm K_0,\bm Q)
\equiv
A\,G_{\rm cc}^{\,n'l';ln}(\bm K_0,\bm Q,0),
\label{eq:appC-Xg-area-scaled-G}
\end{equation}
This redefinition does not remove the normalization-area dependence of the
microscopic Hamiltonian~\eqref{eq:final-projected-Xg-Hamiltonian}; rather, it transfers that dependence from
the coupling vertex to an explicit prefactor $A^{-1}$. Thus, $\mathcal G_{\rm exch}$
carries units of energy times area. The explicit factor $A^{-1}$ is retained
at the Hamiltonian level and cancels only after the exciton-gas momentum sum
is taken in the thermodynamic limit, as shown in App.~\ref{app:exciton-gas-mean-field}.

\paragraph{Electron--hole exchange.}
In addition to the carrier--carrier contribution, the microscopic
exciton--exciton interaction contains an electron--hole exchange channel
associated with the long-range interband Coulomb interaction.  The
corresponding exciton-basis matrix element is conventionally evaluated by
neglecting its weak dependence on the incoming exciton center-of-mass
momenta~\cite{erkensten2021exciton}.  In the present work we retain this
dependence, since the fixed-$\bm K_0$ projection constrains only the
transferred center-of-mass momentum and leaves both the retained momentum
$\bm K_0$ of the tagged exciton and the gas-exciton momentum $\bm Q$
otherwise unrestricted.  We also reinstate independent internal states on
the incoming and outgoing excitonic lines.

We consider the fixed-$\bm K_0$ process $(n,\bm K_0)+(l,\bm Q)\longrightarrow (n',\bm K_0)+(l',\bm Q)$ for which the transferred exciton center-of-mass momentum vanishes. We rewrite the constituent momenta in terms of the exciton center-of-mass
momentum $\bm K$ and relative momentum $\bm k$. Since the hole of momentum
$\bm k_h$ is a missing valence electron of momentum $\bm k_v=-\bm k_h$, the
exciton operator~\eqref{eq:exciton-valence-form} gives
\begin{equation}
    \bm k_e=\bm k+\alpha_e\bm K,
    \qquad
    \bm k_v=\bm k-\alpha_h\bm K,
    \qquad
    \bm k_h=-\bm k_v=\alpha_h\bm K-\bm k,
    \label{eq:appA-xx-eh-constituent-momenta}
\end{equation}
so that $\bm K=\bm k_e-\bm k_v=\bm k_e+\bm k_h$. The interband exchange
kernel below is evaluated at electron-minus-valence-electron differences, so
it is $\bm k_v$ that enters $\bm p_1$ and $\bm p_2$.
There are two electron-hole exchange orientations.  In the first one, the
valence electron belonging to the tagged-exciton line is exchanged with the
electron belonging to the exciton gas line.  If $\bm k$ and $\bm k'$ denote the
relative momenta on the tagged- and exciton gas lines, respectively, the
microscopic momentum entering the interband Coulomb-exchange kernel is $\bm p_1
    =
    \left(\bm k'+\alpha_e\bm Q\right)
    -
    \left(\bm k-\alpha_h\bm K_0\right)
    =
    \bm k'-\bm k
    +\alpha_e\bm Q+\alpha_h\bm K_0$. The opposite orientation exchanges the electron on the tagged-exciton line
with the valence electron on the exciton gas line and therefore probes $\bm p_2=\left(\bm k+\alpha_e\bm K_0\right)-
\left(\bm k'-\alpha_h\bm Q\right)=\bm k-\bm k'+\alpha_e\bm K_0+\alpha_h\bm Q$. 
    
We denote by $\bar V_{\bm p}^{\xi}$ the long-range interband
Coulomb-exchange kernel in the selected valley $\xi$.  Using the
interband electric-dipole matrix element $\bm d_{cv}^{\xi}$~\eqref{eq:bloch-interband-dipole}, it can be written within the band-edge approximation as
\begin{equation}
    \bar V_{\bm p}^{\xi}
    =
    \frac{V_{\bm p}}{e^2}
    \left|\bm p\cdot\bm d_{cv}^{\xi}\right|^2 \,
    \label{eq:appA-xx-eh-kernel}
\end{equation}
where $V_{\bm p}$ is given by Eq.~\eqref{eq:RK-momentum-space}. Within the band-edge approximation adopted here, this kernel is real and even, $\bar V_{-\bm p}^{\xi}=\bar V_{\bm p}^{\xi}=\left[\bar V_{\bm p}^{\xi}\right]^*$. 

To allow the interaction to change the internal state of either exciton, we
introduce the transition density
\begin{equation}
    \rho_{n'n}(\bm k)
    \equiv
    \varphi_{n'}^*(\bm k)\varphi_n(\bm k).
    \label{eq:appA-xx-eh-transition-density}
\end{equation}

The two microscopic electron--hole exchange orientations then give,
respectively,
\begin{subequations}
\begin{align}
    \mathcal E_{\rm eh}^{(1)\,n'l';ln}
    (\bm K_0,\bm Q)
    &=
    \sum_{\bm k,\bm k'}
    \bar V_{
        \bm k'-\bm k+\alpha_e\bm Q+\alpha_h\bm K_0
    }^{\xi}\,
    \rho_{n'n}(\bm k)\,
    \rho_{l'l}(\bm k')\,,
    \label{eq:appA-xx-eh-first-orientation}
    \\
    \mathcal E_{\rm eh}^{(2)\,n'l';ln}
    (\bm K_0,\bm Q)
    &=
    \sum_{\bm k,\bm k'}
    \bar V_{
        \bm k-\bm k'+\alpha_e\bm K_0+\alpha_h\bm Q
    }^{\xi}\,
    \rho_{n'n}(\bm k)\,
    \rho_{l'l}(\bm k').
    \label{eq:appA-xx-eh-second-orientation}
\end{align}
\end{subequations}
Both exchange orientations are already contained in the microscopic
electron--hole exchange interaction.  The corresponding contribution to
the internal-state-resolved exciton--exciton vertex is therefore their sum,
\begin{equation}
\label{eq:appA-xx-eh-full}
\begin{split}
    G_{\rm eh}^{\,n'l';ln}
    (\bm K_0,\bm Q,\bm 0)
    &=
    \mathcal E_{\rm eh}^{(1)\,n'l';ln}
    (\bm K_0,\bm Q)
    +
    \mathcal E_{\rm eh}^{(2)\,n'l';ln}
    (\bm K_0,\bm Q)\,,\\
    &=
\sum_{\bm k,\bm k'}
\rho_{n'n}(\bm k)\,
\rho_{l'l}(\bm k')\,
\left[
\bar V_{
    \bm p+\alpha_e\bm Q+\alpha_h\bm K_0
}^{\xi}
+
\bar V_{
    \bm p-\alpha_e\bm K_0-\alpha_h\bm Q
}^{\xi}
\right]\,,
\end{split}
\end{equation}
where in the second line we used the evenness of
$\bar V_{\bm p}^{\xi}$ and defined $\bm p=\bm k'-\bm k$. Equation~\eqref{eq:appA-xx-eh-full} is the electron--hole exchange
coefficient used in the present fixed-$K_0$ projection~\footnote{Analogously to the carrier--carrier contribution introduced in the previous section, Eq.~\eqref{eq:appC-Xg-area-scaled-G}, we make the normalization-area scaling explicit at the Hamiltonian level by defining the rescaled electron--hole exchange coupling $\mathcal G_{\rm eh}\equiv A G_{\rm eh}$. Accordingly, the corresponding total exchange vertex is rescaled as $\mathcal G_{\rm exch}\equiv A G_{\rm exch}$.
}.  

The conventional low-center-of-mass approximation discussed in Ref.~\cite{erkensten2021exciton} follows as an additional
approximation. Introducing two momentum shifts, $\bm\delta_1
    =
    \alpha_e\bm Q+\alpha_h\bm K_0$ and $\bm\delta_2
    =
    \alpha_e\bm K_0+\alpha_h\bm Q$, 
one may neglect the incoming center-of-mass dependence when these shifts are
small compared with the momentum scale over which
$\bar V_{\bm p}^{\xi}$ varies. Therefore, $\bar V_{\bm p+\bm\delta_1}^{\xi}
    \simeq
    \bar V_{\bm p}^{\xi}$ and $\bar V_{\bm p-\bm\delta_2}^{\xi}
    \simeq
    \bar V_{\bm p}^{\xi}$. 
Under this additional approximation,
Eq.~\eqref{eq:appA-xx-eh-full} reduces to
\begin{equation}
    G_{\rm eh}^{\,n'l';ln}
    (\bm K_0,\bm Q,\bm 0)
    \simeq
    2
    \sum_{\bm k,\bm k'}
    \bar V_{\bm k'-\bm k}^{\xi}\,
    \rho_{n'n}(\bm k)\,
    \rho_{l'l}(\bm k')
\label{eq:appA-xx-eh-low-COM}
\end{equation}
and becomes exact within the present effective-mass expression when
$\bm K_0=\bm Q=\bm 0$.  For state-conserving scattering,
$n'=n$ and $m'=m$, the transition densities reduce to
$|\varphi_n|^2$ and $|\varphi_m|^2$, respectively, and
Eq.~\eqref{eq:appA-xx-eh-low-COM} recovers the conventional
zero-transfer monolayer electron--hole exchange structure discussed in Ref.~\cite{erkensten2021exciton}. We emphasize that this low-center-of-mass approximation is not implied by the fixed-$K_0$ projection. 

\section{General rotating-frame Lindblad equation for a driven excitonic three-level system}
\label{app:general_rotating_lindblad}

This appendix derives the generic Lindblad equation for the selected excitonic three-level system coupled to multiple environments. The goal is to show how the configuration-dependent driven Hamiltonian of Sects.~\ref{subsec:three_level_topologies} and~\ref{subsec:explicit_drives} is combined consistently with the reservoir couplings of Sect.~\ref{subsec:int_ham} and the bath-correlation tensors of App.~\ref{app:bath-corr}, with all quantities expressed in a common drive-rotating frame.
\paragraph{Projection onto the selected three-level configuration:}
We start our derivation from the projected fixed-$K_0$ microscopic Hamiltonians derived in Sect.~\ref{subsec:K0-projection}. The total Hamiltonian of the already kinematically restricted subsystem + reservoirs reads as
\begin{equation}
\ho{H}^{(\bm K_0)}(t)=\ho{H}_X^{(\bm K_0)}+\sum_{\alpha}\ho{H}_{\mathcal B_\alpha}+\ho{H}_{\rm drv}^{(\bm K_0)}(t)+\ho{H}^{(\bm K_0)}_{X-ph}+\ho{H}^{(\bm K_0)}_{X-X_g}+\ho{H}^{fluc,(\bm K_0)}_{X-\gamma}\,.
\label{eq:app_H_total_start}
\end{equation}
The projection onto a driven three-level system is performed before
constructing the reduced Lindblad equation. For a configuration \(\mathcal T\) selected where every exciton state of the three-level system already belongs to the fixed-\(K_0\) sector, we define its associated Hilbert space $H_\mathcal{T}$ and projector operator $\hat{P}_\mathcal{T}$ onto this space as
\begin{equation}
\label{eq:spanned-H-P}
H_{\mathcal T}
=
{\rm span}\{
|1\rangle_{\mathcal T},\,
|2\rangle_{\mathcal T},\,
|3\rangle_{\mathcal T}
\},
\qquad
\hat{P}_\mathcal{T}
=
\sum_{a\in \mathcal{T}}
|a\rangle\langle a| \,,
\end{equation}
where it follows that $\hat{P}_\mathcal{T}=\hat{P}_\mathcal{T}^\dagger$. Thus, every microscopic excitonic operator associated with the subsystem, whether entering the free or interaction Hamiltonians, is restricted to the selected three-level system according to
\begin{equation}
    \label{eq:proj-subs-operator}
    \ho{O}_{\mu,\mathcal T}
    =
    \hat P_{\mathcal T}\,\ho{O}_{\mu}\,\hat P_{\mathcal T}\,.
\end{equation}
The same restriction applies to the reduced subsystem density operator, which is defined on the retained three-level subspace, with its trace understood as being taken over this subspace. Only the projected operators \(\ho O_{\mu,\mathcal T}\) enter the
Born--Markov and secular construction below. To avoid overloading the notation, we suppress the configuration label \(\mathcal T\) in what follows, with the understanding that all subsystem operators act within the selected three-level system, while the excitonic states and their couplings to the reservoirs have already been restricted to the fixed-\(K_0\) sector in Sect.~\ref{subsec:K0-projection}. The subsystem and subsystem-reservoir interaction Hamiltonians take the form~\footnote{The fixed-$\bm K_0$ exciton--photon Hamiltonians~\eqref{eq:fixed-K0-X-photon} and the coherent-drive Hamiltonian obtained from their mean-field component are retained at this stage without a rotating-wave approximation. The RWA is introduced only after projection onto the selected three-level system and transformation to the appropriate interaction/drive-rotating representation.}
\begin{subequations}
\begin{align}
\ho{H}_S(t)&=\ho{H}_{0}+\ho{H}_{\rm drv}(t)\,,
\label{eq:app_HS_generic} \\
\ho{H}_{X-\alpha}&=
\sum_{\mu}\,
\ho{A}^{(\alpha)}_{\mu}\otimes \ho{B}^{(\alpha)}_{\mu},
\label{eq:app_HSB_generic}
\end{align}
\end{subequations}
where $\ho{A}^{(\alpha)}_\mu$ acts on the spanned Hilbert space of the projected subsystem, $H_{\mathcal{T}}$~\eqref{eq:spanned-H-P}, and $\ho{B}^{(\alpha)}_\mu$ acts on the reservoir $\alpha$~\cite{BreuerPetruccione,Lidar2019}. The couplings between the subsystem and reservoir, $\mathcal{W}^{\alpha}_\mu$, have been reabsorbed in the definition of the reservoir operator and thus $\ho{B}^{(\alpha)}_\mu\to \mathcal{W}^{(\alpha)}_\mu \ho{B}^{(\alpha)}_\mu$.

\paragraph{Interaction picture:}
Defining the free Hamiltonian
\begin{equation}
\ho{H}_{\rm free}=\ho{H}_{0}+\sum_\alpha \ho{H}_{B^\alpha},
\label{eq:app_Hfree}
\end{equation}
and the interaction-picture transformation $\tilde{\mathcal{O}}(t)=e^{+\frac{i}{\hbar}\ho{H}_{\rm free} t}\,\hat{\mathcal{O}}\,e^{-\frac{i}{\hbar}\ho{H}_{\rm free} t}$, the total density operator obeys
\begin{equation}
\frac{d\tilde\rho(t)}{dt}
=
-\frac{i}{\hbar}
\left[
\tilde{\mathcal{H}}_{\rm drv}(t)+\sum_\alpha \tilde{\mathcal{H}}_{X-\alpha}(t),\;
\tilde\rho(t)
\right].
\label{eq:app_chi_firstIP}
\end{equation}

\paragraph{Born--Markov reduction in the interaction picture:}
We first perform the reservoir reduction in the interaction picture introduced above, before applying the final drive-rotating transformation. Defining the residual subsystem--bath coupling
\begin{equation}
\tilde{\mathcal V}(t)
=
\sum_\alpha
\tilde{\mathcal H}_{X-\alpha}(t)
=
\sum_{\alpha,\mu}\,
\tilde{\mathcal A}^{(\alpha)}_\mu(t)
\otimes
\tilde{\mathcal B}^{(\alpha)}_\mu(t),
\label{eq:app_Vtilde}
\end{equation}
and assuming that the reservoirs remain stationary and mutually uncorrelated, i.e. $\rho_B
=
\rho_{\rm ph}\otimes\rho_\gamma\otimes\rho_{Xg},$,
the reduced density operator
$\tilde\rho_X(t)={\rm Tr}_B\,\tilde\rho(t)$
satisfies, within the Born--Markov approximation,
\begin{equation}
\frac{d\tilde\rho_X(t)}{dt}
=
-\frac{i}{\hbar}
\left[
\tilde{\mathcal H}_{\rm drv}(t),
\tilde\rho_X(t)
\right]
-\frac{1}{\hbar^2}
\int_0^\infty d\tau\;
{\rm Tr}_B
\left[
\tilde{\mathcal V}(t),
\left[
\tilde{\mathcal V}(t-\tau),
\tilde\rho_X(t)\otimes\rho_B
\right]
\right].
\label{eq:app_BornMarkov_rot}
\end{equation}
Only the fluctuating subsystem--reservoir coupling $\tilde{\mathcal V}$ enters the second-order kernel; the coherent drive is retained explicitly in the first-order subsystem dynamics and is not treated perturbatively in the reservoir coupling. In the bare-transition Born--Markov construction adopted here, the reservoir-induced terms are resolved with respect to the free subsystem Hamiltonian $\hat{\mathcal H}_0$, see Eq.~\eqref{eq:app_Hfree}. Consequently, the frequencies sampled by the bath spectra are the physical Bohr frequencies of the selected three-level system, independently of the subsequent drive-rotating transformation.

Because the reservoirs are independent, their cross-correlators vanish and the dissipative correction decomposes additively,
\begin{equation}
\frac{d\tilde\rho_X(t)}{dt}
=
-\frac{i}{\hbar}
\left[
\tilde{\mathcal H}_{\rm drv}(t),
\tilde\rho_X(t)
\right]
+
\sum_\alpha
\tilde{\mathcal K}_\alpha[\tilde\rho_X(t)],
\label{eq:app_Kalpha_sum}
\end{equation}
where
\begin{align}
\tilde{\mathcal K}_\alpha[\tilde\rho_X]
=
-\frac{1}{\hbar^2}
\int_0^\infty d\tau
\sum_{\mu\nu}
\Big\{&
\mathcal C^{(\alpha)}_{\mu\nu}(\tau)
\left[
\tilde{\mathcal A}^{(\alpha)}_\mu(t),
\tilde{\mathcal A}^{(\alpha)}_\nu(t-\tau)
\tilde\rho_X(t)
\right]+
\mathcal C^{(\alpha)}_{\nu\mu}(-\tau)
\left[
\tilde\rho_X(t)
\tilde{\mathcal A}^{(\alpha)}_\nu(t-\tau),
\tilde{\mathcal A}^{(\alpha)}_\mu(t)
\right]
\Big\},
\label{eq:app_Kalpha_corr}
\end{align}
and the corresponding bath correlation function is given by
\begin{equation}
\mathcal C^{(\alpha)}_{\mu\nu}(\tau)
=
\left\langle
\tilde{\mathcal B}^{(\alpha)}_\mu(\tau)
\hat{\mathcal B}^{(\alpha)}_\nu(0)
\right\rangle_{B^\alpha}\,.
\label{eq:app_Bmunu_again}
\end{equation}
Accordingly, the projected microscopic coupling strengths are already contained in the definition of the bath correlation. 
\paragraph{Bohr-frequency decomposition and secular approximation:}
The interaction-picture subsystem operators are decomposed into eigenoperators of the free subsystem Hamiltonian,
\begin{equation}
\tilde{\mathcal A}^{(\alpha)}_\mu(t)
=
\sum_{\varpi}
e^{-i\varpi t}\,
\mathcal A^{(\alpha)}_\mu(\varpi),
\qquad
\varpi\in\{0,\pm\omega^{\rm B}_{ab}\},\qquad\omega^{\rm B}_{ab}
=
\frac{|\epsilon_a-\epsilon_b|}{\hbar}>0
\label{eq:app_AOmega}
\end{equation}
where $\omega^{\rm B}_{ab}$ denotes the physical Bohr-frequency magnitude associated with a
transition within the retained three-level system. Accordingly,
$\varpi\in\{0,\pm\omega^{\rm B}_{ab}\}$ is a signed subsystem Bohr
frequency and should not be confused with either a drive frequency or
a drive detuning. Positive and negative values of $\varpi$ correspond
to the two opposite frequency sectors associated with the same physical
transition. The eigenoperators~\eqref{eq:app_AOmega} satisfy 
\begin{equation}
\left[
\hat{\mathcal H}_0,
\mathcal A^{(\alpha)}_\mu(\varpi)
\right]
=
-\hbar\varpi\,
\mathcal A^{(\alpha)}_\mu(\varpi).
\label{eq:app_AOmega_eigenoperator}
\end{equation}
Before the secular approximation, insertion of the decomposition
\eqref{eq:app_AOmega_eigenoperator} into the Born--Markov equation~\eqref{eq:app_Kalpha_sum}
generates terms with the generic frequency structure
\begin{equation}
\sum_{\varpi,\varpi'}
e^{i(\varpi'-\varpi)t}
\sum_{\mu,\nu}
\Gamma_{\mu\nu}^{(\alpha)}(\varpi)
\left[
A_{\nu}^{(\alpha)}(\varpi)\widetilde{\rho}_X
A_{\mu}^{(\alpha)\dagger}(\varpi')
-
A_{\mu}^{(\alpha)\dagger}(\varpi')
A_{\nu}^{(\alpha)}(\varpi)\widetilde{\rho}_X
\right]
+\mathrm{H.c.}
\label{eq:appB-presecular-frequency-structure}
\end{equation}
The one-sided Fourier transform of the bath correlator, including the
projected subsystem--reservoir coupling strengths, defines the spectral
correlation tensor~\cite{BreuerPetruccione},
\begin{equation}
\Gamma_{\mu\nu}^{(\alpha)}(\varpi)
=
\frac{1}{\hbar^2}
\int_0^\infty d\tau\,
e^{i\varpi\tau}
C_{\mu\nu}^{(\alpha)}(\tau)
\equiv
\frac{1}{2}\gamma_{\mu\nu}^{(\alpha)}(\varpi)
+
iS_{\mu\nu}^{(\alpha)}(\varpi).
\label{eq:appB-spectral-correlation}
\end{equation}

We adopt the full secular approximation throughout this work. Terms
with $\varpi\neq\varpi'$ in
Eq.~\eqref{eq:appB-presecular-frequency-structure} oscillate at the
difference frequency $\varpi'-\varpi$ and are neglected when this
oscillation is fast compared with the dissipative evolution. Only terms
belonging to the same Bohr-frequency block, $\varpi=\varpi'$, are therefore retained. This distinction is particularly relevant when a given Bohr frequency
is degenerate. If two distinct subsystem transitions $\mu\neq\nu$
possess the same Bohr frequency, they belong to the same secular block,
and the corresponding off-diagonal coefficients
$\gamma_{\mu\nu}^{(\alpha)}(\varpi)$ and
$S_{\mu\nu}^{(\alpha)}(\varpi)$ must in general be retained. By
contrast, when the nonzero Bohr-frequency sectors are nondegenerate,
each physical transition belongs to a separate secular block and
cross terms between distinct transition legs are removed.

The latter situation applies to the three-level systems selected in
Sect.~\ref{subsub:selected-top}. From the WSe$_2$/hBN spectrum summarized in Table~\ref{tab:appD_states}, the
retained photon-coupled transition frequencies are pairwise
nondegenerate within each configuration. In the retained $\Xi$ system, the interband optical
transition is distinct from the $1s\leftrightarrow2p$
transition of energy $113.19\,\mathrm{meV}$.
In the retained $V$ system, the two interband transition
frequencies differ by $124.16\,\mathrm{meV}/\hbar$. Both separations exceed every retained radiative
width by many orders of magnitude, so the quantitative secular criterion
$|\varpi-\varpi'|\gg\gamma$ is satisfied with a very large margin. Hence, no
two distinct retained photon-coupled legs belong to the same nonzero
Bohr-frequency sector. The full secular approximation therefore removes
their mutual cross terms.

The zero-frequency sector $\varpi=0$ is conceptually distinct: several
diagonal subsystem operators may belong to this same block, so
off-diagonal correlation coefficients within the zero-frequency block
are not eliminated by secularization alone. Their presence or absence
must instead be determined from the reservoir-specific projected
coupling, as done in App.~\ref{app:bath-corr}.

With the $\varpi\neq\varpi'$ sectors discarded, the interaction-picture
reduced equation takes the Lindblad form
\begin{equation}
\frac{d\tilde\rho_X}{dt}
=
-\frac{i}{\hbar}
\left[
\tilde{\mathcal H}_{\rm drv}(t)
+
\sum_\alpha
\mathcal H^{(\alpha)}_{LS},
\tilde\rho_X
\right]
+
\sum_\alpha
\mathcal D^{(\alpha)}[\tilde\rho_X],
\label{eq:app_master_interaction}
\end{equation}
where the Lamb shift Hamiltonian and the dissipator are given by respectively
\begin{subequations}
\begin{align}
\mathcal H^{(\alpha)}_{LS}
&=
\hbar
\sum_{\varpi}
\sum_{\mu\nu}
S^{(\alpha)}_{\mu\nu}(\varpi)\,
\mathcal A^{(\alpha)\dagger}_\mu(\varpi)\,
\mathcal A^{(\alpha)}_\nu(\varpi),
\label{eq:app_HLS_interaction} \\
\mathcal D^{(\alpha)}[\tilde\rho_X]
&=
\sum_{\varpi}
\sum_{\mu\nu}
\gamma^{(\alpha)}_{\mu\nu}(\varpi)
\Bigg[
\mathcal A^{(\alpha)}_\nu(\varpi)\,
\tilde\rho_X\,
\mathcal A^{(\alpha)\dagger}_\mu(\varpi)-
\frac{1}{2}
\left\{
\mathcal A^{(\alpha)\dagger}_\mu(\varpi)
\mathcal A^{(\alpha)}_\nu(\varpi),
\tilde\rho_X
\right\}
\Bigg].
\label{eq:app_D_interaction}
\end{align}
\end{subequations}

\paragraph{Final drive-rotating frame.}
The reduced equation above is written in the interaction
picture with respect to $\hat H_0$. The unitary
$\hat U_{\mathcal T}(t)$ introduced in Sect.~\ref{subsec:unified_three_level_ham} instead
defines the transformation from the Schr\"odinger
picture through
$\bar\rho_X=\hat U_{\mathcal T}^\dagger
\rho_X\hat U_{\mathcal T}$.
To connect the two representations, we define
\begin{equation}
\hat W_{\mathcal T}(t)
=
\hat U_{\mathcal T}^\dagger(t)e^{-i\hat H_0t/\hbar},
\qquad
\bar\rho_X(t)
=
\hat W_{\mathcal T}(t)\widetilde\rho_X(t)
\hat W_{\mathcal T}^\dagger(t).
\end{equation}

The coherent Hamiltonian therefore transforms as
\begin{equation}
\begin{split}
    \bar H_{\mathcal T}(t)
&=
\hat W_{\mathcal T}(t)\widetilde H_{\mathrm{drv}}(t)
\hat W_{\mathcal T}^\dagger(t)
+i\hbar\left(\frac{d\,{\hat W}_{\mathcal T}(t)}{dt}\right)
\hat W_{\mathcal T}^\dagger(t)\,,\\
&=
\hat U_{\mathcal T}^\dagger(t)
[\hat H_0+\hat H_{\mathrm{drv}}(t)]
\hat U_{\mathcal T}(t)
-i\hbar\hat U_{\mathcal T}^\dagger(t)
\dot{\hat U}_{\mathcal T}(t).
\end{split}
\end{equation}
Applying the rotating-wave approximation to the
coherent drive and subtracting the irrelevant scalar
$\epsilon_1 I$ gives the Hamiltonians of Sects.~\ref{subsec:unified_three_level_ham}
and~\ref{subsub:selected-top}. The operators entering the interaction-picture
reservoir terms transform according to
\begin{equation}
\bar A_\mu^{(\alpha)}(\varpi,t)
=
\hat W_{\mathcal T}(t)A_\mu^{(\alpha)}(\varpi)
\hat W_{\mathcal T}^\dagger(t).
\end{equation}
Importantly, this transformation changes only the representation of the subsystem transition operators; it does not change the frequency argument of the reservoir spectral tensors. The quantities
$\gamma^{(\alpha)}_{\mu\nu}(\varpi)$
and
$S^{(\alpha)}_{\mu\nu}(\varpi)$
therefore remain evaluated at the physical Bohr frequencies
$\varpi\in\{0,\pm\omega^{\rm B}_{ab}\}$,
rather than at the drive detunings.

The Lamb-shift Hamiltonian and dissipator in the drive-rotating frame are consequently~\footnote{For the transformations considered here,
$\hat W_{\mathcal T}(t)$ is diagonal. Writing
$\hat W_{\mathcal T}(t)
=\sum_a e^{-i\chi_a(t)}\sigma_{aa}$ gives
\[
\hat W_{\mathcal T}\sigma_{ab}
\hat W_{\mathcal T}^\dagger
=
e^{-i[\chi_a(t)-\chi_b(t)]}\sigma_{ab}.
\]
The diagonal operators are invariant. Within each
nondegenerate secular transition block, the phase
cancels between a jump operator and its Hermitian
conjugate in both the dissipator and the Lamb-shift
Hamiltonian. These reservoir contributions therefore
carry no residual explicit time dependence in the
final rotating frame.}
\begin{subequations}
\begin{align}
\bar{\mathcal H}^{(\alpha)}_{LS}(t)
&=
\hbar
\sum_{\varpi}
\sum_{\mu\nu}
S^{(\alpha)}_{\mu\nu}(\varpi)\,
\bar{\mathcal A}^{(\alpha)\dagger}_\mu(\varpi,t)\,
\bar{\mathcal A}^{(\alpha)}_\nu(\varpi,t)\,,
\label{eq:app_HLS_alpha_general} \\
\bar{\mathcal D}^{(\alpha)}_t[\bar\rho_X]
&=
\sum_{\varpi}
\sum_{\mu\nu}
\gamma^{(\alpha)}_{\mu\nu}(\varpi)
\Bigg[
\bar{\mathcal A}^{(\alpha)}_\nu(\varpi,t)\,
\bar\rho_X\,
\bar{\mathcal A}^{(\alpha)\dagger}_\mu(\varpi,t)-
\frac{1}{2}
\left\{
\bar{\mathcal A}^{(\alpha)\dagger}_\mu(\varpi,t)
\bar{\mathcal A}^{(\alpha)}_\nu(\varpi,t),
\bar\rho_X
\right\}
\Bigg]\,.
\end{align}
\end{subequations}

The resulting reduced equation in the common drive-rotating frame is
\begin{equation}
\frac{d\bar\rho_X}{dt}
=
-\frac{i}{\hbar}
\left[
\bar{\mathcal H}^{(\mathrm{RWA})}_{\mathcal T}(t)
+
\sum_\alpha
\bar{\mathcal H}^{(\alpha)}_{LS},
\bar\rho_X
\right]
+
\sum_\alpha
\bar{\mathcal D}^{(\alpha)}[\bar\rho_X].
\label{eq:app_final_master_general}
\end{equation}
Equation~\eqref{eq:app_final_master_general} is the general rotating-frame Lindblad equation for a given selected configuration.

\subsection{Common structural properties and consistency checks of the Lindblad Equation}
\label{app:consistency}
The component equations for the two retained configurations
$\mathcal T\in\{\Xi,V\}$, Eqs.~\eqref{eq:Xi-component-system} and ~\eqref{eq:V-component-system}. In the following, the configuration label on the
density-matrix elements and transition rates is suppressed;
all quantities refer to the selected three-level system, and
$\gamma_{a\to b}^{(\gamma)}=0$ whenever the corresponding
photon-induced transition is absent.

First, summing the three population equations gives
\begin{equation}
\dot{\bar\rho}_{11}
+\dot{\bar\rho}_{22}
+\dot{\bar\rho}_{33}=0,
\label{eq:common-trace-check}
\end{equation}
because every coherent-transfer term and every
photon-induced gain/loss term cancels pairwise. Hence,
conservation of probability follows, i.e. $\operatorname{Tr}\bar\rho_{\mathcal T}(t)=1$
whenever
$\operatorname{Tr}\bar\rho_{\mathcal T}(0)=1$.
Complex conjugation of each coherence equation also
reproduces the equation for the transposed matrix element,
so Hermiticity is preserved. The three populations and
three complex coherences therefore determine the full
density matrix; normalization leaves eight independent
real variables.

Second, when the coherent fields are switched off, the
populations obey the classical rate equations
\begin{equation}
\dot{\bar\rho}_{aa}
=
\sum_{b\neq a}
\left[
\gamma_{b\to a}^{(\gamma)}\bar\rho_{bb}
-
\gamma_{a\to b}^{(\gamma)}\bar\rho_{aa}
\right],
\qquad a=1,2,3.
\label{eq:common-undriven-populations}
\end{equation}
The first term describes population arriving at $\ket{a}$,
whereas the second describes population leaving it.
This form makes probability conservation explicit and
applies to each configuration with its retained photon-coupled
legs. Switching off the coherent fields leaves the exciton-gas
mean field intact. Its diagonality in the retained basis
implies
$[\hat H_{\mathrm{MF}},\bar\rho]_{aa}=0$,
so it produces no direct population transfer, although
its differential energy shifts remain in the coherence
equations. Any dependence of the photon rates on the
renormalized transition frequencies is contained in
their coefficients.

Finally, define the total photon-induced rate leaving
state $\ket{a}$ as
\begin{equation}
\Gamma_a^{\mathrm{out}}
=
\sum_{c\neq a}\gamma_{a\to c}^{(\gamma)}.
\label{eq:common-outgoing-rate}
\end{equation}
Within the independent, nondegenerate secular transition
blocks retained here, the dissipative contribution to
every coherence is
\begin{equation}
\left.\dot{\bar\rho}_{ab}\right|_{\mathrm{diss}}
=
-\frac12
\left(
\Gamma_a^{\mathrm{out}}+\Gamma_b^{\mathrm{out}}
\right)\bar\rho_{ab},
\qquad a\neq b.
\label{eq:common-coherence-damping}
\end{equation}
Thus, a coherence is directly damped by processes leaving
either of its endpoint states. Processes leaving the
third state do not contribute directly to its damping,
although they can affect its evolution indirectly
through the populations and coherent couplings.
In particular, the absence of a direct dipole leg
between two states does not by itself protect their
coherence from decay through other retained transitions.

\subsection{Component equations of the retained three-level systems}
\label{app:component-equations}
This subsection collects the component form of the reduced dynamics of Sect.~\ref{sec:lindblad_three_level} for the two retained three-level systems; the renormalized one- and two-photon detunings are those of Sect.~\ref{sec:VE-renormalization}.

\subsubsection{Ladder \texorpdfstring{$\Xi$}{Xi} system}
For the selected ladder system,
\begin{equation}
\begin{split}
H_\Xi&
=\operatorname{span}\{\ket{1},\ket{2},\ket{3}\}
\\&\equiv
\operatorname{span}\{\ket{0},\ket{1s,\bm K_0},\ket{2p,\bm K_0}\},    
\end{split}
\label{eq:Xi-manifold-components}
\end{equation}
the sequential legs $\ket{1}\leftrightarrow\ket{2}$ and
$\ket{2}\leftrightarrow\ket{3}$ are driven by an interband optical
field and an intra-excitonic THz/mid-IR field, respectively.  The direct
interband transition $\ket{1}\leftrightarrow\ket{3}$ is absent at
electric-dipole order because $\ket{3}\equiv\ket{2p,\bm K_0}$ is
interband-dark within the envelope-function approximation.  With
$\sigma_{ab}=\ket{a}\bra{b}$, the coherent Hamiltonian in the rotating
frame is 
\begin{equation}
\begin{split}
    \bar{\mathcal{H}}_{\Xi,\mathrm{eff}}&^{(\mathrm{RWA})}(t)
=+\hbar\ren\Delta_{12}\sigma_{22}
+\hbar\ren\delta_\Xi\sigma_{33}\\
&+\frac{\hbar}{2}\left[
\Omega_{12}(t)e^{i\phi_{12}}\sigma_{21}
+\Omega_{23}(t)e^{i\phi_{23}}\sigma_{32}
+\mathrm{h.c.}
\right].
\end{split}
\label{eq:Xi-Hamiltonian-components}
\end{equation}
Here the renormalized one-photon and two-photon detunings are respectively
\begin{subequations}
\label{eq:Xi-renormalized-detunings-components}
\begin{align}
\ren\Delta_{12}
&=\Delta_{12}+\frac{\Sigma_{2}^{(X_g,0)}}{\hbar},
\label{eq:Xi-renormalized-detuning-12-components}\\
\ren\Delta_{23}
&=\Delta_{23}
+\frac{\Sigma_{3}^{(X_g,0)}-\Sigma_{2}^{(X_g,0)}}{\hbar},
\label{eq:Xi-renormalized-detuning-23-components}\\
\ren\delta_\Xi
&\equiv\ren\Delta_{12}+\ren\Delta_{23}
=\delta_\Xi+\frac{\Sigma_{3}^{(X_g,0)}}{\hbar},
\label{eq:Xi-two-photon-detuning-components}
\end{align}
\end{subequations}
where $\delta_\Xi=\Delta_{12}+\Delta_{23}$. The photon dissipator is
\begin{align}
\bar{\mathcal D}_{\Xi}^{(\gamma)}[\bar\rho_\Xi]
={}&\gamma_{2\to1}^{(\gamma)}
\mathcal L_{\sigma_{12}}[\bar\rho_\Xi]
+\gamma_{1\to2}^{(\gamma)}
\mathcal L_{\sigma_{21}}[\bar\rho_\Xi]
\notag\\
&+\gamma_{3\to2}^{(\gamma)}
\mathcal L_{\sigma_{23}}[\bar\rho_\Xi]
+\gamma_{2\to3}^{(\gamma)}
\mathcal L_{\sigma_{32}}[\bar\rho_\Xi]\,.
\label{eq:Xi-dissipator-components}
\end{align}
The reduced dynamics therefore obeys
\begin{equation}
\dot{\bar\rho}_\Xi
=-\frac{i}{\hbar}
\left[
\bar{\mathcal{H}}_{\Xi,\mathrm{eff}}^{(\mathrm{RWA})}(t),
\bar\rho_\Xi
\right]
+\bar{\mathcal D}_{\Xi}^{(\gamma)}[\bar\rho_\Xi].
\label{eq:Xi-master-components}
\end{equation}
Thus, the three populations and
three independent coherences of the reduced density matrix consequently satisfy
\begin{subequations}
\label{eq:Xi-component-system}
\begin{align}
\dot{\bar\rho}_{11}
={}&\frac{i}{2}\Big[
\Omega_{12}(t)e^{i\phi_{12}}\bar\rho_{12}
-\Omega_{12}(t)e^{-i\phi_{12}}\bar\rho_{21}
\Big]
-\gamma_{1\to2}^{(\gamma)}\bar\rho_{11}
+\gamma_{2\to1}^{(\gamma)}\bar\rho_{22},
\label{eq:Xi-component-11}\\
\dot{\bar\rho}_{22}
={}&\frac{i}{2}\Big[
\Omega_{12}(t)e^{-i\phi_{12}}\bar\rho_{21}
-\Omega_{12}(t)e^{i\phi_{12}}\bar\rho_{12}
+\Omega_{23}(t)e^{i\phi_{23}}\bar\rho_{23}
-\Omega_{23}(t)e^{-i\phi_{23}}\bar\rho_{32}
\Big]
\notag\\
&+\gamma_{1\to2}^{(\gamma)}\bar\rho_{11}
-\left[
\gamma_{2\to1}^{(\gamma)}+\gamma_{2\to3}^{(\gamma)}
\right]\bar\rho_{22}
+\gamma_{3\to2}^{(\gamma)}\bar\rho_{33},
\label{eq:Xi-component-22}\\
\dot{\bar\rho}_{33}
={}&\frac{i}{2}\Big[
\Omega_{23}(t)e^{-i\phi_{23}}\bar\rho_{32}
-\Omega_{23}(t)e^{i\phi_{23}}\bar\rho_{23}
\Big]
+\gamma_{2\to3}^{(\gamma)}\bar\rho_{22}
-\gamma_{3\to2}^{(\gamma)}\bar\rho_{33},
\label{eq:Xi-component-33}\\
\dot{\bar\rho}_{12}
={}&+i\ren\Delta_{12}\bar\rho_{12}
-\frac{i}{2}\Omega_{12}(t)e^{-i\phi_{12}}
\left(\bar\rho_{22}-\bar\rho_{11}\right)
+\frac{i}{2}\Omega_{23}(t)e^{i\phi_{23}}\bar\rho_{13}
\notag\\
&-\frac{1}{2}\left[
\gamma_{1\to2}^{(\gamma)}+\gamma_{2\to1}^{(\gamma)}
+\gamma_{2\to3}^{(\gamma)}
\right]\bar\rho_{12},
\label{eq:Xi-component-12}\\
\dot{\bar\rho}_{23}
={}&+i\ren\Delta_{23}\bar\rho_{23}
-\frac{i}{2}\Omega_{12}(t)e^{i\phi_{12}}\bar\rho_{13}
-\frac{i}{2}\Omega_{23}(t)e^{-i\phi_{23}}
\left(\bar\rho_{33}-\bar\rho_{22}\right)
\notag\\
&-\frac{1}{2}\left[
\gamma_{2\to1}^{(\gamma)}+\gamma_{2\to3}^{(\gamma)}
+\gamma_{3\to2}^{(\gamma)}
\right]\bar\rho_{23},
\label{eq:Xi-component-23}\\
\dot{\bar\rho}_{13}
={}&+i\ren\delta_\Xi\bar\rho_{13}
-\frac{i}{2}\Omega_{12}(t)e^{-i\phi_{12}}\bar\rho_{23}
+\frac{i}{2}\Omega_{23}(t)e^{-i\phi_{23}}\bar\rho_{12}
\notag\\
&-\frac{1}{2}\left[
\gamma_{1\to2}^{(\gamma)}+\gamma_{3\to2}^{(\gamma)}
\right]\bar\rho_{13}.
\label{eq:Xi-component-13}
\end{align}
\end{subequations}

\subsubsection{\texorpdfstring{$V$}{V} system}
For the selected $V$ system,
\begin{equation}
\begin{split}
    H_V
&=\operatorname{span}\{\ket{1},\ket{2},\ket{3}\}\,,\\
&\equiv
\operatorname{span}\{\ket{0},\ket{1s,\bm K_0},\ket{2s,\bm K_0}\},
\end{split}
\label{eq:V-manifold-components}
\end{equation}
the two interband optical fields drive the legs
$\ket{1}\leftrightarrow\ket{2}$ and
$\ket{1}\leftrightarrow\ket{3}$.  There is no direct
electric-dipole coupling between the two excited states.

With
$\sigma_{ab}=\ket{a}\bra{b}$, the coherent Hamiltonian in the rotating
frame is
\begin{equation}
\begin{split}
    \bar{\mathcal{H}}_{V,\mathrm{eff}}&^{(\mathrm{RWA})}(t)
=+\hbar\ren\Delta_{12}\sigma_{22}
+\hbar\ren\Delta_{13}\sigma_{33}\\
&+\frac{\hbar}{2}\left[
\Omega_{12}(t)e^{i\phi_{12}}\sigma_{21}
+\Omega_{13}(t)e^{i\phi_{13}}\sigma_{31}
+\mathrm{h.c.}
\right]\,.
\end{split}
\label{eq:V-Hamiltonian-components}
\end{equation}
Here, the mean-field-renormalized
one and two photon detunings are
\begin{subequations}
\label{eq:V-renormalized-detunings}
\begin{align}
\ren\Delta_{12}
&\equiv \omega_{L,12}-\ren\omega_{12}
=\Delta_{12}+\frac{\Sigma_2^{(X_g,0)}}{\hbar},
\\
\ren\Delta_{13}
&\equiv \omega_{L,13}-\ren\omega_{13}
=\Delta_{13}+\frac{\Sigma_3^{(X_g,0)}}{\hbar},\\
\ren\delta_V
&\equiv\ren\Delta_{13}-\ren\Delta_{12}
=\delta_V+
\frac{\Sigma_{3}^{(X_g,0)}-\Sigma_{2}^{(X_g,0)}}{\hbar}
\label{eq:V-Raman-mismatch-components}
\end{align}
\end{subequations}

The dissipator generated by the photon fluctuations is
\begin{align}
\bar{\mathcal D}_{V}^{(\gamma)}[\bar\rho_V]
={}&\gamma_{2\to1}^{(\gamma)}\mathcal L_{\sigma_{12}}[\bar\rho_V]
+\gamma_{1\to2}^{(\gamma)}\mathcal L_{\sigma_{21}}[\bar\rho_V]
\notag\\
&+\gamma_{3\to1}^{(\gamma)}\mathcal L_{\sigma_{13}}[\bar\rho_V]
+\gamma_{1\to3}^{(\gamma)}\mathcal L_{\sigma_{31}}[\bar\rho_V],
\label{eq:V-dissipator-components}
\end{align}
where
$\mathcal L_L[\rho]=L\rho L^\dagger-	\frac12\{L^\dagger L,\rho\}$.
The reduced dynamics therefore obeys
\begin{equation}
\dot{\bar\rho}_V
=-\frac{i}{\hbar}
\left[
\bar{\mathcal{H}}_{V,\mathrm{eff}}^{(\mathrm{RWA})}(t),
\bar\rho_V
\right]
+\bar{\mathcal D}_{V}^{(\gamma)}[\bar\rho_V].
\label{eq:V-master-components}
\end{equation}
The components of the reduced density matrix, three populations and
three independent coherences, satisfy the following set of ODEs
\begin{subequations}
\label{eq:V-component-system}
\begin{align}
\dot{\bar\rho}_{11}
={}&\frac{i}{2}\Big[
\Omega_{12}(t)e^{i\phi_{12}}\bar\rho_{12}
-\Omega_{12}(t)e^{-i\phi_{12}}\bar\rho_{21}
+\Omega_{13}(t)e^{i\phi_{13}}\bar\rho_{13}
-\Omega_{13}(t)e^{-i\phi_{13}}\bar\rho_{31}
\Big]
\notag\\
&-\left(\gamma_{1\to2}^{(\gamma)}
+\gamma_{1\to3}^{(\gamma)}\right)\bar\rho_{11}
+\gamma_{2\to1}^{(\gamma)}\bar\rho_{22}
+\gamma_{3\to1}^{(\gamma)}\bar\rho_{33},
\label{eq:V-component-11}\\
\dot{\bar\rho}_{22}
={}&\frac{i}{2}\Big[
\Omega_{12}(t)e^{-i\phi_{12}}\bar\rho_{21}
-\Omega_{12}(t)e^{i\phi_{12}}\bar\rho_{12}
\Big]
+\gamma_{1\to2}^{(\gamma)}\bar\rho_{11}
-\gamma_{2\to1}^{(\gamma)}\bar\rho_{22},
\label{eq:V-component-22}\\
\dot{\bar\rho}_{33}
={}&\frac{i}{2}\Big[
\Omega_{13}(t)e^{-i\phi_{13}}\bar\rho_{31}
-\Omega_{13}(t)e^{i\phi_{13}}\bar\rho_{13}
\Big]
+\gamma_{1\to3}^{(\gamma)}\bar\rho_{11}
-\gamma_{3\to1}^{(\gamma)}\bar\rho_{33},
\label{eq:V-component-33}\\
\dot{\bar\rho}_{12}
={}&+i\ren\Delta_{12}\bar\rho_{12}
-\frac{i}{2}\Omega_{12}(t)e^{-i\phi_{12}}
\left(\bar\rho_{22}-\bar\rho_{11}\right)
-\frac{i}{2}\Omega_{13}(t)e^{-i\phi_{13}}\bar\rho_{32}
\notag\\
&-\frac{1}{2}\left[
\gamma_{1\to2}^{(\gamma)}+\gamma_{1\to3}^{(\gamma)}
+\gamma_{2\to1}^{(\gamma)}
\right]\bar\rho_{12},
\label{eq:V-component-12}\\
\dot{\bar\rho}_{13}
={}&+i\ren\Delta_{13}\bar\rho_{13}
-\frac{i}{2}\Omega_{12}(t)e^{-i\phi_{12}}\bar\rho_{23}
-\frac{i}{2}\Omega_{13}(t)e^{-i\phi_{13}}
\left(\bar\rho_{33}-\bar\rho_{11}\right)
\notag\\
&-\frac{1}{2}\left[
\gamma_{1\to2}^{(\gamma)}+\gamma_{1\to3}^{(\gamma)}
+\gamma_{3\to1}^{(\gamma)}
\right]\bar\rho_{13},
\label{eq:V-component-13}\\
\dot{\bar\rho}_{23}
={}&+i\ren\delta_V\bar\rho_{23}
-\frac{i}{2}\Omega_{12}(t)e^{i\phi_{12}}\bar\rho_{13}
+\frac{i}{2}\Omega_{13}(t)e^{-i\phi_{13}}\bar\rho_{21}
\notag\\
&-\frac{1}{2}\left[
\gamma_{2\to1}^{(\gamma)}+\gamma_{3\to1}^{(\gamma)}
\right]\bar\rho_{23}.
\label{eq:V-component-23}
\end{align}
\end{subequations}

\section{Bath correlators, decay-rate and Lamb-shift matrices}
\label{app:bath-corr}

This appendix collects the reservoir correlation functions, decay-rate matrices, and Lamb-shift coefficients that enter the Lindblad equation~\eqref{eq:app_final_master_general}. From the spectral correlation tensor~\eqref{eq:appB-spectral-correlation}, the decay-rate and Lamb-shift matrices can be written as~\cite{BreuerPetruccione,Lidar2019}
\begin{subequations}
 \begin{align}     
\label{eq:gamma_FT_app}
\gamma^{(\alpha)}_{\mu\nu}(\varpi)&=\frac{1}{\hbar^2}\int_{-\infty}^{+\infty}dt\,e^{i\varpi t}\,\mathcal C^{(\alpha)}_{\mu\nu}(t),
\\
\label{eq:S_PV_app}
S^{(\alpha)}_{\mu\nu}(\varpi)&=\frac{1}{2\pi}\,\mathcal{PV}\!\int_{-\infty}^{+\infty}d\varpi'\,
\frac{\gamma^{(\alpha)}_{\mu\nu}(\varpi')}{\varpi-\varpi'}\,,
 \end{align}  
\end{subequations}
where $\mathcal{PV}$ denotes the Cauchy principal value. Stationarity and Hermiticity of $\rho_{B_\alpha}$ imply that $\gamma^{(\alpha)}(\varpi)$ is Hermitian and positive semidefinite for all Bohr frequencies~\cite{BreuerPetruccione,Lidar2019}.

\subsection{Phonon reservoir}
\label{app-subsec:phonon}
As demonstrated in App.~\ref{app:K0-projection-phonons}, the interacting $X-ph$ Hamiltonian~\eqref{eq:X-ph_Ham} vanishes identically in the fixed-$\bm K_0$ limit after projection onto an excitonic three-level system with fixed center-of-mass momentum $K_0$. Hence, in the
fixed-$\bm K_0$ limit, acoustic phonon channel produces neither a
Markovian pure-dephasing rate nor a dynamical Lamb shift so
\begin{equation}
\bar{\mathcal{D}}_{\mathcal{T}}^{(\mathrm{ph})}=0,
\qquad
\bar{\mathcal{H}}_{\mathrm{LS},\mathcal{T}}^{(\mathrm{ph})}=0.
\label{eq:phonon_vanishing_q0}
\end{equation}
This vanishing is a consequence of the fixed-$\bm K_0$ limit, not a statement about real
monolayer semiconductors~\cite{selig2016excitonic,himami2018}: there, finite-$q$ exciton--phonon
scattering, phonon sidebands, intervalley channels and disorder-assisted scattering broaden the
excitons (Sect.~\ref{subsec:scope}), and App.~\ref{app:dephasing} estimates their effect on the
$\Xi$ results.

\subsection{Photon reservoir}
\label{app:photon-reservoir}
We now consider the fluctuation sector of the quantized electromagnetic
field. Contrary to the unprojected three-dimensional photon continuum,
the fixed-$\bm K_0$ projection derived in App.~\ref{app:K0-projection} fixes the in-plane photon
momentum while leaving its out-of-plane component unrestricted. For
intra-excitonic transitions the projection imposes
$\bm s_\parallel=\bm 0$, whereas for interband transitions it imposes
$\bm s_\parallel=\bm K_0$. Consequently, the photon reservoir entering
the Born--Markov construction is reduced to a one-dimensional continuum
over $s_\perp$.

We further distinguish the polarization structure of the two frequency
sectors entering the selected three-level systems. The interband optical
sector is coupled to the full transverse-polarization photon reservoir
compatible with $\bm s_\parallel=\bm K_0$. By contrast, for the
intra-excitonic THz/mid-IR sector we retain a single in-plane linear
polarization $\bm e_d$, common to the coherent drive and the corresponding
quantum fluctuations. Since $\bm s_\parallel=\bm 0$ for every
intra-excitonic leg, $\bm e_d$ is transverse to all surviving photon modes
$\vec s=(\bm 0,s_\perp)$. We assume that the polarization-selective environment suppresses the
orthogonal polarization channel without appreciably altering the electromagnetic mode spectrum seen by the retained polarization over the relevant intra-excitonic frequencies. The spectral density of the retained THz/mid-IR transitions can therefore be taken to be the same as in the fixed-$\bm K_0$ continuum derived below.

\paragraph{Projected fluctuation coupling:} After separating the coherent field amplitudes, already included in
$\ho{H}_{\rm drv}$~\eqref{eq:Hdrv_def}, from the quantum fluctuations, the projected
intra-excitonic and interband couplings of Eqs.~\eqref{eq:projected-intraband-length-gauge} and~\eqref{eq:projected-interband-length-gauge} can be written in a common form. It is convenient to label each photon-coupled transition by an oriented leg $\mu=(h_\mu,\ell_\mu)$, where $h_\mu$ and $l_\mu$ denote, respectively, the higher- and lower-energy states of the corresponding transition,
and to define 
\begin{equation}
\label{eq:jump-photon}
    \hat L_\mu=\ket{\ell_\mu}\bra{h_\mu}\,,\qquad\,{\omega^{\rm B}_{\mu}} =\frac{\epsilon_{h_\mu}-\epsilon_{l_\mu}}{\hbar}>0
\end{equation}
In addition, we denote the vector $\bm D_\mu$ as the electric-dipole matrix element defined as
\begin{equation}
\bm D_\mu =
\begin{cases}
\bm D^{\rm intra}_{h_\mu\ell_\mu}, & \mu\in{\rm intra},\\[1mm]
\bm D^{\rm inter}_{h_\mu\ell_\mu}(\bm K_0), & \mu\in{\rm inter},
\end{cases}
\qquad
\bm s_{\parallel,\mu} =
\begin{cases}
\bm 0, & \mu\in{\rm intra},\\[1mm]
\bm K_0, & \mu\in{\rm inter}.
\end{cases}
\end{equation}

The fluctuation part of both projected photon couplings can therefore be
written compactly as
\begin{equation}
\hat H_{X-\gamma}^{(\mathrm{fluc})}
=
-
\sum_\mu
\left(
\hat L_\mu^\dagger\bm D_\mu
+
\hat L_\mu\bm D_\mu^*
\right)
\cdot
\delta\hat{\bm E}_{\mu},
\label{eq:appC-fluc-coupling}
\end{equation}
where
$\delta\hat{\bm E}_{\mu}
=
\delta\hat{\bm E}^{(+)}_{\mu}
+
\delta\hat{\bm E}^{(-)}_{\mu}$.
The positive- and negative-frequency reservoir sectors are subsequently
separated by the Bohr-frequency decomposition and secular construction
of App.~\ref{app:general_rotating_lindblad}.

For a quantization volume $V=AL_\gamma$, the positive-frequency
fluctuation field associated with a photon-coupled leg $\mu$ is written as
\begin{equation}
\delta\hat{\bm E}^{(+)}_{\mu}(t)
=
i
\sum_{s_\perp}
\sum_{\lambda\in\mathcal P_\mu}
\sqrt{
\frac{
\hbar\omega_{\bm s_{\parallel,\mu},s_\perp}
}{
2\epsilon_0AL_\gamma
}
}
\,
\bm e_{\bm s_{\parallel,\mu},s_\perp,\lambda}
\,
\delta\hat a_{\bm s_{\parallel,\mu},s_\perp,\lambda}
e^{-i\omega_{\bm s_{\parallel,\mu},s_\perp}t},
\label{eq:appC-fluct-field}
\end{equation}
where
\begin{equation}
\mathcal P_\mu
=
\begin{cases}
\{d\},
&
\mu\in\mathrm{intra},
\\[1mm]
\{1,2\},
&
\mu\in\mathrm{inter}.
\end{cases}
\label{eq:appC-pol-sectors}
\end{equation}
For an intra-excitonic leg,
$\bm s_{\parallel,\mu}=\bm0$ and
$\bm e_{\bm0,s_\perp,d}\equiv\bm e_d$ is the selected in-plane
linear polarization. For an interband leg,
$\bm s_{\parallel,\mu}=\bm K_0$ and
$\lambda=1,2$ labels the complete pair of transverse photon
polarizations.

The fluctuation reservoir is assumed stationary, with
\begin{align}
\left\langle
\delta\hat a_{\vec s,\lambda}
\delta\hat a_{\vec s',\lambda'}^\dagger
\right\rangle_\gamma
&=
\delta_{\vec s,\vec s'}
\delta_{\lambda\lambda'}
\left[n_\gamma(\omega_{\vec s})+1\right],
\\
\left\langle
\delta\hat a_{\vec s,\lambda}^\dagger
\delta\hat a_{\vec s',\lambda'}
\right\rangle_\gamma
&=
\delta_{\vec s,\vec s'}
\delta_{\lambda\lambda'}
n_\gamma(\omega_{\vec s}),
\label{eq:appC-photon-correlators}
\end{align}
where
$n_\gamma(\omega)
=
[\exp(\beta_\gamma\hbar\omega)-1]^{-1}$.
\paragraph{Momentum-constrained bath correlator.}
The polarization contraction entering the bath correlator depends on whether the photon-coupled leg is intra-excitonic or interband. For the polarization-selected intra-excitonic sector,
\begin{equation}
\bm P^{\mathrm{intra}}
=
\bm e_d\bm e_d^{\,T},
\label{eq:appC-P-intra}
\end{equation}
whereas the interband sector retains the complete transverse-polarization
sum. Since the retained dipoles lie in the monolayer plane, its relevant
in-plane block remains
\begin{equation}
\left[
\bm P^{(\rm inter)}
\right]_{ij}
=
\delta_{ij}
-
\frac{c^2}{\omega^2}
s_{\parallel,i}s_{\parallel,j},
\qquad
i,j\in\{x,y\}.
\label{eq:appC-P-inter}
\end{equation}
Accordingly, for transition legs belonging to the same photon sector,
the dipole contraction entering the spectral matrix is
\begin{equation}
F_{\mu\nu}^{(\gamma)}
=
\begin{cases}
\left(
\bm D_\mu^*\cdot\bm e_d
\right)
\left(
\bm e_d\cdot\bm D_\nu
\right),
&
\mu,\nu\in\mathrm{intra},
\\[2mm]
\bm D_\mu^*
\cdot
\bm P^{(\rm inter)}(\omega,\bm K_0)
\cdot
\bm D_\nu,
&
\mu,\nu\in\mathrm{inter}.
\end{cases}
\label{eq:appC-F-hybrid}
\end{equation}
where the intra and interband transition dipoles are given by Eqs.~\eqref{eq:intra-exciton-polarization-general} and~\eqref{eq:projected-inter-dipole}. The selected interband and intra-excitonic transition frequencies are
nondegenerate, so mixed interband--intra-excitonic correlation terms do
not contribute to the full-secular generator derived in App.~\ref{app:general_rotating_lindblad}.

Substituting the field expansion~\eqref{eq:appC-fluct-field} into the reservoir
correlator~\eqref{eq:app_Bmunu_again} and using Eq.~\eqref{eq:appC-F-hybrid}, the polarization sum above gives,
before taking the continuum limit,
\begin{equation}
\begin{split}
C_{\mu\nu}^{(\gamma)}(\tau)
={}&
\delta_{\bm s_{\parallel,\mu},\bm s_{\parallel,\nu}}
\frac{\hbar}{2\epsilon_0 A L_{\gamma}}
\sum_{s_{\perp}}
\omega_{\bm s_{\parallel,\mu},s_{\perp}}\,
F_{\mu\nu}^{(\gamma)}\,
\Big\{
[n_{\gamma}(\omega_{\bm s_{\parallel,\mu},s_{\perp}})+1]
e^{-i\omega_{\bm s_{\parallel,\mu},s_{\perp}}\tau}
+
n_{\gamma}(\omega_{\bm s_{\parallel,\mu},s_{\perp}})
e^{+i\omega_{\bm s_{\parallel,\mu},s_{\perp}}\tau}
\Big\}.
\end{split}
\label{eq:proj-phot-correlator}
\end{equation}
The Kronecker delta follows directly from the photon-mode correlators:
transition legs belonging to different fixed in-plane momentum sectors
couple to independent photon modes. The fixed-$\bm K_0$ projection fixes $\bm s_{\parallel}$ but leaves the
perpendicular momentum $s_{\perp}$ continuous. We therefore use $\sum_{s_{\perp}}
\longrightarrow\,\left(L_{\gamma}/[2\pi]\right)\int_{-\infty}^{+\infty} ds_{\perp}$, so that the arbitrary perpendicular quantization length $L_{\gamma}$
cancels. Thus, after some algebra and using the on-shell condition for the photon field, the time dependent continuum photon correlator is
\begin{equation}
\begin{split}
C_{\mu\nu}^{(\gamma)}(\tau)
={}&
\delta_{\bm s_{\parallel,\mu},\bm s_{\parallel,\nu}}
\frac{\hbar}{2\pi\epsilon_0 A c}
\int_{\omega_{c,\mu}}^{\infty} d\omega\,
\frac{\omega^2}
{\sqrt{\omega^2-\omega_{c,\mu}^2}}\,
F_{\mu\nu}(\omega,\bm s_{\parallel,\mu})
\,
\Big\{
[n_{\gamma}(\omega)+1]e^{-i\omega\tau}
+
n_{\gamma}(\omega)e^{+i\omega\tau}
\Big\}.
\end{split}
\end{equation}
The lower limit is fixed kinematically by the prescribed in-plane photon momentum $\omega_{c,\mu}=c|\bm s_{\parallel,\mu}|$. Thus, $\omega_{c,\mu}$ is the minimum photon frequency compatible with the fixed in-plane momentum sector. In particular, $\omega_{c,\mu}=0$ for an intra-excitonic leg, where $\bm s_{\parallel,\mu}=\bm 0$, whereas $\omega_{c,\mu}=c|\bm K_0|$ for an interband leg.

It is useful to collect the dipole contraction, the momentum-constrained
photon density of states, and the kinematic threshold into the
positive-frequency spectral matrix
\begin{equation}
\label{eq:photon-spectral}
J_{\mu\nu}^{(\gamma)}
(\omega;\bm s_{\parallel,\mu})
=
\delta_{\bm s_{\parallel,\mu},\bm s_{\parallel,\nu}}
\frac{\omega^2}
{\hbar\epsilon_0 A c
\sqrt{\omega^2-\omega_{c,\mu}^2}}\,
F_{\mu\nu}^{(\gamma)}\,
\Theta(\omega-\omega_{c,\mu}).
\end{equation}
With this definition the correlation function~\eqref{eq:proj-phot-correlator} can equivalently be written
as
\begin{equation}
    \label{eq:proj-phot-correlator-2}
C_{\mu\nu}^{(\gamma)}(\tau)
=
\frac{\hbar^2}{2\pi}
\int_{0}^{\infty}d\omega\,
J_{\mu\nu}^{(\gamma)}
(\omega;\bm s_{\parallel,\mu})
\Big\{
[n_{\gamma}(\omega)+1]e^{-i\omega\tau}
+
n_{\gamma}(\omega)e^{+i\omega\tau}
\Big\}.
\end{equation}
The microscopic electric-dipole matrix elements are therefore already
contained in $J_{\mu\nu}^{(\gamma)}$, and consequently in the
decay rate and Lamb-shift coefficients constructed from it below.

\paragraph{Decay-rate matrix:} Given a spectral function, it is possible to calculate directly the decay rate matrix~\cite{BreuerPetruccione,Lidar2019}. Using Eq.~\eqref{eq:photon-spectral}, the photon decay rate matrix is
\begin{equation}
    \gamma_{\mu\nu}^{(\gamma)}(\varpi)
    =
    \mathcal J_{\mu\nu}^{(\gamma)}
    (|\varpi|;\bm s_{\parallel,\mu})
    \begin{cases}
        n_\gamma(\varpi)+1,
        & \varpi>0,
        \\[1mm]
        n_\gamma(|\varpi|),
        & \varpi<0.
    \end{cases}
    \label{eq:photon-fixedK-gamma}
\end{equation}
The photon spectral density~\eqref{eq:photon-spectral} is, in
general, a matrix in the space of photon-coupled transition legs. Within the full secular approximation, terms coupling distinct legs
$\mu\neq\nu$ therefore average to zero, and only the diagonal spectral
elements enter the Lindblad equation. Hence, we abbreviate the notation by denoting $J_\mu^{(\gamma)}
(\omega;\bm s_{\parallel,\mu})
\equiv
J_{\mu\mu}^{(\gamma)}
(\omega;\bm s_{\parallel,\mu})$. For an oriented leg
$\mu=(h_\mu,\ell_\mu)$, with
${\omega^{\rm B}_{\mu}}=(\epsilon_{h_\mu}-\epsilon_{\ell_\mu})/\hbar>0$,
the corresponding downward and upward rates are
\begin{subequations}
\begin{align}
\gamma_{h_\mu\to\ell_\mu}^{(\gamma)}
&\equiv
\gamma_{\mu\mu}^{(\gamma)}(+{\omega^{\rm B}_{\mu}})
=
J_\mu^{(\gamma)}
\bigl({\omega^{\rm B}_{\mu}};\bm s_{\parallel,\mu}\bigr)
\left[n_\gamma({\omega^{\rm B}_{\mu}})+1\right],
\\
\gamma_{\ell_\mu\to h_\mu}^{(\gamma)}
&\equiv
\gamma_{\mu\mu}^{(\gamma)}(-{\omega^{\rm B}_{\mu}})
=
J_\mu^{(\gamma)}
\bigl({\omega^{\rm B}_{\mu}};\bm s_{\parallel,\mu}\bigr)
n_\gamma({\omega^{\rm B}_{\mu}}).
\end{align}
\label{eq:photon-up-down}
\end{subequations}
The first rate describes spontaneous plus stimulated emission, whereas
the second describes thermal absorption. Thus, the downward and upward
processes sample, respectively, the $+{\omega^{\rm B}_{\mu}}$ and $-{\omega^{\rm B}_{\mu}}$
sectors of the decay-rate matrix, while the spectral density is evaluated
at the positive transition-frequency magnitude ${\omega^{\rm B}_{\mu}}$. The rates~\eqref{eq:photon-up-down} satisfy the detailed balance condition, namely $\left[\gamma_{\ell_\mu\to h_\mu}^{(\gamma)}\right]/\left[\gamma_{h_\mu\to\ell_\mu}^{(\gamma)}\right]=e^{-\beta_\gamma\hbar{\omega^{\rm B}_{\mu}}}$. 

For an intra-excitonic leg
$\mu=(h_\mu,\ell_\mu)$,
$\bm s_{\parallel,\mu}=\bm0$, and the polarization contraction is
\begin{equation}
F_{\mu\mu}^{(\gamma,\mathrm{intra})}
=
\left|
\bm e_d\cdot\bm D_\mu^{\rm intra}
\right|^2.
\end{equation}
For the retained $\Xi$ system, we choose the real
combination $\lvert p_d\rangle$ of the degenerate $p$ states
whose intraband transition dipole is aligned with the
retained polarization,
\begin{equation}
\bm D_{p_ds}^{\rm intra}
=
D_{p_ds}^{\rm intra}\bm e_d .
\end{equation}
For this retained state, $F_{\mu\mu}^{(\gamma,\mathrm{intra})}
=
\left|
D_\mu^{\rm intra}
\right|^2$ and hence, the corresponding diagonal spectral density is
\begin{equation}
J_\mu^{(\gamma,\mathrm{intra})}(\omega)
=
\frac{
\left|
\bm D_\mu^{\rm intra}
\right|^2
}{
\hbar\epsilon_0Ac
}
\,\omega ,
\label{eq:appC-intra-J-pol}
\end{equation}
where the second equality applies to the retained polarization-aligned
state. The orthogonal member of the $p$ doublet is neither populated nor
mixed by the retained intra-excitonic photon fluctuations.~\footnote{The additional orthogonal real combination of the degenerate $p$ state, $\ket{p_\perp}$ has an associated dipole $\bm D_{p_\perp s}^{\rm intra}$ which lies in a direction orthogonal to the selected THz polarized beam and therefore $F_{p_\perp p_\perp}^{(\gamma,\mathrm{intra})}
=
F_{p_d p_\perp}^{(\gamma,\mathrm{intra})}
=
F_{p_\perp p_d}^{(\gamma,\mathrm{intra})}
=
0.$}.

Accordingly, the downward and upward intra-excitonic rates are
\begin{subequations}
\label{eq:intra-radiative-rate}
\begin{align}
\gamma_{h_\mu\to\ell_\mu}^{(\gamma,\mathrm{intra})}
&=
\frac{
|\bm D_\mu^{\mathrm{intra}}|^2{\omega^{\rm B}_{\mu}}
}{
\hbar\epsilon_0Ac
}
\left[n_\gamma({\omega^{\rm B}_{\mu}})+1\right],
\label{eq:intra-radiative-rate-down}
\\
\gamma_{\ell_\mu\to h_\mu}^{(\gamma,\mathrm{intra})}
&=
\frac{
|\bm D_\mu^{\mathrm{intra}}|^2{\omega^{\rm B}_{\mu}}
}{
\hbar\epsilon_0Ac
}
n_\gamma({\omega^{\rm B}_{\mu}}).
\label{eq:intra-radiative-rate-up}
\end{align}
\end{subequations}
where ${\omega^{\rm B}_{h\ell}}=(\epsilon_h-\epsilon_\ell)/\hbar>0$.
The appearance of the normalization area in
Eq.~\eqref{eq:intra-radiative-rate} is a consequence of the strict
fixed-$\bm K_0$ restriction. The internal dipole
$\bm D_{\mu}^{\rm intra}$ is independent
of $A$, whereas the projection retains only the single in-plane photon
momentum sector $\bm s_\parallel=\bm 0$. The perpendicular photon
density of states cancels $L_\gamma$ but no two-dimensional
$\bm s_\parallel$ phase-space factor remains to cancel $A^{-1}$.
Since $A$ is used here solely as a plane-wave normalization area,
\begin{equation}
    \lim_{A\rightarrow\infty}
    \gamma_{h\to\ell}^{(\gamma,\mathrm{intra})}
    =
    0\qquad
    \lim_{A\rightarrow\infty}
    \gamma_{\ell\to h}^{(\gamma,\mathrm{intra})}
    =
    0.
    \label{eq:photon-intra-thermodynamic-limit}
\end{equation}
This result should not be interpreted as the absence of
intra-excitonic radiative relaxation in a finite or momentum-resolved
system. Rather, it is the radiative rate into the unique
zero-in-plane-recoil photon sector retained by the strict
fixed-$\bm K_0$ model. Finite center-of-mass localization or
momentum-relaxing processes restore a finite range of in-plane photon
momenta and lie outside the present reduction. For comparison, the same
internal dipole radiating with in-plane recoil allowed, into all directions
and both polarizations of free space, decays at zero temperature at
$\gamma^{\rm fs}_{h\to\ell}=|\bm D_\mu^{\rm intra}|^2(\omega^{\rm B}_\mu)^3/(3\pi\epsilon_0\hbar c^3)$,
independently of $A$. For the retained $2p\to1s$ channel
($|\bm D^{\rm intra}|=0.854\,a_B$, $\hbar\omega^{\rm B}=113.19$~meV) this gives
$\hbar\gamma^{\rm fs}=0.37$~neV, against $1.11$~neV from
Eq.~\eqref{eq:intra-radiative-rate-down} at the baseline area
$A=9.62~\mu$m$^2$ of Sect.~\ref{sec:results}. The two coincide at
$A=3\lambda^2/4\pi=28.6~\mu$m$^2$ ($\lambda=10.95~\mu$m), and both correspond to
a radiative lifetime of the order of a microsecond ($0.59$ and $1.76~\mu$s).
The difference between them, and the frequency dependence $\omega$ instead of
$\omega^3$ that enters item~\ref{item6} of Sect.~\ref{sec:lindblad_three_level}, are
consequences of the fixed-$\mathbf K_0$ limit; lifting them requires a finite-momentum treatment (Sect.~\ref{subsec:scope}). Importantly, this thermodynamic suppression concerns the fluctuating
polarization-selected intra-excitonic photon reservoir only; the coherent
THz/mid-IR drive remains finite because its electric-field amplitude is
prescribed independently of the single-photon quantization volume. Within
the high-polarization-extinction, weak-density-of-states-modification limit
adopted here, the polarization selection does not alter the $A^{-1}$
scaling or the magnitude of the retained intra-excitonic rate.

For an interband photon-coupled leg
$\mu=(n,0)$ - which labels the oriented leg
$|n,\bm K_0\rangle\rightarrow|0\rangle$ - the fixed-$\bm K_0$ projection instead requires
$\bm s_{\parallel,\mu}=\bm K_0$. Using the diagonal spectral density
$J_{\mu}^{(\gamma)}\equiv J_{\mu\mu}^{(\gamma)}$ retained after the
full secular approximation, the interband contribution becomes
\begin{equation}
\begin{split}
    J_{\mu}^{(\gamma,\mathrm{inter})}
(\omega;\bm K_0)
&=
\frac{|\varphi_n(\bm 0)|^2}{\hbar\epsilon_0 c}
\frac{
\omega^2|\bm d_{cv}^{\,\xi}|^2
-
c^2|\bm d_{cv}^{\,\xi}\cdot\bm K_0|^2
}{
\sqrt{\omega^2-c^2K_0^2}
}
\Theta(\omega-cK_0)
\\
&=
g_n^{(\gamma)}
\frac{
\omega^2-\tfrac{1}{2}c^2K_0^2
}{
\sqrt{\omega^2-c^2K_0^2}
}
\Theta(\omega-cK_0),
\qquad
g_n^{(\gamma)}
=
\frac{
|\varphi_n(\bm 0)|^2
|\bm d_{cv}^{\,\xi}|^2
}{
\hbar\epsilon_0 c
}.
\end{split}
\label{eq:appC-inter-spectral}
\end{equation}
where $K_0=|\bm K_0|$. In order to get this expression, we use the projected interband dipole form factor~\eqref{eq:projected-inter-dipole-local} and the fact that $|\bm d_{cv}^{\,\xi}\cdot\bm K_0|^2
=K_0^2|\bm d_{cv}^{\,\xi}|^2/2$ where $\bm d_{cv}^{\,\xi}$ is given by Eq.~\eqref{eq:bloch-interband-dipole}. The interband spectral density
\eqref{eq:appC-inter-spectral} is independent of both $A$ and
$L_\gamma$.

The downward and
upward rates are obtained by sampling the diagonal decay-rate function
at $+{\omega^{\rm B}_{\mu}}$ and $-{\omega^{\rm B}_{\mu}}$, respectively. Thus,
\begin{subequations}
\begin{align}
\gamma_{n\to0}^{(\gamma,\mathrm{inter})}
&\equiv
\gamma_{\mu\mu}^{(\gamma)}(+{\omega^{\rm B}_{\mu}})
=
g_n^{(\gamma)}
\frac{
{\omega^{\rm B}_{\mu}}^2-\tfrac{1}{2}c^2K_0^2
}{
\sqrt{{\omega^{\rm B}_{\mu}}^2-c^2K_0^2}
}
\left[n_\gamma({\omega^{\rm B}_{\mu}})+1\right]
\Theta({\omega^{\rm B}_{\mu}}-cK_0),
\label{eq:appC-inter-radiative-rate-down}
\\
\gamma_{0\to n}^{(\gamma,\mathrm{inter})}
&\equiv
\gamma_{\mu\mu}^{(\gamma)}(-{\omega^{\rm B}_{\mu}})
=
g_n^{(\gamma)}
\frac{
{\omega^{\rm B}_{\mu}}^2-\tfrac{1}{2}c^2K_0^2
}{
\sqrt{{\omega^{\rm B}_{\mu}}^2-c^2K_0^2}
}
n_\gamma({\omega^{\rm B}_{\mu}})
\Theta({\omega^{\rm B}_{\mu}}-cK_0).
\label{eq:appC-inter-radiative-rate-up}
\end{align}
\label{eq:appC-inter-radiative-rate}
\end{subequations}
where ${\omega^{\rm B}_{\mu}}\equiv{\omega^{\rm B}_{n0}}=
\left(\epsilon_n-\epsilon_0\right)/\hbar>0$. The downward transition
$|n,\bm K_0\rangle\rightarrow|0\rangle$ contains spontaneous and
stimulated emission, whereas the upward transition
$|0\rangle\rightarrow|n,\bm K_0\rangle$ describes thermal absorption.
Notice that ${\omega^{\rm B}_{\mu}}={\omega^{\rm B}_{n0}}>0$ in both final expressions: the
transition direction is encoded by whether
$\gamma_{\mu\mu}^{(\gamma)}(\varpi)$ is evaluated in its
$+{\omega^{\rm B}_{\mu}}$ or $-{\omega^{\rm B}_{\mu}}$ sector.

The fixed in-plane momentum therefore produces the usual radiative
light-cone condition ${\omega^{\rm B}_{\mu}}>cK_0.$ Outside the light cone, both on-shell interband transition rates vanish, i.e. $\gamma_{n\to0}^{(\gamma,\mathrm{inter})}
=
\gamma_{0\to n}^{(\gamma,\mathrm{inter})}
=
0$ when ${\omega^{\rm B}_{\mu}}<cK_0.$ 
At the ideal light-cone boundary ${\omega^{\rm B}_{\mu}}=cK_0$, the
one-dimensional photon density of states produces the familiar threshold
singularity already visible in
Eq.~\eqref{eq:appC-inter-spectral}. The rates~\eqref{eq:appC-inter-radiative-rate} should therefore not be used exactly at this boundary,
where finite linewidth, finite momentum width, or other broadening
mechanisms become relevant.

\paragraph{Absolute normalization at $\bm K_0=\bm 0$.} The absolute magnitude
of Eq.~\eqref{eq:appC-inter-radiative-rate} fixes the normalization of the
entire photon sector in a single number. At
$\bm K_0=\bm 0$ and $T_\gamma\to0$ the rate collapses to
$\gamma^{(\gamma,\mathrm{inter})}_{n\to0}
 =g^{(\gamma)}_n\,\omega^{\rm B}_{n0}$.
Using the solver output of Table~\ref{tab:appD_states},
$|\varphi_{1s}(\bm 0)|^2=0.2842\,a_B^{-2}
 =2.01\times10^{17}\,\mathrm{m^{-2}}$, together with the band-edge dipole of
Eq.~\eqref{eq:bloch-interband-dipole},
$|\bm d^{\,\xi}_{cv}|^2=e^2\hbar^2/(2\mu E_g)
 =2.59\times10^{-57}\,\mathrm{C^2m^2}$ for $\mu=0.20\,m_0$ and
$E_g=1.89$~eV, one obtains the dimensionless coupling
$g^{(\gamma)}_{1s}=1.85\times10^{-3}$ and, with
$\hbar{\omega^{\rm B}_{1s,0}}=E_g-E_b(1s)=1728.4$~meV,
\begin{equation}
\gamma^{(\gamma,\mathrm{inter})}_{1s\to0}
= 4.87\times10^{12}\,\mathrm{s^{-1}},
\qquad
\tau_{1s}=0.205\;\mathrm{ps}.
\label{eq:1s-lifetime-check}
\end{equation}
This rate, $\hbar\gamma^{(\gamma,\mathrm{inter})}_{1s\to0}=3.21$~meV, follows
from the Wannier envelope and the band-edge dipole alone, because the
interband spectral density~\eqref{eq:appC-inter-spectral} is independent of
$A$ and $L_\gamma$. It is the rate of a monolayer radiating into vacuum: the
photon modes of Eq.~\eqref{eq:quantized-electric-field} are those of free
space, so the dielectric environment of a real sample is not included, and
the dipole is that of the two-band model of
Eq.~\eqref{eq:bloch-interband-dipole}. Both omissions matter on their own. A
sheet embedded in a thick homogeneous dielectric of refractive index $n$
radiates at a rate reduced by approximately $1/n$, about a factor of two for
hBN, while in a thin hBN/substrate stack the thicknesses of the layers tune the
radiative lifetime by up to an order of magnitude through the Purcell
effect~\cite{Fang2019}. We therefore use
Eq.~\eqref{eq:1s-lifetime-check} as the reference value of the vacuum-embedded
model rather than as a sample-specific prediction. It is consistent with the
microscopic radiative broadening of the $1s$ exciton in hBN-encapsulated
WSe$_2$, which Brem \emph{et al.} compute to be temperature independent and
about $2.5$~meV (full width at half maximum, i.e.\ $\hbar\gamma$ in the present
convention)~\cite{Brem_intrinsic}: the vacuum value lies within about $30\%$
of it, whereas dividing by $n_{\rm hBN}$ alone would give $\approx1.5$~meV,
further from it. The agreement is partly a cancellation between the $1/n$
factor and the stack and dipole effects that the model omits. Because the
reduced dynamics depends on the interband rate only through dimensionless
combinations---$\Omega/\gamma^{(\gamma,\mathrm{inter})}_{1s\to0}$,
$\sigma\gamma^{(\gamma,\mathrm{inter})}_{1s\to0}$ and the adiabaticity
parameter~\eqref{eq:adiabaticity-A}---the results of Sect.~\ref{sec:results}
carry over to a sample with a different radiative rate by rescaling the quoted
energies and times with it at fixed values of these combinations; only
comparisons with scales that do not depend on it, namely the selectivity
limits, the ionization threshold and the exchange shifts, change. The corresponding $2s$ rate is
$\gamma^{(\gamma,\mathrm{inter})}_{2s\to0}
 =5.48\times10^{11}\,\mathrm{s^{-1}}$ ($\tau_{2s}=1.82$~ps), which is the
vertex loss carried by the upper state of the $V$ system.

\paragraph{Lamb shift and free-space vacuum renormalization.} The photon Lamb-shift matrix follows from Eq.~\eqref{eq:S_PV_app}. In terms of the photon spectral function~\eqref{eq:photon-spectral}, it reads as follows
\begin{equation}
\begin{split}
    S_{\mu\nu}^{(\gamma)}(\varpi)
    &=
    \frac{1}{2\pi}
    \mathcal{PV}\,
    \int_{\omega_{c,\mu}}^\infty
    d\omega\,
    \mathcal J_{\mu\nu}^{(\gamma)}
    (\omega;\bm s_{\parallel,\mu})
    \,
    \left[
        \frac{n_\gamma(\omega)+1}{\varpi-\omega}
        +
        \frac{n_\gamma(\omega)}{\varpi+\omega}
    \right].
\end{split}
\label{eq:photon-lamb-unrenormalized}
\end{equation}
It is useful to separate this expression into its zero-temperature
free-space and thermal parts, $S_{\mu\nu}^{(\gamma)}(\varpi)
    =
    S_{\mu\nu}^{(\gamma,\mathrm{vac})}(\varpi)
    +
    S_{\mu\nu}^{(\gamma,\mathrm{th})}(\varpi)$, 
    where 
\begin{subequations}
\label{eq:photon-lamb}
    \begin{align}
        S_{\mu\nu}^{(\gamma,\mathrm{vac})}(\varpi)
    &=
    \frac{1}{2\pi}
    \mathcal{PV}
    \int_{\omega_{c,\mu}}^\infty
    d\omega\,
    \frac{
        \mathcal J_{\mu\nu}^{(\gamma)}
        (\omega;\bm s_{\parallel,\mu})
    }{
        \varpi-\omega
    },
    \label{eq:photon-vacuum-lamb}
    \\
    S_{\mu\nu}^{(\gamma,\mathrm{th})}(\varpi)
    &=
    \frac{\varpi}{\pi}
    \mathcal{PV}
    \int_{\omega_{c,\mu}}^\infty
    d\omega\,
    \frac{
        \mathcal J_{\mu\nu}^{(\gamma)}
        (\omega;\bm s_{\parallel,\mu})
        n_\gamma(\omega)
    }{
        \varpi^2-\omega^2
    } .
    \label{eq:photon-thermal-lamb}
    \end{align}
\end{subequations}
The vacuum contribution in Eq.~\eqref{eq:photon-vacuum-lamb} is ultraviolet sensitive when the
present low-energy electric-dipole description is extrapolated to arbitrarily
large photon frequencies. Indeed, within this approximation
$J^{(\gamma)}_{\mu\nu}(\omega)\propto\omega$ asymptotically, so that
$J^{(\gamma)}_{\mu\nu}(\omega)/(\varpi-\omega)$ approaches a constant at
large $\omega$. The resulting zero-temperature free-space self-energy
therefore depends on photon frequencies outside the range of validity of the
effective excitonic Hamiltonian. Its absolute value cannot be predicted
consistently without specifying a microscopic ultraviolet completion of the
theory. Rather than introducing an additional phenomenological cutoff, we
fix this undetermined vacuum contribution by a renormalization condition. To state this condition precisely, let $\hat H_X^{(W)}$ denote the matter
Hamiltonian constructed from the quasiparticle gap and the Wannier
eigenenergies used throughout this work. These energies follow from the
effective electron--hole problem and, in particular, the Wannier calculation - see App.~\ref{app:rayleigh-ritz} -
does not include the electromagnetic vacuum self-energy of Eq.~\eqref{eq:photon-vacuum-lamb}.
We nevertheless choose $\hat H_X^{(W)}$ as the reference Hamiltonian of the
effective open-system theory at zero photon temperature in free space. This
amounts to introducing a counterterm that cancels the explicit vacuum
self-energy within the effective theory,
\begin{equation}
\ho{H}_{\rm ct}^{(\gamma)}
=
-\ho{H}_{\rm LS}^{(\gamma,\mathrm{vac})},
\qquad
\ho{H}_X^{(W)}
+\ho{H}_{\rm ct}^{(\gamma)}
+\ho{H}_{\rm LS}^{(\gamma,\mathrm{vac})}
=
\ho{H}_X^{(W)} .
\label{eq:counterterm}
\end{equation}
Equation~\eqref{eq:counterterm} should therefore be understood as a matching, or
renormalization, condition: the exciton energies entering the effective
Lindblad model are defined to coincide with the Wannier-solver reference
energies at $T_\gamma=0$ in free space. It does not modify the Wannier
equation, shift its eigenvalues, or imply that the Wannier solver already
contains a QED vacuum correction. Nor does it assert that the microscopic
vacuum self-energy vanishes. Rather, it states that its undetermined
zero-temperature free-space value is absorbed into the definition of the
reference parameters, so that only photon-induced shifts relative to this
reference are retained explicitly. Accordingly, the finite photon-induced Lamb-shift correction is identified with the thermal contribution measured relative to the zero-temperature free-space reference, Eq.~\eqref{eq:photon-thermal-lamb}. 

For a nondegenerate leg $\mu=(h_\mu,\ell_\mu)$, the corresponding
thermal Lamb-shift Hamiltonian contributes
\begin{equation}
\begin{split}
    \hat H_{{\rm LS},\mu}^{(\gamma,\mathrm{th})}
    &=
    \hbar
    S_\mu^{(\gamma,\mathrm{th})}(+{\omega^{\rm B}_{\mu}})
    \ket{h_\mu}\bra{h_\mu}
    \\
    &\quad+
    \hbar
    S_\mu^{(\gamma,\mathrm{th})}(-{\omega^{\rm B}_{\mu}})
    \ket{\ell_\mu}\bra{\ell_\mu}.
\end{split}
\label{eq:photon-single-leg-lamb}
\end{equation}
Since the thermal contribution in
Eq.~\eqref{eq:photon-thermal-lamb} is odd in its Bohr-frequency argument,
$S_\mu^{(\gamma,\mathrm{th})}(-\varpi)
=-S_\mu^{(\gamma,\mathrm{th})}(+\varpi)$,
the resulting thermal photon-induced shift of the subsystem transition
frequency $\omega^{\rm B}_{\mu}$~\eqref{eq:jump-photon} is 

\begin{equation}
\begin{split}
    \delta{\omega^{\rm B}_{\mu}}^{(\gamma)}
    &=
    S_\mu^{(\gamma,\mathrm{th})}(+{\omega^{\rm B}_{\mu}})
    -
    S_\mu^{(\gamma,\mathrm{th})}(-{\omega^{\rm B}_{\mu}})
    \\
    &=
    \frac{2{\omega^{\rm B}_{\mu}}}{\pi}
    \mathcal{PV}
    \int_{\omega_{c,\mu}}^\infty
    d\omega\,
    \frac{
        \mathcal J_\mu^{(\gamma)}(\omega)
        n_\gamma(\omega)
    }{
        {\omega^{\rm B}_{\mu}}^2-\omega^2
    } .
\end{split}
\label{eq:photon-transition-lamb}
\end{equation}

For a retained intra-excitonic oriented leg $\mu$, whose transition dipole
is aligned with the selected polarization $\bm e_d$, the associated thermal
Lamb-shift contribution reads as
\begin{equation}
    S_{\mu}^{(\gamma,\mathrm{th-intra})}({\omega^{\rm B}_{\mu}})
    =
    \frac{
        |\bm D_{\mu}^{\rm intra}|^2
        {\omega^{\rm B}_{\mu}}
    }{
        \pi\hbar\epsilon_0 A c
    }
    \mathcal{PV}
    \int_0^\infty
    d\omega\,
    \frac{
        \omega\, n_\gamma(\omega)
    }{
        {\omega^{\rm B}_{\mu}}^2-\omega^2
    },
    \label{eq:photon-intra-lamb}
\end{equation}
This expression carries the same $A^{-1}$ scaling as the corresponding
fixed-$\bm K_0$ radiative rate and therefore vanishes in the
thermodynamic plane-wave limit. Eq.~\eqref{eq:photon-intra-lamb} can be evaluated analytically when the photon distribution is taken to be the Bose-Einstein distribution $n_{\gamma}(x)
=\left[e^{x}-1\right]^{-1}$ where $x=\hbar\omega/[k_{\rm B}T_{\gamma}]$.  Using the standard Binet integral representation of the digamma function $\psi(z)$~\footnote{The digamma function is defined by $\psi(z)=\Gamma'(z)/\Gamma(z)$, where $\Gamma(z)$ denotes the Gamma function. The Binet representation of $\psi(z)$ is given by Eq.~(6.3.21) of Ref.~\cite{abramowitz}.}, one finds that the thermal photon intra-excitonic Lamb shift reads as 
\begin{equation}
    \label{eq:ther-BE-lamb-photon}
    S_{\mu}^{(\gamma,\mathrm{th-intra})}({\omega^{\rm B}_{\mu}})
=
\frac{
|\bm D_{\mu}^{\rm intra}|^2{\omega^{\rm B}_{\mu}}
}{
2\pi\hbar\epsilon_0 A c
}
\,
\left\{{\rm Re}\,\psi(1+i\eta_{{\omega^{\rm B}_{\mu}}})-\ln\eta_{{\omega^{\rm B}_{\mu}}}\right\}\,
\qquad
\eta_{{\omega^{\rm B}_{\mu}}}=\frac{\hbar|{\omega^{\rm B}_{\mu}}|}{2\pi k_{\rm B}T_\gamma}\,.
\end{equation}

From the previous expression and using Eq.~\eqref{eq:photon-transition-lamb}, the corresponding shift of the intra-excitonic transition frequency is
\begin{equation}
\delta{\omega^{\rm B}_{\mu}}^{(\gamma,\rm{intra})}
=
\frac{
|\bm D_{\mu}^{\rm intra}|^2{\omega^{\rm B}_{\mu}}
}{
\pi\hbar\epsilon_0 A c
}
\,
\left\{{\rm Re}\,\psi(1+i\eta_{{\omega^{\rm B}_{\mu}}})-\ln\eta_{{\omega^{\rm B}_{\mu}}}\right\}.
\end{equation}
Thus, the intra-excitonic thermal Lamb shift can be evaluated without a
numerical principal-value integration. It retains the same $A^{-1}$ scaling
as the corresponding strict-fixed-$\bm K_0$ radiative rate and therefore
vanishes in the thermodynamic plane-wave limit.

After equating~\eqref{eq:appC-inter-spectral} into Eq.~\eqref{eq:S_PV_app}, one obtains the following interband Lamb shift matrix 
\begin{equation}
\begin{split}
S_n^{(\gamma,\mathrm{th-inter})}(\varpi)
&=
\frac{g_n^{(\gamma)}\varpi}{\pi}
\,\mathcal{PV}\,
\int_{cK_0}^{\infty}d\omega\,
\frac{
\left(
\omega^2-\frac{1}{2}c^2K_0^2
\right)n_\gamma(\omega)
}{
\sqrt{\omega^2-c^2K_0^2}
\left(\varpi^2-\omega^2\right)
}\, 
\\
&=
\frac{g_n^{(\gamma)}\varpi}{\pi}
\,
\mathcal I(y,x_c),
\end{split}
\label{eq:photon-inter-lamb}
\end{equation}
where
\begin{equation}
\mathcal I(y,x_c)
=\mathcal{PV} \int_0^\infty du\,
\frac{
u^2+\frac{1}{2}x_c^2
}{
\sqrt{u^2+x_c^2}\,
\left(e^{\sqrt{u^2+x_c^2}}-1\right)
\left(y^2-x_c^2-u^2\right)
}.
\end{equation}
In the previous expression we introduce the dimensionless variables $y=\beta_\gamma\hbar|\varpi|,\,
x_c=\beta_\gamma\hbar cK_0$ and $u=\beta_\gamma\hbar c|s_\perp|$. Interestingly, Eq.~\eqref{eq:photon-inter-lamb} provides an analytical expression in the limit when $K_0\to0$. This limit yields the following expression
\begin{equation}
\left.
S_\mu^{(\gamma,\mathrm{th})}({\omega^{\rm B}_{\mu}})
\right|_{K_0=0}
=
\frac{
g_n^{(\gamma)}{\omega^{\rm B}_{\mu}}
}{2\pi}
\,
\left\{{\rm Re}\,\psi(1+i\eta_{{\omega^{\rm B}_{\mu}}})-\ln\eta_{{\omega^{\rm B}_{\mu}}}\right\},
\end{equation}
and the corresponding interband transition-frequency shift is
\begin{equation}
\left.
\delta{\omega^{\rm B}_{\mu}}^{(\gamma)}
\right|_{K_0=0}
=
\frac{
g_n^{(\gamma)}{\omega^{\rm B}_{\mu}}
}{\pi}
\,
\left\{{\rm Re}\,\psi(1+i\eta_{{\omega^{\rm B}_{\mu}}})-\ln\eta_{{\omega^{\rm B}_{\mu}}}\right\}.
\end{equation}
For optical transitions at ordinary temperatures,
$\eta_{{\omega^{\rm B}_{\mu}}}\gg1$, and using that
\[
\left\{{\rm Re}\,\psi(1+i\eta_{\omega^{\rm B}})-\ln\eta_{\omega^{\rm B}}\right\}
\approx
\frac{1}{12\eta_{\omega^{\rm B}}^2}
+\mathcal O(\eta_{\omega^{\rm B}}^{-4}).
\]
The thermal interband Lamb shift is therefore strongly suppressed as
$(k_{\rm B}T_\gamma/\hbar{\omega^{\rm B}_{\mu}})^2$. Notice that this suppression is
algebraic rather than simply proportional to the resonant occupation
$n_\gamma({\omega^{\rm B}_{\mu}})$, because the Lamb shift also samples off-resonant
thermal photon modes. Although the off-shell correction can remain formally
nonzero outside the radiative light cone, its magnitude is consequently
negligible for the optical transitions and parameter regime considered here.

The applicability of the asymptotic estimate is controlled by
$\eta_{\omega^{\rm B}_\mu}=\hbar\omega^{\rm B}_\mu/2\pi k_BT_\gamma$ and
therefore depends on the adopted temperature. At the baseline $T_\gamma=4$~K
both retained legs are deep in the large-$\eta$ regime,
$\eta=798$ for the interband leg at $\hbar\omega^{\rm B}_\mu\simeq1.7$~eV and
$\eta=52.3$ for the intraband leg at $113.19$~meV, so the bracket is
$1/12\eta^2$ to leading order, i.e. $3.1\times10^{-5}$ for the intraband leg.
The intraband thermal Lamb shift is then suppressed twice over, algebraically
through $(k_BT_\gamma/\hbar\omega^{\rm B}_\mu)^2$ and through the $A^{-1}$
prefactor of Eq.~\eqref{eq:photon-intra-lamb}, and vanishes in the
thermodynamic plane-wave limit. This is what justifies setting it to zero in
the reduced dynamics. At the elevated comparison temperatures the algebraic
part of that argument fails for the intraband leg --- $\eta=2.72$ at $77$~K and
$\eta=0.697$ at $300$~K, where the expansion is not applicable and the full
expression~\eqref{eq:photon-intra-lamb} must be used --- while the $A^{-1}$
suppression is unaffected, since it is kinematic rather than thermal.

Finally, under the full secular approximation, the finite thermal
photon Lamb-shift contribution in a selected three-level system is
diagonal,
\begin{equation}
\hat H_{\rm LS}^{(\gamma,\mathrm{th})}
=
\hbar\sum_{a=1}^{3}
\delta_a^{(\gamma,\mathrm{th})}
\ket{a}\bra{a},
\label{eq:photon-total-lamb}
\end{equation}
where
\begin{equation}
\delta_a^{(\gamma,\mathrm{th})}
=
\sum_{\mu:\,a=h_\mu}
S_\mu^{(\gamma,\mathrm{th})}(+{\omega^{\rm B}_{\mu}})
+
\sum_{\mu:\,a=\ell_\mu}
S_\mu^{(\gamma,\mathrm{th})}(-{\omega^{\rm B}_{\mu}}).
\label{eq:photon-level-lamb}
\end{equation}
Only differences of the level shifts enter the driven dynamics. Thus, for
a driven leg $a\leftrightarrow b$, the photon bath would shift the
corresponding transition frequency by
$\delta_a^{(\gamma,\mathrm{th})}
-\delta_b^{(\gamma,\mathrm{th})}$ and therefore modify the associated
one- and two-photon detunings. The configuration-dependent realization of these
relations is given in Sect.~\ref{sec:lindblad_three_level}.

Eqs.~\eqref{eq:photon-total-lamb} and~\eqref{eq:photon-level-lamb} are displayed here to make explicit the complete contribution generated by the fluctuations of the photon reservoir. At the baseline adopted in this work these two contributions are negligible for
different reasons, and it is worth separating them. The interband shift is
algebraically suppressed as $(k_BT_\gamma/\hbar\omega^{\rm B}_\mu)^2$ and is
negligible at any temperature considered here. The intra-excitonic shift is
suppressed algebraically as well at $T_\gamma=4$~K, where
$\eta_{\omega^{\rm B}_\mu}=52.3$, and is in addition proportional to $A^{-1}$,
so that it vanishes in the thermodynamic plane-wave limit in which the reduced
model is defined. The second statement is a consequence of the selected mode
model rather than an estimate, and it is the one that survives at elevated
temperature, where the algebraic part of the argument does not. We therefore neglect
$\hat H_{\rm LS}^{(\gamma,\mathrm{th})}$ in the numerical Lindblad
calculations, while retaining the corresponding photon-induced
emission and absorption rates. Accordingly, in the numerical implementation
the photon-induced corrections to the one- and two-photon detunings are set
to zero.

\subsection{Exciton reservoir}
\label{app:exciton-gas-mean-field}

We now specialize to the dilute intravalley exciton gas introduced in
Sect.~\ref{sec:multi-env-model}.  In the present work, the exciton gas is retained to first order in its
one-body density matrix, i.e. through the coherent exchange mean field
generated by the strict fixed-$\bm K_0$, zero-transfer sector.  This
choice keeps the density-dependent coherent renormalization that survives
the thermodynamic fixed-$\bm K_0$ limit. Exciton-gas fluctuations, finite-$\bm q$
corrections and the linewidths they produce are not included [contribution~(iv) of
Sect.~\ref{subsec:scope}]; their effect on the $\Xi$ results is estimated in App.~\ref{app:dephasing}.

For a stationary incoherent exciton gas we take the one-body density matrix to be
diagonal in the free exciton basis,
\begin{equation}
\left\langle
\hat X_{l',\bm Q}^\dagger
\hat X_{l,\bm Q'}
\right\rangle_{X_g}
=
\delta_{l'l}\delta_{\bm Q,\bm Q'}\,N_{l,\bm Q},
\label{eq:appC-Xg-incoherent-one-body}
\end{equation}
with
$N_{l,\bm Q}
=\langle\hat X_{l,\bm Q}^\dagger\hat X_{l,\bm Q}\rangle_{X_g}$.
Taking the reservoir expectation value of
Eq.~\eqref{eq:final-projected-Xg-Hamiltonian} therefore gives the coherent
mean-field Hamiltonian
\begin{equation}
    \ho{H}_{\rm MF}^{(X_g,0)}
    =
    \sum_{n',n}
    \Sigma_{n'n}^{(X_g,0)}(\bm K_0)
    |n';\bm K_0\rangle\langle n;\bm K_0|,
    \label{eq:appC-Xg-MF-H}
\end{equation}
where
\begin{equation}
\begin{split}
    \Sigma_{n'n}^{(X_g,0)}(\bm K_0)
    &=\frac{1}{A}\,
    \sum_{l,\mathbf{Q}}
    \mathcal{G}_{\rm exch}^{\,n'l;ln}
    (\bm K_0,\bm Q,\bm0)\,
    N_{l,\bm Q}\,,\\
    &= \sum_l\,\int \frac{d^2Q}{(2\pi)^2}\,\mathcal{G}_{\rm exch}^{\,n'l;ln}
    (\bm K_0,\bm Q,\bm0)\,
    N_{l,\bm Q}\,,
\end{split}
    \label{eq:appC-Xg-self-energy-discrete}
\end{equation}
is the fixed-$K_0$ exciton-gas self-energy matrix. In Eqs.~\eqref{eq:appC-Xg-MF-H} and~\eqref{eq:appC-Xg-self-energy-discrete}, we make use of the area-rescaled $X-X$ vertex~\eqref{eq:appC-Xg-area-scaled-G} and considered a continuum distribution for the exciton gas modes. With the Hermitian prescription adopted in App.~\ref{app:xx-internal-state-exchange}
for the off-shell part of the carrier-exchange vertex, this self-energy is
Hermitian. The conduction--valence Coulomb channel contributes to it only
through carrier exchange. Its direct (Hartree) part belongs to the
combination that cancels between the three Coulomb channels at zero
momentum transfer, whereas its exchange part is the attractive term
$\mathcal E_{\rm cc,a}$ of Eq.~\eqref{eq:appA-xx-cc-att}, which is of the
same order in $A^{-1}$ as the like-carrier term and of opposite sign. The
carrier exchange, with both parts, and the interband electron--hole
exchange are the contributions that survive. The components of the self-energy matrix should not be understood as dephasing rates. Moreover,  when the crystal ground
state $|0\rangle$ belongs to a selected three-level system, its row and
column in Eq.~\eqref{eq:appC-Xg-MF-H} vanish because the
exciton--exciton interaction acts only in sectors containing a tagged
exciton.

\paragraph{Exciton gas mean-field modification of the control detunings.}
The coherent mean field Hamiltonian and its associated self-energy, Eqs.~\eqref{eq:appC-Xg-MF-H} and~\eqref{eq:appC-Xg-self-energy-discrete}, enters the driven three-level dynamics
primarily through density-dependent corrections to the exciton energies. The diagonal elements of $\Sigma_{nn}^{(X_g,0)}$ in Eq.~\eqref{eq:appC-Xg-self-energy-discrete} produce density-dependent coherent energy shifts while the off-diagonal elements $\Sigma_{n'n}^{(X_g,0)}$ are, in general, coherent exchange-induced mixing matrix elements. If appreciable off-diagonal mean-field matrix elements are present, the effective Hamiltonian is no longer diagonal in the selected Wannier
exciton basis, and its eigenstates become mean-field-dressed linear combinations of those basis states within the retained three-level system.

We first consider the case in which the off-diagonal elements of the self-energy matrix~\eqref{eq:appC-Xg-self-energy-discrete}  are negligible. In this case, the effect of the mean-field exciton gas can be represented by the renormalized energy level of the tagged exciton
\begin{equation}
    \ren\epsilon_n(\bm K_0)
    =
    \epsilon_n(\bm K_0)
    + \Sigma_{nn}^{(X_g,0)}(\bm K_0)\,.
    \label{eq:appC-Xg-diagonal-renormalized-energy}
\end{equation}
where $\Sigma_{nn}^{(X_g,0)}(\bm K_0)=0$ when $|n\rangle=|0\rangle$. Consequently, the signed subsystem transition frequency
$\omega_{ab}=(\epsilon_a-\epsilon_b)/\hbar$ --- whose magnitude is the Bohr
frequency $\omega^{\rm B}_{ab}$ used in the spectral densities of
App.~\ref{app:photon-reservoir} --- is shifted according to
\begin{equation}
\ren{\omega}_{ab}
=
\omega_{ab}
+
\frac{
\Sigma_{aa}^{(Xg,0)}
-
\Sigma_{bb}^{(Xg,0)}
}{\hbar}\,,\qquad \omega_{ab}=\frac{\epsilon_a-\epsilon_b}{\hbar}.
\end{equation}
Thus, with the detuning convention
$\Delta_{ab}=\omega_{L,ab}-\omega_{ab}$ adopted in the main text, the corresponding one-photon detuning becomes
\begin{equation}
\label{eq:renorm-detuning-Xg}
\ren{\Delta}_{ab}^{(Xg)}
=
\Delta_{ab}
-
\frac{
\Sigma_{aa}^{(Xg,0)}
-
\Sigma_{bb}^{(Xg,0)}
}{\hbar}.
\end{equation}
Only differences of diagonal mean-field shifts are therefore relevant for
the driven dynamics.

The same result directly determines the density-dependent displacement of
the Raman conditions. For the ladder configuration,
$\delta_\Xi=\Delta_{12}+\Delta_{23}$,
\begin{equation}
\ren{\delta}_{\Xi}^{(Xg)}
=
\delta_\Xi
+
\frac{
\Sigma_{33}^{(Xg,0)}
-
\Sigma_{11}^{(Xg,0)}
}{\hbar},
\end{equation}
whereas for the $V$ configuration,
$\delta_V=\Delta_{13}-\Delta_{12}$,
\begin{equation}
\ren{\delta}_{V}^{(Xg)}
=
\delta_V
+
\frac{
\Sigma_{33}^{(Xg,0)}
-
\Sigma_{22}^{(Xg,0)}
}{\hbar}.
\end{equation}
Hence the two-photon resonance is controlled by the differential mean-field
shift of the two states whose relative coherence defines the corresponding
Raman condition. In the $\Xi$ and $V$ systems the crystal ground state
$|1\rangle=|0\rangle$ carries no tagged exciton, and therefore
$\Sigma_{11}^{(Xg,0)}=0$. The exchange mean field exciton gas can therefore displace the EIT or
STIRAP resonance condition linearly with the exciton-gas density. Prior to any analysis, it is then needed to compare its magnitude with the Rabi frequencies, dissipative linewidths, and the allowed two-photon detuning window. In the following, we consider two scenarios in which the off-diagonal terms of Eq.~\eqref{eq:appC-Xg-self-energy-discrete} can either be neglected or must be retained. 

Let us now consider the case in which symmetry-allowed off-diagonal
mean-field elements $\Sigma_{ab}^{(Xg,0)}$, $a\neq b$, are not negligible.
Their effect cannot, in general, be reduced to a detuning correction. The
tagged-exciton Hamiltonian~\eqref{eq:Exc-ham} is then replaced by
\begin{equation}
\ho{H}_{X,\rm d}^{(\bm K_0)}
=
\ho{H}_X^{(\bm K_0)}
+
\ho{H}_{\rm MF}^{(X_g,0)},
\end{equation}
which must be diagonalized before defining the corresponding transition
frequencies and drive matrix elements,
\begin{equation}
\ho{H}_{X,\rm d}^{(\bm K_0)}
|a^{\rm d}\rangle
=
\epsilon^{\rm d}_a
|a^{\rm d}\rangle .
\label{eq:eff-X-eigenvalue}
\end{equation}
We call the eigenbasis $\{|a^{\rm d}\rangle\}$ the \emph{dressed basis}, and
the superscript ${\rm d}$ marks quantities expressed in it. It follows from a
time-independent change of basis within the retained three-level system and is
unrelated to the drive-rotating frame of Sect.~\ref{subsec:unified_three_level_ham}.

The free Hamiltonian $\ho{H}_X^{(\bm K_0)}$ is already diagonal in the
Wannier basis obtained following App.~\ref{app:rayleigh-ritz}. Hence, no new
Wannier problem is solved: the mean-field dressing is diagonalized only
within the retained excitonic subspace spanned by
$\{|b,\bm K_0\rangle\}$.  Writing
\[
|a^{\rm d}\rangle
=
\sum_b U_{ba}|b,\bm K_0\rangle ,
\]
and replacing it in Eq.~\eqref{eq:eff-X-eigenvalue}, the dressed energies and mixing coefficients satisfy
\begin{equation}
\label{eq:eigenvalue}
\sum_b
\left[
\epsilon_c(\bm K_0)\delta_{cb}
+
\Sigma_{cb}^{(Xg,0)}
\right]
U_{ba}
=
\epsilon^{\rm d}_a \, U_{ca}.
\end{equation}
Thus, $|a^{\rm d}\rangle$ is a mean-field-dressed linear combination of
the original Wannier exciton states associated to a given three-level configuration $\mathcal{T}$.

The driven dynamics can be expressed in the dressed basis, with
transition frequencies
\[
\omega^{\rm d}_{ab}
=
\frac{\epsilon^{\rm d}_a-\epsilon^{\rm d}_b}{\hbar},
\]
and transformed drive matrix elements, e.g.
\[
{\bm D}^{\rm d}_{ab}
=
\langle a^{\rm d}|\hat{\bm D}|b^{\rm d}\rangle .
\]
When the off-diagonal elements vanish, $U$ is the identity and the dressed
energies reduce to the renormalized energies of
Eq.~\eqref{eq:VE-renormalized-energies}, $\epsilon^{\rm d}_a=\ren\epsilon_a$;
otherwise they differ at second order,
$\epsilon^{\rm d}_a\simeq\ren\epsilon_a+\sum_{b\neq a}|\Sigma^{(Xg,0)}_{ab}|^2/(\ren\epsilon_a-\ren\epsilon_b)$,
which is why the two are denoted differently. An appreciable off-diagonal exchange mean field can therefore modify both
the transition frequencies and the Rabi couplings, rather than merely shift the detunings.

In App.~\ref{app:offdiag}, we perform a representative estimate to assess the relevance of the off-diagonal self-energy contributions. For the $\mathrm{WSe_2}/\mathrm{hBN}$ material considered here, we find that the off-diagonal elements of the self-energy~\eqref{eq:appC-Xg-self-energy-discrete} can be safely neglected for all selected three-level systems. 

\paragraph{Absence of dissipative effects within the exciton-gas mean-field approximation.} The Hamiltonian Eq.~\eqref{eq:appC-Xg-MF-H} generates deterministic unitary dynamics. Accordingly, retaining only this contribution does not induce dissipative corrections.  A diagonal mean field changes relative phases and therefore renormalizes the one- and two-photon detunings, while an off-diagonal mean field coherently mixes internal exciton states. Thus, within the present approximation and regardless of the selected excitonic three-level configuration $\mathcal{T}$, there are no dephasing or Lamb-shift contributions associated with this mean field 
\begin{equation}
    \mathcal D^{(X_g)}_{\mathcal{T}}=0\,,\qquad \ho{H}_{\rm{LS},\mathcal{T}}^{(X_g)}=0.
    \label{eq:dissipative-Xgas}
\end{equation}
These conclusions are a statement about the scope of the present model, not a claim that exciton--exciton interactions are intrinsically nondissipative.  Connected gas fluctuations, finite-$\bm q$ scattering, finite center-of-mass coherence width, collision broadening, disorder, and other mechanisms that generate a
finite bath correlation time are deferred to a momentum-resolved extension study. 

\subsection{Omitted intra-excitonic channels and the photon temperature}
\label{app:omitted-channels}
This subsection gives the estimates behind item~\ref{item6} of Sect.~\ref{sec:lindblad_three_level} and its $V$ counterpart in Sect.~\ref{sec:V-component-equations}. In the retained $\Xi$ system the target $2p$ lies $10.97$~meV
\emph{below} the $2s$ state, to which it is dipole-connected with
$|d_{2s,2p}|=2.740\,a_B$, i.e. $3.21$ times the retained $|d_{1s,2p}|$
(Table~\ref{tab:appD_dipoles}). Both the retained $2p\to1s$ channel and this
omitted $2p\leftrightarrow2s$ channel are intra-excitonic and therefore share
the $A^{-1}$ scaling of Eq.~\eqref{eq:photon-intra-thermodynamic-limit}, so
their ratio is $A$-independent. The larger dipole of the omitted channel
almost exactly compensates its smaller frequency,
\begin{equation}
\frac{|d_{2s,2p}|^2\,\omega^{\rm B}_{2s,2p}}
     {|d_{1s,2p}|^2\,\omega^{\rm B}_{2p,1s}}
= 10.30\times0.0969 = 0.998,
\label{eq:omitted-channel-ratio}
\end{equation}
so that, within the fixed-$\mathbf K_0$ photon reservoir, the two spectral
densities agree to $0.2\%$. This near equality is a property of the
fixed-$\mathbf K_0$ limit, whose one-dimensional photon continuum makes both
spectral densities linear in frequency [Eq.~\eqref{eq:appC-intra-J-pol}]. With
in-plane recoil allowed, each channel would radiate into the full
three-dimensional photon phase space at the free-space rate
$|d|^2\omega^3/(3\pi\epsilon_0\hbar c^3)$, and the
ratio~\eqref{eq:omitted-channel-ratio} would become
$10.30\times0.0969^3=9.4\times10^{-3}$. As for the rates themselves (App.~\ref{app:photon-reservoir}), only a finite-momentum
treatment would lift it. Within the present reduction the two
channels differ only in thermal occupation, and it is this that fixes the
temperature at which the reduction is usable. Writing the ratio of the omitted
upward channel to the retained downward one as
$0.998\,n_\gamma(10.97\,\mathrm{meV})/[1+n_\gamma(113.19\,\mathrm{meV})]$, one
obtains $1.5\times10^{-14}$ at $T_\gamma=4$~K, $0.24$ at $77$~K and $1.86$ at
$300$~K; with the free-space frequency dependence the same ratios would be
$1.4\times10^{-16}$, $2.2\times10^{-3}$ and $1.8\times10^{-2}$. The
fixed-$\mathbf K_0$ reservoir therefore requires a low photon temperature, and
we adopt $T_\gamma=4$~K as the baseline throughout this
work, where the omitted channel is exponentially dead.

Here $T_\gamma$ is the temperature of the photon modes that couple to
the sample, which need not coincide with that of the lattice. In the strict
fixed-$\bm K_0$ model the retained intra-excitonic reservoir consists of the
$s_\parallel=0$ modes, i.e.\ radiation propagating along the normal, which is
also the direction of the optical access of a cryostat. If the optical path
transmitted room-temperature radiation at $\hbar\omega=10.97$~meV
($\lambda\simeq113~\mu$m, $2.65$~THz), the occupation of that mode would be set
by the $300$~K environment and the omitted $2p\to2s$ channel would return to the
ratio $1.86$ quoted above, independently of the lattice temperature and of $A$.
The $4$~K baseline therefore presupposes that the far-infrared radiation reaching
the sample is thermalized with the cold stage, for instance by cold shields and a
cold filter that blocks the far infrared while transmitting the $113.19$~meV
control tone. The same exposure is harmless for the retained $2p\to1s$ channel,
for which $n_\gamma(113.19~{\rm meV},300~{\rm K})=1.3\times10^{-2}$.

The $V$ truncation omits one channel, the intra-excitonic $2s\to2p$ transition
at $10.97$~meV out of the upper driven state, which is $V$'s counterpart to the $\Xi$ leak above. Its
status is the opposite of that leak, and the difference is kinematic rather
than thermal. In $\Xi$ the omitted $2p\leftrightarrow2s$ channel competes with
a retained channel that is also intra-excitonic, so both carry the $A^{-1}$
scaling of Eq.~\eqref{eq:photon-intra-thermodynamic-limit} and their ratio is
$A$-independent --- which is why that leak cannot be argued away by the choice
of normalization area, and why it is temperature that controls it. In $V$ the
omitted $2s\to2p$ channel carries $A^{-1}$ while the retained exit of the same
state, the interband $2s\to0$ decay, is $A$-independent. Within the strict
fixed-$\bm K_0$ plane-wave model the omitted channel is therefore suppressed
relative to the retained one by the same factor that suppresses every
intra-excitonic rate, and vanishes in the thermodynamic limit. The $V$
three-level truncation is in this specific sense better controlled than the
$\Xi$ one, at every temperature; the comparison between the two three-level systems is
accordingly not biased in $\Xi$'s favor by the truncation itself.

\section{Rayleigh--Ritz Variational Solver for 2D Exciton States}
\label{app:rayleigh-ritz}
 
We describe the numerical method used to compute the exciton non-hydrogenic Rydberg series, and validate it against exact results for the two-dimensional Coulomb problem and against experimental data from WSe$_2$/hBN~\cite{Stier2018}. In this appendix, the main equations are quoted in natural units (lengths in $a_B$, energies in $E_0$, etc.), while the numerical results in the tables are given in conventional physical units (meV, nm, etc.), except where a natural unit is stated explicitly (e.g. radii in $a_B$, etc). The solver code and the post-processing and configuration-scoring code of Sect.~\ref{sec:top_manifold} was written with the assistance of Claude Opus 4.8 and 5.5 models to the authors' specification, and was verified by the authors against the exact two-dimensional Coulomb series (Table~\ref{tab:coulomb}) and the magneto-optical data of Ref.~\cite{Stier2018} (Table~\ref{tab:wse2}).  
 
\subsection{Variational ansatz}
 
The exciton envelope function in a two-dimensional semiconductor satisfies the radial Schr\"odinger equation (in natural units $E_0 = \mu\kappa_{\mathrm{coul}}^2/\hbar^2$, $a_B = \hbar^2/\mu\kappa_{\mathrm{coul}}$):
\begin{equation}
  \left[-\frac{1}{2}\left(\frac{d^2}{dr^2}+\frac{1}{r}\frac{d}{dr}
  -\frac{m^2}{r^2}\right)+V(r)\right]\psi=E\,\psi,
  \label{eq:radial}
\end{equation}
where $m$ is the angular momentum quantum number, $V(r)$ is the
electron--hole interaction potential, and
$\kappa_{\mathrm{coul}}=e^2/(4\pi\varepsilon_0\,\kappa_{\mathrm{env}})$
is the environmentally screened Coulomb coupling constant.
The code supports two potentials:
the Coulomb potential $V_{\mathrm{C}}(r)=-1/r$, and the Rytova--Keldysh
potential appropriate for quasi-2D semiconductors,
\begin{equation}
  V_{\mathrm{K}}(r)=-\frac{\pi}{2r_0^{\mathrm{eff}}}
  \left[H_0\!\!\left(\frac{r}{r_0^{\mathrm{eff}}}\right)
  -Y_0\!\!\left(\frac{r}{r_0^{\mathrm{eff}}}\right)\right],
  \label{eq:keldysh}
\end{equation}
where $H_0$ and $Y_0$ are the Struve and Bessel functions of the
second kind, and $r_0^{\mathrm{eff}}=r_0/(\kappa_{\mathrm{env}}\,a_B)$
is the effective screening length in natural units.
 
We expand each eigenstate in a basis of 2D hydrogenic radial
functions~\cite{yang1991} spanning \emph{several} variational length
scales at once. We take the basis to be the union of hydrogenic functions over
an even-tempered ladder of scales $\{\lambda_s\}_{s=1}^{N_s}$,
\begin{equation}
  \psi_n(r)=\sum_{s=1}^{N_s}\sum_{j=0}^{N_f-1}
  c_{n,sj}\,R_j^{(m)}(r;\lambda_s),
  \label{eq:ansatz}
\end{equation}
with $N_f$ radial functions per scale and
\begin{equation}
  R_j^{(m)}(r;\lambda)=A_{j}^{(m)}\,\xi^{|m|}\,e^{-\xi/2}
  \,L_{j}^{(2|m|)}(\xi),\qquad \xi=\frac{2r}{\nu_j\,\lambda},
  \label{eq:basis}
\end{equation}
where $\nu_j=j+|m|+\tfrac{1}{2}$ and $L_j^{(\alpha)}$ are the generalized
Laguerre polynomials. Throughout, $\psi_n(r)$ denotes the \emph{radial}
factor normalized as $\int_0^\infty|\psi_n|^2 r\,dr=1$, while the full
two-dimensional envelope used in the main text is
$\varphi_n(\bm r)=\psi_n(r)\,e^{im\theta}/\sqrt{2\pi}$, so that
$\int d^2\bm r\,|\varphi_n|^2=1$ and
$|\varphi_n(0)|^2=|\psi_n(0)|^2/2\pi$. $A_j^{(m)}$ is the normalization constant such
that $\int_0^\infty |R_j|^2\,r\,dr=1$. For $\lambda=1$ and the
Coulomb potential these functions are the exact 2D hydrogen
eigenstates.
 
\subsection{Generalized eigenvalue problem}
 
Substituting the multi-scale ansatz~\eqref{eq:ansatz} into the
Rayleigh--Ritz variational principle yields a \emph{single} generalized
eigenvalue problem
\begin{equation}
  \mathbf{H}\,\mathbf{c}_n = E_n\,\mathbf{O}\,\mathbf{c}_n,
  \label{eq:gep}
\end{equation}
where $H_{ij}=\langle R_i|\hat{H}|R_j\rangle$ and
$O_{ij}=\langle R_i|R_j\rangle$ are the Hamiltonian and overlap
matrices in the combined basis, computed by composite Gauss--Legendre
quadrature over an extended, piecewise radial grid. Because the basis
mixes several scales it is genuinely non-orthogonal ($\mathbf{O}\neq
\mathbf{I}$) and mildly ill-conditioned: nearby scales produce
near-linearly-dependent functions. We therefore diagonalize
$\mathbf{O}=\mathbf{V}_O\,\mathrm{diag}(s_\alpha)\,\mathbf{V}_O^{\mathsf
T}$, discard the directions whose eigenvalues fall below a small
relative threshold (guarding against variational collapse), and apply
symmetric (L\"owdin) orthogonalization
$\mathbf{U}=\mathbf{V}_{O,\mathrm{kept}}\,\mathrm{diag}(s_\alpha^{-1/2})$
to reduce Eq.~\eqref{eq:gep} to a standard eigenproblem
$\mathbf{U}^{\mathsf T}\mathbf{H}\mathbf{U}\,\mathbf{d}_n=E_n\mathbf{d}_n$,
$\mathbf{c}_n=\mathbf{U}\mathbf{d}_n$. The resulting eigenvectors are
$\mathbf{O}$-orthonormal, $\mathbf{c}_a^{\mathsf T}\mathbf{O}\,
\mathbf{c}_b=\delta_{ab}$, so the eigenstates form a genuine
orthonormal set---essential for the transition dipole matrix elements
and density form factors that enter the Lindbladian.
 
The scale ladder is even-tempered and auto-calibrated. A master scale
$\lambda_g$ is obtained by minimizing the $m{=}0$ ground-state energy,
$\lambda_g=\arg\min_\lambda E_0^{(m=0)}(\lambda)$, and the ladder is
placed geometrically over $[\,0.10\,\lambda_g,\,1.40\,\lambda_g\,]$.
The $m{=}0$ ground state provides a robust material length scale;
calibrating in-sector for $|m|\ge 2$ is unreliable, because there the
ground energy is nearly flat in $\lambda$ and the minimizer drifts to
the search boundary. Centering every angular-momentum sector on the
same master scale, with the deliberately wide span above, yields stable
bound series for all $m$. The pseudocode for the solver is given below.
\vspace{-2mm}
\begin{algorithm}[H]
\footnotesize
\caption{Multi-scale Rayleigh--Ritz solver}
\begin{algorithmic}[1]
\Require Potential $V(r)$, quantum number $m$; $N_f$ functions/scale, $N_s$ scales
\State Calibrate master scale $\lambda_g=\arg\min_\lambda E_0^{(m=0)}(\lambda)$
\State Build even-tempered ladder $\{\lambda_s\}$ over $[0.10\,\lambda_g,\,1.40\,\lambda_g]$
\State Assemble basis $\{R_j^{(m)}(r;\lambda_s):\,j=0,\ldots,N_f{-}1;\ s=1,\ldots,N_s\}$
\State Compute $H_{ij}$, $O_{ij}$ by composite Gauss--Legendre quadrature
\State Diagonalize $\mathbf{O}$; keep subspace with $s_\alpha>$ threshold
\State Form $\mathbf{U}=\mathbf{V}_{O,\mathrm{kept}}\,\mathrm{diag}(s_\alpha^{-1/2})$
\State Solve $\mathbf{U}^{\mathsf T}\mathbf{H}\mathbf{U}\,\mathbf{d}_n=E_n\mathbf{d}_n$; set $\mathbf{c}_n=\mathbf{U}\mathbf{d}_n$
\State \Return eigenvalues $\{E_n\}$ and $\mathbf{O}$-orthonormal eigenvectors $\{\mathbf{c}_n\}$
\end{algorithmic}
\end{algorithm}

\subsection{Validation I: 2D Coulomb benchmark}
 
For the Coulomb potential $V(r)=-1/r$, the exact 2D eigenvalues are
$E_n=-1/2(n-\tfrac{1}{2})^2$ with principal quantum number
$n=n_r+|m|+1$. Table~\ref{tab:coulomb} shows the solver output for
$m{=}0$ states using an even-tempered ladder that includes the exact
Coulomb scale $\lambda=1$. All eigenvalues are reproduced with relative
errors of order $10^{-10}$---limited by quadrature and the overlap
cutoff rather than by machine precision---and never fall below the exact
values, confirming the absence of variational collapse. The
$\mathbf{O}$-orthonormalized eigenvectors satisfy
$\max_{ab}|\mathbf{c}_a^{\mathsf T}\mathbf{O}\mathbf{c}_b-\delta_{ab}|
\lesssim 10^{-10}$.
 
\begin{table}[h!]
\centering
\caption{2D Coulomb benchmark ($m{=}0$), multi-scale basis with an
even-tempered ladder including $\lambda=1$. Energies in units of $E_0$.}
\label{tab:coulomb}
\footnotesize
\setlength{\tabcolsep}{3.5pt}
\begin{tabular}{@{}cccccc@{}}
\toprule
$n$ & $n_r$ & $\nu$ & $E_n^{\mathrm{exact}}$ & $E_n^{\mathrm{num}}$
& Rel.\ error \\
\midrule
 1 &  0 & 0.5 & $-2.000\,000\,0$ & $-2.000\,000\,0$
    & $2\times10^{-11}$ \\
 2 &  1 & 1.5 & $-0.222\,222\,2$ & $-0.222\,222\,2$
    & $2\times10^{-12}$ \\
 3 &  2 & 2.5 & $-0.080\,000\,0$ & $-0.080\,000\,0$
    & $5\times10^{-12}$ \\
 4 &  3 & 3.5 & $-0.040\,816\,3$ & $-0.040\,816\,3$
    & $7\times10^{-13}$ \\
 5 &  4 & 4.5 & $-0.024\,691\,4$ & $-0.024\,691\,4$
    & $4\times10^{-12}$ \\
 6 &  5 & 5.5 & $-0.016\,528\,9$ & $-0.016\,528\,9$
    & $2\times10^{-12}$ \\
 7 &  6 & 6.5 & $-0.011\,834\,3$ & $-0.011\,834\,3$
    & $3\times10^{-11}$ \\
 8 &  7 & 7.5 & $-0.008\,888\,9$ & $-0.008\,888\,9$
    & $2\times10^{-11}$ \\
 9 &  8 & 8.5 & $-0.006\,920\,4$ & $-0.006\,920\,4$
    & $2\times10^{-11}$ \\
10 &  9 & 9.5 & $-0.005\,540\,2$ & $-0.005\,540\,2$
    & $8\times10^{-12}$ \\
\bottomrule
\end{tabular}
\end{table}

\subsection{Validation II: WSe$_2$/hBN exciton Rydberg series}
\label{app:validation}
 
We compare the solver against the magneto-optical measurements of
Stier \textit{et al.}~\cite{Stier2018}, who resolved the $1s$--$4s$
Rydberg series of the neutral A-exciton in hBN-encapsulated monolayer
WSe$_2$ under magnetic fields up to 65\,T. Before presenting the
comparison, we describe how the reference values are obtained from
the original paper, and distinguish carefully between experimental
observables, derived experimental quantities, and theoretical results.
 
\paragraph{Experimental observables from Ref.~\cite{Stier2018}:}
The primary measured quantities are the absorption peak positions
(transition energies) of the $ns$ exciton states in the zero-field
optical spectrum. These are reported in the main text and
Figs.~1b--c and 2a of Ref.~\cite{Stier2018}:
$E_{\mathrm{trans}}(1s) = 1.723\,$eV (stated in text, p.\,2),
$E_{\mathrm{trans}}(2s) = 1.853\,$eV (stated in text, p.\,2, and
described as ``130\,meV higher in energy'' than the $1s$).
The $3s$ transition energy $E_{\mathrm{trans}}(3s)\approx 1.875\,$eV
is not quoted explicitly but follows from the statement (p.\,3) that
the $3s$ state lies ``152\,meV above the $1s$ ground state,''
i.e.\ $1.723+0.152=1.875\,$eV. 
 
\paragraph{Energy spacings (model-independent):}
Because the spacings
$\Delta E_{n,n'}=E_{\mathrm{trans}}(n's)-E_{\mathrm{trans}}(ns)$
are direct differences of measured peak positions, they are the most
model-independent experimental observables and do not depend on
any fitted parameters. The values used in Table~\ref{tab:wse2} are:
$2s{-}1s = 1.853-1.723 = 130\,$meV (stated explicitly on pp.\,2--4
of Ref.~\cite{Stier2018}),
$3s{-}2s = 1.875-1.853 = 22\,$meV (stated on p.\,4: ``experimentally-measured
separations of 130\,meV and 22\,meV'').
 
\paragraph{Absolute binding energies (derived):}
The binding energies
$E_b(ns) = E_{\mathrm{gap}} - E_{\mathrm{trans}}(ns)$ require the
free-particle bandgap, which Stier \textit{et al.}\ extract by fitting
the Rydberg series with their Keldysh model. They report (p.\,4 and
Fig.~3b) $E_{\mathrm{gap}}\approx 1.890\,$eV for hBN/WSe$_2$/hBN.
This yields:
$E_b(1s) = 1.890-1.723 = 167\,$meV,
$E_b(2s) = 1.890-1.853 = 37\,$meV, and 
$E_b(3s) = 1.890-1.875 = 15\,$meV.
Because $E_{\mathrm{gap}}$ is itself a model-dependent
quantity---extracted from a global fit rather than measured
directly---any systematic error in it shifts all binding energies by a
common offset. As a result, the binding energies are less robust
benchmarks than the energy spacings.
 
\paragraph{Theoretical values from Ref.~\cite{Stier2018}:}
In addition to the experimental data, Stier \textit{et al.}\ report
their own numerical solution of the Schr\"odinger equation with the
Keldysh potential (Eq.~\ref{eq:keldysh}), using the same parameters
we adopt ($\mu=0.20\,m_0$, $\kappa_{\mathrm{env}}=4.5$,
$r_0=4.5\,$nm). Their theoretical binding energies, obtained by a
finite-difference method on a real-space grid (Supplemental Material
of Ref.~\cite{Stier2018}), are:
$E_b^{\,\mathrm{th}}(1s) = 161\,$meV (stated on p.\,4),
$E_b^{\,\mathrm{th}}(3s) \approx 15.7\,$meV (read from Fig.~3b at
$\kappa=4.5$). The $2s$ theoretical binding energy is not given
individually, but the theoretical $1s{-}2s$ separation of
124\,meV (stated on p.\,4) yields
$E_b^{\,\mathrm{th}}(2s) \approx 161-124 = 37\,$meV. The
theoretical $2s{-}3s$ separation of 21.3\,meV (stated on p.\,4)
gives $E_b^{\,\mathrm{th}}(3s) \approx 37-21.3 = 15.7\,$meV,
consistent with the value read from Fig.~3b. 
These theoretical values serve as a benchmark for our numerics:
any deviation in the column ``$\Delta_{\mathrm{num}}$'' of
Table~\ref{tab:wse2} reflects differences in the two numerical
implementations of the \textit{same} Keldysh Hamiltonian, not
limitations of the physical model.
 
\paragraph{Material parameters:}
Following Ref.~\cite{Stier2018}, we use the exciton reduced mass
$\mu=0.20\,m_0$ (experimentally determined from the high-field shifts
of the $3s/4s$ excitons; see their Fig.~2b), the environmental
dielectric constant $\kappa_{\mathrm{env}}=4.5$ 
, and the WSe$_2$ intrinsic screening length
$r_0=4.5\,$nm (from \textit{ab initio} calculations~\cite{berkelbach2013}).
These yield $a_B=1.19\,$nm, $E_0=269\,$meV, and
$r_0^{\mathrm{eff}}=0.84\,a_B$.
No parameters are adjusted to fit the data. The individual carrier masses do not enter the Wannier problem, which depends
only on $\mu$; they enter only through the total mass $M$ and the mass
fractions $\alpha_{e,h}$ of Eqs.~\eqref{eq:X-carriertrans} and~\eqref{eq:appA-xx-cc-oriented}. For these quantities we adopt the first-principles carrier masses
$m_e=0.39\,m_0$ and $m_h=0.51\,m_0$ at the $K$ point of monolayer
WSe$_2$~\cite{Jin2014}, giving $M=0.90\,m_0$, $\alpha_e=0.433$ and
$\alpha_h=0.567$, while retaining the measured reduced mass $\mu=0.20\,m_0$.
The reduced mass implied by these carrier masses, $0.22\,m_0$, falls within the
range $0.16$--$0.22\,m_0$ spanned by first-principles calculations with~\cite{Kormanyos2015}
and without~\cite{Jin2014} spin--orbit coupling. Because only $M$ and
$\alpha_{e,h}$ are taken from Ref.~\cite{Jin2014}, this theoretical uncertainty
enters only the validity estimate of App.~\ref{app:offdiag}; none of the
tabulated quantities depend on it.
 
\paragraph{Results:}
For these parameters the auto-calibrated ladder spans
$\lambda_s\in[0.27,\,3.76]\,a_B$; the bound states settle at scales of
order a few $a_B$, larger than the Coulomb $a_B$, reflecting the
weakened (non-Coulombic) attraction at short range due to dielectric
screening from the 2D material.
Table~\ref{tab:wse2} compares our computed binding energies with both
the theoretical Keldysh-model values from Ref.~\cite{Stier2018}
(column ``$\Delta_{\mathrm{num}}$,'' testing our \textit{numerics})
and with the experimental values (column ``$\Delta_{\mathrm{model}}$,''
testing the \textit{physical model}).

\begin{table}[t]
\centering
\setlength{\tabcolsep}{8pt}
\renewcommand{\arraystretch}{1.2}
\caption{Validation of our multi-scale Rayleigh--Ritz numerical
code of the WSe$_2$/hBN Rytova--Keldysh problem. $\Delta_{\mathrm{num}}=\text{Code}-\text{Stier}$
is the numerics-vs-numerics difference between our algorithm and the independent finite-difference
Keldysh values of Stier \emph{et al.}~\cite{Stier2018}; $\Delta_{\mathrm{model}}=
\text{Code}-\text{Expt.}$ is the difference against the measured values; and
$w_{\mathrm{hyd}}$ is the overlap of each eigenstate with its best single
2D-hydrogenic function. Energy spacings are model-independent (they require no
absolute-energy reference).}
\label{tab:wse2}
\begin{tabular}{l r r r r r r}
\hline\hline
 & Code & Stier (th.) & $\Delta_{\rm num}$ & Expt. & $\Delta_{\rm model}$ &
 $w_{\rm hyd}$ \\
 & (meV) & (meV) & (meV) & (meV) & (meV) & \\
\hline
\multicolumn{7}{l}{\emph{Binding energies}} \\
$1s$ & $161.59$ & $161$  & $+0.6$ & $167$ & $-5.4$ & $0.995$ \\
$2s$ & $37.43$  & $37$   & $+0.4$ & $37$  & $+0.4$ & $0.972$ \\
$3s$ & $16.13$  & $15.7$ & $+0.4$ & $15$  & $+1.1$ & $0.973$ \\
\hline
\multicolumn{7}{l}{\emph{Energy spacings (model-independent)}} \\
$2s{-}1s$ & $124.2$ & $124$  & $+0.2$ & $130$ & $-5.8$ & --- \\
$3s{-}2s$ & $21.3$  & $21.3$ & $0.0$  & $22$  & $-0.7$ & --- \\
\hline\hline
\end{tabular}
\end{table}
 
\paragraph{Discussion:}
The comparison can be read at two levels.
 
\textit{Numerical accuracy} ($\Delta_{\mathrm{num}}$).---The column
$\Delta_{\mathrm{num}}$ compares our multi-scale Rayleigh--Ritz
eigenvalues with the independent finite-difference solution of the same
Keldysh Hamiltonian by Stier \textit{et al.}~\cite{Stier2018}. Across
the $1s$--$3s$ series the two agree to $\le 0.6\,$meV
($\Delta_{\mathrm{num}}=+0.6,\,+0.4,\,+0.4\,$meV), confirming that the
variational and finite-difference solutions of the same model coincide
at sub-meV accuracy.
 
\textit{Model accuracy} ($\Delta_{\mathrm{model}}$).---The column
$\Delta_{\mathrm{model}}$ compares our Keldysh-model results with
experimental binding energies. The excited states now agree closely
with experiment---the $2s$ and $3s$ to within $\sim\!1\,$meV---while the
$1s$ underbinds by $5.4\,$meV. The
residual on the $1s$ reflects contributions not captured by the
Wannier--Keldysh framework of Eq.~\eqref{eq:keldysh}---polaron
renormalization, Berry-curvature corrections, and short-range
exchange---which most strongly deepen the tightly bound ground state.

\textit{Energy spacings}.---The spacings provide the cleanest test,
being free of the $E_{\mathrm{gap}}$ extraction. Our $2s{-}1s$ and
$3s{-}2s$ spacings reproduce Stier's \emph{theoretical} Keldysh
spacings to $0.2$ and $0.0\,$meV, an independent confirmation of the
numerics. Against \emph{experiment}, the dominant $2s{-}1s$ spacing lies
$5.8\,$meV ($4.5\%$) below the measured $130\,$meV; because Stier's own
Keldysh theory shows the same offset, this discrepancy is a limitation
of the Rytova--Keldysh \emph{model}, not of the numerics. 
 
The last column of Table~\ref{tab:wse2} reports $w_{\mathrm{hyd}}$, the
squared overlap of each eigenstate with the best single hydrogenic
function at its own scale. All states are well described by a single
hydrogenic function ($w_{\mathrm{hyd}}\approx 0.97$--$1.00$); the strong
apparent mixing seen with a single \emph{shared} scale was a
basis artifact, not intrinsic non-hydrogenic character. Because the
multi-scale eigenstates form an $\mathbf{O}$-orthonormal set, the
transition dipole matrix elements and density form factors entering the
Lindbladian are evaluated directly in this common orthonormal basis,
with no residual non-orthogonality to correct.

\paragraph{Convergence of the retained states.} Table~\ref{tab:wse2}
certifies the $ns$ series against an independent finite-difference solution,
but the three-level systems retained in Sect.~\ref{sec:top_manifold} also use the $2p$
state and the splittings derived from it. We therefore verified these
separately by varying every discretization parameter of the scheme about its
default: the number of radial functions per scale ($N_f=5$--$8$), the number
of scales ($N_s=5$--$8$), the quadrature order ($2000$--$8000$ points), and
the overlap cutoff ($10^{-8}$--$10^{-12}$). Across all thirteen settings the
retained binding energies vary by less than $2.3\times10^{-5}$~meV, the
derived splittings $\Delta_\Xi=E_b(1s)-E_b(2p)$ and
$\Delta_V=E_b(1s)-E_b(2s)$ by less than $2.3\times10^{-5}$~meV, and the
matrix elements $|\varphi_{1s}(\bm 0)|^2$ and $|d_{1s,2p}|$ by less than
$10^{-4}$ and $10^{-6}$ in absolute value, respectively. The $2p$ state is
thus converged to the same level as the $1s$, and every quantity entering
the reduced dynamics is converged far beyond the precision at which it is
quoted. This is a statement about the numerics only: the model error against
experiment, $5.4$~meV on the $1s$ (Table~\ref{tab:wse2}), is larger by five
orders of magnitude and is unaffected by these parameters.

\subsection{Exciton eigenstates and inputs to the configuration classification}
\label{app:eigenstate_inputs}
 
Table~\ref{tab:appD_states} lists the eleven lowest exciton eigenstates together
with the per-state quantities that the configuration screening of
Sect.~\ref{sec:top_manifold} consumes. These quantities are the binding energy $E_b$, which
determines the leg splittings $\Delta E$; the interband
brightness proxy
$|\varphi_n(0)|^2=|\psi_n(0)|^2/(2\pi)$, which is nonzero
only for $s$ states; the mean-square radius
$\langle r^2\rangle_n$, which determines the size penalty
$\Pi_n$; and the normalized brightness
$\normq B_n=B_n/\max_k B_k$. All quantities are in natural units
($a_B=1.19$~nm, $E_0=269$~meV, Rytova--Keldysh screening length
$r_0^{\rm eff}=0.84\,a_B$).
 
The remaining input is the set of intraband (THz/mid-IR) dipole magnitudes
$|d_{nm}|=|e\langle n|\bm r|m\rangle|$ of Table~\ref{tab:appD_dipoles}. By the
$\Delta m=\pm1$ selection rule these connect only $s$- and $p$-like states, so
they form the $6\times5$ block shown, with all other pairs vanishing; note the
strong near-diagonal (adjacent-shell) elements and the weak off-diagonal ones.
The scores of Sect.~\ref{sec:top_manifold} use the normalized dipoles
$\normq D_{nm}=|d_{nm}|/\max_{i<j}|d_{ij}|$ [Eq.~\eqref{eq:normalized_primitives}], with
$\max|d_{ij}|=32.29\,a_B$ (the $6s$--$6p$ element)~\footnote{In Table~\ref{tab:appD_dipoles} the dipoles are listed in the $|m|=1$ eigenbasis,
$|d_{nm}| = |I_{nm}|/\sqrt{2}$ with
$I_{nm}=\int \psi_n(r)\,r\,\psi_m(r)\,r\,dr$.}. Together,
Tables~\ref{tab:appD_states} and~\ref{tab:appD_dipoles} constitute the complete
input to the configuration-dependent scores, so that the candidate rankings of
Tables~\ref{tab:rank_ladder_wse2}-\ref{tab:rank_v_wse2} can be reproduced from these solver outputs alone.

\begin{table*}[t]
\centering
\setlength{\tabcolsep}{9pt}
\renewcommand{\arraystretch}{1.2}
\caption{The eleven lowest exciton eigenstates of
hBN-encapsulated monolayer WSe$_2$, from the multi-scale Rayleigh--Ritz solver
of App.~\ref{app:rayleigh-ritz} ($E_0=269$~meV, $a_B=1.19$~nm,
$r_0^{\rm eff}=0.84\,a_B$). Columns are the binding energy $E_b$, mean radius
$\langle r\rangle$, mean-square radius $\langle r^2\rangle$, interband
brightness $|\varphi_n(0)|^2=|\psi_n(0)|^2/2\pi$, its normalized value
$\normq B_n$, the size penalty $\Pi_n$ [Eq.~\eqref{eq:size_penalty}], and the
optical character.}
\label{tab:appD_states}
\begin{tabular}{r l c r r r c c c l}
\hline\hline
$n$ & State & $(n_r,m)$ & $E_b$ (meV) & $\langle r\rangle$ ($a_B$) &
$\langle r^2\rangle$ ($a_B^2$) & $|\varphi_n(0)|^2$ & $\normq B_n$ & $\Pi_n$ &
Character \\
\hline
$0$  & $1s$ & $(0,0)$    & $161.59$ & $1.186$  & $1.99$    & $2.84\times10^{-1}$ & $1.0000$ & $0.5000$ & bright ($s$) \\
$1$  & $2p$ & $(0,\pm1)$ & $48.39$  & $3.663$  & $16.48$   & $0$                 & $0.0000$ & $0.1078$ & dark ($p$)   \\
$2$  & $2s$ & $(1,0)$    & $37.43$  & $5.401$  & $34.27$   & $2.98\times10^{-2}$ & $0.1050$ & $0.0549$ & bright ($s$) \\
$3$  & $3p$ & $(1,\pm1)$ & $19.09$  & $10.109$ & $118.10$  & $0$                 & $0.0000$ & $0.0166$ & dark ($p$)   \\
$4$  & $3s$ & $(2,0)$    & $16.13$  & $12.552$ & $179.40$  & $8.45\times10^{-3}$ & $0.0297$ & $0.0110$ & bright ($s$) \\
$5$  & $4p$ & $(2,\pm1)$ & $10.10$  & $19.543$ & $433.22$  & $0$                 & $0.0000$ & $0.0046$ & dark ($p$)   \\
$6$  & $4s$ & $(3,0)$    & $8.92$   & $22.693$ & $579.95$  & $3.47\times10^{-3}$ & $0.0122$ & $0.0034$ & bright ($s$) \\
$7$  & $5p$ & $(3,\pm1)$ & $6.23$   & $31.976$ & $1150.47$ & $0$                 & $0.0000$ & $0.0017$ & dark ($p$)   \\
$8$  & $5s$ & $(4,0)$    & $5.64$   & $35.831$ & $1438.67$ & $1.75\times10^{-3}$ & $0.0062$ & $0.0014$ & bright ($s$) \\
$9$  & $6p$ & $(4,\pm1)$ & $4.22$   & $47.409$ & $2518.52$ & $0$                 & $0.0000$ & $0.0008$ & dark ($p$)   \\
$10$ & $6s$ & $(5,0)$    & $3.89$   & $51.969$ & $3018.29$ & $1.00\times10^{-3}$ & $0.0035$ & $0.0007$ & bright ($s$) \\
\hline\hline
\end{tabular}
\end{table*}
 
\begin{table}[t]
\centering
\setlength{\tabcolsep}{11pt}
\renewcommand{\arraystretch}{1.2}
\caption{Intraband (THz/mid-IR) dipole magnitudes $|d_{nm}|=|e\langle n|\bm r|m
\rangle|$ connecting $s$-like (rows) and $p$-like (columns) states of
Table~\ref{tab:appD_states}; all other pairs vanish by $\Delta m=\pm1$. The
normalized dipoles are $\normq D_{nm}=|d_{nm}|/\max_{i<j}|d_{ij}|$ with
$\max|d_{ij}|=32.29\,a_B$ (the $6s$--$6p$ element).}
\label{tab:appD_dipoles}
\begin{tabular}{l r r r r r}
\hline\hline
$|d_{nm}|$ ($a_B$) & $2p$ & $3p$ & $4p$ & $5p$ & $6p$ \\
\hline
$1s$ & $0.854$ & $0.266$ & $0.146$  & $0.096$  & $0.070$  \\
$2s$ & $2.740$ & $2.783$ & $0.858$  & $0.474$  & $0.317$  \\
$3s$ & $0.005$ & $7.156$ & $5.676$  & $1.704$  & $0.931$  \\
$4s$ & $0.010$ & $0.039$ & $13.548$ & $9.545$  & $2.797$  \\
$5s$ & $0.009$ & $0.053$ & $0.137$  & $21.925$ & $14.392$ \\
$6s$ & $0.007$ & $0.041$ & $0.130$  & $0.289$  & $32.290$ \\
\hline\hline
\end{tabular}
\end{table}

\subsection{Negligibility of the off-diagonal exciton--gas exchange self-energy}
\label{app:offdiag}
Following the discussion in App.~\ref{app:exciton-gas-mean-field}, we show numerically that, for the selected three-level systems of the $\mathrm{WSe_2}/\mathrm{hBN}$ material considered here, the projected exciton-gas mean-field self-energy~\eqref{eq:appC-Xg-self-energy-discrete} is effectively diagonal in the retained Wannier eigenbasis.

The relevant criterion is not whether an off-diagonal element is strictly nonzero, but whether it produces appreciable mixing between the retained Wannier states of a given selected configuration. We therefore compare each off-diagonal self-energy matrix element with the corresponding separation between the diagonal mean-field-renormalized energies by defining
\begin{equation}
  \chi_{ab}
  =
  \frac{\bigl|\Sigma^{(Xg,0)}_{ab}\bigr|}
  {\bigl|\,\epsilon_a+\Sigma^{(Xg,0)}_{aa}
  -\epsilon_b-\Sigma^{(Xg,0)}_{bb}\,\bigr|}\,,
  \qquad a\neq b .
  \label{eq:eta}
\end{equation}
A value $\chi_{ab}\ll 1$ indicates that mixing through the corresponding $a\leftrightarrow b$ channel is perturbatively small and may be neglected. If this condition holds for all off-diagonal channels within the retained three-level system, the Wannier states remain excellent approximate eigenstates of the exciton-gas mean-field-renormalized Hamiltonian.

\paragraph{Approximations of the representative estimate.} The self-energy matrix~\eqref{eq:appC-Xg-self-energy-discrete} considers that the mean-field exciton gas occupation number $N_{l,\bm Q}$~\eqref{eq:appC-Xg-incoherent-one-body} is a stationary, internally incoherent bath, so its one-body density matrix is diagonal in the internal-state and momentum labels. In addition, we shall assume that $N_{l,\bm Q}$ is rotationally invariant with no net angular-momentum polarization,
$N_{l,\mathbf Q}\equiv N_l(|\mathbf Q|)$. Writing each envelope as
$\varphi_l(\mathbf{k})=R_l(k)\,e^{i m_l\theta_{\mathbf k}}$ - being $m_l$ the angular momentum of the $l$-th exciton energy level -, every populated
internal state contributes an \emph{isotropic} momentum density
$|\varphi_l(\mathbf k)|^2=R_l(k)^2$, so the total gas density is isotropic
irrespective of which internal states are occupied. The tagged transition density
carries the phase $e^{i(m_n-m_{n'})\theta_{\mathbf k}}$, and together with the
rotationally invariant exchange kernels and neglecting mass fraction recoils in the envelope arguments of Eqs.~\eqref{eq:appA-xx-cc-oriented} and~\eqref{eq:appA-xx-eh-full}, the angular integrals at $\mathbf K_0=0$ 
enforce
\begin{equation}
  \Sigma^{(Xg,0)}_{n'n}\propto \delta_{m_{n'},m_n},
  \label{eq:selrule}
\end{equation}
i.e. the exchange mean field is block-diagonal in the envelope angular momentum.
We stress that Eq.~\eqref{eq:selrule} follows from the incoherence and isotropy of
the exciton gas alone. Equation~\eqref{eq:selrule} settles both retained three-level systems at
once. For the ladder $\Xi=\{|0\rangle,|1s\rangle,|2p\rangle\}$ the only
excitonic off-diagonal element, $\Sigma_{1s,2p}$, connects $m=0$ to $|m|=1$ and
vanishes identically, so the $2\times2$ excitonic block is exactly diagonal.
For $V=\{|0\rangle,|1s\rangle,|2s\rangle\}$ the single admissible element is
$\Sigma_{1s,2s}$ ($m=0$), which is nonzero.

The selection rule~\eqref{eq:selrule} fixes the \emph{structure} of $\Sigma^{(Xg,0)}$~\eqref{eq:appC-Xg-self-energy-discrete}; its
\emph{magnitude} is set by the internal-state occupations $N_m$. Since $N_m/N_{1s}\lesssim 1\%$ for $m\neq 1s$, we retain only the first $1s$ contribution such that $N_{1s}\approx N_X$~\footnote{In
quasi-equilibrium these are Boltzmann suppressed by the intra-excitonic gaps,
which for WSe$_2$/hBN exceed $100$~meV (the smallest, $1s\!\to\!2p$, is
$113$~meV from Table~\ref{tab:appD_states}). The excited-internal fraction $\sim e^{-113\,\mathrm{meV}/k_BT}$ is $\approx1\%$
at room temperature and is $\mathcal{O}(10^{-143})$ at the adopted $4$~K
baseline, so at the baseline $N_{1s}=N_X$ to any accuracy of interest among the internal
states of the bright intravalley envelope model, and the retention of the $1s$
contribution alone is exact within that model; spin- and momentum-dark species
lie outside it and are discussed in Sect.~\ref{sec:lindblad_three_level}, item~\ref{item4}; we retain that
contribution to the vertices, the correction from thermally populated excited
internal states being at the percent level and immaterial to the conclusions
below.}. 

\paragraph{Evaluation.}
The exchange exciton-exciton interaction is the only contribution to the self-energy matrix~\eqref{eq:appC-Xg-incoherent-one-body}. Moreover, in the low-center-of-mass sector the
self-energy is linear in the exciton gas density, $\Sigma^{(Xg,0)}_{n'n}=n_X\, \zeta_{n'n}$
with $n_X=\int d^2Q\,N(|\mathbf Q|)/(2\pi)^2$. The reduced vertex separates into
carrier--carrier and electron--hole contributions,
$\zeta_{n'n}=\zeta^{cc}_{n'n}+\zeta^{eh}_{n'n}$, with the carrier--carrier part itself the
sum of a like-carrier and an electron--hole Coulomb term,
$\zeta^{cc}_{n'n}=\zeta^{cc,{\rm r}}_{n'n}+\zeta^{cc,{\rm a}}_{n'n}$, all built solely from the variational
envelopes~\eqref{eq:ansatz}, entering through their order-$|m|$ Hankel transforms
$R_n(k)=\int_0^\infty\! dr\,r\,\psi_n(r)\,J_{|m_n|}(kr)$.  

Both exchange channels are evaluated at $\bm K_0=\bm 0$ and in the
low-center-of-mass limit $\bm Q\to\bm 0$, so that the vertices are
independent of the gas momentum and
$\Sigma^{(Xg,0)}_{n'n}=n_X\,\zeta_{n'n}$. For the electron--hole channel this
is the approximation~\eqref{eq:appA-xx-eh-low-COM}; for the
carrier--carrier channel it is the $\bm\Delta=\bm K_0-\bm Q\to\bm 0$ limit
of Eq.~\eqref{eq:appA-xx-cc-oriented}, i.e.
Eq.~\eqref{eq:appA-xx-cc-zero-transfer}. The step is controlled when the
shifts $\alpha_{e,h}\bm\Delta$ are small compared with the momentum width of
the envelopes. With $\bm K_0=0$ the scale of $\bm\Delta$ is the thermal gas momentum
$Q_{\rm th}=\sqrt{2Mk_BT_X}/\hbar$, with $T_X$ the gas temperature, taken equal
to the lattice temperature. Since $\alpha_h>\alpha_e$ for WSe$_2$ (App.~\ref{app:validation}), the
larger shift is $\alpha_hQ_{\rm th}$, which reaches $0.09$, $0.41$, $0.58$ and
$0.82$ of the $1s$ momentum scale $\langle k\rangle a_B=0.646$ at $T_X=4$, $77$,
$150$ and $300$~K respectively (the electron shift $\alpha_eQ_{\rm th}$ reaches
$0.07$, $0.32$, $0.44$ and $0.62$). At the adopted baseline $T_X=4$~K the shift
is $9\%$ of the $1s$ momentum scale and the reduction is well controlled; it
remains usable, at the level of a $40\%$ shift, up to liquid-nitrogen
temperature and becomes marginal at room temperature, where a
momentum-resolved evaluation of Eq.~\eqref{eq:appA-xx-cc-oriented} under
$\int d^2Q\,N(|\bm Q|)$ would be required. The baseline is therefore the same
one at which the three-level truncation is controlled (Sect.~\ref{sec:lindblad_three_level},
item~\ref{item6}); the two independent conditions select it jointly.

The carrier--carrier exchange~\eqref{eq:appA-xx-cc-symmetrized} is evaluated in real space, where the kernel is finite:
\begin{equation}
\label{eq:Xicc}
\begin{split}
  \zeta^{cc,{\rm r}}_{n'n}&=2\,(2\pi)\,\delta_{m_{n'},m_n}\!\int_0^\infty\!\! dr\,r\,V(r)\,
     \Big[\mathcal F^{(|m_n|)}_{n',1s}(r)\,\mathcal F^{(|m_n|)}_{1s,n}(r)\,+\,\mathcal F^{(0)}_{n',n}(r)\,\mathcal F^{(0)}_{1s,1s}(r)\Big],\\
  \mathcal F^{(m)}_{ab}(r)&=\int_0^\infty\!\! dk\,k\, R_a(k)\,R_b(k)J_m(kr),
\end{split}
\end{equation}
where $V(r)$ is the Rytova--Keldysh kernel~\eqref{eq:keldysh} and the factor
$2$ collects the carrier--carrier and valence--valence channels, which
contribute equally. The two terms in the bracket are the two contractions of
the exchange vertex derived in App.~\ref{app:xx-internal-state-exchange}: the
first cross-pairs the tagged and gas envelopes and is the one retained in the
conventional single-internal-state treatment, the second pairs the envelopes
of each exciton with themselves. They enter with equal weight, as
Eq.~\eqref{eq:appA-xx-cc-zero-transfer} shows, and only their sum is
Hermitian. The Kronecker delta is the selection rule~\eqref{eq:selrule},
which both terms produce independently through their angular integrals; it is
written explicitly here because the radial integral alone does not enforce
it. Because $V(r)$ in Eq.~\eqref{eq:Xicc} is the Rytova--Keldysh kernel, which
is minus the repulsive carrier--carrier interaction, the positive prefactor
reproduces the factor $-2$ of Eq.~\eqref{eq:appA-xx-cc-zero-transfer}:
$\zeta^{cc,{\rm r}}_{n'n}$ is negative because of the fermionic exchange
sign, not because the line is attractive. The electron--hole Coulomb term is
evaluated in the single-momentum form~\eqref{eq:appA-xx-cc-att-wannier}, with
the Hermitian prescription of App.~\ref{app:xx-internal-state-exchange},
\begin{equation}
\zeta^{cc,{\rm a}}_{n'n}=2\,(2\pi)\,\delta_{m_{n'},m_n}\!\int_0^\infty\!\!dk\,k\,
R_{n'}(k)\,R_n(k)\,R_{1s}(k)^2\Big[k^2+E_b(1s)+\tfrac12\big(E_b(n)+E_b(n')\big)\Big],
\label{eq:Xica}
\end{equation}
in natural units, with binding energies in $E_0$; its diagonal elements are
positive definite. The electron--hole (interband)
exchange, whose kernel $\propto|q|$ is regular, is kept in momentum space,
\begin{equation}
\label{eq:Xieh}
\begin{split}
  \zeta^{eh}_{n'n}&=2\!\int_0^\infty\!\!dk\,k\!\int_0^\infty\!\!dk'\,k'\,
   R_{n'}(k)\,R_{n}(k)\ R_{1s}(k')^2\,w_0(k,k'),\\
  w_0(k,k')&=\tfrac{E_0}{4\,E_g}\!\int\! d\phi\,\frac{|k-k'|}{1+r_0^{\rm eff} |k-k'|}\,,
\end{split}
\end{equation}
where we used explicitly the band-edge
interband dipole~\eqref{eq:bloch-interband-dipole} and the Rytova–Keldysh potential~\eqref{eq:keldysh}. The variational eigenvectors carry an arbitrary overall sign, which does not
affect the diagonal vertices but does affect the off-diagonal ones. We fix it
by requiring $R_n(k)>0$ as $k\to0$ for every $n$. 

The numerical evaluation of the matrix elements~\eqref{eq:Xicc} and~\eqref{eq:Xieh} rests only on the Wannier solver output and is checked at five levels.
(i) The Hankel transforms are unitary, $\int_0^\infty dk\,k\, R_n(k)^2=1$
to better than $10^{-5}$ for every retained state, and to $10^{-9}$ on the
composite Gauss--Legendre momentum grid used for the convergence check~(iv).
(ii) The self-energy matrix is real in this basis - the Wannier envelopes and the exchange kernels being real - so its Hermiticity reduces to the symmetry $\Sigma_{ab}=\Sigma_{ba}$; for the like-carrier vertex~\eqref{eq:Xicc} this is an
analytic property of the two contractions, and for the electron--hole term~\eqref{eq:Xica}
it holds by the Hermitian prescription of App.~\ref{app:xx-internal-state-exchange}. (iii) the symmetry-forbidden element
$\Sigma_{1s,2p}$ vanishes numerically, consistent with
Eq.~\eqref{eq:selrule}. (iv) The radial quadrature is converged: the vertices
are unchanged to four decimal places under independent refinement of the
radial grid ($4000$--$8000$ points, $r_{\max}=400$--$800\,a_B$) and of the
momentum grid ($4000$--$8000$ points, $k_{\max}=60$--$100\,a_B^{-1}$). The carrier--carrier channel~\eqref{eq:Xicc} is
evaluated in real space so that the
integrable Coulomb singularity $|k-k'|^{-1}$  of the momentum representation never has to be sampled on a grid; the price is that the form factor
$\mathcal F^{(m)}_{ab}(r)$ inherits a slowly decaying oscillatory tail from
the finite momentum cutoff, so the radial quadrature must either be truncated
at the physical support of the integrand or use a momentum grid fine enough
to suppress that tail. (v) The code reproduces the two-dimensional hydrogenic
limit. With the bare Coulomb kernel ($r_0^{\rm eff}\to0$), for which
$E_b(1s)=2E_0$ and the envelopes are analytic, it returns
$\zeta^{cc,{\rm r}}_{1s,1s}=-19.0762$ and $\zeta^{cc,{\rm a}}_{1s,1s}=+25.1327$,
so that $\zeta^{cc,{\rm r}}_{1s,1s}+\zeta^{cc,{\rm a}}_{1s,1s}=+6.0566\,E_0a_B^2$;
the analytic values are $-315\pi^3/512$, $8\pi$ and
$8\pi-315\pi^3/512=6.05661$, a relative difference of $5\times10^{-6}$.
With $\lambda_{2D}=a_B/2$, i.e.\ $E_b(1s)\lambda_{2D}^2=E_0a_B^2/2$, the same
vertex is $6.057\,E_b(1s)\lambda_{2D}^2$ per exchanged carrier species, the
quantity given at $q=0$ by Fig.~2 and Eq.~(22) of Ref.~\cite{ciuti1998}. Their
Monte Carlo value read from that figure, $(2/\pi)^2\times14.7\simeq5.94$, is
$2\%$ lower. Per exchanged carrier species the like-carrier and
electron--hole parts are $-19.08$ and $+25.13\,E_b(1s)\lambda_{2D}^2$, in
agreement with the two partial exchange integrals, $\simeq19$ and $\simeq-25$
in the same units, of Ref.~\cite{Glazov2009}, up to the sign convention of
that work. The like-carrier lines alone would give the opposite sign,
$-19.08\,E_0a_B^2$.

\paragraph{Results for WSe$_2$/hBN.}
Table~\ref{tab:offdiag} lists the reduced vertices obtained from the eleven-state
solver output of Table~\ref{tab:appD_states}. 
Because $\chi_{ab}$ is a ratio of self-energies to level spacings, it is
controlled by the single physical scale $\Sigma_{1s}\equiv\Sigma^{(Xg,0)}_{1s,1s}$ (the $1s$ shift, positive in the present convention) and is otherwise independent of the overall normalization
\begin{equation}
\label{eq:eta_approx}
  \chi_{ab}\simeq
  \frac{|\zeta_{ab}|}{|\epsilon_a-\epsilon_b|}\,\frac{\Sigma_{1s}}{\zeta_{1s,1s}}
  \;\;+\;O\!\left(\Sigma_{1s}^2\right).
\end{equation}
Since $\zeta_{1s,1s}>0$, the reference shift $\Sigma_{1s}$ is positive, a blue
shift, and Table~\ref{tab:eta} is tabulated against $n_X$ together with
$\Sigma_{1s}$. Across the validated range $n_X\le2\times10^{11}$~cm$^{-2}$ one
finds $\chi_{1s,2s}\le0.023$ with the Hermitian prescription and at most
$0.046$ for either of the in and out forms, and the criterion $\chi=0.1$ is
reached only at $n_X=8.7\times10^{11}$~cm$^{-2}$ ($4.4\times10^{11}$~cm$^{-2}$
for the less favorable form), i.e.\ $n_Xa_B^2\approx0.01$, outside the
dilute window. The retained Wannier states are therefore good approximate
eigenstates of the mean-field renormalized Hamiltonian throughout the regime
of interest, and the dressed basis of Eqs.~\eqref{eq:eigenvalue} need not be
constructed explicitly.

We therefore neglect the off-diagonal element of the $V$ block in the reduced
dynamics. The two leading orders at which it would enter observable quantities
fix the size of what is omitted. At second order it repels the two levels,
widening their separation by
$2|\Sigma^{(Xg,0)}_{1s,2s}|^2/(\epsilon_{2s}-\epsilon_{1s})=0.033$~meV at
$n_X=10^{11}$~cm$^{-2}$, i.e.\ $7.8\%$ of the first-order differential shift of
Sect.~\ref{sec:VE-renormalization}; the ratio scales as $n_X$ and stays below $10\%$ up to
$n_X=1.3\times10^{11}$~cm$^{-2}$. At first order it admixes the envelopes,
changing $\varphi_{2s}(0)$, and hence the $2s$ Rabi coupling, by $3.6\%$ and the
$2s$ interband rate by about $7\%$ at the same density, with $0.4\%$ and $0.7\%$ for the
$1s$. These estimates use the Hermitian prescription for the off-diagonal
vertex; the less favorable of the in and out forms doubles them. Neither
correction alters the structure of the $V$ generator, and neither
affects the screening of Sect.~\ref{sec:top_manifold}, whose rankings are unchanged at the per-cent
level of $\chi$. They set instead the density-dependent accuracy of the $V$
results: better than $10\%$ for $n_X\lesssim1.4\times10^{11}$~cm$^{-2}$, and about
$14\%$ at the top of the window of Table~\ref{tab:eta}. For the ladder $\Xi$ no approximation is involved: $\Sigma^{(Xg,0)}_{1s,2p}\equiv0$
by Eq.~\eqref{eq:selrule}, so its excitonic block is exactly diagonal and the diagonal
treatment is exact to all orders in the mean field.

Two limitations of this estimate should be recorded. First, the vertices are
evaluated in the $\bm\Delta\to\bm 0$ limit discussed above, which is
controlled up to liquid-nitrogen temperature and marginal at $300$~K.
Second, the composite-boson (phase-space-filling) corrections dropped
throughout scale as $n_X\lambda$ with the exchange area
$\lambda=A\sum_{\bm p}|\varphi_{1s}(\bm p)|^4
 =2\pi\int_0^\infty dk\,k\,R_{1s}(k)^4=13.40\,a_B^2=18.98$~nm$^2$; over the
window of Table~\ref{tab:eta} the product $n_X\lambda$ runs from
$1.9\times10^{-3}$ to $3.8\times10^{-2}$, so the neglected Pauli terms are at
the few-per-cent level at the top of that window and grow linearly with
density. The results quoted here should accordingly be read as accurate at
the per-cent level for $n_X\lesssim10^{11}$~cm$^{-2}$.

\begin{table}[t]
\centering
\caption{Reduced exciton--gas exchange vertices $\zeta_{ab}$ (in $E_0 a_B^2$,
$E_0=269$~meV, $a_B=1.19$~nm) for WSe$_2$/hBN at $\mathbf K_0=0$, from the
Wannier envelopes of Table~\ref{tab:appD_states}. Off-diagonal elements not listed vanish by the
selection rule~\eqref{eq:selrule}. Within the carrier--carrier exchange, the
like-carrier term $\zeta^{cc,{\rm r}}$ [Eq.~\eqref{eq:Xicc}] is negative by
the fermionic exchange sign and the electron--hole term $\zeta^{cc,{\rm a}}$
[Eq.~\eqref{eq:Xica}] is positive; the interband electron--hole channel
carries the manifestly positive kernel of Eq.~\eqref{eq:Xieh}. Off-diagonal
entries of $\zeta^{cc,{\rm a}}$ use the Hermitian prescription of
App.~\ref{app:xx-internal-state-exchange}; the in and out forms give
$\zeta_{1s,2s}=7.42$ and $0.09$, and $\zeta_{2p,3p}=3.26$ and $2.00$. The sign convention on the
off-diagonal entries is $R_n(k)>0$ as $k\to0$.}
\label{tab:offdiag}
\begin{tabular}{lcccc}
\hline\hline
$(a,b)$ & $\zeta^{cc,{\rm r}}_{ab}$ & $\zeta^{cc,{\rm a}}_{ab}$ & $\zeta^{eh}_{ab}$ & $\zeta_{ab}$ \\ \hline
$(1s,1s)$ & $-32.24$ & $37.60$ & $0.215$  & $5.57$ \\
$(2s,2s)$ & $-30.97$ & $37.46$ & $0.189$  & $6.68$ \\
$(2p,2p)$ & $-27.40$ & $32.03$ & $0.199$  & $4.83$ \\
$(3p,3p)$ & $-27.96$ & $35.79$ & $0.184$  & $8.01$ \\
$(1s,2s)$ & $-9.92$  & $13.70$ & $-0.022$ & $3.76$ \\
$(2p,3p)$ & $-3.99$  & $6.64$  & $-0.013$ & $2.63$ \\
\hline\hline
\end{tabular}
\end{table}

\begin{table}[t]
\centering
\caption{Mixing diagnostic $\chi_{ab}$, Eq.~\eqref{eq:eta_approx}, versus the
exciton density $n_X$, with the corresponding $1s$ mean-field shift
$\Sigma_{1s}=\Sigma^{(Xg,0)}_{1s,1s}$, for the single admissible
off-diagonal element of each retained three-level system ($V$: $1s$--$2s$, spacing
$124.16$~meV). The ladder $\Xi$ has no admissible off-diagonal element. The
last column gives the range spanned by the in and out forms of the
off-diagonal vertex (App.~\ref{app:xx-internal-state-exchange}).}
\label{tab:eta}
\begin{tabular}{lccc}
\hline\hline
$n_X$ (cm$^{-2}$) & $\Sigma_{1s}$ (meV) & $\chi_{1s,2s}$ ($V$) & in/out range \\ \hline
$1\times10^{10}$   & $0.21$ & $0.0012$ & $0.0000$--$0.0023$ \\
$2.5\times10^{10}$ & $0.53$ & $0.0029$ & $0.0001$--$0.0057$ \\
$5\times10^{10}$   & $1.06$ & $0.0058$ & $0.0001$--$0.0114$ \\
$1\times10^{11}$   & $2.12$ & $0.0115$ & $0.0003$--$0.0228$ \\
$2\times10^{11}$   & $4.24$ & $0.0231$ & $0.0006$--$0.0455$ \\
\hline\hline
\end{tabular}
\end{table}

\subsection{Drive-induced ionization of the \texorpdfstring{$2p$}{2p} exciton}
\label{app:ionization}
The magnitude of the ionization channel of item~\ref{item7} of Sect.~\ref{sec:lindblad_three_level} can be estimated within the Rytova--Keldysh
model of this appendix. To first order in the drive, Fermi's golden rule with the rotating-wave
amplitude of Eq.~\eqref{eq:modelA_legresolved} gives, for the retained polarization-aligned state
$|p_d\rangle$,
\begin{equation}
\begin{split}
\hbar\Gamma_{\rm ion}&=\frac{\pi}{2}\,
\frac{(\hbar\Omega_{23})^2}{|\langle 1s|\mathbf e_d\!\cdot\!\mathbf r|p_d\rangle|^{2}}\,
S_d(E_f),\\
S_d(E)&=\sum_f\big|\langle f|\mathbf e_d\!\cdot\!\mathbf r|p_d\rangle\big|^2\,
\delta(E_f-E),
\end{split}
\label{eq:2p-ionization}
\end{equation}
where the sum runs over the electron--hole scattering states of Eq.~\eqref{eq:radial}, which
the dipole selection rule restricts to $m=0$ and $|m|=2$, and
$E_f=\hbar\omega_{L,23}-E_b(2p)=64.80$~meV. Evaluating $S_d$ with
energy-normalized continuum solutions of Eq.~\eqref{eq:radial}, cross-checked against a box
discretization of the continuum to within $2\%$, gives
$S_d(E_f)=0.141\,a_B^2/E_0$, $98\%$ of it in the $|m|=2$ channel, and hence
\begin{equation}
\begin{split}
\hbar\Gamma_{\rm ion}&=0.304\,\frac{(\hbar\Omega_{23})^2}{E_0}\\
&=1.13\times10^{-3}\,{\rm meV}^{-1}\,(\hbar\Omega_{23})^2 .
\end{split}
\label{eq:2p-ionization-num}
\end{equation}
The rate is independent of $A$, being fixed by $\Omega_{23}$, and is small
against the vertex width: $\hbar\Gamma_{\rm ion}=2.9~\mu$eV
($\Gamma_{\rm ion}^{-1}=0.23$~ns) at the EIT/Autler--Townes crossover
$\hbar\Omega_{23}\simeq1.6$~meV of Sect.~\ref{subsec:explicit_drives}, and $28~\mu$eV
($\Gamma_{\rm ion}^{-1}=23$~ps), i.e.\ $0.9\%$ of
$\hbar\gamma^{(\gamma,{\rm inter})}_{1s\to0}$, at $\hbar\Omega_{23}=5$~meV.
Because the channel acts only on the $2p$ population, the accumulated loss of a
pulse sequence is $P_{\rm ion}=\int dt\,\Gamma_{\rm ion}(t)\,\bar\rho_{33}(t)$,
which we evaluate from the solution of Eq.~\eqref{eq:Xi-master-components} to first order in
$\Gamma_{\rm ion}$, without modifying the generator. For the weak-excitation
conversion of Sect.~\ref{subsec:Xi-transfer}, with Gaussian envelopes of width
$\sigma_\ell=1.1$--$1.9$~ps and delay $(1.0$--$1.5)\,\sigma_\ell$, the loss per
converted $2p$ exciton is $P_{\rm ion}/\bar\rho_{33}(t_f)=0.5$--$1.6\%$ at
$\hbar\Omega_{23}=5$~meV and $0.2$--$0.7\%$ at $3$~meV. Drive-induced ionization is
therefore negligible in the response regime and a sub-percent to percent-level
loss of the converted population, more than thirty times smaller than the
emission from the $1s$ vertex; it bears on the achievable conversion efficiency,
not on the structure of the generator derived here.

\section{Generalities of the numerical solver of the Lindblad dynamics}
\label{app:num_Lindblad}

The reduced dynamics of Eqs.~\eqref{eq:Xi-component-system} and~\eqref{eq:V-component-system}
is integrated in the drive-rotating frame as a linear equation for the
column-stacked density matrix, 
\begin{equation}
\begin{split}
    \dot{\vec\rho}&=\mathcal L(t)\,\vec\rho\,,\\
    \mathcal L(t)&=\mathcal L_0+\sum_\ell\Omega_\ell(t)\,\mathcal K_\ell
\end{split}
\end{equation}
Here $\mathcal L_0$ contains the renormalized detunings and the photon
dissipator~\eqref{eq:generic-photon-dissipator}, and $\mathcal K_\ell$ is the coherent
coupling of leg $\ell$ at unit Rabi frequency. Trace, Hermiticity and
positivity are not built into this representation, so they serve as
independent checks. All quantities are expressed in the natural units of
App.~\ref{app:rayleigh-ritz} ($E_0=269$~meV, $\hbar/E_0=2.45$~fs). The control
inputs are the renormalized detunings $\hbar\ren\Delta$ and the Rabi
energies $\hbar\Omega$; carrier frequencies and field amplitudes are recovered
from them afterwards through Eqs.~\eqref{eq:VE-renormalized-detunings}-\eqref{eq:VE-two-photon-detunings},
\eqref{eq:Omega_inter_def} and~\eqref{eq:Omega_intra_def}. Time-dependent runs use the
explicit eighth-order Dormand--Prince integrator DOP853~\cite{Hairer1993,
Virtanen2020}. Its adaptive step is controlled by relative and absolute
tolerances $10^{-8}$ and $10^{-10}$, tightened to $10^{-10}$ and $10^{-12}$ in
purely coherent runs, where $\bar\rho$ stays of rank one. In the
weak-excitation $\Xi$ runs, whose excited populations are of order $10^{-3}$,
the absolute tolerance is lowered to $10^{-13}$ ($10^{-14}$ in coherent runs). The integration
proceeds piecewise between the envelope breakpoints, with the step capped at
one twentieth of the shortest envelope time. The emitted vertex-photon number
$N_{\rm emit}=\int dt\,\gamma^{(\gamma)}_{2\to1}\bar\rho_{22}$ and the
ionization loss $P_{\rm ion}$ of Sect.~\ref{sec:lindblad_three_level}, item \ref{item7}, are
accumulated as additional components of the same system. Stationary
responses under plateau drive are obtained directly from the null space of the
time-independent generator, $\mathcal L\vec\rho_{\rm ss}=0$ with
${\rm Tr}\,\bar\rho_{\rm ss}=1$, by singular-value decomposition of the
augmented matrix. The null space is one-dimensional in every dissipative run.
Finite plateaus are propagated only to quantify how closely a pulse of given
duration $T_{\rm pulse}$ approaches this steady state.

The solver was validated against known results. The generic generator reproduces the component
equations~\eqref{eq:Xi-component-11}--\eqref{eq:Xi-component-13} and
\eqref{eq:V-component-11}--\eqref{eq:V-component-23} to $3\times10^{-17}$
for arbitrary phases. Along every trajectory, trace and Hermiticity are
preserved to $3\times10^{-15}$ and the smallest eigenvalue of $\bar\rho$ stays
above $-3\times10^{-10}$. With the drives switched off, the populations follow
the rate equations~\eqref{eq:common-undriven-populations} and the coherences decay at the
rates~\eqref{eq:common-coherence-damping} to $5\times10^{-15}$, at 4 and 300~K and
with the exchange mean field included. The weak-probe $\Xi$ susceptibility,
extrapolated to $\Omega_{12}\to0$, agrees with the analytic homogeneous
cascade susceptibility~\cite{GeaBanacloche1995} to better than $3\times10^{-6}$ for
$\hbar\Omega_{23}=0.5$--$5$~meV. The free decay of $\bar\rho_{23}$ after a
short $V$ pulse pair reproduces $\tfrac12(\gamma^{(\gamma)}_{2\to1}+
\gamma^{(\gamma)}_{3\to1})$ to $7\times10^{-7}$, independently of the drive
amplitude, together with the 33.3~fs beat period. In the linear regime the
stationary $V$ coherence reduces to its lowest-order form
$\bar\rho_{23}=\bar\rho_{21}\bar\rho_{13}$. For the counterintuitive $\Xi$
sequence, the coherent Hamiltonian annihilates the dark
state~\eqref{eq:xi-dark-state} to machine precision. In the
weak-excitation regime the conversion efficiency~\eqref{eq:conversion-efficiency}
agrees to $2\times10^{-3}$ of its value with an independent integration of
the first-order amplitude equations, in which the ground-state amplitude is
held at one and the vertex carries the energy
$-i\hbar\gamma^{(\gamma,{\rm inter})}_{1s\to0}/2$; the comparison covers the
baseline, the sequential order, the uncalibrated case~3 and the window
$\hbar\Omega_{23}=3$--$5$~meV, $\sigma=1.1$--$1.9$~ps. The efficiency is
independent of $\Omega_{12}$ to $10^{-3}$. Plateaus of 50--60~ps converge to the
null-space steady state to $10^{-6}$ when the pumping time into the dark state
is short compared with the plateau, i.e.\ for $\hbar\Omega_{23}\gtrsim1$~meV and in
the $V$ response; the narrow window at $\hbar\Omega_{23}=0.5$~meV forms more slowly
(Sect.~\ref{subsec:Xi-response}). A reference run repeated with
the implicit Radau integrator~\cite{Hairer1996} and with QuTiP's
\texttt{mesolve}~\cite{Lambert2026} agrees to $\sim10^{-9}$. 

\paragraph*{Run details of Sects.~\ref{subsec:Xi-response} and~\ref{subsec:V-comparison}.}
On the coherent-population-trapping plateau of Fig.~\ref{fig:R4}(a), the bright
component $\sin^2\theta=2.5\times10^{-3}$ of the ground state drives the vertex
to a peak population $\bar\rho_{22}=1.6\times10^{-4}$; its emission,
$N_{\rm emit}=2.50\times10^{-3}$, returns it until $\bar\rho_{22}=8.7\times10^{-10}$
at the end of the plateau, and the purity stays above $1-1.1\times10^{-6}$. After
the switch-off $\bar\rho_{33}=2.49\times10^{-3}$, $99.7\%$ of
$(\Omega_{12}/\Omega_{23})^2$, and $|\bar\rho_{13}|=0.0499\simeq\Omega_{12}/\Omega_{23}$.
In case~3 the vertex population at the end of the plateau is
$\bar\rho_{22}=4.6\times10^{-5}$. The weak-probe response is linear in the probe
to $3\times10^{-3}$, from a comparison with half the probe amplitude. The
absorption at the window center is $4\Gamma_{12}\Gamma_{13}/\Omega_{23}^2$, and
there $|\bar\rho_{13}|=\Omega_{12}/\Omega_{23}$, the dark-state ratio. The window
width is $\Omega_{23}^2/2\Gamma_{12}$ for weak control (FWHM $0.077$~meV at
$\hbar\Omega_{23}=0.5$~meV, $2\%$ below this value) and falls below it as the
control grows ($9\%$ at $1$~meV, $19\%$ at $1.5$~meV). The dispersion at the
center has the slope $d\,{\rm Re}\,\hat\chi/d\ren\Delta_{12}=4\Gamma_{12}/\Omega_{23}^2$,
$6.4$~meV$^{-1}$ at $1$~meV, ten times the two-level value and of opposite sign.
The separation of the absorption maxima grows linearly with $\Omega_{23}$ on
both sides of the crossover, while above $\Gamma_{12}-\Gamma_{13}$ the poles split by
$[\Omega_{23}^2-(\Gamma_{12}-\Gamma_{13})^2]^{1/2}$, which approaches $\Omega_{23}$.
Population is pumped into the dark state at about $\Omega_{23}^2/4\Gamma_{12}$.
In the $V$ run of Fig.~\ref{fig:R5}(a)--(c), $\bar\rho_{22}$ decays in
$0.21$~ps and $\bar\rho_{33}$ in $1.8$~ps; the fitted decay rate of
$|\bar\rho_{23}|$ reproduces $\Gamma^{V}_{23}$ to $10^{-6}$, and $11$ beat periods
elapse within $T_2$. Without the dissipator the pulses leave
$\bar\rho_{22}=\bar\rho_{33}=|\bar\rho_{23}|=0.050$; in case~3 the mean field
detunes the two legs by $2.1$ and $2.5$~meV; the purity stays above $0.997$.
The extrapolated beat amplitude agrees with the lowest-order prediction to
within $2$--$7\%$ at $1$~meV and $0.5$--$2\%$ at $0.5$~meV; the deviation grows
with the pulse area ($0.35$--$0.69$~rad) and does not affect $T_2$. It reflects
the excitation these pulses create, $\max_t(1-\bar\rho_{11})=0.04$--$0.11$ at
$1$~meV and $0.010$--$0.029$ at $0.5$~meV, which at $1$~meV allows truncation
errors of up to about $10\%$, consistent with that deviation; the $0.5$~meV runs
are accurate to about $2\%$. Under the weak plateau drive of
Fig.~\ref{fig:R5}(e) the Lorentzian fit of ${\rm Im}\,\hat\chi_{13}$ has a
residual of $10^{-7}$, the $0\to1s$ absorption is flat to $8\times10^{-4}$ over
the scan of $\ren\delta_V$, and $\bar\rho_{23}\simeq\bar\rho_{21}\bar\rho_{13}$
holds to $1\%$, so that $|\bar\rho_{23}|^2$ has the same half-width.

\paragraph*{Use of AI tools.} The code of this appendix---the propagator and
steady-state solver, the scripts that execute the runs of
Sect.~\ref{sec:results} and reduce their output to the reported quantities
(conversion efficiencies, loss budgets, linewidths and coherence times), and the plotting
scripts of Figs.~\ref{fig:R1}--\ref{fig:R5}---was written with the assistance
of Claude Opus 4.8 and 5.5 models, which also executed the
runs under the authors' direction. The
authors specified the equations to be implemented, Eqs.~\eqref{eq:Xi-component-system} and~\eqref{eq:V-component-system}, the
parameter set, the protocols and the numerical tolerances, and reviewed the
code. Its output was verified through the checks listed above, which compare
it with analytic limits and with independently written integrators (the Radau
integrator and QuTiP's \texttt{mesolve}).

\section{Sensitivity to pure dephasing and population decay of the \texorpdfstring{$2p$}{2p} exciton}
\label{app:dephasing}

The fixed-$\mathbf K_0$ generator contains no pure dephasing and no population decay of $2p$
beyond its radiative decay (Sect.~\ref{subsec:scope}). To estimate how the $\Xi$ results respond to the
width of $2p$ in a real monolayer, we add these terms to the $\Xi$ generator by
hand, first a pure dephasing,
\begin{equation}
\dot{\bar\rho}\big|_\phi=\gamma_\phi\left(2\hat P_3\bar\rho\hat P_3
-\hat P_3\bar\rho-\bar\rho\hat P_3\right),\qquad \hat P_3=|3\rangle\langle3|,
\label{eq:dephasing-term}
\end{equation}
which damps $\bar\rho_{13}$ and $\bar\rho_{23}$ at the rate $\gamma_\phi$ and
leaves the populations and $\bar\rho_{12}$ unchanged. It describes
fluctuations of the $2p$ energy relative to $|1\rangle$ and $|2\rangle$, such
as those produced by contributions~(i), (iv) and~(v) of
Sect.~\ref{subsec:scope}; $\gamma_\phi$ is not derived here and is treated as
a parameter. A dephasing of $\bar\rho_{13}$ alone, with $\bar\rho_{12}$ and
$\bar\rho_{23}$ undamped, cannot be written in Lindblad form. In the
weak-probe limit only $\bar\rho_{12}$ and $\bar\rho_{13}$ enter the linear
response, so Eq.~\eqref{eq:dephasing-term} is there equivalent to the
replacement $\Gamma^{\Xi}_{13}\to\Gamma^{\Xi}_{13}+\gamma_\phi$.

The runs use the propagator and settings of App.~\ref{app:num_Lindblad} with
Eq.~\eqref{eq:dephasing-term} added. At $\gamma_\phi=0$ they reproduce the
values of Sect.~\ref{subsec:Xi-transfer} to $10^{-15}$; at
$\hbar\gamma_\phi=4$~meV, tolerances ten times tighter change $\eta$ by less
than $2\times10^{-15}$; and the window-center absorption obtained from the
null-space steady state agrees with the closed expression below to
$2\times10^{-4}$.

\paragraph*{Transparency window.} In the weak-probe limit the residual
absorption at the window center is
$\Gamma_{12}(\Gamma^{\Xi}_{13}+\gamma_\phi)/[\Gamma_{12}(\Gamma^{\Xi}_{13}+\gamma_\phi)+\Omega_{23}^2/4]$.
A transparency depth of at least $0.9$ requires
$\hbar\gamma_\phi\le(\hbar\Omega_{23})^2/36\hbar\Gamma_{12}-\hbar\Gamma^{\Xi}_{13}$,
i.e.\ $4.3$, $17$ and $39~\mu$eV at $\hbar\Omega_{23}=0.5$, $1$ and
$1.5$~meV, so the narrowest windows, the best suited to the density
calibration, are the most sensitive. For $\hbar\gamma_\phi=3.96$~meV, half the
measured width of $2p$~\cite{Kusaba2021}, the residual absorption is $0.962$
at $\hbar\Omega_{23}=1$~meV and $0.504$ at $5$~meV.

\begin{figure}[t]
\includegraphics[width=\textwidth]{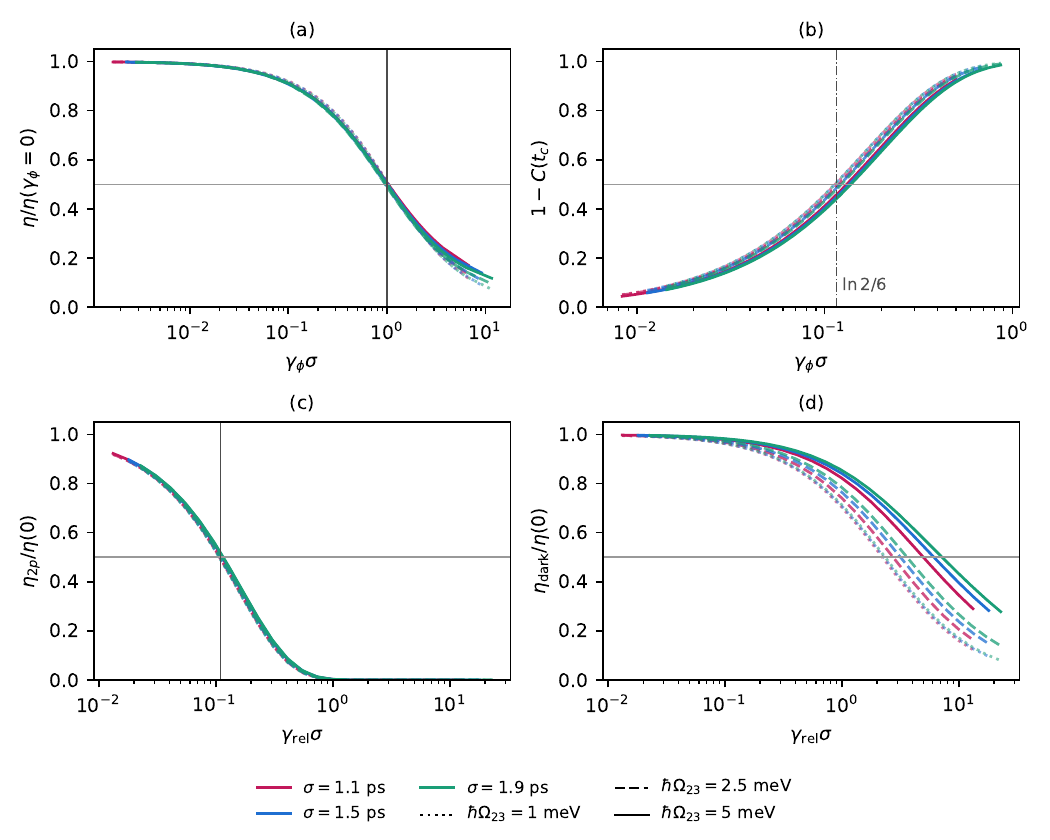}
\caption{Response of the $\Xi$ conversion to a pure dephasing $\gamma_\phi$ of
$2p$, Eq.~\eqref{eq:dephasing-term} [(a), (b)], and to a population decay
$\gamma_{\rm rel}$ of $2p$ into non-radiating states,
Eq.~\eqref{eq:relax-term} [(c), (d)], added outside the fixed-$\mathbf K_0$
generator. Counterintuitive order, $\tau_d=1.25\,\sigma$, weak optical leg;
$\sigma=1.1$ (magenta), $1.5$ (blue) and $1.9$~ps (green);
$\hbar\Omega_{23}=1$ (dotted), $2.5$ (dashed) and $5$~meV (solid).
(a)~Conversion efficiency normalized to its value at $\gamma_\phi=0$, against
$\gamma_\phi\sigma$; vertical line, $\gamma_\phi\sigma=1$.
(b)~Incoherent fraction $1-C$ at $t_c=t_{12}+3\sigma$; dash-dotted line,
$\ln2/6$, the free-decay estimate of the text. (c)~$2p$ fraction $\eta_{2p}$ at $t_f$ and (d)~dark-sector fraction
$\eta_{\rm dark}$, normalized to $\eta$ at zero width, against
$\gamma_{\rm rel}\sigma$ ($\gamma_\phi=0$); vertical line in (c),
$\gamma_{\rm rel}\sigma=0.11$, close to $\ln2/6$, the free decay of $2p$ over
the $6\sigma$ from the optical pulse to $t_f$. Horizontal lines: $1/2$.
The data were computed with the propagator of App.~\ref{app:num_Lindblad} and
the terms of Eqs.~\eqref{eq:dephasing-term} and~\eqref{eq:relax-term}. The
run scripts and the plotting script were written with the assistance of
Claude Opus 5.5 model, which also executed the runs under the authors'
direction; the authors specified the terms, the grids and the checks, and
checked the plotted quantities against the run outputs.}
\label{fig:dephasing}
\end{figure}

\paragraph*{Conversion efficiency.} The conversion runs cover $\sigma=1.1$,
$1.5$ and $1.9$~ps with $\tau_d=1.25\,\sigma$, $\hbar\Omega_{23}=1$, $2.5$ and
$5$~meV, the weak optical leg of Sect.~\ref{subsec:Xi-transfer}, the
counterintuitive order, and $\hbar\gamma_\phi$ from $10^{-3}$ to $4$~meV.
Normalized to its value at $\gamma_\phi=0$, the efficiency depends on
$\gamma_\phi$ and $\sigma$ essentially through $\gamma_\phi\sigma$
[Fig.~\ref{fig:dephasing}(a)]: the nine curves fall to one half at
$\gamma_\phi\sigma=1.01$ on average (range $0.96$--$1.05$), with a slight
decrease as $\sigma$ grows (mean $1.02$ at $\sigma=1.1$~ps, $0.99$ at
$1.9$~ps). Since $\tau_d=1.25\,\sigma$ throughout, $\sigma$ stands here for
the timescale of the pulse pair at fixed $\tau_d/\sigma$. At
$\hbar\gamma_\phi=3.96$~meV the efficiency keeps $7$--$17\%$ of its
$\gamma_\phi=0$ value; at the baseline it falls from $0.479$ to $0.067$. The
counterintuitive order still leads the sequential one there,
$\eta=0.067$ against $4.9\times10^{-3}$ at $\hbar\Omega_{23}=5$~meV (a ratio
of $14$, against $656$ at $\gamma_\phi=0$), but the vertex is no longer kept
empty: the vertex suppression rises from $0.096$ to $1.1$. In the sequential
order the efficiency grows with $\gamma_\phi$ at $\hbar\Omega_{23}=2.5$ and
$5$~meV, because the dephasing interrupts the coherent $1s\leftrightarrow2p$
Rabi cycling of Sect.~\ref{subsec:Xi-transfer}.

\paragraph*{Coherence of the converted population.} With dephasing, the
coherence fraction $C$ of Sect.~\ref{subsec:Xi-transfer} keeps decaying after
the pulses, so its value at the end of the integration depends on how long the
run continues. We therefore evaluate it at $t_c=t_{12}+3\sigma$, after which
both pulses are below $1.1\%$ of their peak. At $\gamma_\phi=0$, $1-C(t_c)$
equals the values of Sect.~\ref{subsec:Xi-transfer} to the digits quoted
there, $1.1\times10^{-3}$ at the baseline and $6.4\times10^{-3}$ in case~3.
With dephasing, the coherence is lost much earlier than the conversion
[Fig.~\ref{fig:dephasing}(b)]: $1-C(t_c)=1/2$ at $\gamma_\phi\sigma=0.123$ on
average (range $0.113$--$0.138$), independent of $\sigma$ at fixed
$\Omega_{23}$ and rising by $15$--$22\%$ from $\hbar\Omega_{23}=1$ to
$5$~meV. This is seven to nine times below the half-point of $\eta$, and at
this dephasing the efficiency still keeps $88$--$91\%$ of its value at
$\gamma_\phi=0$. The value is close to $\ln2/6=0.116$, which follows from the
free decay $e^{-2\gamma_\phi\Delta t}$ of a coherence formed near the peak of
the optical pulse over the remaining $\Delta t=3\sigma$ to $t_c$. The
coherence therefore follows the duration of the protocol, and its half-point
depends on the convention for $t_c$. Under pure dephasing at rates comparable to the
measured width of $2p$, the protocol still transfers part of the excitation
to $2p$, and the counterintuitive order remains favorable, but it delivers
that excitation as population rather than as a coherent superposition with
the crystal ground state. Population decay, below, removes even this.

\paragraph*{Population decay of $2p$.} The measured width of $2p$ may also
come from population decay, for instance by phonon emission into
momentum-dark excitons, the counterpart for $2p$ of contribution~(ii) of
Sect.~\ref{subsec:scope}, or by relaxation into lower states, contribution~(iii). We describe it by a rate
$\gamma_{\rm rel}$ out of $|3\rangle$, either into a non-radiating sink,
\begin{equation}
\dot{\bar\rho}\big|_{\rm rel}=-\frac{\gamma_{\rm rel}}{2}\left(\hat P_3\bar\rho+\bar\rho\hat P_3\right),
\label{eq:relax-term}
\end{equation}
which removes population from the three-level system, $P_{\rm sink}(t)=1-{\rm Tr}\,\bar\rho(t)$,
or back to $|2\rangle$ by the jump operator $\sqrt{\gamma_{\rm rel}}\,|2\rangle\langle3|$,
which preserves the trace. Together with Eq.~\eqref{eq:dephasing-term}, a
homogeneous width of $2p$ has the half-width
$\hbar\gamma_{\rm rel}/2+\hbar\gamma_\phi$; we set it to $3.955$~meV, half the
measured width~\cite{Kusaba2021}, and vary the share of population decay,
$f=\hbar\gamma_{\rm rel}/(2\times3.955~{\rm meV})$, from $0$ (the pure dephasing
above) to $1$. Besides $\eta$, Eq.~\eqref{eq:conversion-efficiency}, we
evaluate at $t_f$ the $2p$ fraction
$\eta_{2p}=\bar\rho_{33}/(\bar\rho_{33}+P_{\rm sink}+N_{\rm emit})$ and the
dark-sector fraction
$\eta_{\rm dark}=(\bar\rho_{33}+P_{\rm sink})/(\bar\rho_{33}+P_{\rm sink}+N_{\rm emit})$,
the share of the excitation that ends in $2p$ or in its relaxation products
rather than being re-emitted from the vertex; without decay into the sink
both equal $\eta$. Since $\bar\rho_{33}(t_f)$ vanishes with population decay,
the vertex suppression is taken here as $\max_t\bar\rho_{22}/\max_t\bar\rho_{33}$,
which equals the definition of Sect.~\ref{subsec:Xi-transfer} without decay.
The runs use the propagator of App.~\ref{app:num_Lindblad} with these terms
added. At $f=0$ they reproduce the pure-dephasing runs above exactly; the
population lost to the sink equals $\gamma_{\rm rel}\int\bar\rho_{33}\,dt$ to
$1.1\times10^{-8}$; the smallest eigenvalue of $\bar\rho$ stays above
$-3\times10^{-16}$ and, for decay back to $1s$, the trace is preserved to
$9\times10^{-16}$; tolerances ten times tighter change the results by less
than $4\times10^{-10}$.

At the measured width and $\sigma=1.5$~ps (Table~\ref{tab:relax}), any share
$f\ge1/4$ leaves no $2p$ population at the end of the pulses
($\eta\le8\times10^{-13}$, in both orders and for both destinations). With
decay into the sink, the dark-sector fraction at $\hbar\Omega_{23}=5$~meV is
$0.114$ at $f=1/4$ and $0.135$ at $f=1$, against $\eta=0.479$ without the
width; the vertex is not emptied (vertex suppression $4.1$ and $12.8$, against
$0.096$); and the sequential order gives more than the counterintuitive one at
every Rabi energy, so the pulse-order signature of dark-state transfer is
reversed. With decay back to $1s$ the relaxed population is re-emitted, and
$\eta_{\rm dark}=\eta\le8\times10^{-13}$. For decay into the sink alone
($f=1$, $\hbar\gamma_{\rm rel}$ from $0$ to $7.91$~meV), the $2p$ fraction,
normalized to $\eta$ at zero width, depends on $\gamma_{\rm rel}$ and $\sigma$
essentially through $\gamma_{\rm rel}\sigma$ and falls to one half at
$\gamma_{\rm rel}\sigma=0.106$--$0.115$ for all nine $(\sigma,\Omega_{23})$
[Fig.~\ref{fig:dephasing}(c)]. Like the half point of the coherence, this
value measures the time between the transfer and the evaluation, here the
$6\sigma$ from the optical pulse to $t_f$. A further run over the same nine
$(\sigma,\Omega_{23})$, $33$ values of $\gamma_{\rm rel}\sigma$ from $10^{-3}$
to $30$ and both pulse orders records both fractions $k\sigma$ after the
second pulse, $k=1,2,3,6$; at $k=6$ it reproduces the runs above exactly. The
$2p$ fraction halves at $\gamma_{\rm rel}\sigma=0.47$--$0.68$, $0.28$--$0.35$
and $0.20$--$0.23$ for $k=1$, $2$, $3$, approaching the free-decay values
$\ln2/k=0.69$, $0.35$, $0.23$ (and $0.116$ for $k=6$). The transfer itself is
affected only at larger rates. At $\hbar\Omega_{23}=5$~meV the pulse-order
contrast of the dark-sector fraction, counterintuitive over sequential, falls
from $149$--$747$ without the width to $10$ at $\gamma_{\rm rel}\sigma=0.12$--$0.18$,
to $2$ at $0.82$--$1.25$ ($\hbar\gamma_{\rm rel}=0.43$--$0.49$~meV) and to $1$
at $2.4$--$4.6$, for $\sigma=1.1$--$1.9$~ps; at contrast $2$ the dark-sector
fraction keeps $82$--$85\%$ of its value at zero width. At $2.5$~meV the
contrast falls to $2$ at $\gamma_{\rm rel}\sigma=0.36$--$0.84$; at $1$~meV it is
below $2$ already without the width ($0.63$--$1.23$;
Sect.~\ref{subsec:Xi-transfer}). The dark-sector fraction changes little after
the pulses and falls to one half only at $\gamma_{\rm rel}\sigma=2.2$--$7.2$
[Fig.~\ref{fig:dephasing}(d)], later for stronger terahertz drive.

\begin{table}[t]
\centering
\caption{Population decay of $2p$ at the measured width
($\hbar\gamma_{\rm rel}/2+\hbar\gamma_\phi=3.955$~meV), $\sigma=1.5$~ps,
decay into the sink, Eq.~\eqref{eq:relax-term}. $\eta(0)$: conversion
efficiency without the width; $\eta_{\rm dark}$: dark-sector fraction in the
counterintuitive (ci) and sequential (seq) orders, for the share $f$ of the
width due to population decay; last column: vertex suppression
$\max_t\bar\rho_{22}/\max_t\bar\rho_{33}$, counterintuitive order, $f=1$. In
all these runs $\eta\le8\times10^{-13}$.}
\label{tab:relax}
\begin{tabular}{cccccc}
\hline\hline
$\hbar\Omega_{23}$ (meV) & $\eta(0)$ & $\eta_{\rm dark}$, ci, $f=1/4$ & $\eta_{\rm dark}$, ci, $f=1$ & $\eta_{\rm dark}$, seq, $f=1$ & vertex suppr. \\ \hline
$1$   & $0.091$ & $0.0086$ & $0.0087$ & $0.016$ & $198$ \\
$2.5$ & $0.327$ & $0.045$  & $0.048$  & $0.085$ & $36$ \\
$5$   & $0.479$ & $0.114$  & $0.135$  & $0.214$ & $12.8$ \\
\hline\hline
\end{tabular}
\end{table}

\section{Spectral selectivity and the window of the \texorpdfstring{$V$}{V} beat}
\label{app:drive-windows}
This appendix derives the bounds on the Gaussian pulse widths quoted in Sect.~\ref{subsubsec:leg_resolved_drive}.

The Gaussian width $\sigma_\ell$ is bounded from below by the requirement
that each tone address only its intended leg. For an envelope
$f^{\rm G}_\ell(t)=\exp[-(t-t_\ell)^2/2\sigma_\ell^2]$ the spectral width is
$\sigma_E=\hbar/\sigma_\ell$, so a competing transition detuned by
$|\Delta|$ from the carrier is suppressed provided
$\sigma_\ell\gtrsim n\hbar/|\Delta|$ for a chosen margin $n$. For the
retained WSe$_2$/hBN three-level systems the binding constraint on the
intra-excitonic leg is the $1s\to3p$ transition, detuned by $29.30$~meV
and carrying $0.31$ of the retained $1s\to2p$ dipole
(Table~\ref{tab:appD_dipoles}), which at $n=3$ requires
$\sigma_\ell\gtrsim0.067$~ps, i.e. a pulse longer than about $0.16$~ps
FWHM. On the interband side, the nearest bright competitor of the $0\to1s$ tone is the $0\to2s$ transition at $124.16$~meV, requiring only $\sigma_\ell\gtrsim0.016$~ps. This bound applies to the optical leg of both retained three-level systems and, in the $V$ system, also ensures that neither optical tone addresses the other driven leg. The tightest condition of the $V$ system is instead set by its upper tone. The $0\to2s$ carrier lies $E_b(2s)-E_b(3s)=21.30$~meV below the bright $0\to3s$ transition, whose interband amplitude is $|\varphi_{3s}(0)/\varphi_{2s}(0)|=0.53$ of the retained one (Table~\ref{tab:appD_states}); the intervening $2p$ state is interband-dark and does not compete. At $n=3$ this requires $\sigma_\ell\gtrsim0.093$~ps, i.e.\ a pulse longer than about $0.22$~ps FWHM, the most restrictive selectivity condition among all retained legs. Since $f^{G}_\ell$ has spectral amplitude $e^{-\sigma_\ell^{2}\Delta^{2}/2\hbar^{2}}$, the criterion $\sigma_\ell\gtrsim n\hbar/|\Delta|$ corresponds to a suppression $e^{-n^{2}/2}$ of the competing transition. Because the $V$ system is driven with the plateau envelope~\eqref{eq:envelope_plateau}, the same condition is carried by its switching edges: the spectrum of each $\tanh$ edge, normalized to its zero-frequency value, is $x/\sinh x$ with $x=\pi|\Delta|\tau_{\rm on(off)}/2\hbar$, and requiring the same suppression $e^{-n^{2}/2}$ gives $\tau_{\rm on},\tau_{\rm off}\gtrsim4.56\,\hbar/|\Delta|=0.14$~ps at $n=3$, while the plateau itself only needs $T_{\rm pulse}\gg2\pi\hbar/|\Delta|\simeq0.19$~ps. All of these bounds remain permissive, so sub-picosecond envelopes already resolve every retained leg spectrally, with the upper $V$ tone setting the tightest requirement. We stress that this settles only the bandwidth condition. The complementary
requirement, that the Rabi energies themselves remain small compared with these
same detunings, $|\Omega|\ll|\Delta|$, is not implied by it and is assessed
separately, since it depends on the field amplitude through
Eq.~\eqref{eq:Omega_inter_def} rather than on the pulse duration. The Rabi energies depend on the choice of sample area $A$, which enters through the normalization of the Wannier envelope and, consequently, through the collective interband matrix element; see Eq.~\eqref{eq:interband-dipole-drive} below. The choice of envelope family is dictated by the configuration and the observable of
interest. We report both envelope families for both configurations. In $V$,
$f^{\rm plateau}_\ell$ addresses the stationary coherent response of the
bright--bright three-level system, while a synchronous Gaussian pair prepares the
excited-state coherence whose free decay is the quantum beat; in $\Xi$, the
ladder structure admits a steady cascade-absorption analysis under
$f^{\rm plateau}_\ell$ and a finite-pulse conversion analysis under $f^{G}_\ell$.

For the $V$ beat the Gaussian width is bounded from both sides. From below,
the selectivity condition of the upper tone requires $\sigma_\ell\gtrsim0.093$~ps.
From above, the pair must end before the coherence it prepares has decayed:
$\bar\rho_{23}$ is damped at
$\tfrac12\big(\gamma^{(\gamma)}_{2\to1}+\gamma^{(\gamma)}_{3\to1}\big)$, Eq.~\eqref{eq:V-component-23}, i.e.\ with a lifetime $T_2^{V}=0.37$~ps, and a free beat is
observable only for $\sigma_\ell\lesssim T_2^{V}/2\simeq0.18$~ps. The admissible
window, $0.093\lesssim\sigma_\ell\lesssim0.18$~ps, spans a factor of two. The
Rabi condition $\hbar\Omega_{13}\ll21.29$~meV further restricts the preparation
to small pulse areas, so that the beat amplitude is perturbative in the drives.
We use $\sigma_\ell=0.12$~ps (FWHM $0.28$~ps), which suppresses the competing
$0\to3s$ transition to $e^{-\sigma_\ell^2\Delta^2/2\hbar^2}\simeq5\times10^{-4}$
and ends within a third of $T_2^{V}$, together with
$\hbar\Omega_{12},\hbar\Omega_{13}\lesssim1$~meV (pulse areas
$\lesssim0.46$~rad). In the drive-rotating frame $\bar\rho_{23}$ evolves only at
$\ren\delta_V$ once the pulses are over; the beat at $\omega_{23}$ is the
laboratory-frame phase restored by the inverse of the drive rotation~\eqref{eq:drive-rotation}. No
comparable restriction arises in $\Xi$, where $\bar\rho_{13}$ is undamped in the
present reduction [Eq.~\eqref{eq:Xi-component-13}].

\bibliography{open_quantum_references}
\end{document}